\RequirePackage[T1]{fontenc}
\documentclass[12pt,a4paper,final]{book}
\usepackage{lmodern}
\usepackage{amsmath}
\usepackage{amssymb}
\usepackage{amsfonts}
\usepackage{bm}
\usepackage{graphicx}
\usepackage{systeme}
\usepackage{enumerate}
\usepackage{comment}
\usepackage{here}
\usepackage{mathrsfs}
\usepackage{url}
\usepackage{braket}
\usepackage{cite}
\usepackage[dvipsnames,svgnames,x11names]{xcolor}
\usepackage[colorlinks,citecolor=DeepSkyBlue4,linkcolor=DodgerBlue4,urlcolor=DodgerBlue3,breaklinks=true,linktoc=all]{hyperref}
\usepackage[top=30truemm,bottom=30truemm,left=25truemm,right=25truemm]{geometry}
\usepackage{physics}
\usepackage{showkeys}
\usepackage{authblk}
\usepackage{tikz}
\usetikzlibrary{shapes.geometric, positioning, arrows.meta}
\usepackage{ytableau}
\allowdisplaybreaks
\usepackage{tabularx}
\usepackage{array}
\usepackage{ragged2e}
\usepackage{CJKutf8}

\usepackage{fancyhdr}
\renewcommand{\chaptermark}[1]{\markboth{\chaptername\ \thechapter}{}}

\fancypagestyle{plain}{%
  \fancyhf{}%
  \fancyhead[LE,RO]{\thepage}%
  \fancyhead[RE,LO]{\rightmark}%
}

\title{\begin{CJK}{UTF8}{ipxm} Foliated-Exotic Duality and Anomaly Inflow\\ in Fracton Quantum Field Theories \\
(フラクトン系の場の量子論における \\
葉層-エキゾチック双対性とアノマリー流入)\end{CJK}}

\author{\begin{CJK}{UTF8}{ipxm} 島村　洲太朗\end{CJK}}
\affil{\textit{Department of Physics, The University of Tokyo, Bunkyo-ku, Tokyo 113-0033, Japan}}

\date{}

\renewcommand{\L}{\mathcal{L}}
\newcommand{\Z}{\mathbb{Z}}
\newcommand{\R}{\mathbb{R}}
\newcommand{\T}{\mathbb{T}}

\usepackage{xcolor}

\usepackage[status=draft,inline,nomargin]{fixme}
\fxusetheme{color}
\FXRegisterAuthor{ko}{ako}{KO}

\FXRegisterAuthor{shu}{ashu}{\color{magenta}SHU}

\begin{document}

\begin{titlepage}
\vspace*{20pt}

	\begin{center}
		\Large Doctoral Dissertation\vspace{2mm}\\
		\Large \begin{CJK}{UTF8}{ipxm}	博士論文 \end{CJK}
	\end{center}
	
	\bigskip

	\begin{center}
	{\Large {
		Foliated-Exotic Duality and Anomaly Inflow\\ in Fracton Quantum Field Theories \\}
		\vspace*{2mm}
			\Large \begin{CJK}{UTF8}{ipxm}
			{（フラクトン系の場の量子論における \\
葉層-エキゾチック双対性とアノマリー流入）} \end{CJK}
	}\\
		\vspace*{50pt}

	\Large A Dissertation Submitted  for  the Degree of Doctor of Philosophy\\
  December 2025\vspace{2mm}\\
	\Large \begin{CJK}{UTF8}{ipxm}	令和 7年12月 博士（理学）申請 \end{CJK}
	
	\vspace*{60pt}
	
	\Large Department of Physics, Graduate School of Science,\\
    The University of Tokyo\vspace{2mm}\\
		\begin{CJK}{UTF8}{ipxm} 東京大学大学院理学系研究科 
        物理学専攻 \end{CJK}

	\vspace*{50pt}
	\Large
	Shutaro Shimamura\vspace{2mm}\\
	 \Large \begin{CJK}{UTF8}{ipxm} 島村  洲太朗 \end{CJK}
	\end{center}

\vspace*{\fill}
\end{titlepage}

\cleardoublepage
\thispagestyle{empty}

\vspace*{50pt}
\begin{center}
	\textbf{Abstract}
\end{center}
Fracton phases are new types of phases of matter characterized by subsystem global symmetry, which is a generalized global symmetry whose symmetry operator is partially topological.
Their continuum low-energy effective descriptions admit two different formulations: an exotic quantum field theory (QFT) using exotic tensor gauge fields, and a foliated QFT constructed from a foliation structure and foliated gauge fields.
For certain fracton QFTs, these two descriptions are equivalent, which is called the foliated-exotic duality.

In this dissertation, we extend the foliated-exotic duality by combining it with the anomaly inflow mechanism for 't Hooft anomalies of subsystem symmetries. This dissertation has two main results.

First, we discuss the exotic and foliated $BF$ theories in 2+1 dimensions, which exhibit the mixed 't Hooft anomaly of $\Z_N \times \Z_N$ subsystem symmetry.
This anomaly is captured by a subsystem symmetry-protected topological (SSPT) phase for $\Z_N \times \Z_N$ subsystem symmetry in one dimension higher.
By extending the foliated-exotic duality in the fractonic $BF$ theory to the SSPT phase, we establish the field correspondences in the SSPT phase and construct the foliated description of the SSPT phase.

Second, we discuss the exotic $\phi$-theory in 2+1 dimensions -- a fractonic gapless scalar field theory, which has the 't Hooft anomaly of $U(1) \times U(1)$ subsystem symmetry.
The anomaly is captured by an SSPT phase for $U(1) \times U(1)$ subsystem symmetry in 3+1 dimensions via the anomaly inflow mechanism.
Extending the foliated-exotic duality to the $\phi$-theory, we establish field correspondences in the $\phi$-theory and construct the foliated $\phi$-theory that is equivalent to the exotic $\phi$-theory.
Furthermore, we also investigate the foliated-exotic duality in the $\hat\phi$-theory, which is dual to the $\phi$-theory, and construct the foliated $\hat\phi$-theory.
These provide the first examples of the foliated-exotic duality in gapless theories.

\cleardoublepage



\setlength{\parskip}{2mm}
\setlength{\abovedisplayskip}{10pt} 
\setlength{\belowdisplayskip}{10pt} 
\numberwithin{equation}{section}

\tableofcontents

\chapter{Introduction}

Certain new types of phases of matter have attracted much attention in recent years.
The phases have characteristic excitations that cannot move in space: \textit{fracton}, or can only move along a certain dimensional submanifold like a line: \textit{lineon}, or a plane: \textit{planon}.
They are called fracton phases (see \cite{Nandkishore:2018sel,Pretko:2020cko,Gromov:2022cxa} for reviews).
Fracton phases were originally motivated by quantum storage \cite{Haah:2011drr} and glassy dynamics \cite{Chamon:2004lew}.

The excitations with characteristic mobility originate from \textit{subsystem global symmetry}, which is one of the generalized global symmetries \cite{Gaiotto:2014kfa}.
Symmetry operators of subsystem symmetry are supported on specific submanifolds such as points, lines or planes.
In this sense, subsystem symmetry resembles higher-form global symmetry\cite{Gaiotto:2014kfa}, the standard generalization of global symmetries where the symmetry operators are supported on arbitrary codimension-$(p+1)$ submanifolds for a $p$-form global symmetry.
However, in contrast to higher-form global symmetries, these submanifolds cannot be deformed in arbitrary directions, meaning that subsystem symmetry is not fully topological.
Then, symmetry operators supported on each submanifold, which cannot be continuously deformed into one another, carry independent charges \cite{Seiberg:2019vrp}.
This type of global symmetry leads to characteristic features such as restricted mobility of excitations and a sub-extensive ground state degeneracy, motivating various research.
The most standard lattice example of fracton phases with subsystem symmetry is described by the X-cube model \cite{Vijay:2016phm}, which will be reviewed in Section \ref{section:fracton_phases}.

Although fracton phases were first studied as lattice models (earlier studies are in \cite{Paramekanti:2002iup,Haah:2011drr,Bravyi:2010jfq,Vijay:2015mka,Vijay:2016phm}), they are also studied as effective continuum quantum field theories (QFTs), which are called fracton QFTs.
The property of subsystem symmetry is inconsistent with relativistic systems, so fracton QFTs lack continuous rotational symmetry, and have discrete rotational symmetry instead.
One way to implement fracton QFTs is by employing exotic tensor gauge fields, which is called \textit{exotic QFTs}\cite{Slagle:2017wrc,Pretko:2018jbi,You:2019ciz,Seiberg:2020bhn,Seiberg:2020wsg,Seiberg:2020cxy,Gorantla:2020xap,Gorantla:2021bda,Gorantla:2020jpy,Geng:2021cmq,Yamaguchi:2021qrx,Yamaguchi:2021xeq,Gorantla:2022eem,Gorantla:2022ssr,Honda:2022shd,Burnell:2021reh,Luo:2022mrj}.
Exotic tensor gauge fields are in the representations of discrete spatial rotational symmetry, making exotic QFTs manifestly invariant under this spatial symmetry.
While exotic QFTs can describe some fracton phases, the QFTs constructed from \textit{foliation} structure also exhibit fractonic features, and such QFTs are called \textit{foliated QFTs} \cite{Slagle:2018swq,Slagle:2020ugk,Hsin:2021mjn,Geng:2021cmq,Ohmori:2022rzz,Spieler:2023wkz,Cao:2023rrb,Ebisu:2023idd,Ebisu:2024cke,Hsin:2023ooo,Hsin:2024eyg,Gorantla:2025fzz}.\footnote{Some lattice models can also be constructed by using foliation structure. They describe the foliated fracton phases \cite{Shirley:2017suz,Shirley:2018nhn,Shirley:2018vtc,Shirley:2019uou,Shirley:2022wri}.}
A foliation is a decomposition of a manifold into an infinite number of submanifolds: \textit{leaves}.
In the foliated QFT context, we consider foliations where the intersections of leaves form a lattice-like structure.
Foliated QFTs contain foliated gauge fields, which are considered as lower form gauge fields on the leaves, and \textit{bulk} gauge fields, which mediate the foliated gauge fields.
Both exotic and foliated fracton QFTs exhibit UV/IR mixing; the low-energy infrared (IR) physics depends on some microscopic quantities in short-distance ultraviolet (UV) physics.
As a manifestation of this, the tensor and foliated gauge fields can have singularities and discontinuities in specific directions \cite{Seiberg:2020bhn,Seiberg:2020wsg,Seiberg:2020cxy,Slagle:2020ugk,Hsin:2021mjn}.
The study of fracton QFTs is expected to broaden our understanding of the mathematical structure of non-relativistic QFT appearing in condensed matter physics.

It was noted in \cite{Hsin:2021mjn} that there are pairs of exotic and foliated QFTs sharing the same subsystem symmetries and related quantities, and thus we expect such a pair to describe the same physics.
In \cite{Ohmori:2022rzz}, the author of the dissertation considered the exotic $BF$ theory \cite{Slagle:2017wrc,Seiberg:2020cxy} and the foliated $BF$ theory \cite{Slagle:2020ugk,Hsin:2021mjn}, each of which was considered as a continuous exotic or foliated QFT description of the X-cube model \cite{Vijay:2016phm}, and demonstrated the equivalence of the two theories.\footnote{These fractonic $BF$ theories are similar to the ordinary $BF$ theory \cite{Maldacena:2001ss,Banks:2010zn,Kapustin:2014gua} and have several analogies with it.}
This correspondence is referred to as \textit{the foliated-exotic duality}, and subsequently, correspondences have been established for other theories as well \cite{Spieler:2023wkz,Cao:2023rrb,Ebisu:2023idd,Ebisu:2024cke,Hsin:2024eyg}.
This structure is specific to fracton phases, and is considered as a new type of duality.
Since exotic and foliated QFTs have different manifest structures, the study of the foliated-exotic duality provides new perspectives on fracton QFTs and may reveal new mathematical relations in exotic gauge fields.

With respect to global symmetry, we consider the anomaly inflow mechanism \cite{Callan:1984sa} for global symmetry.
Let us consider a $D$-dimensional QFT with a global symmetry whose group is $G$. The global symmetry acts on a charged object as a global transformation.
Then, we can couple the global symmetry to a background gauge field $C$ for $G$, and replace the parameter of the global transformation with a local parameter, which is absorbed in the gauge transformation of $C$.
In this situation, the partition function may not be invariant under the local transformation $C \rightarrow C^g$:
\begin{align}
    Z[C^g] = e^{i\alpha(C,g)} Z[C] \,.
\end{align}
If the function $\alpha(C,g)$ cannot be canceled by a local counterterm of $C$ and $g$, the QFT is said to have an 't Hooft anomaly \cite{tHooft:1979rat}.
Due to the non-invariance of the partition function, we cannot sum over the background gauge field $C$; the global symmetry cannot be gauged.
In the anomaly inflow mechanism, an 't Hooft anomaly in a $D$-dimensional QFT is captured by a classical field theory in $D+1$ dimensions.
Then, by coupling the classical field theory in the $(D+1)$-dimensional bulk to the $D$-dimensional QFT on the boundary, the 't Hooft anomaly can be canceled, and the total system becomes invariant.
In relativistic QFTs, the classical field theories are called invertible field theories \cite{Freed:2016rqq} or symmetry-protected topological (SPT) phases \cite{Gu:2009dr,Chen:2010gda}.
In this case, the classification of anomalies is interpreted as the cobordism group classification of SPT phases\cite{Chen:2011pg,Wen:2013oza,Kapustin:2014tfa,Kapustin:2014dxa}.

Then, what about the case of subsystem symmetry? In some exotic theories \cite{Burnell:2021reh,Luo:2022mrj} and simple foliated theories \cite{Hsin:2021mjn}, 't Hooft anomalies of the subsystem symmetries are captured by classical field theories called subsystem symmetry-protected topological (SSPT) phases (Lattice models of SSPT phases appear in \cite{You:2018oai,Devakul:2018fhz}).
However, the relation between 't Hooft anomalies of subsystem symmetry and SSPT phases is obscure.
For example, the foliation structure of the SSPT phase is not canonically determined \cite{Burnell:2021reh,Luo:2022mrj,Okuda:2024azp}. 
This fact obstructs the classification of anomalies of subsystem symmetry.

In this dissertation, we explore the foliated-exotic duality by relating the anomaly inflow mechanism for subsystem symmetry.
Beyond matching known pairs of foliated and exotic theories, we establish the correspondences between exotic tensor gauge fields and foliated gauge fields in a class of fracton QFTs, and construct previously unknown exotic and foliated QFT descriptions, which are equivalent to known fracton QFTs.

In Chapter \ref{chapter:2+1d_bfanomaly_and_3+1d_sspt}, we will discuss the $\Z_N \times \Z_N$ mixed 't Hooft anomaly\footnote{If two global symmetries can be gauged respectively, but cannot be gauged simultaneously, the system is said to have a mixed 't Hooft anomaly.} of subsystem symmetries in the exotic and foliated $BF$ theory in 2+1 dimensions \cite{Seiberg:2020bhn,Ohmori:2022rzz}, and the SSPT phase for $\Z_N \times \Z_N$ subsystem symmetry in 3+1 dimensions that cancels the anomaly.
While the exotic description of the SSPT phase whose spatial rotational symmetry group is $\Z_4$ was known in \cite{Burnell:2021reh,Luo:2022mrj}, the equivalent foliated description was not known, and we construct it by extending the foliated-exotic duality of the fractonic $BF$ theories.
Furthermore, we change the foliation structure of the SSPT phase, and construct the exotic description of the SSPT phase, whose spatial rotational symmetry group is $S_4$, via the foliated-exotic duality. 
This can be seen as a systematic way to construct exotic QFTs with different foliation structures. 
These two SSPT phases with different foliation structures or different spatial rotational symmetries cancel the same anomaly of the 2+1d exotic/foliated $BF$ theory. This fact is considered as a hint for characterizing 't Hooft anomalies of subsystem symmetry.
This discussion is mainly based on the author's paper \cite{Shimamura:2024kwf}.

In Chapter \ref{chapter_phianomaly}, we will consider the SSPT phase for $U(1) \times U(1)$ subsystem symmetry in 3+1 dimensions.
The boundary anomalous theory is the $\phi$-theory in 2+1 dimensions \cite{Seiberg:2020bhn,Gorantla:2021bda}, which is a gapless scalar field theory with mixed 't Hooft anomaly for $U(1) \times U(1)$ subsystem symmetry \cite{Burnell:2021reh}.
While the exotic description of $\phi$-theory was known, the equivalent foliated description was not known, and we have successfully constructed the foliated $\phi$-theory equivalent to the exotic $\phi$-theory by extending the foliated-exotic duality of the SSPT phases and establishing correspondences between the fields.
To our knowledge, this is the first example of the nontrivial foliated description of gapless fracton QFT.
Due to this duality, the $\phi$-theory can be seen as layers of 1+1d compact scalar field theory coupled by the bulk gauge fields from a perspective of the foliated QFT description. 
This description makes the structure of the $\phi$-theory clearer, and allows us to analyze the theory using established tools from standard QFT.
Furthermore, we discuss the $T$-duality-like duality in the $\phi$-theory, which is considered as a fractonic theory version of the ordinary $T$-duality in relativistic compact scalar field theories.
The exotic version of this duality was studied in \cite{Seiberg:2020bhn,Spieler:2024fby}, and we extend it to the foliated version and construct the foliated $\hat{\phi}$-theory equivalent to the exotic $\hat{\phi}$-theory.
These results provide a systematic description of exotic tensor gauge fields and foliated gauge fields for flat foliations, and facilitating an understanding of the general correspondence between these fields.
This discussion is based on the paper \cite{Ohmori:2025fuy}.\footnote{Around the same time as \cite{Ohmori:2025fuy}, the work \cite{Apruzzi:2025mdl}, which independently constructs the foliated $\phi$-theory based on the SymTFT for the subsystem symmetry, was submitted.}

Here is a comment related to dipole symmetry.
In the exotic and foliated QFTs discussed in this dissertation, the tensor gauge fields are symmetric and \textit{hollow}, meaning that they satisfy $A_{ij} = A_{ji}$ and $A_{ii} = 0$, and the components of the foliated gauge fields $A^k_{kij...}$ are gauged.
The subsystem symmetry appearing in the theories is also referred to as dipole symmetry \cite{Seiberg:2020bhn}.
However, in recent studies on dipole symmetry \cite{Radzihovsky:2019jdo,Ebisu:2023idd,Ebisu:2024cke,Ebisu:2024eew,Huang:2023zhp,Han:2024nvu,Bertolini:2025qcy}, the theories typically involve tensor gauge fields where $A_{ii} \neq 0$ and the components of the foliated (or dipole) gauge fields $A^k_{kij...}$ are not gauged.
Regarding scalar field theories, a similar model is studied in the context of dipole symmetry \cite{Huang:2023zhp}.
The subsystem symmetry and the dipole (or multipole) symmetry are also referred to as modulated symmetry, which is the position-dependent global symmetry \cite{Ebisu:2025mtb}.

\subsection*{Organization}

The organization of this dissertation is as follows.

In Chapter \ref{chapter:review}, we will review generalized global symmetries and fracton phases, describing the standard examples.

In Chapter \ref{chapter:exotic_foliated_qft}, we will review exotic tensor gauge fields and foliated gauge fields. Although this chapter is a review, we provide new systematic descriptions of these fields.
This chapter is mainly based on the appendix of the author's paper \cite{Ohmori:2025fuy}.

Chapter \ref{chapter:2+1d_bfanomaly_and_3+1d_sspt} is one of the main parts of this dissertation.
In this chapter, we will discuss the 't Hooft anomaly of $\Z_N \times \Z_N$ subsystem symmetry in the exotic and foliated $BF$ theory in 2+1 dimensions, and construct the foliated-exotic duality in the SSPT phase for $\Z_N \times \Z_N$ subsystem symmetry in 3+1 dimensions that cancels the anomaly.
This chapter is mainly based on the author's paper \cite{Shimamura:2024kwf}.

Chapter \ref{chapter_phianomaly} is another main part of this dissertation.
In this chapter, we will discuss the SSPT phase for $U(1) \times U(1)$ subsystem symmetry in 3+1 dimensions, which cancels the 't Hooft anomaly of $U(1) \times U(1)$ subsystem symmetry in the $\phi$-theory in 2+1 dimensions. We will construct the foliated $\phi$-theory and the foliated-exotic duality in the $\phi$-theory. Furthermore, we will extend the $T$-duality-like duality in the exotic $\phi$-theory to the foliated version and construct the foliated $\hat{\phi}$-theory equivalent to the exotic $\hat{\phi}$-theory.
This chapter is based on the author's paper \cite{Ohmori:2025fuy}.

Chapter \ref{chapter:conclusion} is devoted to conclusion and future directions.

\subsection*{Special notes}

This dissertation is based on the following papers.
\begin{center}
\begin{tabularx}{\linewidth}{l>{\raggedright\arraybackslash}X}
 \cite{Shimamura:2024kwf} &  S. Shimamura,  ``Anomaly of subsystem symmetries in exotic and foliated $BF$ theories,''  \href{https://link.springer.com/article/10.1007/JHEP06(2024)002}{JHEP \textbf{06} (2024) 002}, \href{https://arxiv.org/abs/2404.10601}{arXiv:2404.10601 [cond-mat.str-el]}.\\ 
 \vspace*{-7pt} & \\
 \cite{Ohmori:2025fuy} & K. Ohmori and S. Shimamura,  ``Gapless foliated-exotic duality,''  \href{https://link.springer.com/article/10.1007/JHEP09(2025)049}{JHEP \textbf{09} (2025) 049}, \href{https://arxiv.org/abs/2504.10835}{arXiv:2504.10835 [cond-mat.str-el]}.\\
\end{tabularx}
\end{center}

\chapter{Symmetry and Fracton Phases}
\label{chapter:review}

In this chapter, we review (generalized) global symmetry and fracton phases, which are important concepts in this dissertation.
We will introduce some standard examples to illustrate these concepts, and many aspects of the theories presented in later chapters can be understood through analogy with these examples.

In this dissertation, we take Euclidean spacetime and the coordinate is written as $(x^0, x^k)$, where $x^0$ is the time component and $x^k \ (k = 1,2,...)$ are spatial components.

\section{Generalized Global Symmetry}
\label{section:generalized_symmetry}

Global symmetry is a powerful tool to analyze physical systems.
Ordinary global symmetry leads to conservation laws and the Ward identity via Noether's theorem and constrains the dynamics of systems. 
Moreover, concepts related to symmetry, such as spontaneous symmetry breaking, gauging, and anomalies, can deepen our understanding of physical phenomena.
In QFT textbooks, (internal) global symmetry appears as a transformation acting on local fields, and the transformation forms a continuous group $G$.
However, in recent years, the notion of global symmetry has been generalized beyond this conventional concept, which is called generalized global symmetry \cite{Gaiotto:2014kfa}.
As a result, we can now apply symmetry-related concepts across a broader range of contexts.
In this section, we review the generalized global symmetry and related concepts --- gauging, the 't Hooft anomaly \cite{tHooft:1979rat} and the anomaly inflow mechanism \cite{Callan:1984sa}.

\subsection{Global Symmetry}

We consider a $D$-dimensional relativistic\footnote{Since we consider the Euclidean spacetime, "relativistic" means that the theory is invariant under $SO(D)$ rotation.} QFT with an ordinary global symmetry. The symmetry acts on local operator (field) $\mathcal{O}(x)$ as
\begin{align}
  \mathcal{O}(x) \rightarrow \mathcal{O}^g(x) \,,
\end{align}
where the symmetry transformation forms a group, and $g$ belongs to the continuous Lie group $G$.
The Noether theorem states that for each continuous symmetry, there exists a one-form conserved current $J$ satisfying the conservation law
\begin{align}
  d \ast J  = (\partial_\mu J^\mu) d^Dx = 0 \,. \label{rev:cons_law}
\end{align}
In quantum theory, the conservation law is understood as the operator formalism or in the correlation functions, which is called the Ward identity.
The conserved charge $Q[\Sigma_{t}]$ is defined by
\begin{align}
  Q[\Sigma_{t}] = \int_{\Sigma_{t}} \ast J = \int_{x^0 = t} d^{D-1}x \, J^0 \,,
\end{align}
where $\Sigma_{t}$ is a $(D-1)$-dimensional time slice at time $x^0 = t$.
From the conservation law, the charge $Q[\Sigma_{t}]$ is conserved:
\begin{align}
  Q[\Sigma_{t_1}] - Q[\Sigma_{t_2}] = \int_{M_{D}} d \ast J = 0 \,,
\end{align}
where $M_{D}$ is the $D$-dimensional region bounded by $\Sigma_{t_1}$ and $\Sigma_{t_2}$.
The symmetry operator is defined by
\begin{align}
  U_g[\Sigma_{t}] = \exp \left( i \sum_a \theta_a Q^a[\Sigma_{t}] \right) \,,
\end{align}
where $g = \exp\left(i \sum_a \theta_a T^a\right) \in G$ with the generators $T^a$ of the Lie algebra of $G$.\footnote{The conserved current and charge exist for the number of generators $\text{dim} \, G$, but we omit the index $a$ for simplicity.} 
The symmetry operators satisfy the group multiplication law:
\begin{align}
  U_{g_1}[\Sigma_{t}] U_{g_2}[\Sigma_{t}] = U_{g_1 g_2}[\Sigma_{t}] \,.
\end{align}
It acts on the local operator $\mathcal{O}(x)$ as
\begin{align}
  U_g[\Sigma_{t}] \cdot \mathcal{O}(x) = U_g[\Sigma_{t}] \mathcal{O}(x) U_g[\Sigma_{t}]^{-1} = \mathcal{O}^g(x) \,,
\end{align}
where $x$ is on the time slice $\Sigma_{t}$, from the Ward identity
\begin{align}
  \langle d \ast J (x) \, \mathcal{O}(x_1) \dots \rangle = -i  \delta(x - x_1) \langle \delta_\text{g} \mathcal{O}(x_1) \dots \rangle \,,
\end{align}
where $\dots$ represent other operators inserted away from $x$, and $\delta_\text{g} \mathcal{O}(x)$ is the infinitesimal variation of $\mathcal{O}(x)$ under the symmetry transformation.
If we write the local operator $\mathcal{O}(x)$ in a representation $R$ of $G$ as $\mathcal{O}_a(x) \ (a = 1,2,...,\mathrm{dim} R)$, the symmetry action is expressed as
\begin{align}
  \mathcal{O}_a^g(x) = \sum_{b} \rho_{ab}^R(g) \mathcal{O}_b(x) \,,
\end{align}
where $\rho^R(g)$ is the representation matrix of $g$ in the representation $R$. The local operator $\mathcal{O}(x)$ is called a charged object if it transforms non-trivially under the symmetry transformation.
When $G = U(1)$, for example, the symmetry operator is expressed as 
\begin{align}
  U_\alpha[\Sigma_{t}] = \exp \left( i \alpha \int_{\Sigma_{t}} \ast J \right) \,,
\end{align}
where $\alpha$ is a $2\pi$-periodic real parameter: $e^{i\alpha} \in U(1)$, and it acts on the charged local operator $\mathcal{O}(x)$ with charge $q \in \Z$ as
\begin{align}
  U_\alpha [\Sigma_{t}] \cdot \mathcal{O}(x) = e^{i q \alpha} \mathcal{O}(x) \,.
\end{align}
When the symmetry group is discrete, the conserved current $J$ and the conserved charge $Q[\Sigma_{t}]$ are not defined, but the symmetry operator $U_g[\Sigma_{t}]$ is still defined by the group transformation.

In this derivation, we do not have to specify the support of the symmetry operator as time slice due to the rotational invariance. Then, we take the closed $(D-1)$-dimensional manifold $\Sigma_{D-1}$ as the support of the symmetry operator:
\begin{align}
  U_g[\Sigma_{D-1}] = \exp \left( i \sum_a \theta_a Q^a[\Sigma_{D-1}] \right) \,.
\end{align}
From the conservation law \eqref{rev:cons_law}, the symmetry operator is topological; it is invariant under continuous deformation of the manifold $\Sigma_{D-1}$ as long as it does not cross any operator insertions:
\begin{align}
  U_g[\Sigma_{D-1}] =  U_g[\Sigma'_{D-1}]\,.
\end{align}
When $\Sigma_{D-1}$ extends in the time direction, we sometimes call the symmetry operator a defect.
Then, when $\Sigma_{D-1}$ surrounds the local operator $\mathcal{O}(x)$, the symmetry operator acts on $\mathcal{O}(x)$ as
\begin{align}
  U_g[\Sigma_{D-1}] \cdot \mathcal{O}(x) = \mathcal{O}^g(x) \,.
\end{align}
This fact means that the conservation law becomes the topologicalness of the symmetry operator.
Therefore, conversely, we can consider any topological operator as the symmetry operator.
This is the idea of generalized global symmetry \cite{Gaiotto:2014kfa}.
If symmetry operators are topological operators defined on a codimension-$(p+1)$ manifold $\Sigma_{D-p-1}$, the symmetry is called a $p$-form global symmetry (higher-form symmetry). Then, the charged object is a $p$-dimensional extended operator.
If symmetry operators are topological operators that do not necessarily have an inverse and satisfy the categorical multiplication law, the symmetry is called a non-invertible symmetry (See for review \cite{Shao:2023gho}).
If symmetry operators are partially topological, the symmetry is called a subsystem symmetry, which is the main topic of this dissertation.

\subsection{Gauging and 't Hooft Anomaly}

Next, we consider coupling the global symmetry to the background gauge field $C$ for the symmetry $G$.
The partition function of the QFT with the background gauge field is denoted as
\begin{align}
  Z[C] = \int d \phi \, e^{-S[\phi] + i \int_{M_D} C \wedge \ast J} \,,
\end{align}
where $\phi$ represents all the dynamical fields in the QFT, $S[\phi]$ is the action, and $M_D$ is the spacetime manifold.
For the $p$-form global symmetry, the background gauge field is a $(p+1)$-form gauge field.
When the symmetry group $G$ is discrete, we can insert the defect network of the symmetry operator $U_g[\Sigma]$ as the background flat gauge field. 
The symmetry transformation acts on the charged objects by a global parameter. Then, the parameter is promoted to the local gauge parameter when we couple to the background gauge field. Under the transformation, the background gauge field $C$ transforms as
\begin{align}
  C \sim C^g \,.
\end{align}
In this situation, if the partition function is invariant under the transformation: $Z[C^g] = Z[C]$, then we can sum over the background gauge field $C$:
\begin{align}
  Z_\text{gauged} = \int dC \, Z[C] \,,
\end{align}
where the integral is taken over all gauge equivalence classes of $C \sim C^g$. Then, the symmetry operator becomes trivial: $U_g[\Sigma] = 1$, and the global symmetry is said to be gauged. This procedure is called gauging.

On the other hand, the partition function may not be invariant under the transformation:
\begin{align}
  Z_\text{anom.}[C^g] = e^{i \int_{M_{D}} \beta(C,g)} Z_\text{anom.}[C] \,,
\end{align}
where $\beta(C,g)$ is a phase depending on the background gauge field $C$ and the gauge transformation $g$. If the function $\beta(C,g)$ cannot be canceled by a local counterterm of $C$ and $g$, we cannot sum over the background gauge field $C$.
Then, the QFT is said to have an 't Hooft anomaly \cite{tHooft:1979rat}.
The 't Hooft anomaly is an obstruction to gauging the global symmetry $G$, so if we gauged the symmetry $G$, the theory would become inconsistent and have the gauge anomaly \cite{Adler:1969gk,Bell:1969ts}.

This 't Hooft anomaly is known to be canceled by combining the QFT with a classical field theory in one dimension higher. This mechanism is called the anomaly inflow mechanism \cite{Callan:1984sa}. In recent years, it has been understood that the classical field theories are invertible field theories \cite{Freed:2016rqq}, which is low-energy effective field theory of symmetry-protected topological (SPT) phases \cite{Gu:2009dr,Chen:2010gda}.\footnote{In this dissertation, we also use the term SPT phases to refer to the invertible field theories.}
The SPT phases for $G$ are gapped phases that are connected to the trivial phase without quantum phase transition, but cannot be connected to the trivial phase without breaking the symmetry $G$ \cite{Chen:2011pg}.
The classification of anomalies is considered to be the cobordism group classification of SPT phases \cite{Wen:2013oza,Kapustin:2014tfa,Kapustin:2014dxa}.
The important property of the SPT phases is that they are almost trivial in the bulk but non-trivial on the boundary.
For the effective Lagrangian of an SPT phase $L_\text{SPT}[C]$, the partition function
\begin{align}
  Z_\text{SPT}[C] = e^{- \int_{M_{D+1}} L_\text{SPT}[C] } \,,
\end{align}  
is invariant if the theory is placed on a closed manifold $M_{D+1}$. However, if the manifold $M_{D+1}$ has a boundary $B_{D} = \partial M_{D+1}$, it transforms as
\begin{align}
    Z_\text{SPT}[C^g] = e^{-i \int_{B_{D}} \beta(C,g)} Z_\text{SPT}[C] \,.
\end{align}
Therefore, by putting the anomalous theory on the boundary $B_{D}$ and considering the combined system of the SPT phase in the bulk and the anomalous QFT on the boundary as
\begin{align}
  Z_\text{total}[C] = Z_\text{SPT}[C] Z_\text{anom.}[C] \,,
\end{align}
this partition function is invariant under the gauge transformation:
\begin{align}
  Z_\text{total}[C^g] = Z_\text{total}[C] \,.
\end{align}
Thus, the 't Hooft anomaly in the $D$-dimensional QFT is captured by the SPT phase in $D+1$ dimensions. This is the anomaly inflow mechanism.

\subsection{Examples of Generalized Global Symmetry}

Here, we provide some standard examples of global symmetry.

\subsubsection*{Compact scalar field theory in $d+1$ dimensions}

We review the free compact scalar field theory in $d+1$ dimensions. The $d=1$ case is widely known as the simplest conformal field theory, and we will revisit it in Section \ref{subsection:construction_of_boundary_theory}.
This theory has the $U(1)$ zero-form momentum symmetry and the $U(1)$ $(d-1)$-form winding symmetry.
This $U(1) \times U(1)$ symmetry has a mixed 't Hooft anomaly, where we cannot gauge both symmetries simultaneously.
The anomaly is captured by the SPT phase for $U(1) \times U(1)$ symmetry in $(d+1)+1$ dimensions.
In this section, we consider the $(d+1)$-dimensional torus $M_{d+1} = \T^{d+1}$ as spacetime.

The Lagrangian of the compact scalar field theory is given by
\begin{align}
\begin{split}
  L_{\phi} &= \frac{R^2}{2} d\phi \wedge \ast d\phi \\
  &= \frac{R^2}{2} \left[ (\partial_0 \phi)^2 + \sum_{i=1}^d  (\partial_i \phi)^2 \right] d^{d+1}x \,, \label{comp_lag}
\end{split}
\end{align}
where $R$ is a parameter with mass dimension $(d-1)/2$, and the real scalar field $\phi$ has the periodicity $\phi \sim \phi + 2\pi$.
The dynamical gauge transformation of $\phi$ is
\begin{align}
  \phi \sim \phi + 2\pi w \,, \label{gauge_eq_phi}
\end{align}
where $w$ is an integer.
The equation of motion is given by
\begin{align}
  d \ast d \phi = 0 \,.
\end{align}
From this equation, the current 
\begin{align}
  J = i R^2 d \phi \,,
\end{align}
is conserved:
\begin{align}
  d \ast J = 0 \,.
\end{align}
This current corresponds to the $U(1)$ zero-form momentum symmetry.
The symmetry operator is
\begin{align}
  U_\alpha [\Sigma_{d}] = \exp \left( i \alpha \int_{\Sigma_{d}} \ast J \right) \,,
\end{align}
where $\alpha$ is $2\pi$-periodic parameter: $e^{i\alpha} \in U(1)$.
The symmetry transformation on $\phi$ is
\begin{align}
  \phi \rightarrow \phi + \alpha \,,
\end{align}
so the gauge-invariant charged operator is
\begin{align}
  V_q[x] = e^{i q \phi(x)} \,,
\end{align}
where $q \in \Z$ is the charge of the operator.
This theory also has another conserved current
\begin{align}
  \hat{J} = \frac{(-1)^d}{2\pi} \ast d \phi \,,
\end{align}
which satisfies the conservation law\footnote{In $(d+1)$-dimensional Euclidean spacetime, we have $\ast\ast = (-1)^{n(d+1-n)}$ for an $n$-form field.}
\begin{align}
  d \ast \hat{J} = 0 \,.
\end{align}
The current $\hat{J}$ is $d$-form, and this corresponds to the $U(1)$ $(d-1)$-form winding symmetry.
The symmetry operator is
\begin{align}
  \hat{U}_{\hat{\alpha}} [\Sigma_{1}] = \exp \left( i \hat{\alpha} \int_{\Sigma_{1}} \ast \hat{J} \right) \,, \label{winding_sym_op1}
\end{align}
 where $\hat{\alpha}$ is $2\pi$-periodic parameter: $e^{i\hat{\alpha}} \in U(1)$.
 When the closed manifold $\Sigma_{1}$ is the line $\mathcal{C}^i$ along the $x^i$ direction, the symmetry operator becomes
\begin{align}
\begin{split}
  \hat{U}_{\hat{\alpha}} [\mathcal{C}^i] &= \exp \left( i \frac{\hat{\alpha}}{2\pi} \int_{\mathcal{C}^i} dx^i\, \partial_i \phi   \right) \\
  &= e^{ i \hat{\alpha} w_{\phi}^i} \,,
\end{split}
\end{align}
where $w_{\phi}^i \in \Z$ is the winding number of $\phi$ along the $x^i$ direction.
The charged object is a $(d-1)$-dimensional defect, which is expressed by using the dual $(d-1)$-form gauge field $\hat{A}$ introduced below.

Then, we consider a duality of the compact scalar field theory \eqref{comp_lag}. The duality is known as the $T$-duality in $d=1$ case \cite{Buscher:1987sk,Buscher:1987qj}.
The Lagrangian is also written as
\begin{align}
\begin{split}
  L_{\phi} &= \frac{R^2}{2} E \wedge \ast E + \frac{i}{2\pi} \hat{E} \wedge  (d \phi - E )  \,, \label{comp_lag_dual}
\end{split}
\end{align}
where $E$ is a dynamical one-form field, and $\hat{E}$ is a dynamical $d$-form field. Integrating out $\hat{E}$, we have 
\begin{align}
  E = d \phi \,, \label{eq_of_motion_E_hat}
\end{align}
and we obtain the original Lagrangian \eqref{comp_lag}.
On the other hand, integrating out $E$, we have
\begin{align}
  E = \frac{i}{2\pi R^2} \ast \hat{E} \,, \label{eq_of_motion_E}
\end{align}
so the Lagrangian becomes
\begin{align}
\begin{split}
  L_{\phi} &= \frac{1}{8 \pi^2 R^2} \hat{E} \wedge \ast \hat{E} + \frac{i}{2\pi} \hat{E} \wedge d \phi  \,. \label{comp_lag_dual2}
\end{split}
\end{align}
Then, integrating out $\phi$, the equation of motion is
\begin{align}
  d \hat{E} = 0 \,,
\end{align}
and due to the gauge equivalence \eqref{gauge_eq_phi}, the field $\hat{E}$ satisfies the quantization condition
\begin{align}
  \int_{\Sigma_{d}} \frac{\hat{E}}{2\pi} \in \Z \,.
\end{align}
Therefore, we can write $\hat{E}$ locally as
\begin{align}
  \hat{E} = d \hat{A} \,, \label{eq_of_motion_phi}
\end{align}
where $\hat{A}$ is a $(d-1)$-form gauge field with the gauge transformation
\begin{align}
  \hat{A} \sim \hat{A} + d \hat{\lambda} \,,
\end{align}
where $\hat{\lambda}$ is a $U(1)$ $(d-2)$-form gauge parameter.\footnote{For $d=1$, we can write $\hat{E}$ locally as
\begin{align}
  \hat{E} = d \hat{\phi} \,,
\end{align}
where $\hat{\phi}$ is a zero-form gauge field with the gauge transformation
\begin{align}
  \hat{\phi} \sim \hat{\phi} + 2\pi \hat{w} \,,
\end{align}
where $\hat{w}$ is an integer and we can regard $\hat{w}$ as a exterior derivative of a $(-1)$-form gauge parameter $\hat{\chi}$ that satisfies $d\hat{\chi} = \hat{w}$.}
Then, the Lagrangian \eqref{comp_lag_dual2} becomes
\begin{align}
\begin{split}
  L_{\hat{A}} &= \frac{1}{8 \pi^2 R^2} d \hat{A} \wedge \ast d \hat{A} \,. \label{comp_lag_dual3}
\end{split}
\end{align}
From the relations \eqref{eq_of_motion_E_hat}, \eqref{eq_of_motion_E}, and \eqref{eq_of_motion_phi}, we have the relation
\begin{align}
  d \phi = \frac{i}{2\pi R^2} \ast d \hat{A} \,.
\end{align}
In this dual description, the currents of the $U(1)$ zero-form momentum symmetry and the $U(1)$ $(d-1)$-form winding symmetry are expressed as
\begin{align}
  J &= - \frac{1}{2\pi} \ast d \hat{A} \,, \\
  \hat{J} &= \frac{i}{(2\pi R)^2} d \hat{A} \,.
\end{align}
Therefore, the momentum symmetry and the winding symmetry are exchanged under this duality.
Then, the charged object for the $U(1)$ $(d-1)$-form winding symmetry \eqref{winding_sym_op1} is expressed as
\begin{align}
  \hat{V}_{\hat{q}}[\Sigma_{d-1}] = \exp \left( i \hat{q} \int_{\Sigma_{d-1}} \hat{A} \right) \,,
\end{align}
where $\hat{q} \in \Z$ is the charge of the operator.

Next, we consider coupling to the background gauge fields for the $U(1)$ zero-form momentum symmetry and the $U(1)$ $(d-1)$-form winding symmetry.
The background gauge field for the $U(1)$ zero-form momentum symmetry is a one-form gauge field $C$ with the background gauge transformation
\begin{align}
  C \sim C + d \gamma \,,
\end{align}
where $\gamma$ is a $U(1)$ background zero-form gauge parameter.
The background gauge field for the $U(1)$ $(d-1)$-form winding symmetry is a $d$-form gauge field $\hat{C}$ with the background gauge transformation
\begin{align}
  \hat{C} \sim \hat{C} + d \hat{\gamma} \,,
\end{align}
where $\hat{\gamma}$ is a $U(1)$ background $(d-1)$-form gauge parameter.
When coupled to the background gauge fields, the gauge field $\phi$ obtains the background gauge transformation
\begin{align}
  \phi \sim \phi + \gamma \,.
\end{align}
The compact scalar field theory coupled to the background gauge fields is given by
\begin{align}
\begin{split}
  L_{\phi}\left[C,\hat{C}\right] &= \frac{R^2}{2} (d\phi - C) \wedge \ast (d\phi - C) + \frac{i}{2\pi} \hat{C} \wedge (d \phi - C) \,. \label{comp_lag_bg}
\end{split}
\end{align}
Under the background gauge transformations, the Lagrangian transforms as
\begin{align}
  \delta_\text{g} L_{\phi}\left[C,\hat{C}\right] = - \frac{i}{2\pi} d \hat{\gamma} \wedge C \,,
\end{align}
so the partition function transforms as
\begin{align}
  Z_{\phi}\left[C + d\gamma,\hat{C} + d \hat{\gamma}\right] = \exp \left( \int_{M_{d+1}} \frac{i}{2\pi} d \hat{\gamma} \wedge C \right) Z_{\phi}\left[C,\hat{C}\right] \,.
\end{align}
This phase cannot be canceled by a local counterterm of $C$, $\hat{C}$, $\gamma$, and $\hat{\gamma}$ in $d+1$ dimensions, so the $U(1)$ zero-form momentum symmetry and the $U(1)$ $(d-1)$-form winding symmetry have the mixed 't Hooft anomaly.
This anomaly is the same as the quantum anomaly of free fermion theory in $1+1$ dimensions, which is known as the chiral anomaly \cite{Adler:1969gk,Bell:1969ts}.

This anomaly is captured by the SPT phase for $U(1) \times U(1)$ symmetry in $(d+1)+1$ dimensions, which is given by the effective Lagrangian
\begin{align}
  L_\text{SPT}\left[C,\hat{C}\right] =   \frac{i}{2\pi} \hat{C} \wedge d C \,, \label{spt_compact_scalar}
\end{align}
where $C$ and $\hat{C}$ are background gauge fields in $(d+1)+1$ dimensions.
Under the background gauge transformations, the Lagrangian transforms as
\begin{align}
  \delta_\text{g} L_\text{SPT}\left[C,\hat{C}\right] = d \left[ (-1)^d \frac{i}{2\pi} d \hat{\gamma} \wedge C \right] \,.
\end{align}
On a closed manifold, the action is gauge invariant.
However, when we consider the manifold $N_{(d+1)+1} = \T^{d+1} \times \R_{x^{d+1} \geq 0}$ with a boundary $\partial N_{(d+1)+1} = (-1)^d \T^{d+1} = (-1)^d M_{d+1}$, under the background gauge transformations, the action transforms as
\begin{align}
\begin{split}
  \delta_\text{g} S_\text{SPT}\left[C,\hat{C}\right] 
  &= \int_{(-1)^d M_{d+1}} (-1)^d \frac{i}{2\pi} d \hat{\gamma} \wedge C \\
  &= \int_{M_{d+1}}  \frac{i}{2\pi} d \hat{\gamma} \wedge C \,.
\end{split}
\end{align}
Thus, the partition function transforms as
\begin{align}
  Z_\text{SPT}\left[C + d\gamma,\hat{C} + d \hat{\gamma}\right] =  \exp \left( - \int_{M_{d+1}} \frac{i}{2\pi} d \hat{\gamma} \wedge C \right) Z_\text{SPT}\left[C,\hat{C}\right]\,,
\end{align}
which cancels the anomaly of the compact scalar field theory on the boundary $M_{d+1}$. This is an example of the anomaly inflow mechanism.

\subsubsection*{Ordinary $BF$ theory in 2+1 dimensions}

We review the ordinary $BF$ theory in 2+1 dimensions \cite{Maldacena:2001ss,Banks:2010zn,Kapustin:2014gua}, which is a topological quantum field theory (TQFT).
This theory has the $\Z_N$ electric one-form symmetry and the $\Z_N$ magnetic one-form symmetry with the mixed 't Hooft anomaly.
The $BF$ theory is the low energy effective field theory of the $\Z_N$ Toric code in 2+1 dimensions with anyon excitations \cite{Kitaev:1997wr}.
This theory is similar to the fractonic $BF$ theories introduced in later sections, and many aspects of the fractonic $BF$ theories can be understood through analogy with this theory.
In this section, we consider the $(2+1)$-dimensional torus $M_{2+1} = \T^{2+1}$ as spacetime.

The Lagrangian of the $BF$ theory in 2+1 dimensions is given by
\begin{align}
  L_{BF} = \frac{iN}{2\pi} \hat{b} \wedge d a \,, \label{rev:bf_lag}
\end{align}
where $N$ is a positive integer, and $a$ and $\hat{b}$ are $U(1)$ one-form gauge fields.
Their gauge transformations are
\begin{align}
  a &\sim a + d \lambda \,, \\
  \hat{b} &\sim \hat{b} + d \hat{\lambda} \,,
\end{align}
where $\lambda$ and $\hat{\lambda}$ are $U(1)$ zero-form gauge parameters.
The equations of motion are
\begin{align}
  \frac{iN}{2\pi} d a = 0 \,, \\
  \frac{iN}{2\pi} d \hat{b} = 0 \,,
\end{align}
and due to the gauge equivalence, the fields satisfy the quantization conditions
\begin{align}
  \int_{\Sigma_{1}} a \in \frac{2\pi}{N} \Z \,, \label{rev:quantization_a}\\ 
  \int_{\Sigma_{1}} \hat{b} \in \frac{2\pi}{N} \Z \,, \label{rev:quantization_hatb}
\end{align}
where $\Sigma_{1}$ is any closed one-dimensional manifold.
Then, $a$ and $\hat{b}$ are $\Z_N$ one-form gauge fields.\footnote{Strictly, to define the Lagrangian \eqref{rev:bf_lag}, we need to take into account the global structure of the gauge fields. The detailed calculation is given in \cite{Cordova:2019jnf}.} 
This theory is dual to the theory described by the Lagrangian
\begin{align}
  L_{\Z_N} = \frac{i}{2\pi} \hat{E} \wedge ( d \phi - N a) \,,
\end{align}
where $\phi$ is a compact scalar field (zero-form gauge field), and $\hat{E}$ is a two-form field.
The compact scalar field $\phi$ has the gauge transformation
\begin{align}
  \phi \sim \phi + N \gamma + 2\pi w \,,
\end{align}
where $w$ is an integer.
This means that $U(1)\setminus\Z_N$ in the gauge group $U(1)$ is Higgsed and the gauge group is broken to $\Z_N$.
Integrating out $\phi$ and dualizing $\hat{E}$ to the one-form gauge field $\hat{b}$, we obtain the $BF$ theory \eqref{rev:bf_lag}.

This theory has $\Z_N$ electric and magnetic one-form symmetries. The symmetry operator of the $\Z_N$ electric one-form symmetry is
\begin{align}
  U_{m}[\Sigma_{1}] &= \exp \left( i m \int_{\Sigma_{1}} \hat{b} \right) \,,
\end{align}
where $m = 0,1,\ldots,N-1$ and $\Sigma_{1}$ is a closed one-dimensional manifold.
The charged object is
\begin{align}
  W_{n}[\hat{\Sigma}_{1}] = \exp \left( i n \int_{\hat{\Sigma}_{1}} a \right) \,,
\end{align}
where $n = 0,1,\ldots,N-1$ and $\hat{\Sigma}_{1}$ is a closed one-dimensional manifold. These operators satisfy $U_{1}[\Sigma_{1}]^N = W_{1}[\hat{\Sigma}_{1}]^N = 1$, so they are $\Z_N$ operators. 
On the other hand, the symmetry operator of the $\Z_N$ magnetic one-form symmetry is $W_{n}[\hat{\Sigma}_{1}]$ and the charged object is $U_{m}[\Sigma_{1}]$.
They satisfy the following commutation relation:
\begin{align}
  U_{m}[\Sigma_{1}] W_{n}[\hat{\Sigma}_{1}] = e^{ - 2\pi i m n \, \text{Link}(\Sigma_{1},\hat{\Sigma}_{1})/N } W_{n}[\hat{\Sigma}_{1}] U_{m}[\Sigma_{1}] \,,
\end{align}
where $\text{Link}(\Sigma_{1},\hat{\Sigma}_{1}) \in \Z$ is the linking number of $\Sigma_{1}$ and $\hat{\Sigma}_{1}$.
Therefore, when the $\Sigma_{1}$ surrounds the defect on $\hat{\Sigma}_{1}$, the symmetry operator $U_{m}[\Sigma_{1}]$ acts on the charged object $W_{n}[\hat{\Sigma}_{1}]$ as
\begin{align}
  U_{m}[\Sigma_{1}] \cdot W_{n}[\hat{\Sigma}_{1}] = e^{ - 2\pi i m n /N } W_{n}[\hat{\Sigma}_{1}] \,,
\end{align}
and similarly for the magnetic one-form symmetry.
The defects $W_{n}[\hat{\Sigma}_{1}]$ and $U_{m}[\Sigma_{1}]$ represent the worldlines of the charged particles, which are anyons.

On the torus $M_{2+1} = \T^{2+1}$, there are two one-dimensional non-contractible closed loops in space, and they satisfy
\begin{align}
  U_{1}[\mathcal{C}^1] W_{1}[\mathcal{C}^2] = e^{ - 2\pi i /N }  W_{1}[\mathcal{C}^2] U_{1}[\mathcal{C}^1] \,.
\end{align}
Then, the symmetries are spontaneously broken, and the ground state degeneracy is $N^2$.

Next, we consider coupling to the background gauge fields for the $\Z_N$ electric and magnetic one-form symmetries.
We introduce $U(1)$ background two-form gauge fields $C$ and $\hat{C}$.
Their background gauge transformations are
\begin{align}
  C &\sim C + d \gamma \,, \\
  \hat{C} &\sim \hat{C} + d \hat{\gamma} \,,
\end{align}
where $\gamma$ and $\hat{\gamma}$ are $U(1)$ background one-form gauge parameters.
To restrict the gauge fields to $\Z_N$, we add to the Lagrangian the terms $\frac{iN}{2\pi}\hat{\chi}  dC $ and $\frac{iN}{2\pi}\chi  d\hat{C} $, where $\chi$ and $\hat{\chi}$ are $U(1)$ zero-form gauge fields.
Then, the $BF$ theory coupled to the background gauge fields is given by
\begin{align}
  L_{BF}\left[C,\hat{C}\right] = \frac{iN}{2\pi}\left[ \hat{b} \wedge (d a - C) - \hat{C} \wedge a \right] + \frac{iN}{2\pi}\hat{\chi}  dC + \frac{iN}{2\pi}\chi  d\hat{C} \,. \label{rev:bf_lag_bg}
\end{align}
The gauge fields $a$ and $\hat{b}$ obtain the background gauge transformations
\begin{align}
  a &\sim a + \gamma \,, \\
  \hat{b} &\sim \hat{b} + \hat{\gamma} \,,
\end{align}
and the dynamical fields $\chi$ and $\hat{\chi}$ have the dynamical gauge transformations
\begin{align}
  \chi &\sim \chi - \lambda \,, \\
  \hat{\chi} &\sim \hat{\chi} - \hat{\lambda} \,.
\end{align}
Under the background gauge transformations, the Lagrangian transforms as
\begin{align}
  \delta_\text{g} L_{BF}\left[C,\hat{C}\right] = \frac{iN}{2\pi} \left[ - \hat{\gamma} \wedge C - (\hat{C} + d\hat{\gamma} )  \wedge \gamma \right] \,, \label{rev:spt_gauge_transf2}
\end{align}
which indicates the mixed 't Hooft anomaly of the $\Z_N \times \Z_N$ symmetry.

This anomaly is captured by the SPT phase for $\Z_N \times \Z_N$ symmetry in $3+1$ dimensions, which is given by the effective Lagrangian
\begin{align}
  L_\text{SPT}\left[C,\hat{C}\right] = \frac{iN}{2\pi} \hat{\beta} \wedge dC + \frac{iN}{2\pi} \beta \wedge d\hat{C} +   \frac{iN}{2\pi} \hat{C} \wedge C \,, \label{spt_bf}
\end{align}
where $C$ and $\hat{C}$ are background gauge fields in $3+1$ dimensions, and $\beta$ and $\hat{\beta}$ are  $U(1)$ dynamical one-form gauge fields that restrict $C$ and $\hat{C}$ to $\Z_N$ gauge fields.
The dynamical gauge transformations of $\beta$ and $\hat{\beta}$ are
\begin{align}
  \beta &\sim \beta + d s \,, \\
  \hat{\beta} &\sim \hat{\beta} + d \hat{s} \,,
\end{align}
where $s$ and $\hat{s}$ are $U(1)$ zero-form gauge parameters. When we put the SPT phase on a manifold with a boundary, we set the gauge parameters $s$ and $\hat{s}$ to zero on the boundary.
$\beta$ and $\hat{\beta}$ also have the background gauge transformations
\begin{align}
  \beta &\sim \beta - \gamma \,, \\
  \hat{\beta} &\sim \hat{\beta} - \hat{\gamma} \,.
\end{align}
Then, under the background gauge transformations, the Lagrangian transforms as
\begin{align}
  \delta_\text{g} L_\text{SPT}\left[C,\hat{C}\right] = d \left\{ \frac{iN}{2\pi} \left[ \hat{\gamma} \wedge C + (\hat{C} + d\hat{\gamma} )  \wedge \gamma \right] \right\} \,. 
\end{align}
Therefore, when we consider the manifold $N_{3+1} = \T^{2+1} \times \R_{x^3 \geq 0}$ with a boundary $\partial N_{3+1} = (-1)^{2} \T^{2+1} = M_{2+1}$, under the background gauge transformations, the action transforms as
\begin{align}
  \delta_\text{g} S_\text{SPT}\left[C,\hat{C}\right] 
  &= \int_{ M_{2+1}} \frac{iN}{2\pi} \left[ \hat{\gamma} \wedge C + (\hat{C} + d\hat{\gamma} )  \wedge \gamma \right]  \,,
\end{align}
which cancels the anomaly of the $BF$ theory \eqref{rev:spt_gauge_transf2} on the boundary $M_{2+1}$.

\section{Fracton Phases}
\label{section:fracton_phases}

\subsection{Fracton Phases and Subsystem Symmetry}

Fracton phases are new types of phases of matter that have characteristic excitations called fractons (See for reviews \cite{Nandkishore:2018sel,Pretko:2020cko,Gromov:2022cxa}).
Fractons are immobile particles that cannot move freely in space.
The fracton phases sometimes have other particles called lineons and planons, which can move only along a line or a plane, respectively.
These particles with restricted mobility is originated from the subsystem symmetry, which is a partially topological global symmetry.
Since the subsystem symmetry operators are topological only along certain submanifolds, they cannot be deformed and moved freely in space, and the operators on different submanifolds have independent charges \cite{Seiberg:2019vrp}.
From this property, the fracton phases have a huge ground state degeneracy that grows exponentially with linear system size, not with the volume.

In the earlier studies, the fracton phases have attracted much attention as exactly solvable spin models from the point of view of glassy dynamics and quantum memory \cite{Chamon:2004lew,Haah:2011drr,Bravyi:2010jfq}.
The Haah's code \cite{Haah:2011drr} has fracton excitations at the corners of the fractal operators, and the subsystem symmetry are defined on the fractal submanifolds.
On the other hand, there are fracton models that have fracton excitations at the corners of membrane operators, and the subsystem symmetry are defined on rigid lines or planes \cite{Chamon:2004lew,Bravyi:2010jfq}.
As for gapless models, they do not have fracton excitations, but they have subsystem symmetry \cite{Paramekanti:2002iup,Seiberg:2020bhn,Seiberg:2020wsg}.
Thus, we refer to models with subsystem symmetries as fracton phases in this dissertation.
The most well-known example of fracton phases with subsystem symmetries is the X-cube model \cite{Vijay:2016phm}, which has $\Z_2 \times \Z_2$ subsystem symmetry in 3+1 dimensions.
In Section \ref{section:rev_xcube_model}, we review the X-cube model, and see the properties of the fracton models: subsystem symmetry, excitations with restricted mobility and a huge ground state degeneracy.

The fracton phases with subsystem symmetries have been studied not only in lattice models but also in continuum quantum field theories --- Fracton QFTs \cite{Slagle:2017wrc,Pretko:2018jbi,You:2019ciz,Seiberg:2020bhn,Seiberg:2020wsg,Seiberg:2020cxy,Gorantla:2020xap,Gorantla:2021bda,Gorantla:2020jpy,Geng:2021cmq,Yamaguchi:2021qrx,Yamaguchi:2021xeq,Gorantla:2022eem,Gorantla:2022ssr,Honda:2022shd,Burnell:2021reh,Luo:2022mrj,Slagle:2018swq,Slagle:2020ugk,Hsin:2021mjn,Geng:2021cmq,Ohmori:2022rzz,Spieler:2023wkz,Cao:2023rrb,Ebisu:2023idd,Ebisu:2024cke,Hsin:2023ooo,Hsin:2024eyg,Gorantla:2025fzz}.
Fracton QFT is non-relativistic QFT that has subsystem symmetries, so it has discrete rotational symmetry.
Study of fracton QFTs is important to understand universal properties of fracton phases and general aspects of QFTs themselves.
As explained in Introduction, they are classified into two types: exotic QFTs and foliated QFTs, and the duality in two types of fracton QFTs \cite{Ohmori:2022rzz} is the main topic of this dissertation.
 
In Section \ref{section:rev_fractonic_bf_3+1}, we review an example of fracton QFT --- the exotic $BF$ thoery in 3+1 dimensions \cite{Slagle:2017wrc,Seiberg:2020cxy,Ohmori:2022rzz}. This theory is considered as the low-energy effective field theory of the $\Z_N$ X-cube model, which is the $\Z_N$ generalization of the X-cube model with $\Z_N \times \Z_N$ subsystem symmetry.

\subsection{The X-cube Model}
\label{section:rev_xcube_model}

In this section, we review the X-cube model \cite{Vijay:2016phm}, which is a prototypical example of fracton phases with subsystem symmetry.
Here we consider the X-cube model defined on a three-dimensional cubic lattice with periodic boundary conditions.

In this section, we denote the lattice spacing as $a$, and the lattice sites are given by $(n^1, n^2, n^3)$ with integers $n^1,n^2,n^3$.
The system size in the $x^i$ direction is $aL^i$ with a positive integer $L^i$.

We assign two-dimensional Hilbert space $\mathcal{H}_e = \left\{ \ket{0}_e, \ket{1}_e \right\}$ and the $\Z_2$ operators $A_e$ and $B_e$ on each link $e$ of the cubic lattice.
The operators satisfy the relations
\begin{align}
  A_e^2 = B_e^2 = 1 \,,\\
  A_e B_e  = - B_e A_e \,,
\end{align}
and act on the vectors in $\mathcal{H}_e$ as 
\begin{align}
  A_e \ket{0}_e &= \ket{0}_e \,, \\
  A_e \ket{1}_e &= -\ket{1}_e \,, \\
  \ket{1}_e &= B_e \ket{0}_e \,, \\
  \ket{0}_e &= B_e \ket{1}_e \,.
\end{align}
Thus they are the Pauli operators.
The total Hilbert space is given by the tensor product of the Hilbert spaces on all links $\mathcal{H} = \bigotimes_e \mathcal{H}_e$.
We have the gauge transformation at each site as
\begin{align}
  B_{(n^1 + 1/2,n^2,n^3)} &\sim B_{(n^1 + 1/2,n^2,n^3)} \Lambda_{(n^1,n^2,n^3)}^1 \Lambda_{(n^1+1,n^2,n^3)}^1 \,, \\
  B_{(n^1,n^2 + 1/2,n^3)} &\sim B_{(n^1,n^2 + 1/2,n^3)} \Lambda_{(n^1,n^2,n^3)}^2 \Lambda_{(n^1,n^2+1,n^3)}^2 \,, \\
  B_{(n^1,n^2,n^3 + 1/2)} &\sim B_{(n^1,n^2,n^3 + 1/2)} \Lambda_{(n^1,n^2,n^3)}^3 \Lambda_{(n^1,n^2,n^3+1)}^3 \,,
\end{align}
where $\Lambda_{s}^k$ is a $\Z_2$ gauge parameter satisfying $\Lambda_{s}^1 \Lambda_{s}^2 \Lambda_{s}^3 = 1$ on each site $s$, and we label each link by the midpoint of the two sites it connects.

\begin{figure}[H]
   \begin{center}
    \includegraphics[width=0.8\hsize]{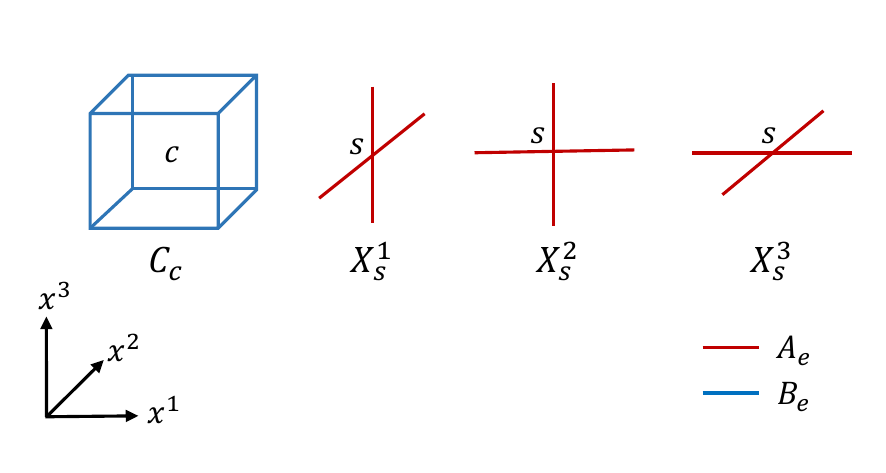} 
    \end{center}
    \vspace{-1cm}
      \caption{The cube operator $C_c$ (left) and the cross operator $X_s^k$ (right) in the X-cube model. The red and blue links represent the operators $A_e$ and $B_e$, respectively.}
    \label{fig:xcube_operators}
\end{figure}

The Hamiltonian of the X-cube model is given by
\begin{align}
  H = - \sum_{c:\text{cube}}  C_c  - \sum_{\substack{s:\text{site} \\ k = 1,2,3}} X_s^k  \,, \label{rev:xcube_ham}
\end{align}
where we define the cube operator $C_c$ on each cube $c$ as
\begin{align}
\begin{split}
  &C_{c = (n^1+1/2,n^2+1/2,n^3+1/2)} \\
  &=  B_{(n^1+1/2,n^2,n^3)}   B_{(n^1+1,n^2+1/2,n^3)}  B_{(n^1+1/2,n^2+1,n^3)}  B_{(n^1,n^2+1/2,n^3)}\\
  &\times B_{(n^1,n^2 ,n^3+1/2)}   B_{(n^1+1,n^2,n^3+1/2)}  B_{(n^1 + 1,n^2 +1,n^3+1/2)}  B_{(n^1,n^2+1,n^3+1/2)} \\
  &\times B_{(n^1+1/2,n^2,n^3+1)}  B_{(n^1+1,n^2+1/2,n^3+1)}  B_{(n^1+1/2,n^2+1,n^3+1)}    B_{(n^1,n^2+1/2,n^3+1)}\,,
\end{split}
\end{align}
and the Hermitian cross operator $X_s^k$ on each site $s$ as
\begin{align}
  X_{s=(n^1,n^2,n^3)}^1 &=  A_{(n^1,n^2+1/2,n^3)}  A_{(n^1,n^2,n^3+ 1/2)} A_{(n^1,n^2-1/2,n^3)} A_{(n^1,n^2,n^3-1/2)}  \,, \\
  X_{s=(n^1,n^2,n^3)}^2 &=  A_{(n^1+1/2,n^2,n^3)}  A_{(n^1,n^2,n^3+ 1/2)}    A_{(n^1-1/2,n^2,n^3)}  A_{(n^1,n^2,n^3-1/2)}  \,, \\ 
  X_{s=(n^1,n^2,n^3)}^3 &=  A_{(n^1+1/2,n^2,n^3)}  A_{(n^1,n^2+ 1/2,n^3)}    A_{(n^1-1/2,n^2,n^3)}  A_{(n^1,n^2-1/2,n^3)}  \,,
\end{align}
as in Figure \ref{fig:xcube_operators}.
They commute with each other, so the model is exactly solvable.
A ground state $\ket{GS}$ satisfies
\begin{align}
  C_c \ket{GS} =  \ket{GS} \,, \\
  X_s^k \ket{GS} = \ket{GS} \,.
\end{align}
By considering the independent constraints among the cube operators and the cross operators, we can calculate the ground state degeneracy.
The ground state degeneracy is $2^{2L^1 + 2L^2 + 2L^3 - 3}$ \cite{Vijay:2016phm}, which grows exponentially with the linear system size.

\begin{figure}[H]
  \centering
  \begin{minipage}{0.45\hsize}
    \centering
    \includegraphics[width=\hsize]{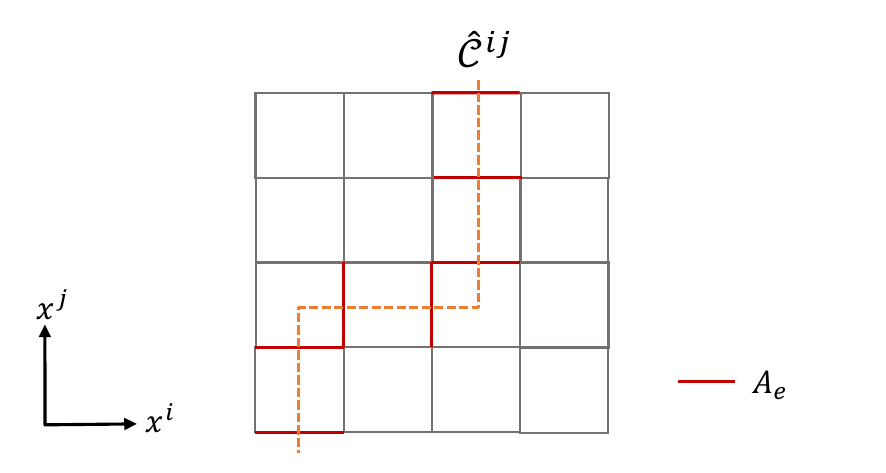}
    \caption{The orange dashed loop is the zigzagging closed loop $\hat{\mathcal{C}}^{ij}$ in the $x^i x^j$-plane. The red links represent the operator $A_e$.}
    \label{fig:dipsym}
  \end{minipage}
  \hspace{0.5cm}
  \begin{minipage}{0.45\hsize}
    \centering
    \includegraphics[width=\hsize]{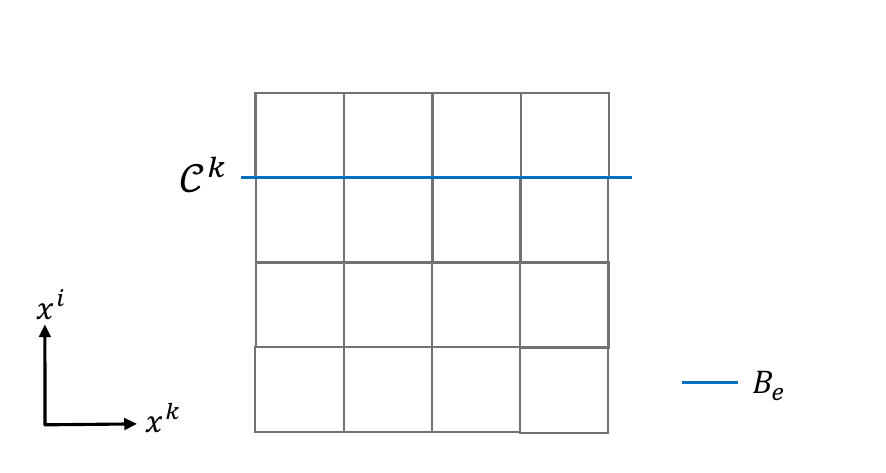}
    \caption{The blue line is the closed line $\mathcal{C}^{k}$ in the $x^k$ direction. The blue links also represent the operator $B_e$.}
    \label{fig:tensym}
  \end{minipage}
\end{figure}

The X-cube model has $\Z_2 \times \Z_2$ subsystem symmetry.
The charge of the $\Z_2$ dipole symmetry is
\begin{align}
  W^{k}[\hat{\mathcal{C}}^{ij}] = \prod_{e \in \hat{\mathcal{C}}^{ij}} A_{e} \,,
\end{align}
where $\hat{\mathcal{C}}^{ij}$ is a zigzagging closed one-dimensional loop on the dual lattice links in the $x^i x^j$-plane with fixed $x^k$ (see Figure \ref{fig:dipsym}).\footnote{The charge of the $\Z_2$ dipole symmetry cannot be deformed freely even on the $x^i x^j$-plane unless we demand the Gauss law $X_s^k = 1$ on each site $s$. Thus, this is called unconstrained symmetry \cite{Seiberg:2020cxy}.}
The charge of the $\Z_2$ tensor symmetry is
\begin{align}
  V^{k}[\mathcal{C}^{k}] = \prod_{e \in \mathcal{C}^{k}} B_{e} \,,
\end{align}
where $\mathcal{C}^{k}$ is a closed one-dimensional line on the lattice links in the $x^k$ direction with fixed $x^i$ and $x^j$ (see Figure \ref{fig:tensym}).
These operators commute with the Hamiltonian \eqref{rev:xcube_ham}, so they are conserved charges.
They satisfies the following commutation relation:
\begin{align}
  W^{1}[\hat{\mathcal{C}}^{23}] V^{2}[\mathcal{C}^{2}] &= - V^{2}[\mathcal{C}^{2}] W^{1}[\hat{\mathcal{C}}^{23}] \,, \\
  W^{1}[\hat{\mathcal{C}}^{23}] V^{3}[\mathcal{C}^{3}] &= - V^{3}[\mathcal{C}^{3}] W^{1}[\hat{\mathcal{C}}^{23}] \,,  \\
  W^{1}[\hat{\mathcal{C}}^{23}] V^{1}[\mathcal{C}^{1}] &=  V^{1}[\mathcal{C}^{1}] W^{1}[\hat{\mathcal{C}}^{23}] \,,
\end{align}
similarly for other directions, when the loop $\hat{\mathcal{C}}^{ij}$ and the line $\mathcal{C}^{k}$ cross each other once.

\begin{figure}[H]
  \centering
  \begin{minipage}{0.45\hsize}
    \centering
    \includegraphics[width=\hsize]{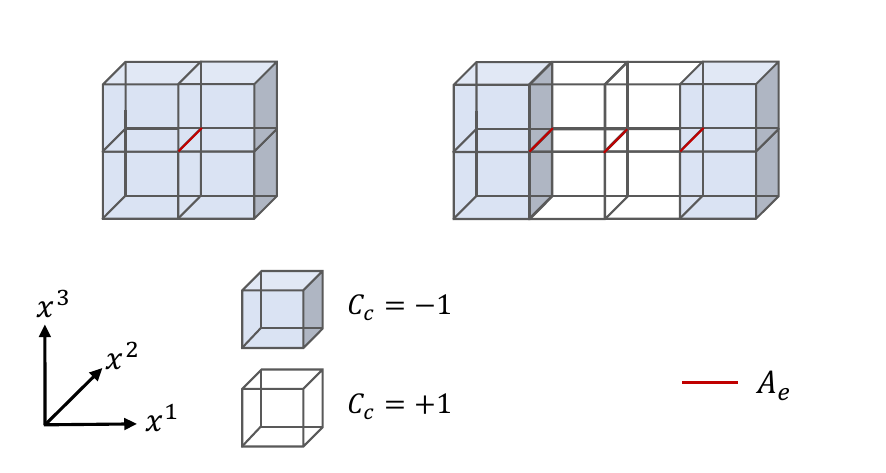}
    \caption{The fracton excitation created by acting with $A_e$. The red links represent $A_e$ and the colored cubes correspond to the fractons.}
    \label{fig:xcube_fracton}
  \end{minipage}
  \hspace{0.5cm}
  \begin{minipage}{0.45\hsize}
    \centering
    \includegraphics[width=\hsize]{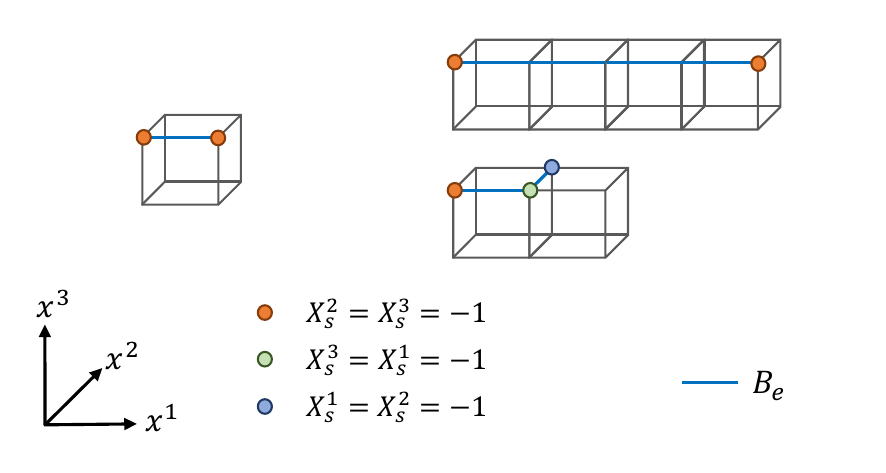}
    \caption{The lineon excitations created by acting with $B_e$. The blue links represent $B_e$ and the colored sites correspond to the lineons.}
    \label{fig:xcube_lineon}
  \end{minipage}
\end{figure}

Let us consider a state $A_{e} \ket{GS}$.
This state is an excited state with four excitations, whose energy is $+1$, at the four cubes that share the link $e$, since it satisfies
\begin{align}
  C_{c} A_{e} \ket{GS} = - A_{e} \ket{GS} \,,
\end{align}
where $c$ are the four cubes sharing the link $e$ (see Figure \ref{fig:xcube_fracton}).
This excitation is called a fracton that cannot be moved individually.
The dipole of two fractons can move only in a plane by acting with $A_{e'}$ on adjacent links $e'$ without creating additional excitations, so it is called a planon.
We also consider another state $B_{e} \ket{GS}$.
This state is an excited state with two excitations, whose energy is $+2$, at the two ends of the link $e$, since it satisfies
\begin{align}
  X_{s}^k B_{e} \ket{GS} = - B_{e} \ket{GS} \,,
\end{align}
where $s$ are the two sites at the ends of the link $e$ and $k$ is the other directions than the direction of the link $e$ (see Figure \ref{fig:xcube_lineon}).
This excitation can move only along the line of the link $e$ by acting with $B_{e'}$ on adjacent links $e'$ without creating additional excitations, so it is called a lineon.

The $\Z_N$ generalization of the X-cube model is defined by placing $N$-dimensional Hilbert space on each link of the cubic lattice \cite{Seiberg:2020cxy}.

\subsection{Fractonic BF Theories in 3+1 Dimensions}
\label{section:rev_fractonic_bf_3+1}

In this section, we review the fractonic $BF$ theory in 3+1 dimensions \cite{Slagle:2017wrc,Seiberg:2020cxy}, which is considered as the low-energy effective field theory of the $\Z_N$ X-cube model.
The fractonic $BF$ theory here is constructed by exotic tensor gauge fields in the representations of the spatial discrete rotational symmetry $S_4$, which is 90-degree rotational symmetry of the cube.
Therefore, we call the theory the exotic $BF$ theory.
The exotic tensor gauge fields are explained in detail in Chapter \ref{chapter:exotic_foliated_qft}.
This theory has $\Z_N \times \Z_N$ subsystem symmetries as the lattice X-cube model.

The Lagrangian of the exotic $BF$ theory in 3+1 dimensions is given by
\begin{align}
  \L = -\frac{iN}{2\pi} \left[ \frac{1}{3} \sum_{(i,j,k)} \hat{B}^{k(ij)}_0 (2 \partial_k A_{ij} - \partial_i A_{jk} - \partial_j A_{ki})  -  \sum_{(i,j)} \hat{B}^{ij} (\partial_0 A_{ij} - \partial_i \partial_j A_0 )  \right] \,, \label{rev:exotic_bf_lag}
\end{align}
where $N$ is a positive integer, and the summations are taken over
$(i,j,k) = (1,2,3),$ \\
\noindent  $(2,3,1),(3,1,2)$.
The fields $A_0$, $A_{12}$, $A_{23}$ and $A_{31}$ are the components of the $U(1)$ exotic tensor gauge field $\bm{A} = (A_0, A_{ij})$, and the fields $\hat{B}^{1(23)}_0$, $\hat{B}^{2(31)}_0$, $\hat{B}^{3(12)}_0$, $\hat{B}^{12}$, $\hat{B}^{23}$ and $\hat{B}^{31}$ are the components of another $U(1)$ exotic tensor gauge field $\hat{\bm{B}} = (\hat{B}^{k(ij)}_0, \hat{B}^{ij})$. They belong to the representations of spatial rotational symemtry $S_4$. The $U(1)$ exotic tensor gauge field $\bm{A} = (A_0, A_{ij})$ is in the representation $(\bm{1},\bm{3}')$ of $S_4$, where $A_{ij}$ is a symmetric tensor field with $i \neq j$.
The $U(1)$ exotic tensor gauge field $\hat{\bm{B}} = (\hat{B}^{k(ij)}_0, \hat{B}^{ij})$ is in the representation $(\bm{2},\bm{3}')$ of $S_4$, where $\hat{B}^{k(ij)}_0$ is a tensor field with $i \neq j, j \neq k, k \neq i$ that is symmetric for the indices $(i,j)$ and satisfies $\hat{B}^{i(jk)}_0 + \hat{B}^{j(ki)}_0 + \hat{B}^{k(ij)}_0 = 0$, and $\hat{B}^{ij}$ is a symmetric tensor field with $i \neq j$. These representations will be explained in Section \ref{section:exotic_qft_tensor_gauge_fields}. The gauge transformations of the exotic tensor gauge fields are
\begin{align}
  A_0 &\sim A_0 + \partial_0 \Lambda \,, \\
  A_{ij} &\sim A_{ij} + \partial_i \partial_j \Lambda  \,, \\
  \hat{B}^{k(ij)}_0 &\sim \hat{B}^{k(ij)}_0 +  \partial_0 \hat{\Lambda}^{k(ij)} \,, \\
  \hat{B}^{ij} &\sim \hat{B}^{ij} + \partial_k \hat{\Lambda}^{k(ij)} \,,
\end{align}
where $\Lambda$ is a $U(1)$ gauge parameter in $\bm{1}$, and $\hat{\Lambda}^{k(ij)}$ is a tensor gauge parameter in $\bm{2}$.
The equations of motion are
\begin{align}
  \frac{iN}{2\pi} (2 \partial_k A_{ij} - \partial_i A_{jk} - \partial_j A_{ki}) &= 0 \,, \\
  \frac{iN}{2\pi} (\partial_0 A_{ij} - \partial_i \partial_j A_0 ) &= 0 \,, \\
  \frac{iN}{2\pi} (\partial_0 \hat{B}^{ij} - \partial_k \hat{B}^{k(ij)}_0)  &= 0 \,, \\
  \frac{iN}{2\pi} \sum_{(i,j)} \partial_i \partial_j \hat{B}^{ij}  &= 0 \,. 
\end{align}
The left-hand sides of these equations are gauge-invariant form, which appear as $d_{\text{e}} \bm{A}$ and $d_{\text{e}} \hat{\bm{B}}$. From an analogy with ordinary $BF$ theory, the tensor gauge fields are considered to be restricted to $\Z_N$ gauge fields due to the gauge equivalence and the equations of motion.

Next, we discuss global symmetries.
The subsystem symmetry operators are partially topological gauge-invariant operators.
Since the fracton theory is not fully rotational invariant, the time and space directions must be treated in different manners even in Euclidean spacetime. This fact implies that we have two types of symmetries: space-like symmetry and time-like symmetry \cite{Gorantla:2022eem}. A space-like symmetry has a charged operator in space, and the corresponding symmetry operator acts on the charged operator. On the other hand, a time-like symmetry has a time-like charged defect and the corresponding symmetry operator in space can remotely detect the time-like defect.\footnote{In relativistic Euclidean QFT, which has the full rotational spacetime symmetry, time-like symmetry and space-like symmetry are not distinguished.}
The $\Z_N$ dipole symmetry is a space-like subsystem symmetry, whose symmetry operator is
\begin{align}
  W_m^k \left[ [x^k_1, x^k_2] \times \mathcal{C}^{ij} \right] &= \exp \left[ i m \int^{x^k_2}_{x^k_1} dx^k \oint_{C^{ij}} \left(dx^i  A_{ki} + dx^j A_{jk} \right) \right] \,, \label{rev:exotic_bf_dipole_sym_op}
\end{align}
where $m = 0,1,\ldots,N-1$, and $\mathcal{C}^{ij}$ is a closed one-dimensional loop in the $x^i x^j$-plane. This operator is defined on a strip $[x^k_1, x^k_2] \times \mathcal{C}^{ij}$ with a fixed width long the $x^k$ direction, which can be moved in the $x^0$ direction from the equations of motion, so this is a symmetry.
In addition, $\mathcal{C}^{ij}$ can be deformed in the $x^i x^j$-plane, so the operator is topological in the $x^i x^j$-plane, but cannot be deformed in the $x^k$ directions. Then, for separated intervals, the operators are independent and have different charges, which is the property of subsystem global symmetry.
The $\Z_N$ tensor symmetry is another space-like subsystem symmetry, whose symmetry operator is
\begin{align}
  V_m^k [ \mathcal{C}^k] = \exp \left[ i m \oint_{\mathcal{C}^k} dx^k \, \hat{B}^{ij}  \right] \,, \label{rev:exotic_bf_tensor_sym_op}
\end{align}
where $m = 0,1,\ldots,N-1$, and $\mathcal{C}^k$ is a closed one-dimensional line along the $x^k$ direction. This operator is defined on the line $\mathcal{C}^k$, which can be moved in the $x^0$ direction from the equations of motion, so this is also a symmetry.
In addition, $\mathcal{C}^k$ cannot be deformed in the space directions, so the operator is not topological in space.
The symmery operators satisfy the following commutation relation:
\begin{align}
  W_m^1 \left[ [x^1_1, x^1_2] \times \mathcal{C}^{23} \right] V_n^2 [ \mathcal{C}^2] = e^{ 2\pi i m n I([x^1_1, x^1_2] \times \mathcal{C}^{23},\mathcal{C}^2) /N } V_n^2 [ \mathcal{C}^2] W_m^1 \left[ [x^1_1, x^1_2] \times \mathcal{C}^{23} \right] \,, \\
  W_m^1 \left[ [x^1_1, x^1_2] \times \mathcal{C}^{23} \right] V_n^3 [ \mathcal{C}^3] = e^{ 2\pi i m n I([x^1_1, x^1_2] \times \mathcal{C}^{23},\mathcal{C}^3) /N } V_n^3 [ \mathcal{C}^3] W_m^1 \left[ [x^1_1, x^1_2] \times \mathcal{C}^{23} \right] \,,
\end{align}
similarly for other directions, where $I([x^k_1, x^k_2] \times \mathcal{C}^{ij},\mathcal{C}^l)$ is the intersection number of the strip $[x^k_1, x^k_2] \times \mathcal{C}^{ij}$ and the line $\mathcal{C}^l$.
Thus, the charged objects of the dipole symmetry are the line operators $V_m^k [ \mathcal{C}^k]$ and the charged objects of the tensor symmetry are the strip operators $W_m^k \left[ [x^k_1, x^k_2] \times \mathcal{C}^{ij} \right]$.

We can also consider the time-like subsystem symmetries.
The charged object of the $\Z_N$ electric symmetry is the defect
\begin{align}
  F_n \left[ \mathcal{C}^0 \right] &= \exp \left[ i n \oint_{\mathcal{C}^0}  dx^0 A_0  \right] \,, \label{rev:exotic_bf_dipole_charged_obj}
\end{align}
where $n = 0,1,\ldots,N-1$, and $\mathcal{C}^0$ is a closed one-dimensional loop in the time direction.
The $\mathcal{C}^0$ cannot be deformed in space directions, so this represents the fracton.
The time-like symmetry operator of the $\Z_N$ electric symmetry is the cage operator 
\begin{align}
\begin{split}
  T_m \left[ \mathcal{C}^{\text{cage}}  \right] &= \exp \left[ i m \int^{x^1_2}_{x^1_1} dx^1 \Delta_{23} \hat{B}^{23}(x^2_1,x^2_2,x^3_1,x^3_2) \right] \\
  & \times \exp \left[ i m \int^{x^2_2}_{x^2_1} dx^2 \Delta_{31} \hat{B}^{31}(x^3_1,x^3_2,x^1_1,x^1_2,) \right] \\
  &  \times \exp \left[ i m \int^{x^3_2}_{x^3_1} dx^3 \Delta_{12} \hat{B}^{12}(x^1_1,x^1_2,x^2_1,x^2_2) \right] \\
  &= \exp \left[ i m \int_{\mathcal{B}} d x^1 d x^2 d x^3  \left( \partial_1 \partial_2 \hat{B}^{12} + \partial_2 \partial_3 \hat{B}^{23} + \partial_3 \partial_1 \hat{B}^{31} \right)  \right]  \,,
\end{split} 
\end{align}
where $m = 0,1,\ldots,N-1$, and $\mathcal{C}^{\text{cage}}$ is a cage, which is composed of the twelve edges of a cuboid $\mathcal{B} = [x^1_1, x^1_2] \times [x^2_1, x^2_2] \times [x^3_1, x^3_2]$ in space.
The $\Delta_{ij}$ means the difference
\begin{align}
  \Delta_{ij} \hat{B}^{ij} (x^i_1, x^i_2, x^j_1, x^j_2) = \hat{B}^{ij}(x^i_2,x^j_2) -  \hat{B}^{ij}(x^i_2,x^j_1) -  \hat{B}^{ij}(x^i_1,x^j_2) + \hat{B}^{ij}(x^i_1,x^j_1) \,,
\end{align}
where we omit the other arguments $x^0$ and $x^k$.
This cage operator can detect the fracton defect:
\begin{align}
  T_m \left[ \mathcal{C}^{\text{cage}} \right] \cdot F_n \left[ \mathcal{C}^0 \right] = e^{ 2\pi i m n /N } F_n \left[ \mathcal{C}^0 \right]  \,,
\end{align}
where $\mathcal{C}^0$ is inside the cuboid $\mathcal{B}$.
The charged object of the $\Z_N$ magnetic symmetry is the defect
\begin{align}
  V_{n,\text{lin}}^k [ \mathcal{C}^{0k}] = \exp \left[ i n \oint_{\mathcal{C}^{0k}} \left(dx^0 \hat{B}^{k(ij)}_0 +  dx^k  \hat{B}^{ij}  \right)\right] \,, \label{rev:exotic_bf_tensor_charged_obj}
\end{align}
where $n = 0,1,\ldots,N-1$, and $\mathcal{C}^{0k}$ is a closed one-dimensional loop in the $x^0 x^k$-plane.
The $\mathcal{C}^{0k}$ cannot be deformed in other space directions, so this represents the lineon.
The time-like symmetry operator of the $\Z_N$ magnetic symmetry is the belt operator
\begin{align}
\begin{split}
  U_m^{[ij]k} \left[ [x^k_1, x^k_2] \times \mathcal{C}^{ij} \right] &= \exp \left[ i m \int^{x^j_2}_{x^j_1} dx^j \int^{x^k_2}_{x^k_1} dx^k \Delta_i A_{jk}(x^i_1,x^i_2)   \right] \\
  &\times \exp \left[ i m \int^{x^k_2}_{x^k_1} dx^k \int^{x^i_2}_{x^i_1} dx^i  \Delta_j A_{ki}(x^j_1,x^j_2) \right] \\ 
  &= \exp \left[ i m \int_{\mathcal{B}} dx^1 dx^2 dx^3 \left( \partial_i A_{jk} - \partial_j A_{ki} \right) \right] \,, \label{rev:exotic_bf_belt_op}
\end{split}
\end{align}
where $m = 0,1,\ldots,N-1$, and $\mathcal{C}^{ij,\text{rect}}$ is the edges of a rectangle $[x^i_1, x^i_2] \times [x^j_1, x^j_2]$ in the $x^i x^j$-plane, so $[x^k_1, x^k_2] \times \mathcal{C}^{ij,\text{rect}}$ is a belt, which is composed of the four faces of a cuboid $\mathcal{B} = [x^1_1, x^1_2] \times [x^2_1, x^2_2] \times [x^3_1, x^3_2]$ in space that are orthogonal to the $x^i$ and $x^j$ directions.
The $\Delta_{i}$ means the difference
\begin{align}
  \Delta_{i} A_{jk} (x^i_1, x^i_2) = A_{jk}(x^i_2) - A_{jk}(x^i_1)  \,,
\end{align}
where we omit the other arguments $x^0$, $x^j$ and $x^k$.
The belt operators satisfy
\begin{align}
  U_m^{[12]3} U_m^{[23]1} U_m^{[31]2} = 1 \,.
\end{align}
This belt operator can detect the lineon defect:
\begin{align}
  U_m^{[ij]k} \left[ [x^k_1, x^k_2] \times \mathcal{C}^{ij} \right] \cdot V_{n,\text{lin}}^i [ \mathcal{C}^{0i}] &= e^{ -2\pi i m n  /N } V_{n,\text{lin}}^i [ \mathcal{C}^{0i}]  \,, \\
  U_m^{[ij]k} \left[ [x^k_1, x^k_2] \times \mathcal{C}^{ij} \right] \cdot V_{n,\text{lin}}^j [ \mathcal{C}^{0j}] &= e^{ 2\pi i m n  /N } V_{n,\text{lin}}^j [ \mathcal{C}^{0j}]  \,, \\
  U_m^{[ij]k} \left[ [x^k_1, x^k_2] \times \mathcal{C}^{ij} \right] \cdot V_{n,\text{lin}}^k [ \mathcal{C}^{0k}] &=  V_{n,\text{lin}}^k [ \mathcal{C}^{0k}]  \,.
\end{align}

The exotic $BF$ theory is equivalent to the foliated $BF$ theory in 3+1 dimensions \cite{Slagle:2018swq,Slagle:2020ugk}, which is the first example of the foliated-exotic duality \cite{Ohmori:2022rzz}. 
In addition, the subsystem symmetries have the mixed 't Hooft anomaly, which is captured by the SSPT phase for $\Z_N \times \Z_N$ subsystem symmetries in $4+1$ dimensions \cite{Burnell:2021reh}, but we do not deal with them in this dissertation.
Instead, we will discuss the 2+1d case, which is simpler, of the 't Hooft anomaly of the fractonic exotic/foliated $BF$ theory in 2+1 dimensions and the corresponding SSPT phase in 3+1 dimensions in Chapter \ref{chapter:2+1d_bfanomaly_and_3+1d_sspt}.

\chapter{Exotic Tensor Gauge Fields and Foliated Gauge Fields}
\label{chapter:exotic_foliated_qft}

In this chapter, we review exotic tensor gauge fields and foliated gauge fields, which are used in fracton QFTs.
We provide a unified treatment of these fields, offering novel insights and a systematic description.
This chapter is mainly based on the appendix of the author's paper \cite{Ohmori:2025fuy}, but we have made revisions in several points.

\section{Exotic QFT and Tensor Gauge Fields}
\label{section:exotic_qft_tensor_gauge_fields}

\begin{figure}[H]
   \begin{center}
    \includegraphics[width=1\hsize]{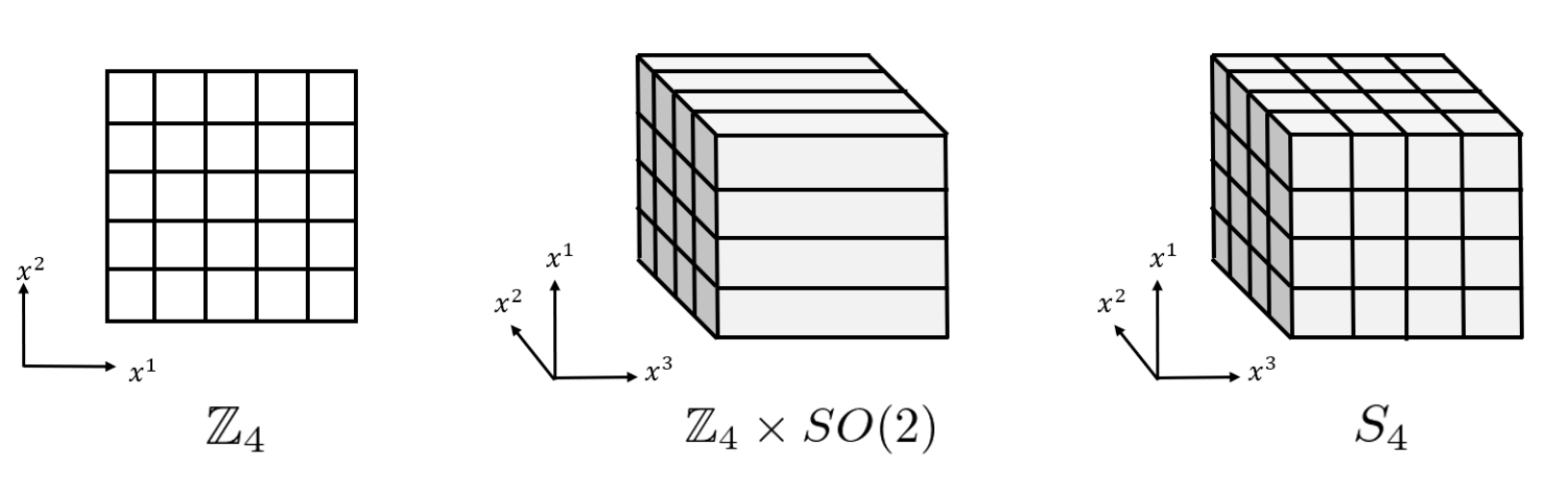} 
    \end{center}
    \vspace{-1cm}
      \caption{The discrete spatial rotational symmetry groups $Z_4$ (left), $\Z_4 \times SO(2)$ (middle) and $S_4$ (right) and figures invariant under the rotations.}
      \label{fig_rep}
\end{figure}

Fractonic theories do not have the full spatial rotational symmetry, but instead, have discrete spatial rotational symmetry in particular directions.
Some fractonic QFTs admit a description by exotic QFTs with exotic tensor gauge fields
\cite{Pretko:2018jbi,You:2019ciz,Slagle:2017wrc,Seiberg:2020bhn,Seiberg:2020wsg,Seiberg:2020cxy,Gorantla:2020xap,Gorantla:2021bda,Gorantla:2020jpy,Geng:2021cmq,Yamaguchi:2021qrx,Yamaguchi:2021xeq,Burnell:2021reh,Gorantla:2022eem,Gorantla:2022ssr,Honda:2022shd}.
The tensor gauge fields are the gauge fields in the representations of the discrete rotational symmetry.
In this dissertation, we mainly deal with the following theories:
\begin{itemize}
\item Theories in 2+1 dimensions with the 90-degree rotational symmetry for $(x^1,x^2)$ whose symmetry group is the cyclic group $\Z_4$
  \item Theories in 3+1 dimensions with the 90-degree rotational symmetry for $(x^1,x^2)$ and the continuous rotational symmetry for $(x^0,x^3)$ whose symmetry group is isomorphic to $\Z_4 \times SO(2)$
  \item Theories in 3+1 dimensions with the 90-degree rotational symmetry for $(x^1,x^2,x^3)$ whose symmetry group is the octahedral group, which is isomorphic to the symmetric group $S_4$
\end{itemize}
Therefore, we consider the representations of $\Z_4$ and $S_4$. See Figure \ref{fig_rep} for illustrations of these groups.

The 90-degree rotations in two-dimensional space generate the cyclic group $\Z_4$. The irreducible representations of $\Z_4$ are one-dimensional representations $\bm{1}_n \ (n = 0,\pm1,2)$.
The 90-degree rotation $e^{i\frac{\pi}{2}}$ acts on the representations in $\bm{1}_n$ as $e^{i\frac{n\pi}{2}}$.
We write the tensor using $SO(2)$ vector indices as $F$ for $\bm{1}_0$, $F_{12}=F_{21}$ for $\bm{1}_2$. Under the 90-degree rotation, the tensors transform as
\begin{align}
    F &\rightarrow F \,, \\
    F_{12} &\rightarrow - F_{12} \,.
\end{align}
These transformations can be understood %
as follows.
$F_{12}$ is the component of a hollow ($F_{ii}=0$) symmetric tensor $F_{12}\, dx^1 \odot dx^2$,\footnote{The symbol $\odot$ represents the symmetric tensor $dx^1 \odot dx^2 = \frac{1}{2} (dx^1 \otimes dx^2 + dx^2 \otimes dx^1)$, while the wedge product $\wedge$ means $dx^1 \wedge dx^2 = \frac{1}{2} (dx^1 \otimes dx^2 - dx^2 \otimes dx^1)$.}
so under the rotation $x^1 \rightarrow x'^1 = x^2$, $x^2 \rightarrow x'^2 = -x^1$, it transforms as
\begin{align}
\begin{split}
    F_{12} \, dx^1 \odot dx^2 &= F'_{12} \, dx'^1 \odot dx'^2 \\
    &= - F'_{12} \, dx^2 \odot dx^1 \,,
\end{split}
\end{align}
and thus, we have $F'_{12} = - F_{21} = -F_{12}$.

The 90-degree rotations in three-dimensional space generate the octahedral group that is isomorphic to the symmetric group $S_4$. The irreducible representations of $S_4$ are classified as the following tensors (see Appendix in \cite{Seiberg:2020wsg,Seiberg:2020cxy}):
\begin{itemize}
    \item $(F)$ in $\bm{1}$ 
    \item $(F_{ijk})$ in $\bm{1'}$, totally symmetric in the indices $i$, $j$, $k$
    \item $(F_{[ij]k}) = (F_{i(jk)})$ in $\bm{2}$, satisfying $F_{[ij]k} = -F_{[ji]k}$, $F_{[12]3} + F_{[23]1} + F_{[31]2} = 0$, $F_{i(jk)} = F_{i(kj)}$ and $F_{1(23)} + F_{2(31)} + F_{3(12)} = 0$ 
    \item $(F_i)$ in $\bm{3}$
    \item $(F_{ij})$ in $\bm{3'}$, satisfying $F_{ij} = F_{ji}$
\end{itemize}
The indices are the $SO(3)$ vector indices, where $i, j, k$  are $1$ or $2$ or $3$ and different from one another. The two bases of irreducible representation $\bm{2}$ are related as
\begin{align}
    F_{k(ij)} &= F_{[ki]j} - F_{[jk]i} \,,   \label{rep2rel1} \\
    F_{[ij]k} &= \frac{1}{3} ( F_{i(jk)} - F_{j(ki)}) \,. \label{rep2rel2}
\end{align}

We consider $U(1)$ exotic tensor gauge fields.
A feature of the tensor gauge fields is that they can have some delta function singularities and step function discontinuities, which is a consequence of the UV/IR mixing in fracton phases \cite{Seiberg:2020bhn,Seiberg:2020wsg,Seiberg:2020cxy}.\footnote{The allowed singularities and discontinuities are determined so that the action has up to a $O(a^{-1})$ dependence where $a$ is the UV cutoff length. }
As examples of the exotic tensor gauge fields, we describe the fields in (2+1)-dimensional $\Z_4$-invariant fractonic theories, (3+1)-dimensional $\Z_4 \times SO(2)$-invariant fractonic theories, and (3+1)-dimensional $S_4$-invariant fractonic theories.\footnote{We wanted to formulate the symmetric tensor form and the coupling product corresponding wedge product in differential forms, but we could not find a satisfactory way to do so. We leave it for future work.}

\subsection{Exotic Tensor Gauge Fields in \texorpdfstring{$\Z_4$}{Z4} representations in 2+1 Dimensions}
\label{subsection:exotic_tensor_gauge_fields_z4_2d}

First, we describe the exotic tensor gauge fields used in theories in 2+1 dimensions with the 90-degree rotational symmetry for $(x^1,x^2)$ whose symmetry group is $\Z_4$.

\subsubsection*{$U(1)$ tensor zero-form gauge fields}

We consider a compact scalar field $\phi$ in the representation $\bm{1}_0$ with the periodicity $\phi \sim \phi + 2\pi$.
Unlike ordinary compact scalar fields, the field $\phi$ has a gauge transformation
\begin{align}
    \phi &\sim \phi + 2\pi w^1 + 2\pi w^2 \,,
\end{align}
where $w^1 = w^1(x^1)$ and $w^2 = w^2(x^2)$ are integer-valued gauge parameters.
We let $w^k (x^k) \ (k=1,2)$ have step function discontinuities in the $x^k$ direction.
Under the 90-degree rotation, they transform as $w^1(x^1) \rightarrow w'^1(x'^1 = x^2) = w^2(x^2)$ and $w^2 (x^2) \rightarrow w'^2(x'^2 = -x^1) = w^1(x^1)$, and then, $w(x^1,x^2) = w^1(x^1) + w^2(x^2) = w'^1(x'^1) + w'^2(x'^2) = w'(x'^1,x'^2)$.\footnote{They also have the gauge transformations
$w^1(x^1) \sim w^1(x^1) + w_\text{const.}$ and
$w^2(x^2) \sim w^2(x^2) - w_\text{const.}$,
where $w_\text{const.}$ is a constant integer-valued gauge parameter.
}
This compact scalar $\phi$ is similar to the $U(1)$ zero-form gauge field, so we call $\phi$ a $U(1)$ tensor zero-form gauge field $\bm{\phi} = (\phi)$.
Another tensor zero-form gauge field is a compact scalar $\bm{\hat\phi} = (\hat{\phi}^{12})$ in the representation $\bm{1}_2$.
The field $\hat{\phi}^{12}$ has a gauge transformation
\begin{align}
    \hat{\phi}^{12} &\sim \hat{\phi}^{12} + 2\pi \hat{w}^{1} - 2\pi \hat{w}^2  \,,
\end{align}
where $\hat{w}^k = \hat{w}^k(x^k)\ (k=1,2)$ has the same properties as $w^k(x^k)$.\footnote{Note that $\hat{w}^k\ (k=1,2)$ have the different gauge transformations from $w^k\ (k=1,2)$: $\hat{w}^1(x^1) \sim \hat{w}^1(x^1) + \hat{w}_\text{const.}$ and $\hat{w}^2(x^2) \sim \hat{w}^2(x^2) + \hat{w}_\text{const.}$, where $\hat{w}_\text{const.}$ is a constant integer-valued gauge parameter.} Under the 90-degree rotation, $\hat{w}^{12}(x^1,x^2) = \hat{w}^{1}(x^1) - \hat{w}^{2}(x^2)$ transforms in $\bm{1}_2$ as $\hat{w}^{12}(x^1,x^2) \rightarrow \hat{w}'^{12}(x'^1,x'^2) = \hat{w}'^{1}(x'^1) - \hat{w}'^{2}(x'^2) = \hat{w}^{2}(x^2) - \hat{w}^{1}(x^1) = - \hat{w}^{12}(x^1,x^2)$.
We define the exterior derivatives of $\bm
{\phi} = (\phi)$ and $\bm{\hat\phi} = (\hat\phi^{12})$ as
\begin{align}
    d_\text{e} \bm{\phi} &= (\partial_0 \phi, \partial_1 \partial_2 \phi ) \,, \\
    d_\text{e} \bm{\hat\phi} &= (\partial_0 \hat\phi^{12}, \partial_1 \partial_2 \hat\phi^{12} ) 
\end{align}
in the representation $(\bm{1}_0,\bm{1}_2)$ and $(\bm{1}_2,\bm{1}_0)$, which are gauge invariant.

\subsubsection*{$U(1)$ tensor one-form gauge fields}

Next, we consider $U(1)$ tensor one-form gauge fields.
These fields are $\bm{A}=(A_0,A_{12})$ in $(\bm{1}_0,\bm{1}_2)$ and $\bm{\hat{A}}=(\hat{A}^{12}_0, \hat{A})$ in $(\bm{1}_2,\bm{1}_0)$.
Their gauge transformations are 
\begin{align}
    A_0 &\sim A_0 + \partial_0 \Lambda \,,  \\
    A_{12} &\sim A_{12} + \partial_1\partial_2 \Lambda \,. 
\end{align}
and
\begin{align}
    \hat{A}^{12}_0 &\sim \hat{A}^{12}_0 + \partial_0  \hat{\Lambda}^{12} \,,  \\
    \hat{A} &\sim \hat{A} + \partial_1 \partial_2  \hat{\Lambda}^{12} \,,
\end{align}
where $\bm{\Lambda}= (\Lambda)$ and $\bm{\hat\Lambda} = (\hat{\Lambda}^{12})$ are $U(1)$ tensor zero-form gauge parameters in the $\bm{1}_0$ and $\bm{1}_2$, respectively, and they have their own gauge transformations.
We can rewrite the gauge transformations as
\begin{align}
    \bm{A} &\sim \bm{A} + d_\text{e} \bm{\Lambda} \,, \\
    \bm{\hat{A}} &\sim \bm{\hat{A}} + d_\text{e} \bm{\hat{\Lambda}} \,.
\end{align}
We define the exterior derivatives of $\bm{A}$ and $\bm{\hat{A}}$ as
\begin{align}
    d_\text{e} \bm{A} &= (\partial_0 A_{12} - \partial_1 \partial_2 A_0) \,, \\
    d_\text{e} \bm{\hat{A}} &= (\partial_0 \hat A  - \partial_1 \partial_2 \hat A^{12}_0) \,,
\end{align}
in $\bm{1}_2$ and $\bm{1}_0$, which are gauge invariant.

\subsubsection*{$U(1)$ tensor two-form gauge fields}

We can also consider $U(1)$ tensor two-form gauge fields in 2+1 dimensions, although we do not deal with them in this dissertation.
These fields are $\bm{B} = (B_{012})$ in $\bm{1}_2$ and $\bm{\hat{B}} = (\hat{B}_0)$ in $\bm{1}_0$.
Their gauge transformations are
\begin{align}
    B_{012} &\sim B_{012} + \partial_0 \Gamma_{12} - \partial_1 \partial_2 \Gamma_0 \,, \\
    \hat{B}_0 &\sim \hat{B}_0 + \partial_0 \hat{\Gamma} - \partial_1 \partial_2 \hat{\Gamma}^{12}_0 \,,
\end{align}
where $\bm{\Gamma}=(\Gamma_0, \Gamma_{12})$ and $\bm{\hat{\Gamma}}= (\hat{\Gamma}^{12}_0,\hat{\Gamma})$ are $U(1)$ tensor one-form gauge parameters in the $(\bm{1}_0,\bm{1}_{2})$ and $(\bm{1}_{2},\bm{1}_0)$, respectively, and they have their own gauge transformations.
We can also rewrite the gauge transformations in terms of the exterior derivatives as
\begin{align}
    \bm{B} &\sim \bm{B} + d_\text{e} \bm{\Gamma} \,, \\
    \bm{\hat{B}} &\sim \bm{\hat{B}} + d_\text{e} \bm{\hat{\Gamma}} \,.
\end{align}
Exterior derivatives of $\bm{B}$ and $\bm{\hat{B}}$ are assumed to be zero: $d_\text{e} \bm{B} = 0$ and $d_\text{e} \bm{\hat{B}} = 0$.
Note that we can check that $(d_\text{e})^2$ always vanishes.

\subsection{Exotic Tensor Gauge Fields in \texorpdfstring{$\Z_4 \times SO(2)$}{Z4 × SO(2)} representations in 3+1 Dimensions}
\label{subsection:exotic_tensor_gauge_fields_z4so2_3d}

Next, we describe the exotic tensor gauge fields used in theories in 3+1 dimensions with the 90-degree rotational symmetry for $(x^1,x^2)$ and the continuous rotational symmetry for $(x^0,x^3)$ whose symmetry group is $\Z_4 \times SO(2)$. This case is a straightforward extension of the (2+1)-dimensional case.

\subsubsection*{$U(1)$ tensor zero-form gauge fields}

The $U(1)$ tensor zero-form gauge fields are compact scalars $\bm{\phi} = (\phi)$ and $\bm{\hat\phi} = (\hat\phi^{12})$. They are in the $\Z_4$ representations $\bm{1}_0$ and $\bm{1}_2$, respectively, and in the scalar in the $SO(2)$ representation. Their gauge transformations are the same as those in the (2+1)-dimensional case.
The exterior derivatives of $\bm{\phi} = (\phi)$ and $\bm{\hat\phi} = (\hat\phi^{12})$ are defined as
\begin{align}
    d_\text{e} \bm{\phi} &= (\partial_0 \phi, \partial_1 \partial_2 \phi, \partial_3 \phi) \,, \\
    d_\text{e} \bm{\hat\phi} &= (\partial_0 \hat\phi^{12}, \partial_1 \partial_2 \hat\phi^{12}, \partial_3 \hat\phi^{12}) 
\end{align}
in the representation $(\bm{1}_0,\bm{1}_2, \bm{1}_0)$ and $(\bm{1}_2,\bm{1}_0, \bm{1}_2)$ for the $\Z_4$ representation, and $(\partial_0 \phi, \partial_3 \phi)$ and $(\partial_0 \hat\phi^{12}, \partial_3 \hat\phi^{12})$ are vectors in the $SO(2)$ representation.

\subsubsection*{$U(1)$ tensor one-form gauge fields}

The $U(1)$ tensor one-form gauge fields are $\bm{A}=(A_0,A_{12},A_3)$ and $\bm{\hat{A}}=(\hat{A}^{12}_0, \hat{A},\hat{A}^{12}_3)$. They are in the same representations as $d_\text{e} \bm{\phi}$ and $d_\text{e} \bm{\hat\phi}$, respectively.
Their gauge transformations are 
\begin{align}
    A_0 &\sim A_0 + \partial_0 \Lambda \,,  \\
    A_{12} &\sim A_{12} + \partial_1\partial_2 \Lambda \,, \\
    A_3 &\sim A_3 + \partial_3 \Lambda \,,
\end{align}
and
\begin{align}
    \hat{A}^{12}_0 &\sim \hat{A}^{12}_0 + \partial_0  \hat{\Lambda}^{12} \,,  \\
    \hat{A} &\sim \hat{A} + \partial_1 \partial_2  \hat{\Lambda}^{12} \,, \\
    \hat{A}^{12}_3 &\sim \hat{A}^{12}_3 + \partial_3  \hat{\Lambda}^{12} \,,
\end{align}
where $\bm{\Lambda}= (\Lambda)$ and $\bm{\hat\Lambda} = (\hat{\Lambda}^{12})$ are $U(1)$ tensor zero-form gauge parameters and they have their own gauge transformations.
We can rewrite the gauge transformations as
\begin{align}
    \bm{A} &\sim \bm{A} + d_\text{e} \bm{\Lambda} \,, \\
    \bm{\hat{A}} &\sim \bm{\hat{A}} + d_\text{e} \bm{\hat{\Lambda}} \,.
\end{align}
We define the exterior derivatives of $\bm{A}$ and $\bm{\hat{A}}$ as
\begin{align}
    d_\text{e} \bm{A} &= (\partial_0 A_{12} - \partial_1 \partial_2 A_0,\partial_3 A_{12} - \partial_1 \partial_2 A_3, \partial_0 A_3 - \partial_3 A_0)  \,, \\
    d_\text{e} \bm{\hat{A}} &= (\partial_0 \hat{A}  - \partial_1 \partial_2 \hat{A}^{12}_0, \partial_3 \hat{A}  - \partial_1 \partial_2 \hat{A}^{12}_3, \partial_0 \hat{A}^{12}_3  - \partial_3 \hat{A}^{12}_0)  \,,
\end{align}
which are gauge invariant. $d_\text{e} \bm{A}$ and $d_\text{e} \bm{\hat{A}}$ are in the $\Z_4$ representations $(\bm{1}_2,\bm{1}_2,\bm{1}_0)$ and $(\bm{1}_0,\bm{1}_0,\bm{1}_2)$, respectively, and $\partial_0 A_3 - \partial_3 A_0$ and $\partial_0 \hat{A}^{12}_3  - \partial_3 \hat{A}^{12}_0$ are anti-symmetric tensors in the $SO(2)$ representation.

\subsubsection*{$U(1)$ tensor two-form gauge fields}

The $U(1)$ tensor two-form gauge fields are $\bm{B} = (B_{012}, B_{312}, B_{03})$ and $\bm{\hat{B}} = (\hat{B}_0, \hat{B}_3, \hat{B}_{03}^{12})$. They are in the same representations as $d_\text{e} \bm{A}$ and $d_\text{e} \bm{\hat{A}}$, respectively.
Their gauge transformations are
\begin{align}
    B_{012} &\sim B_{012} + \partial_0 \Gamma_{12} - \partial_1 \partial_2 \Gamma_0 \,, \\
    B_{312} &\sim B_{312} + \partial_3 \Gamma_{12} - \partial_1 \partial_2 \Gamma_3 \,, \\
    B_{03} &\sim B_{03} + \partial_0 \Gamma_{3} - \partial_3 \Gamma_0 \,,
\end{align}
and
\begin{align}
    \hat{B}_0 &\sim \hat{B}_0 + \partial_0 \hat{\Gamma} - \partial_1 \partial_2 \hat{\Gamma}^{12}_0 \,, \\
    \hat{B}_3 &\sim \hat{B}_3 + \partial_3 \hat{\Gamma} - \partial_1 \partial_2 \hat{\Gamma}^{12}_3 \,, \\
    \hat{B}^{12}_{03} &\sim \hat{B}^{12}_{03} + \partial_0 \hat{\Gamma}^{12}_3 - \partial_3  \hat{\Gamma}^{12}_0 \,, 
\end{align}
where $\bm{\Gamma}=(\Gamma_0, \Gamma_{12}, \Gamma_0)$ and $\bm{\hat{\Gamma}}= (\hat{\Gamma}^{12}_0, \hat{\Gamma}, \hat{\Gamma}^{12}_3)$ are $U(1)$ tensor one-form gauge parameters and they have their own gauge transformations.
We can also rewrite the gauge transformations in terms of the exterior derivatives as
\begin{align}
    \bm{B} &\sim \bm{B} + d_\text{e} \bm{\Gamma} \,, \\
    \bm{\hat{B}} &\sim \bm{\hat{B}} + d_\text{e} \bm{\hat{\Gamma}} \,.
\end{align}
We define the exterior derivatives of $\bm{B}$ and $\bm{\hat{B}}$ as
\begin{align}
    d_\text{e} \bm{B} &= (\partial_0 B_{312} - \partial_3 B_{012} + \partial_1 \partial_2 B_{03})  \,, \\
    d_\text{e} \bm{\hat{B}} &= (\partial_0 \hat{B}_3 - \partial_3 \hat{B}_0 + \partial_1 \partial_2 \hat{B}^{12}_{03})  \,,
\end{align}
which are gauge invariant. $d_\text{e} \bm{B}$ and $d_\text{e} \bm{\hat{B}}$ are in the $\Z_4$ representation $\bm{1}_0$, and they are scalars in the $SO(2)$ representation.

\subsubsection*{$U(1)$ tensor three-form gauge fields}

In 3+1 dimensions, we can also consider $U(1)$ tensor three-form gauge fields $\bm{C} = (C_{0123})$ and $\bm{\hat{C}} = (\hat{C}^{12}_{03}) $ in the same representations as $d_\text{e} \bm{B}$ and $d_\text{e} \bm{\hat{B}}$. Their gauge transformations are
\begin{align}
    C_{0123} &\sim C_{0123} + \partial_0 \Pi_{312} - \partial_3 \Pi_{012} + \partial_1 \partial_2 \Pi_{03} \,, \\
    \hat{C}^{12}_{03} &\sim \hat{C}^{12}_{03} + \partial_0 \hat{\Pi}_3 - \partial_3  \hat{\Pi}_0 + \partial_1 \partial_2 \hat{\Pi}^{12}_{03} \,,
\end{align}
where $\bm{\Pi} = (\Pi_{012}, \Pi_{312}, \Pi_{03})$ and $\bm{\hat{\Pi}} = (\hat{\Pi}_0, \hat{\Pi}_3, \hat{\Pi}^{12}_{03})$ are $U(1)$ tensor two-form gauge parameters and they have their own gauge transformations. We also rewrite the gauge transformations in terms of the exterior derivatives as
\begin{align}
    \bm{C} &\sim \bm{C} + d_\text{e} \bm{\Pi} \,, \\
    \bm{\hat{C}} &\sim \bm{\hat{C}} + d_\text{e} \bm{\hat{\Pi}} \,.
\end{align}
Exterior derivatives of $\bm{C}$ and $\bm{\hat{C}}$ are assumed to be zero: $d_\text{e} \bm{C} = 0$ and $d_\text{e} \bm{\hat{C}} = 0$. Then, we can check that $(d_\text{e})^2$ always vanishes.

\subsection{Exotic Tensor Gauge Fields in \texorpdfstring{$S_4$}{S4} representations in 3+1 Dimensions}
\label{subsection:exotic_tensor_gauge_fields_s4_3d}

We also describe the exotic tensor gauge fields used in theories in 3+1 dimensions with the 90-degree rotational symmetry for $(x^1,x^2,x^3)$ whose symmetry group is $S_4$.

The exotic tensor gauge fields with $S_4$ have several types, depending on which representations they are projected to. When the exterior derivative of $\bm{\phi} = (\phi)$  is $d_\text{e} \bm{\phi} = (\partial_0 \phi, \partial_1 \partial_2 \partial_3 \phi)$ in $(\bm{1}, \bm{1'})$, the exotic tensor gauge fields are similar to the (2+1)-dimensional case \cite{Gorantla:2020xap}, so we do not explain them here.
We describe the case where the exterior derivative of $\bm{\phi} = (\phi)$ in $\bm{1}$ is $d_\text{e} \bm{\phi} = (\partial_0 \phi, \partial_i \partial_j \phi)$ in $(\bm{1}, \bm{3'})$ and the exterior derivative of $\bm{\hat\phi} = (\hat\phi^{k(ij)})$ in $\bm{2}$ is $d_\text{e} \bm{\hat\phi} = (\partial_0 \hat\phi^{k(ij)}, \partial_k \hat\phi^{k(ij)})$ in $(\bm{2}, \bm{3'})$ \cite{Seiberg:2020wsg,Seiberg:2020cxy}.

\subsubsection*{$U(1)$ tensor zero-form gauge fields}

The $U(1)$ tensor zero-form gauge fields are compact scalars $\bm{\phi} = (\phi)$ in the representation $\bm{1}$ and $\bm{\hat\phi} = (\hat\phi^{k(ij)}) = (\hat\phi^{[ij]k})$ in the representation $\bm{2}$. From the relations \eqref{rep2rel1} and \eqref{rep2rel2}, we express $\bm{\hat\phi}$ as either $\bm{\hat\phi}^{i(jk)}$ or $\bm{\hat\phi}^{[ij]k}$:
\begin{align}
    \hat{\phi}^{k(ij)} &= \hat{\phi}^{[ki]j} - \hat{\phi}^{[jk]i} \,,  \\
    \hat{\phi}^{[ij]k} &= \frac{1}{3} ( \hat{\phi}^{i(jk)} - \hat{\phi}^{j(ki)}) \,.
\end{align}
Their gauge transformations are
\begin{align}
    \phi &\sim \phi + 2\pi w^1 + 2\pi w^2 + 2\pi w^3 \,, \\
    \hat{\phi}^{k(ij)} &\sim \hat{\phi}^{k(ij)} + 2\pi \hat{w}^i - 2\pi \hat{w}^j \,, \\
    \hat{\phi}^{[ij]k} &\sim \hat{\phi}^{[ij]k} + \frac{2\pi}{3} (\hat{w}^1 + \hat{w}^2 + \hat{w}^3) - 2\pi \hat{w}^k \,,
\end{align}
where $w^k = w^k(x^k)\ (k=1,2,3)$ and $\hat{w}^k = \hat{w}^k(x^k)\ (k=1,2,3)$ has the same properties as the (2+1)-dimensional case. The exterior derivatives of $\bm
{\phi}$ and $\bm{\hat\phi}$ are defined as
\begin{align}
    d_\text{e} \bm{\phi} &= (\partial_0 \phi, \partial_i \partial_j \phi) \,, \\
    d_\text{e} \bm{\hat\phi} &= (\partial_0 \hat\phi^{k(ij)}, \partial_k \hat\phi^{k(ij)}) \\
    &= (\partial_0 \hat\phi^{[ij]k}, \partial_k \hat\phi^{[ki]j} -  \partial_k\hat\phi^{[jk]i}) \,,
\end{align}
in the representation $(\bm{1}, \bm{3'})$ and $(\bm{2}, \bm{3'})$, respectively, which are gauge invariant.

\subsubsection*{$U(1)$ tensor one-form gauge fields}

The $U(1)$ tensor one-form gauge fields are $\bm{A}=(A_0,A_{ij})$ in $(\bm{1}, \bm{3'})$ and $\bm{\hat{A}}=(\hat{A}^{k(ij)}_0, \hat{A}^{ij}) = (\hat{A}^{[ij]k}_0, \hat{A}^{ij})$ in $(\bm{2}, \bm{3'})$. Their gauge transformations are
\begin{align}
    A_0 &\sim A_0 + \partial_0 \Lambda \,,  \\
    A_{ij} &\sim A_{ij} + \partial_i \partial_j \Lambda \,,
\end{align}
and
\begin{align}
    \hat{A}^{k(ij)}_0 &\sim \hat{A}^{k(ij)}_0 + \partial_0  \hat{\Lambda}^{k(ij)} \,,  \\
    \hat{A}^{ij} &\sim \hat{A}^{ij} + \partial_k \hat{\Lambda}^{k(ij)} \,,
\end{align}
or equivalently,
\begin{align}
    \hat{A}^{[ij]k}_0 &\sim \hat{A}^{[ij]k}_0 + \partial_0  \hat{\Lambda}^{[ij]k} \,,  \\
    \hat{A}^{ij} &\sim \hat{A}^{ij} + \partial_k \hat{\Lambda}^{[ki]j} - \partial_k  \hat{\Lambda}^{[jk]i}\,,
\end{align}
where $\bm{\Lambda}= (\Lambda)$ in $\bm{1}$ and $\bm{\hat\Lambda} = (\hat{\Lambda}^{k(ij)}) = (\hat{\Lambda}^{[ij]k})$ in $\bm{2}$ are $U(1)$ tensor zero-form gauge parameters and they have their own gauge transformations. We can rewrite the gauge transformations as
\begin{align}
    \bm{A} &\sim \bm{A} + d_\text{e} \bm{\Lambda} \,, \\
    \bm{\hat{A}} &\sim \bm{\hat{A}} + d_\text{e} \bm{\hat{\Lambda}} \,.
\end{align}
We define the exterior derivatives of $\bm{A}$ and $\bm{\hat{A}}$ as
\begin{align}
    d_\text{e} \bm{A} &= (\partial_0 A_{ij} - \partial_i \partial_j A_0, 2 \partial_k A_{ij} - \partial_i A_{jk} - \partial_j A_{ki} ) \\
    &= (\partial_0 A_{ij} - \partial_i \partial_j A_0, \partial_i A_{jk} - \partial_j A_{ki})  \,, \\
    d_\text{e} \bm{\hat{A}} &= (\partial_0   \hat A^{ij} - \partial_k \hat{A}^{k(ij)}_0, \partial_1 \partial_2 \hat{A}^{12} + \partial_2 \partial_3 \hat{A}^{23} + \partial_3 \partial_1 \hat{A}^{31}  )\\
    &= (\partial_0   \hat A^{ij} - \partial_k \hat{A}^{[ki]j}_0 + \partial_k \hat{A}^{[jk]i}_0 , \partial_1 \partial_2 \hat{A}^{12} + \partial_2 \partial_3 \hat{A}^{23} + \partial_3 \partial_1 \hat{A}^{31}  )  \,,
\end{align}
in the representation $(\bm{3'}, \bm{2})$ and $(\bm{3'}, \bm{1})$, respectively, which are gauge invariant. Note that the $\bm{2}$ component of $d_\text{e} \bm{A}$ are
\begin{align}
    (d_\text{e} \bm{A})_{k(ij)} &=  2 \partial_k A_{ij} - \partial_i A_{jk} - \partial_j A_{ki}  \,, \\(d_\text{e} \bm{A})_{[ij]k} &=  \partial_i A_{jk} - \partial_j A_{ki} \,.
\end{align}

\subsubsection*{$U(1)$ tensor two-form gauge fields}

The $U(1)$ tensor two-form gauge fields are $\bm{B} = (B_{0ij}, B_{k(ij)}) = (B_{0ij}, B_{[ij]k})$ in $(\bm{3'}, \bm{2})$ and $\bm{\hat{B}} = (\hat{B}^{ij}_0, \hat{B})$ in $(\bm{3'}, \bm{1})$. Their gauge transformations are
\begin{align}
    B_{0ij} &\sim B_{0ij} + \partial_0 \Gamma_{ij} - \partial_i \partial_j \Gamma_0 \,, \\
    B_{k(ij)} &\sim B_{k(ij)} + 2 \partial_k \Gamma_{ij} - \partial_i \Gamma_{jk} - \partial_j \Gamma_{ki} \,,
\end{align}
or equivalently,
\begin{align}
    B_{0ij} &\sim B_{0ij} + \partial_0 \Gamma_{ij} - \partial_i \partial_j \Gamma_0 \,, \\
    B_{[ij]k} &\sim B_{[ij]k} + \partial_i \Gamma_{jk} - \partial_j \Gamma_{ki} \,,
\end{align}
and
\begin{align}
    \hat{B}^{ij}_0 &\sim \hat{B}^{ij}_0 + \partial_0 \hat{\Gamma}^{ij} - \partial_k \hat{\Gamma}^{k(ij)}_0 \,, \\
    &\sim \hat{B}^{ij}_0 + \partial_0 \hat{\Gamma}^{ij} - \partial_k \hat{\Gamma}^{[ki]j}_0 + \partial_k \hat{\Gamma}^{[jk]i}_0 \,, \\
    \hat{B} &\sim \hat{B} + \partial_1 \partial_2 \hat{\Gamma}^{12} + \partial_2 \partial_3 \hat{\Gamma}^{23} + \partial_3 \partial_1 \hat{\Gamma}^{31}  \,,
\end{align}
where $\bm{\Gamma}=(\Gamma_0, \Gamma_{ij})$ in $(\bm{1},\bm{3'})$ and $\bm{\hat{\Gamma}}= (\hat{\Gamma}^{k(ij)}_0, \hat{\Gamma}) = (\hat{\Gamma}^{[ij]k}_0, \hat{\Gamma})$ in $(\bm{2},\bm{3'})$ are $U(1)$ tensor one-form gauge parameters and they have their own gauge transformations. We can also rewrite the gauge transformations in terms of the exterior derivatives as
\begin{align}
    \bm{B} &\sim \bm{B} + d_\text{e} \bm{\Gamma} \,, \\
    \bm{\hat{B}} &\sim \bm{\hat{B}} + d_\text{e} \bm{\hat{\Gamma}} \,.
\end{align}
We define the exterior derivatives of $\bm{B}$ and $\bm{\hat{B}}$ as
\begin{align}
    d_\text{e} \bm{B} &= (\partial_0 B_{k(ij)} - 2 \partial_k B_{0ij} + \partial_i B_{0jk} + \partial_j B_{0ki}) \\
    &= (\partial_0 B_{[ij]k} - \partial_i B_{0jk} + \partial_j B_{0ki})  \,, \\
    d_\text{e} \bm{\hat{B}} &= (\partial_0 \hat{B} - \partial_1 \partial_2 \hat{B}^{12}_0 - \partial_2 \partial_3 \hat{B}^{23}_0 - \partial_3 \partial_1 \hat{B}^{31}_0 ) \,,
\end{align}
in the representation $\bm{2}$ and $\bm{1}$, respectively, which are gauge invariant. Note that the $\bm{2}$ component of $d_\text{e} \bm{B}$ are
\begin{align}
    (d_\text{e} \bm{B})_{k(ij)} &= \partial_0 B_{k(ij)} - 2 \partial_k B_{0ij} + \partial_i B_{0jk} + \partial_j B_{0ki} \,, \\
    (d_\text{e} \bm{B})_{[ij]k} &= \partial_0 B_{[ij]k} - \partial_i B_{0jk} + \partial_j B_{0ki} \,.
\end{align}

\subsubsection*{$U(1)$ tensor three-form gauge fields}

The $U(1)$ tensor three-form gauge fields are $\bm{C} = (C_{0k(ij)}) = (C_{0[ij]k})$ in $\bm{2}$ and $\bm{\hat{C}} = (\hat{C})$ in $\bm{1}$. Their gauge transformations are
\begin{align}
    C_{0k(ij)} &\sim C_{0k(ij)} + \partial_0 \Pi_{k(ij)} - 2 \partial_k \Pi_{0ij} + \partial_i \Pi_{0jk} + \partial_j \Pi_{0ki} \,, 
\end{align}
or equivalently,
\begin{align}
    C_{0[ij]k} &\sim C_{0[ij]k} + \partial_0 \Pi_{[ij]k} - \partial_i \Pi_{0jk} + \partial_j \Pi_{0ki} \,,
\end{align}
and
\begin{align}
    \hat{C} &\sim \hat{C} + \partial_0 \hat{\Pi} - \partial_1 \partial_2 \hat{\Pi}^{12}_0 - \partial_2 \partial_3 \hat{\Pi}^{23}_0 - \partial_3 \partial_1 \hat{\Pi}^{31}_0 \,,
\end{align}
where $\bm{\Pi} = (\Pi_{0ij}, \Pi_{k(ij)}) = (\Pi_{0ij}, \Pi_{[ij]k})$ in $(\bm{3'}, \bm{2})$ and $\bm{\hat{\Pi}} = (\hat{\Pi}_0^{ij},\hat{\Pi})$ in $(\bm{3'},\bm{1})$ are $U(1)$ tensor two-form gauge parameters and they have their own gauge transformations. We also rewrite the gauge transformations in terms of the exterior derivatives as
\begin{align}
    \bm{C} &\sim \bm{C} + d_\text{e} \bm{\Pi} \,, \\
    \bm{\hat{C}} &\sim \bm{\hat{C}} + d_\text{e} \bm{\hat{\Pi}} \,.
\end{align}
Exterior derivatives of $\bm{C}$ and $\bm{\hat{C}}$ are assumed to be zero: $d_\text{e} \bm{C} = 0$ and $d_\text{e} \bm{\hat{C}} = 0$. Then, we can check that $(d_\text{e})^2$ always vanishes.




\section{Foliated QFT and Foliated Gauge Fields}
\label{section:foliated_qft_foliated_gauge_fields}

Although this section is based on \cite{Slagle:2020ugk,Hsin:2021mjn, Ohmori:2022rzz, Shimamura:2024kwf}, we present a reorganized formulation that provides a new perspective.

\begin{figure}[H]
   \begin{center}
    \includegraphics[width=1\hsize]{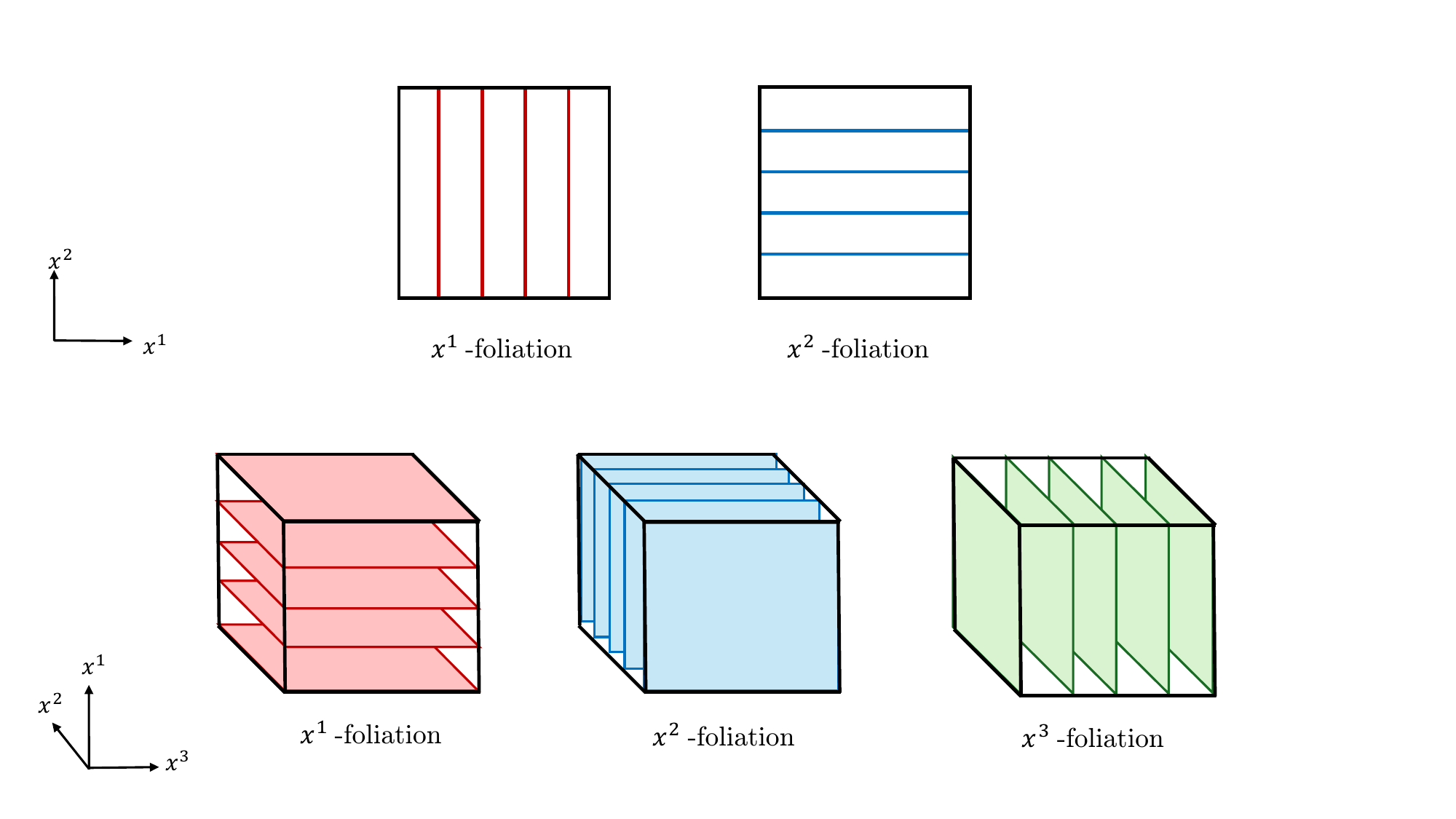} 
    \end{center}
    \vspace{-1cm}
      \caption{The flat foliation structures in 2+1 and 3+1 dimensions. The figure shows only a finite number of leaves, but in reality each foliation has infinitely many leaves.}
      \label{fig_foli}
\end{figure}

We can construct fractonic QFTs by using the foliation structure.
A foliation is a decomposition of a manifold into a stack of an infinite number of submanifolds, each of which we call a \textit{leaf}.
We consider codimension-one foliations specified by the one-form foliation field $e$.
The foliation field $e$ is nonzero everywhere and must satisfy $e \wedge de = 0$.
Given a foliation field $e$, we can define the foliation structure as the set of the codimension-one submanifolds orthogonal to $e$.
The foliation field $e$ has a gauge transformation $e \rightarrow f e$, where $f$ is a zero-form parameter.
QFTs coupled to the foliation structures on the spacetime are called foliated QFTs.
A foliation is called flat when $de = 0$.
In this dissertation, we focus on the particular flat foliation fields that are aligned with the coordinate axes and fix the gauge by setting $e^k = d x^k\ (k =1,2,...)$, where $x^k$ denotes a spatial coordinate.
In 2+1 dimensions, we consider the two simultaneous foliations $e^1 = dx^1$ and $e^2 = dx^2$ where the leaves are the lines orthogonal to the $x^1$ direction and the lines orthogonal to the $x^2$ direction.
In 3+1 dimensions, we consider the two simultaneous foliations $e^1 = dx^1$ and $e^2 = dx^2$ and the three simultaneous foliations $e^1 = dx^1$, $e^2 = dx^2$ and $e^3 = dx^3$ where the leaves are the planes orthogonal to the $x^k\ (k=1,2,3)$ directions (Fig. \ref{fig_foli}).

Next, we introduce $U(1)$ foliated gauge fields indexed by $k$.
They can be considered as $U(1)$ gauge fields on the leaves orthogonal to the $x^k$ direction.
We also introduce $U(1)$ gauge fields that mediate foliated gauge fields, which are called \textit{bulk} gauge fields.\footnote{The word “bulk” in the bulk gauge field has nothing to do with the word “bulk” in the bulk SSPT phase.}
We also refer to a pair of a foliated gauge field and a bulk gauge field as a foliated gauge field as well.
As in the case of the exotic tensor gauge fields, the foliated gauge fields can have delta function singularities and step function discontinuities.
In the following, we consider a foliated QFT in 2+1 dimensions with two flat foliations $e^1 = dx^1$ and $e^2 = dx^2$. The (3+1)-dimensional case is similar to the (2+1)-dimensional case.
We have two types of foliated gauge fields: type-$A$ foliated gauge fields and type-$B$ foliated gauge fields.
Type-$A$ foliated gauge fields are ($n$+1)-form gauge fields $\tilde{A}^k$ satisfying $\widetilde{A}^k \wedge e^k = 0$, and then we 
write $\widetilde{A}^k = A^k \wedge e^k$ locally, where $A^k$ is some $n$-form gauge field.
In addition to the gauge transformations $A^k$ inherits from $\widetilde{A}^k$, it has an additional gauge transformation $A^k \sim A^k + \sigma^k\wedge e^k$, with which we set the component $A^k_{i_1,\cdots,i_n}$ to zero when $e^k = dx^k$ and one of $i_p$ is $k$.

\subsection{Foliated Gauge fields in 2+1 Dimensions}
\label{subsection:foliated_gauge_fields}

\subsubsection*{$U(1)$ type-$A$ foliated (0+1)-form gauge field}

We consider a one-form gauge field $\Phi^k e^k\ (k =1,2)$ whose gauge transformation is
\begin{align}
    \Phi^k e^k \sim \Phi^k e^k + 2\pi d W^k \,, 
\end{align}
where $W^k$ is an integer-valued gauge parameter and it can have step function discontinuities in the $x^k$ direction, so $dW^k = \partial_k W^k e^k$.
We call this field a $U(1)$ type-$A$ foliated (0+1)-form gauge field.
In addition, we consider zero-form gauge field $\Phi$ whose gauge transformation is
\begin{align}
    \Phi \sim \Phi  + 2\pi W^1 + 2\pi W^2 \,. 
\end{align}
We call this field a $U(1)$ type-$A$ bulk zero-form gauge field.
Then, the set of fields $\bm{\Phi} = (\Phi^k e^k , \Phi)$ is also called $U(1)$ type-$A$ foliated (0+1)-form gauge field.\footnote{In two foliations case in 2+1 dimensions, the set of fields $\bm{\Phi} = (\Phi^k e^k , \Phi)$ means $\bm{\Phi} = (\Phi^1 e^1, \Phi^2 e^2 , \Phi)$.}
We define the exterior derivatives of $\bm{\Phi}$ as
\begin{align}
    d_\text{f} \bm{\Phi} = \left(d\Phi^k \wedge e^k , d \Phi - \sum^{2}_{k=1} \Phi^k e^k \right) \,, 
\end{align}
which is gauge invariant.\footnote{
Mathematically, the resulting complex is the mapping cone construction known in the context of homological algebra. About the mapping cone, see a standard textbook, e.g.\ \cite{Weibel_1994}.
}

\subsubsection*{$U(1)$ type-$A$ foliated (1+1)-form gauge field}

Similarly, we can consider $U(1)$ type-$A$ foliated (1+1)-form gauge field. The foliated fields are the two-form gauge fields $A^k\wedge e^k\  (k =1,2)$ with gauge transformations
\begin{align}
    A^k\wedge e^k \sim A^k \wedge e^k + d\lambda^k \wedge e^k\,, 
\end{align}
where $\lambda^k\, e^k$ is a $U(1)$ type-$A$ foliated (0+1)-form gauge parameter and it has its own gauge transformation. 
The bulk field is the one-form gauge field $a$ with gauge transformations
\begin{align}
    a \sim a + d\lambda - \sum^{2}_{k=1} \lambda^k\, dx^k \,,
\end{align}
where $\lambda$ is a $U(1)$ type-$A$ bulk zero-form gauge parameter and has its own gauge transformation.  
We define the exterior derivatives of $\bm{A} = (A^k \wedge e^k, a)$ as
\begin{align}
    d_\text{f} \bm{A} = \left(d A^k \wedge e^k, da + \sum^2_{k=1} A^k \wedge e^k \right) \,,
\end{align}
which is gauge invariant. The gauge transformation of $\bm{A}$ can be written as
\begin{align}
    \bm{A} \sim \bm{A} + d_\text{f} \bm{\lambda} \,,
\end{align}
where $\bm{\lambda} = (\lambda^k e^k , \lambda)$. Therefore, we can derive $(d_\text{f})^2 = 0$.

\subsubsection*{$U(1)$ type-$A$ foliated (2+1)-form gauge field}

We can also consider $U(1)$  type-$A$ foliated (2+1)-form gauge field $\bm{B}= ( B^k \wedge e^k, b)$. $B^k \wedge e^k \ (k=1,2)$ is the (2+1)-form foliated gauge fields and $b$ is the type-$A$ bulk two-form gauge field. Their gauge transformations are
\begin{align}
    B^k \wedge e^k &\sim B^k \wedge e^k + d\gamma^k \wedge e^k \,, \\
    b &\sim b + d\gamma + \sum^{2}_{k=1} \gamma^k \wedge e^k  \,,
\end{align}
or
\begin{align}
    \bm{B} \sim \bm{B} + d_\text{f} \bm{\gamma} \,,
\end{align}
where $\bm{\gamma} = (\gamma^k \wedge e^k, \gamma)$ is a $U(1)$ type-$A$ foliated (1+1)-form gauge parameter. We can define the exterior derivative of $\bm{B}$ as
\begin{align}
    d_\text{f} \bm{B} = \left(0, db - \sum^{2}_{k=1} B^k \wedge e^k \right) \,.
\end{align}
We set the exterior derivative of any three-form gauge fields to zero in 2+1 dimensions, and then $(d_\text{f})^2 = 0$ always holds.\footnote{We can also consider a type-$A$ foliated (3+1)-form gauge field in 2+1 dimensions as $\bm{C} = (0,c)$. In the 3+1 dimensions, we can consider a type-$A$ foliated (3+1)-form gauge field $\bm{C} = (C^k \wedge e^k,c)$ and the similar argument holds. Generally, we can consider  type-$A$ foliated $(n+1)$-form gauge fields to the top form in any dimensions.}

\subsubsection*{$U(1)$ type-$B$ foliated zero-form gauge field}

The other type of foliated gauge field is the type-$B$ foliated gauge field. We consider zero-form gauge field $\hat\Phi^k \ (k = 1,2)$ whose gauge transformation is
\begin{align}
    \hat\Phi^k \sim \hat\Phi^k + 2\pi \hat{W}^k + \hat{\xi} \,,
\end{align}
where $\hat{W}^k$ is an integer-valued gauge parameter and it can have step function discontinuities in the $x^k$ direction, and $\hat{\xi}$ is a zero-form gauge parameter. 
The bulk gauge field is a one-form gauge field $\hat\Phi$ whose gauge transformation
\begin{align}
    \hat\Phi \sim \hat\Phi + d\hat\xi \,.
\end{align}
In addition, we must consider the auxiliary field $\alpha^k e^k \ (k =1,2)$, which is a $U(1)$ type-$A$ foliated (0+1)-form gauge field whose gauge transformation is
\begin{align}
    \alpha^k e^k \sim \alpha^k e^k + 2\pi d\hat{W}^k \,.
\end{align}
Then, the $U(1)$ type-$B$ foliated zero-form gauge field is the set of fields $\bm{\hat{\Phi}} = (\hat\Phi^k, \hat\Phi, \alpha^k e^k)$, and we define the exterior derivative as
\begin{align}
    d_\text{f}\bm{\hat{\Phi}} = (d\hat\Phi^k -\hat\Phi-\alpha^k e^k , d\hat\Phi, d\alpha^k \wedge e^k) \,,
\end{align}
which is gauge invariant. We always couple $d\hat\Phi^k -\hat\Phi-\alpha^k e^k$ to type-$A$ foliated gauge field with $e^k$, so we can ignore the auxiliary fields $\alpha^k e^k$.

\subsubsection*{$U(1)$ type-$B$ foliated one-form gauge field}

Next, we consider the $U(1)$ type-$B$ foliated one-form gauge field. The foliated field is a one-form gauge field $\hat{A}^k\ (k=1,2)$, the bulk field is a two-form gauge field $\hat{a}$, and the auxiliary field $\beta^k\wedge e^k$ is a type-$A$ foliated (1+1)-form gauge field. The gauge transformations of $\bm{\hat{A}} = (\hat{A}^k, \hat{a}, \beta^k\wedge e^k)$ are
\begin{align}
    \hat{A}^k &\sim \hat{A}^k + d \hat\lambda^k - \hat{\lambda} - \zeta^k e^k \,, \\
    \hat{a} &\sim \hat{a} + d\hat\lambda \,, \\
    \beta^k\wedge e^k &\sim \beta^k\wedge e^k + d\zeta^k \wedge e^k \,,
\end{align}
or
\begin{align}
    \bm{\hat{A}} \sim \bm{\hat{A}} + d_\text{f} \bm{\hat\lambda} \,, 
\end{align}
where $\bm{\hat\lambda}= (\hat\lambda^k, \hat\lambda, \zeta^k e^k)$ is a $U(1)$ type-$B$ foliated zero-form gauge parameter.
We define the exterior derivative of $\bm{\hat{A}}$ as
\begin{align}
    d_\text{f} \bm{\hat{A}} = (d\hat{A}^k + \hat{a} + \beta^k \wedge e^k, d\hat{a}, d\beta^k \wedge e^k) \,,
\end{align}
which is gauge invariant, and we also have $(d_\text{f})^2 = 0$. As in the case of type-$B$ foliated zero-form gauge fields, we can ignore the auxiliary fields $\beta^k\wedge e^k$ and the gauge parameter $\zeta^k e^k$. Furthermore, due to the gauge  parameter $\zeta^k e^k$, we can set $\hat{A}^k_k$ to zero. Thus, we have only to consider $\hat{A}^1_0, \hat{A}^1_2, \hat{A}^2_0,$ and $\hat{A}^2_1$ as the components of $\hat{A}^k$.

\subsubsection*{$U(1)$ type-$B$ foliated two-form gauge field}

We can also consider the $U(1)$ type-$B$ foliated two-form gauge field in the same way. The foliated field is a two-form gauge field $\hat{B}^k\ (k=1,2)$, the bulk field is a three-form gauge field $\hat{b}$, and the auxiliary field $\rho^k\wedge e^k$ is type-$A$ foliated (2+1)-form gauge field. The gauge transformations of $\bm{\hat{B}} = (\hat{B}^k, \hat{b},\rho^k\wedge e^k)$ are
\begin{align}
    \hat{B}^k &\sim \hat{B}^k + d \hat\gamma^k + \hat{\gamma} + \sigma^k \wedge e^k \,, \\
    \hat{b} &\sim \hat{b} + d\hat\gamma \,, \\
    \rho^k\wedge e^k &\sim \rho^k\wedge e^k + d\sigma^k \wedge e^k \,,
\end{align}
or
\begin{align}
    \bm{\hat{B}} \sim \bm{\hat{B}} + d_\text{f} \bm{\hat\gamma} \,, 
\end{align}
where $\bm{\hat\gamma}= (\hat\gamma^k, \hat\gamma, \sigma^k \wedge e^k)$ is a $U(1)$ type-$B$ foliated one-form gauge parameter. 
We define the exterior derivative of $\bm{\hat{B}}$ as
\begin{align}
    d_\text{f} \bm{\hat{B}} = (d\hat{B}^k - \hat{b} - \rho^k \wedge e^k, 0 ,0) \,,
\end{align}
which is gauge invariant, and we also have $(d_\text{f})^2 = 0$. We can also ignore the auxiliary field $\rho^k\wedge e^k$ and the gauge parameter $\sigma^k \wedge e^k$. Furthermore, due to the gauge  parameter $\sigma^k \wedge e^k$, we can set $\hat{B}^k_{ki}$ to zero. Thus, we have only to consider $\hat{B}^1_{02}$ and $\hat{B}^2_{01}$ as the components of $\hat{B}^k$.\footnote{We can also consider a type-$B$ foliated three-form gauge field in 2+1 dimensions as $\bm{\hat{C}} = (\hat{C}^k,0,0)$. In the 3+1 dimensions, we can consider a type-$B$ foliated three-form gauge field $\bm{\hat{C}} = (\hat{C}^k,\hat{c},\tau^k \wedge e^k)$ and the similar argument holds. Generally, we can consider type-$B$ foliated $n$-form gauge fields to the top form in any dimensions.}

\subsubsection*{\texorpdfstring{$\Z_4$}{Z4} rotation of foliated gauge fields}
\label{appendix23}

Here we collect how the foliated gauge fields transform under the 90-degree rotation $\Z_4$. We consider the rotation $x'^1 = x^2$, $x'^2 = -x^1$. From the transformations of $e^k = dx^k$, we determine the transformation rules for the type-$A$ foliated gauge field as
\begin{align}
    \Phi^1 &\rightarrow \Phi^2 \,, \\
    \Phi^2 &\rightarrow - \Phi^1 \,, \\
    A^1_2 &\rightarrow - A^2_1 \,, \\
    A^2_1 &\rightarrow - A^1_2 \,, \\
    A^1_0 &\rightarrow  A^2_0 \,, \\
    A^2_0 &\rightarrow - A^1_0 \,, 
\end{align}
and so forth. As for the type-$B$ foliated gauge field, the rules are
\begin{align}
    \hat{\Phi}^1 &\rightarrow \hat{\Phi}^2 \,, \\
    \hat\Phi^2 &\rightarrow  \hat\Phi^1 \,, \\
    \hat{A}^1_2 &\rightarrow -\hat{A}^2_1 \,, \\
    \hat{A}^2_1 &\rightarrow \hat{A}^1_2 \,, \\
    \hat{A}^1_0 &\rightarrow  \hat{A}^2_0 \,, \\
    \hat{A}^2_0 &\rightarrow \hat{A}^1_0 \,, 
\end{align}
and so forth. The bulk fields obey the ordinary rules.

\subsection{Foliated Wedge Product}

We can define the wedge product of type-$A$ and  type-$B$ foliated gauge fields as follows. We let a type-$A$ foliated $(n+1)$-form gauge field be $\bm{A} = (A^k \wedge e^k, a)$ and a type-$B$ foliated $m$-form gauge field be $\bm{\hat{B}} = (\hat{B}^k,\hat{b}, \beta^k \wedge e^k)$, and we define the wedge product of these as\footnote{This definition is different from the one in \cite{Ohmori:2025fuy}.}
\begin{align}
    \bm{A} \wedge_\text{f} \bm{\hat{B}} &= \sum_k A^k \wedge e^k \wedge \hat{B}^k   + (-1)^m \, a \wedge \hat{b} \,, \\
    \bm{\hat{B}} \wedge_\text{f} \bm{A} &= \sum_k  \hat{B}^k \wedge A^k \wedge e^k + (-1)^n \hat{b} \wedge a \,.
\end{align}
Note that the auxiliary fields $\beta^k \wedge e^k$ do not appear. Then, we have the relation
\begin{align}
    \bm{A} \wedge_\text{f} \bm{\hat{B}} = (-1)^{(n+1) m} \bm{\hat{B}} \wedge_\text{f} \bm{A} \,.
\end{align}
Since the exterior derivatives of $\bm{A}$ and $\bm{\hat{B}}$ are
\begin{align}
    d_\text{f} \bm{A} &= \left(d A^k \wedge e^k, da + (-1)^{n+1} \sum_k A^k \wedge e^k \right) \,, \\
    d_\text{f} \bm{\hat{B}} &= \left(d\hat{B}^k + (-1)^{m+1} \hat{b} + (-1)^{m+1} \beta^k \wedge e^k, d\hat{b}, d\beta^k \wedge e^k \right) \,,
\end{align}
we have the Leibniz rule
\begin{align}
\begin{split}
    d ( \bm{A} \wedge_\text{f} \bm{\hat{B}} ) 
    & = \sum_k \left( d A^k \wedge e^k \wedge \hat{B}^k + (-1)^{n+1} A^k \wedge e^k \wedge d\hat{B}^k \right) \\
    & \quad + (-1)^m \left( da \wedge \hat{b} + (-1)^{n} a \wedge d\hat{b} \right) \\
    &= \sum_k d A^k \wedge e^k \wedge \hat{B}^k + (-1)^{m} \left( da  + (-1)^{n+1} \sum_k A^k \wedge e^k \right) \wedge \hat{b} \\
    &\quad + (-1)^{n+1} \left\{ \sum_k A^k \wedge e^k \wedge \left( d\hat{B}^k + (-1)^{m+1} \hat{b} \right) + (-1)^{m+1} a \wedge d\hat{b} \right\} \\ 
    & = d_\text{f} \bm{A} \wedge_\text{f} \bm{\hat{B}} + (-1)^{n+1} \bm{A} \wedge_\text{f} d_\text{f} \bm{\hat{B}} \,,
\end{split}
\end{align}
and similarly,
\begin{align}  
    d (\bm{\hat{B}} \wedge_\text{f} \bm{A} ) = d_\text{f} \bm{\hat{B}} \wedge_\text{f} \bm{A} + (-1)^{m} \bm{\hat{B}} \wedge_\text{f} d_\text{f} \bm{A} \,.
\end{align}
For example, the wedge product of the type-$B$ foliated one-form gauge field $\bm{\hat{C}} = (\hat{C}^k, \hat{c},\beta^k \wedge e^k)$ and the exterior derivative of the type-$A$ foliated (1+1)-form gauge field $\bm{C} = (C^k \wedge e^k, c)$ is
\begin{align}
\begin{split}
    \bm{\hat{C}} \wedge_\text{f} d_\text{f} \bm{C} 
    &= \sum^2_{k = 1} \hat{C}^k \wedge d C^k \wedge e^k + \hat{c} \wedge \left(  dc +  \sum^2_{k = 1} C^k \wedge e^k \right) \\
    &= \sum^2_{k = 1} ( d \hat{C}^k + \hat{c} ) \wedge C^k \wedge e^k - d \hat{c} \wedge c  \\
    &= d_\text{f} \bm{\hat{C}} \wedge_\text{f} \bm{C} \,,
\end{split}
\end{align}
where we have dropped the total derivative term. This form arises in the foliated SSPT Lagrangian.

We can also define the Hodge star of foliated gauge fields. The Hodge star of the type-$A$ foliated $(n+1)$-form gauge field $\bm{A} = ( A^k \wedge e^k, a)$ as
\begin{align}
    \ast \bm{A} = \left(\ast (A^k \wedge e^k), (-1)^{D-n-1} \ast a, 0 \right) \,,
\end{align}
which is the same form as a type-$B$ foliated $(D - n - 1)$-form gauge field where $D$ is the spacetime dimension. The Hodge star of the type-$B$ foliated $m$-form gauge field $\bm{\hat{B}} = (\hat{B}^k,\hat{b}, \beta^k \wedge e^k)$ as
\begin{align}
    \ast \bm{\hat{B}} = \left( e^k \wedge \ast (\hat{B}^k  \wedge e^k), (-1)^{D-m-1}\ast \hat{b} \right) \,.
\end{align}
$\ast \bm{\hat{B}}$ is the same form as a type-$A$ foliated $\{(D - m - 1) + 1\}$-form gauge field. Then, we can define the squares of type-$A$ and type-$B$ foliated gauge fields as
\begin{align}
    ( \bm{A} )^2 &= \bm{A} \wedge_\text{f} \ast \bm{A} = \sum_k A^k  \wedge e^k \wedge \ast ( A^k \wedge e^k ) + a \wedge \ast a  \,, \\
    ( \bm{\hat{B}} )^2 &= \sum_k \bm{\hat{B}} \wedge_\text{f} \ast \bm{\hat{B}} = \hat{B}^k \wedge e^k \wedge \ast (\hat{B}^k \wedge e^k )  + \hat{b} \wedge \ast \hat{b}  \,.
\end{align}
For example, the squares of the exterior derivatives of the type-$A$ foliated (0+1)-form gauge field and type-$B$ foliated zero-form gauge field are
\begin{align}
    ( d_\text{f} \bm{\Phi} )^2 &= \sum^{2}_{k=1} ( d\Phi^k \wedge e^k ) \wedge \ast ( d\Phi^k \wedge e^k ) + \left( d\Phi - \sum^{2}_{k=1} \Phi^k e^k \right) \wedge \ast \left( d\Phi - \sum^{2}_{k=1} \Phi^k e^k \right)  \,, \\
    ( d_\text{f} \bm{\hat{\Phi}} )^2 &= \sum^{2}_{k=1} \left\{ ( d\hat\Phi^k - \hat\Phi ) \wedge e^k \right\} \wedge \ast \left\{ ( d\hat\Phi^k - \hat\Phi ) \wedge e^k \right\} + ( d\hat\Phi ) \wedge \ast ( d\hat\Phi )  \,,
\end{align}
which arise in the foliated Lagrangians of $\phi$-theory and $\hat\phi$-theory.

\chapter{Fractonic \texorpdfstring{$BF$}{BF} Theories in 2+1 Dimensions and Their Anomalies}
\label{chapter:2+1d_bfanomaly_and_3+1d_sspt}

This chapter is mainly based on the work in \cite{Shimamura:2024kwf}. We revise the contents of the work using the insights obtained in \cite{Ohmori:2025fuy}, rewriting it to be more systematic and organized.

In this chapter, we discuss the $\Z_N \times \Z_N$ mixed 't Hooft anomaly of subsystem symmetries in the exotic and foliated $BF$ theory in 2+1 dimensions \cite{Seiberg:2020bhn,Ohmori:2022rzz}. Although they are equivalent, it is easier to couple the exotic $BF$ theory to background tensor gauge fields and construct the 3+1d exotic SSPT phase with two simultaneous foliations that cancels the 't Hooft anomaly. After constructing them, we will assume field correspondences between the exotic and foliated $BF$ theories extending the previous result without background gauge fields \cite{Ohmori:2022rzz}, and construct the 2+1d foliated $BF$ theory coupled to background foliated and bulk gauge fields. In the foliated $BF$ theory, we find that the non-topological defect that describes a fracton is considered as a symmetry-like operator. Next, using the field correspondences, we will construct the foliated form of the 3+1d SSPT phase with two foliations that cancels the 't Hooft anomaly of the foliated $BF$ theory. Finally, we will construct a bulk SSPT phase with three simultaneous foliations. In the foliated form, it is simple to construct the SSPT phase with three foliations from the SSPT phase with two foliations we have constructed. In the exotic form, on the other hand, the relation between the two SSPT phases is not manifest. Here we will construct the exotic form of the SSPT phase with three foliations via the foliated-exotic duality. This can be seen as a systematic way to construct exotic QFTs with different foliation structures. In addition, we will see that these two foliated SSPT phases are connected via a smooth deformation. This fact is considered as a hint for characterizing 't Hooft anomalies of subsystem symmetry.

\begin{figure}[H]
   \begin{center}
    \includegraphics[width=1\hsize]{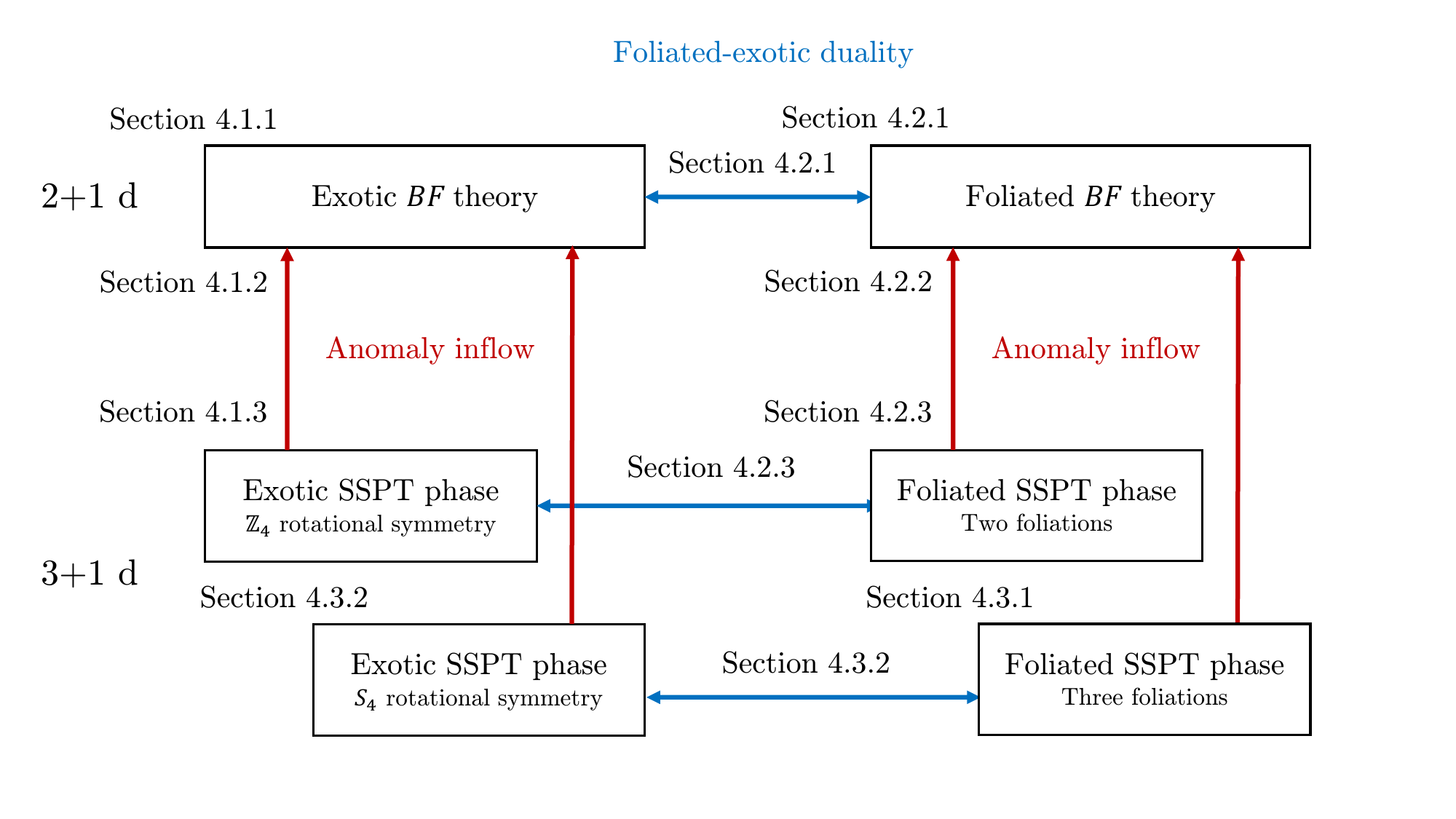} 
    \end{center}
    \vspace{-1cm}
      \caption{The structure of the anomaly inflow and the foliated-exotic dualities.}
      \label{fig_bf}
\end{figure}

The organization of this chapter is as follows and is also summarized in Figure \ref{fig_bf}. 

In Section \ref{section:2+1d_bfanomaly}, we consider the anomaly of the 2+1d exotic $BF$ theory. We will review the 2+1d exotic $BF$ theory and its subsystem symmetries \cite{Seiberg:2020bhn,Ohmori:2022rzz}, and consider the 2+1d exotic $BF$ theory coupled to background tensor gauge fields. Then, we construct the exotic form of the 3+1d SSPT phase with $\Z_4 \times SO(2)$ rotational symmetry that matches the anomaly of the 2+1d exotic $BF$ theory. 

In Section \ref{section:anomaly_foliated_bf}, we will consider the anomaly of the 2+1d foliated $BF$ theory \cite{Ohmori:2022rzz}. First, we review the 2+1d foliated $BF$ theory and the foliated-exotic duality in the foliated and exotic $BF$ theories. In Section \ref{subsection:coupling_background_foliated_gauge_fields}, we expand the field correspondences to the background gauge fields and construct the 2+1d foliated $BF$ theory coupled to background foliated and bulk gauge fields. In Section \ref{subsection:foliated_sspt}, we construct the foliated SSPT phase with two foliations from the exotic SSPT phase with $\Z_4 \times SO(2)$ rotational symmetry, which is one of the main result of the paper \cite{Shimamura:2024kwf}.

\begin{figure}[H]
   \begin{center}
    \includegraphics[width=1\hsize]{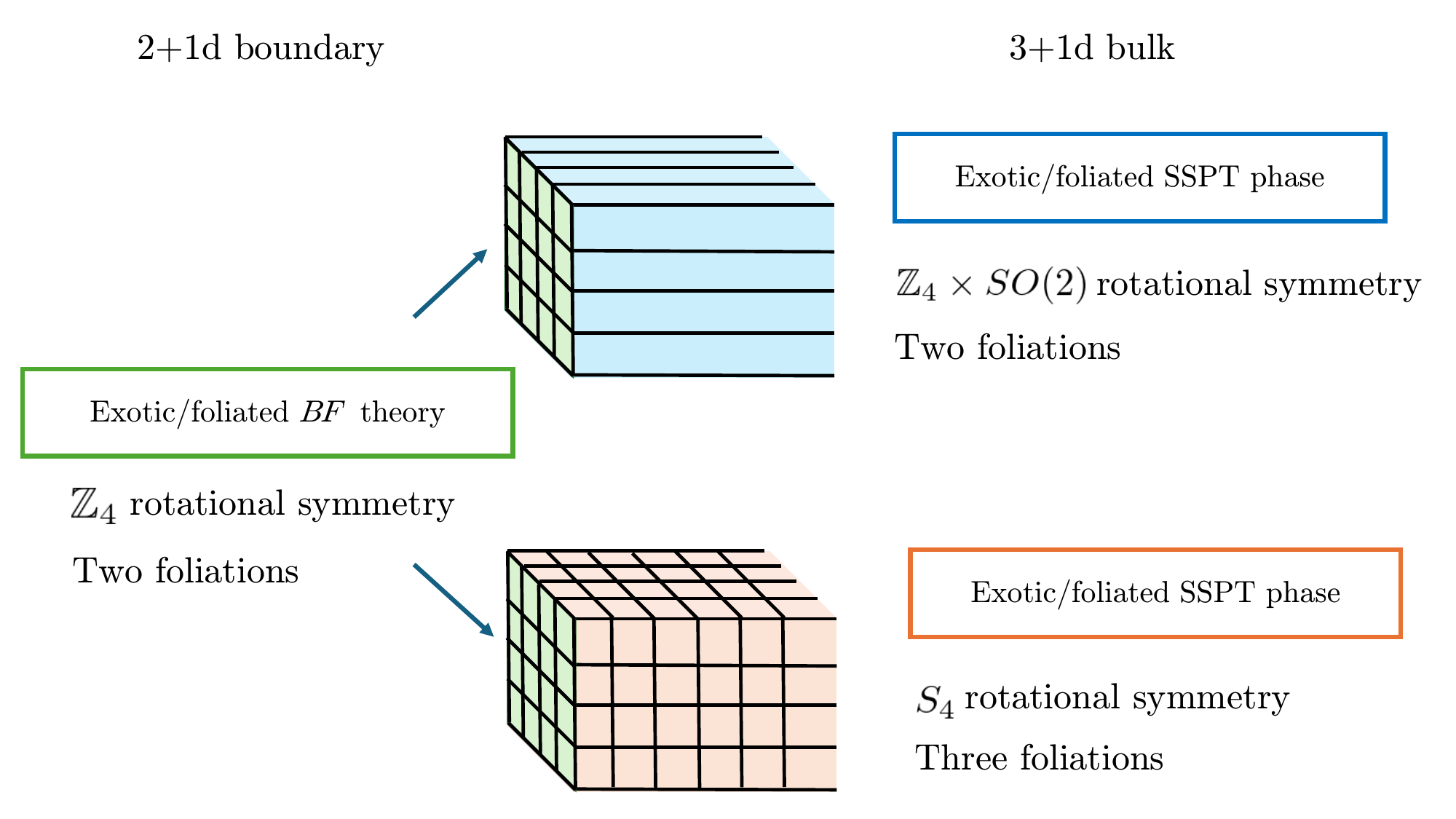} 
    \end{center}
      \caption{The foliation structures and the rotational symmetries in the fractonic $BF$ theory on the boundary and the SSPT phases in the bulk.}
      \label{fig_sspt}
\end{figure}

In Section \ref{section:change_foliation}, we will discuss changing the foliation structure from two foliations to three foliations. The SSPT phase with three foliations also cancels the same anomaly of the 2+1d exotic/foliated $BF$ theory. In Section \ref{subsection:foliated_sspt_3foliations}, we see the change is easily carried out in the foliated SSPT phase. In Section \ref{subsection:exotic_sspt_3foliations}, we convert the SSPT phase with three foliations from the foliated form into the exotic form by assuming field correspondences, and construct the exotic SSPT phase with $S_4$ rotational symmetry, which is the other main result of the paper \cite{Shimamura:2024kwf}. The folaition structures and the rotational symmetries in the fractonic $BF$ theory on the boundary and the SSPT phases in the bulk are showed in Figure \ref{fig_sspt}.

\section{Anomaly in the 2+1d Exotic \texorpdfstring{$BF$}{BF} Theory}
\label{section:2+1d_bfanomaly}

In this section, we review the exotic $BF$ theory in 2+1 dimensions \cite{Seiberg:2020bhn,Ohmori:2022rzz}, which is the low-energy effective QFT of the $\Z_N$ plaquette Ising model \cite{Johnston:2016mbz}, and consider the two types of subsystem symmetries and their mixed 't Hooft anomaly. Due to the anomaly, we cannot gauge both of the subsystem symmetries simultaneously. This anomaly can be canceled by a classical field theory in one dimension higher, which is called a subsystem symmetry-protected topological (SSPT)
phase\cite{You:2018oai,Devakul:2018fhz}. We will consider the mixed 't Hooft anomaly and the SSPT phase associated with the 2+1d exotic $BF$ theory. We basically proceed in parallel with the case of the 3+1d exotic $BF$ theory studied in \cite{Burnell:2021reh}.

\subsection{Exotic \texorpdfstring{$BF$}{BF} Theory and Symmetries}
\label{subsection:exotic_bf_and_symmetries}

We take Euclidean spacetime to be a (2+1)-dimensional torus $\T^{2+1}$ of lengths $l^0$, $l^1$ and $l^2$, and the coordinate $(x^0, x^1, x^2)$ on it. We consider the exotic $BF$ theory, whose rotational symmetry is only the 90 degree ones. Such a theory has exotic tensor gauge fields, each of which is in a representation of the 90 degree rotation group $\Z_4$. The 2+1d exotic $BF$ theory contains a $U(1)$ tensor zero-form gauge field $\bm{\hat\phi} = (\hat{\phi}^{12})$ in $\bm{1}_2$ and a $U(1)$ tensor one-form gauge field $\bm{A} = (A_0,A_{12})$ in $(\bm{1}_0,\bm{1}_2)$. Their gauge transformations are
\begin{align}
    \hat{\phi}^{12} &\sim \hat{\phi}^{12} + 2\pi \hat{w}^1 - 2\pi \hat{w}^2 \,, \label{egauge3} \\
    A_{0} &\sim A_{0} + \partial_0 \Lambda \,, \label{egauge1} \\
    A_{12} &\sim A_{12} + \partial_1\partial_2 \Lambda \,, \label{egauge2} 
\end{align}
where $\hat{w}^k$ is an $x^k$-dependent integer-valued gauge parameter, and $\bm{\Lambda} = (\Lambda)$ is a tensor one-form gauge parameter in $\bm{1}_0$. The gauge parameter $\Lambda$ has its own gauge transformation: $\Lambda \sim \Lambda + 2\pi m^1 + 2\pi m^2$, where $m^k$ is an $x^k$-dependent integer-valued gauge parameter. Due to constant parts of $\hat{w}^k$ and $m^k$, $\hat{\phi}^{12}$ and $\Lambda$ can be regarded as $U(1)$-valued: $\hat{\phi}^{12} \sim \hat{\phi}^{12} + 2\pi$, $\Lambda \sim \Lambda + 2\pi$. These tensor gauge fields and parameters can have particular types of singularities and discontinuities \cite{Seiberg:2020bhn,Ohmori:2022rzz}.

The exotic $BF$ Lagrangian is\footnote{The subscript \text{e} means that the theory is written by using exotic tensor gauge fields, which we call the exotic form. Also, the subscript \text{f}, which will appear later, means that the theory is written by using foliated gauge fields and bulk gauge fields, which we call the foliated form.}
\begin{align}
    \L_{\text{e}} = -\frac{i N}{2\pi}\hat{\phi}^{12}(\partial_0 A_{12} - \partial_{1}\partial_{2} A_0) \,. \label{elagrangian}
\end{align}
where $N$ is a positive integer.

The equations of motion are
\begin{align}
    &\frac{iN}{2\pi} \partial_{1}\partial_{2} \hat{\phi}^{12} = 0 \,, \label{eeom1}\\
    &\frac{iN}{2\pi} \partial_{0} \hat{\phi}^{12} = 0 \,, \label{eeom2} \\
    &\frac{iN}{2\pi} (\partial_0 A_{12} - \partial_{1}\partial_{2} A_0) = 0 \,. \label{eeom3}
\end{align}

Let us discuss global symmetries. The subsystem symmetries are described by the partially topological gauge-invariant defects and operators.
The exotic $BF$ theory has two types of space-like subsystem symmetries. One is the $\Z_N$ electric global symmetry that is generated by the symmetry operator
\begin{align}
    V_1[x] = \exp \left[ i \hat{\phi}^{12} \right]\,. \label{epoint}
\end{align}
This operator satisfies $V_1[x]^N = 1$, so it generates a $\Z_N$ subsystem symmetry.\footnote{We denote the elements of $\Z_N$ by $e^{2\pi im/N} \, (m = 1,2, \ldots, N)$, and then they correspond to the symmetry operators $V_m[x] = V_1[x]^m = \exp \left[ i m \hat{\phi}^{12} \right]$ for the $\Z_N$ electric global symmetry and $W^k_m \left[ [x^k_1,x^k_2] \times \mathcal{C}^j \right]$ for the $\Z_N$ dipole global symmetries given below.}
The charged operators are the strip operators
\begin{align}
    W^1_n \left[ [x^1_1,x^1_2] \times \mathcal{C}^2 \right] = \exp \left[ i n \int^{x^1_2}_{x^1_1} dx^1 \oint_{\mathcal{C}^2}  dx^2 A_{12} \right] \,, \label{sestrip1}\\
    W^2_n \left[ [x^2_1,x^2_2] \times \mathcal{C}^1  \right] = \exp \left[ i n \int^{x^2_2}_{x^2_1} dx^2 \oint_{\mathcal{C}^1}  dx^1 A_{12}  \right]\,, \label{sestrip2}
\end{align}
where $n = 1,2,\ldots, N-1$, and  $\mathcal{C}^j$ is a one-dimensional loop along the $x^j$ direction. Thus, $[x^k_1,x^k_2] \times \mathcal{C}^j$ is a  two-dimensional strip with a fixed width along the $x^k$ direction in the $(x^1,x^2)$-plane.
The operator $W^k_1 \left[ [x^k_1,x^k_2] \times \mathcal{C}^j \right]$ is also a $\Z_N$ operator: $W^k_1 \left[ [x^k_1,x^k_2] \times \mathcal{C}^j \right]^N  = 1$.
The symmetry operator $V_1[x]$ acts on $W^k_n \left[ [x^k_1,x^k_2] \times \mathcal{C}^j \right]$ as
\begin{align}
    V_1[x]\  W^k_n \left[ [x^k_1,x^k_2] \times \mathcal{C}^j \right]\ V_1[x]^{-1} =  e^{-2\pi i n /N} \  W^k_n \left[ [x^k_1,x^k_2] \times \mathcal{C}^j \right]  \ \ ,\quad \text{if} \quad x^k_1 < x^k < x^k_2\,. \label{eactvw}
\end{align}
The $\Z_N$ electric global symmetry is a space-like subsystem symmetry on a zero-dimensional submanifold. For the action on the field, the $\Z_N$ electric global symmetry acts as
\begin{align}
    A_{12} \rightarrow A_{12} + \Gamma_{12} \,,
\end{align}
where $\Gamma_{12}$ satisfies
\begin{align}
     \int^{x^k_2}_{x^k_1} dx^k \oint_{\mathcal{C}^j}  dx^j  \Gamma_{12} \in \frac{2\pi}{N} \Z \,,
\end{align}
so $\Gamma_{12}$ can be written as $\Gamma_{12} = \frac{2\pi}{N}\left( \frac{1}{l^2} \partial_1 n_1(x^1) + \frac{1}{l^1} \partial_2 n_2(x^2)  \right) $, where $n_k(x^k)$ is an $x^k$-dependent integer-valued and single-valued function.

The other space-like symmetries are the $\Z_N$ dipole global symmetries that are generated by the strip operators $W_1^k \left[ [x^k_1,x^k_2] \times \mathcal{C}^j \right]$ in \eqref{sestrip1} and \eqref{sestrip2}.
The charged operator is $V_n[x] = \exp \left[i n \hat\phi^{12}  \right]$, and the actions are
\begin{align}
    W_1^k\left[ [x^k_1,x^k_2] \times \mathcal{C}^j \right]\  V_n[x]\ W^k_1 \left[ [x^k_1,x^k_2] \times \mathcal{C}^j \right]^{-1} =  e^{2\pi i n /N} \  V_n[x]  \ \ ,\quad \text{if} \quad x^k_1 < x^k < x^k_2\,. \label{eactwv}
\end{align}
For the action on the field, the $\Z_N$ dipole global symmetries act as
\begin{align}
    \hat{\phi}^{12} \rightarrow \hat{\phi}^{12} + \hat{\Gamma}^{12} \,,
\end{align}
where $\hat{\Gamma}^{12}$ is valued in $2\pi \Z/N$, so $\hat{\Gamma}^{12}$ can be written as  $\hat{\Gamma}^{12} = \frac{2\pi}{N}\left(n_1(x^1)  + n_2(x^2) \right)$, where $n_k(x^k)$ is an $x^k$-dependent integer-valued and single-valued function.

From the actions \eqref{eactvw} and \eqref{eactwv}, each symmetry operator is the charged operator of the other, which is similar to the symmetries in the ordinary $BF$ theory \cite{Maldacena:2001ss,Banks:2010zn,Kapustin:2014gua}. From this fact, we expect that the two types of subsystem symmetries have a mixed 't Hooft anomaly.

We also have a time-like symmetry \cite{Gorantla:2022eem} whose charged defect describes a fracton, which is called the $\Z_N$ tensor symmetry. The symmetry operator is a quadrupole operator:
\begin{align}
    T_m \left[\mathcal{C}^{12,\text{rect}}(x^1_1,x^1_2,x^2_1,x^2_2)\right] = \exp\left[i m \Delta_{12} \hat{\phi}^{12} (x^1_1,x^1_2,x^2_1,x^2_2) \label{etimelikeop}   \right] \,,
\end{align}
where $\Delta_{12} \hat{\phi}^{12}(x^1_1,x^1_2,x^2_1,x^2_2) = \hat{\phi}^{12}(x^1_2,x^2_2) - \hat{\phi}^{12}(x^1_2,x^2_1) - \hat{\phi}^{12}(x^1_1,x^2_2) + \hat{\phi}^{12}(x^1_1,x^2_1)$, and \\
\noindent  $\mathcal{C}^{12,\text{rect}}(x^1_1,x^1_2,x^2_1,x^2_2)$ is a rectangle whose vertices are the four points above. This operator is a product of the operators $V_m[x] = \exp \left[ i m \hat{\phi}^{12} \right]$ localized at the corners of the rectangle. The charged defect is
\begin{align}
    F_n[\mathcal{C}^0] = \exp \left[ i n \oint_{\mathcal{C}^0} dx^0 A_0 \right] \,, \label{efracton}
\end{align}
where $\mathcal{C}^0$ is a closed one-dimensional loop along the time $x^0$ direction. The deformation of $\mathcal{C}_1$ would break the gauge invariance of the defect, which means that this defect describes a fracton that cannot move alone in space. The operator $T_m \left[\mathcal{C}^{12,\text{rect}}(x^1_1,x^1_2,x^2_1,x^2_2)\right]$ detects the fracton defect $F_n[\mathcal{C}^0]$ as
\begin{align}
    T_m \left[\mathcal{C}^{12,\text{rect}}(x^1_1,x^1_2,x^2_1,x^2_2)\right] \cdot F_n [\mathcal{C}^0] = e^{2\pi i m n /N} F_n[\mathcal{C}^0] \,, \label{tfctvw}
\end{align}
when $\mathcal{C}^{12,\text{rect}}(x^1_1,x^1_2,x^2_1,x^2_2)$ surrounds $\mathcal{C}^0$.\footnote{The edges of a rectangle $\mathcal{C}^{12,\text{rect}}(x^1_1,x^1_2,x^2_1,x^2_2)$ cannot be remotely detected by other operators, but the operator $T[\mathcal{C}^{12,\text{rect}}(x^1_1,x^1_2,x^2_1,x^2_2)]$ is actually an operator on a rectangle. In fracton theory, operators are not necessarily remotely detectable unlike ordinary topological order or topological field theory \cite{Ohmori:2022rzz}} For the action on the field, the $\Z_N$ tensor symmetry acts as
\begin{align}
    A_0 \rightarrow A_0 + \Gamma_0 \,,
\end{align}
where $\Gamma_0$ is a $\Z_N$ field, which satisfies
\begin{align}
     \oint_{\mathcal{C}^0} dx^0  \Gamma_0 \in \frac{2\pi}{N} \Z \,,
\end{align}
so $\Gamma_0$ can be written as $ \Gamma_0 = \frac{2\pi}{N} \frac{1}{l^0} ( n_1(x^1)  + n_2(x^2))$, where $n_k(x^k)$ is an $x^k$-dependent integer-valued and single-valued function\cite{Gorantla:2022eem}.

In addition, we can construct the defect that describes a dipole of fractons separated in the $x^k$ direction. A dipole of fractons can be represented as
\begin{align}
    F_n[\mathcal{C}^0(x^k_1,x^j)]  F_n[\mathcal{C}^0(x^k_2,x^j)]^{-1}  = \exp \left[ i n \oint_{\mathcal{C}^0} dx^0 \int^{x^k_2}_{x^k_1} dx^k \partial_k A_0(x^0,x^k,x^j)  \right] \,,
\end{align}
where $\mathcal{C}^0(x^1,x^2)$ is a closed loop along the $x^0$ direction at a point $(x^1,x^2)$ in space, and $(k,j)=(1,2),(2,1)$. This defect is partially topological, that is, it can be deformed to the strip defect
\begin{align}
    W^k_{n,\text{dip}} \left[ [x^k_1,x^k_2] \times \mathcal{C}^{0j} \right] &= \exp \left[ i n \int^{x^k_2}_{x^k_1} dx^k \oint_{\mathcal{C}^{0j}} \left( dx^0 \partial_k A_0 + dx^j A_{12} \right)  \right] \,, \label{estripk}
\end{align}
where $\mathcal{C}^{0j}$ is a one-dimensional loop in the $(x^0,x^j)$-plane in spacetime.

\subsection{Coupling to the Background Tensor Gauge Fields}
\label{subsection:couple_background_gauge_fields}

In this section, we couple the subsystem symmetries to background gauge fields and replace the parameters of the symmetry actions on the fields with local transformations. Then, the local transformations are absorbed into the gauge transformations of the background gauge fields. To gauge these symmetries, one has to sum over configurations of the gauge fields. If the partition function is not invariant under the gauge transformations, we cannot gauge all the symmetries at the same time, indicating a mixed 't Hooft anomaly. We will see that the partition function of the 2+1d exotic $BF$ theory is indeed not gauge invariant and the subsystem symmetries have a mixed 't Hooft anomaly.

The tensor time-like symmetry $T\left[\mathcal{C}^{12,\text{rect}}(x^1_1,x^1_2,x^2_1,x^2_2)\right]$ and the electric space-like symmetry $V_m[x]$ are coupled to the $U(1)$ tensor two-form gauge field $\bm{C}= (C_{012})$ in $\bm{1}_2$. The dipole symmetries $W_m^k [ [x^k_1,x^k_2] \times \mathcal{C}^j ]$ are coupled to the $U(1)$ tensor one-form gauge field $\bm{\hat{C}}  = (\hat{C}^{12}_0, \hat{C})$ in $(\bm{1}_2, \bm{1}_0)$. We will see later that the background gauge transformations of $\bm{C} = (C_{012})$ and $\bm{\hat{C}} = (\hat{C}^{12}_0, \hat{C})$ are absorbed into the local transformations of $ \bm{A} = (A_0, A_{12})$ and $\bm{\hat{\phi}} =  (\hat{\phi}^{12})$. The tensor gauge field $\bm{A}$ transforms as
\begin{align}
    A_0 &\sim A_0 + \Gamma_0 \,, \label{a0bk} \\
    A_{12} &\sim A_{12} + \Gamma_{12} \,, \label{a12bk}
\end{align}
where $\bm{\Gamma} = (\Gamma_0,\Gamma_{12})$ is a background tensor one-form gauge parameter. Then, the background gauge transformations of $\bm{C} = (C_{012})$ is
\begin{align}
    C_{012} \sim C_{012} +  \partial_0 \Gamma_{12} - \partial_1 \partial_2 \Gamma_0 \,.
\end{align}
The tensor gauge field $\bm{\hat{\phi}}$ transforms as
\begin{align}
    \hat{\phi}^{12} \sim \hat{\phi}^{12} +  \hat{\Gamma}^{12} \,, \label{phi12bk}
\end{align}
where $\bm{\hat{\Gamma}} = (\hat{\Gamma}^{12})$ is a background tensor zero-form gauge parameter. Then, the background gauge transformations of $\bm{\hat{C}} = (\hat{C}^{12}_0, \hat{C})$ are
\begin{align}
    \hat{C}^{12}_0 &\sim \hat{C}^{12}_0 + \partial_0 \hat{\Gamma}^{12} \,, \\
    \hat{C} &\sim \hat{C} + \partial_1 \partial_2 \hat{\Gamma}^{12} \,.
\end{align}

The Lagrangian including the background gauge fields is
\begin{align}
\begin{split}
    \L_{\text{e}} \left[ \bm{C}, \bm{\hat{C}} \right] = &\frac{iN}{2\pi} \left[ -\hat{\phi}^{12} (\partial_0 A_{12} - \partial_1 \partial_2 A_0 -C_{012} ) - \hat{C}^{12}_0 A_{12}  - \hat{C} A_0  \right] \\ 
    &- \frac{iN}{2\pi} \chi (\partial_0 \hat{C} - \partial_1 \partial_2 \hat{C}^{12}_0) - \frac{iN}{2\pi}( 2\pi\hat{h}^{12} ) C_{012} \,, \label{ebfc}
\end{split}
\end{align}
Since the symmetries coupled to the $U(1)$ gauge fields are $\Z_N$ symmetries, we need terms of $\chi$ and $\hat{h}^{12}$ that are dynamical fields, so that the $U(1)$ gauge fields $\bm{C}$ and $\bm{\hat{C}}$ are restricted to $\Z_N$ gauge fields. $\chi$ is a $U(1)$-valued function in $\bm{1}_0$, and $\hat{h}^{12}$ is an integer-valued function in $\bm{1}_2$. They are tensor zero-form gauge fields, but their dynamical gauge transformations are
\begin{align}
    \chi &\sim \chi + \Lambda \,, \\
    \hat{h}^{12} &\sim \hat{h}^{12} + \hat{w}^1 - \hat{w}^2 \,.
\end{align}

Under the background gauge transformations, the Lagrangian transforms as
\begin{align}
\begin{split}
     \delta_\text{g} \L_{\text{e}} \left[ \bm{C}, \bm{\hat{C}} \right] &=  \frac{iN}{2\pi}  \left[ - \hat{\Gamma}^{12} (\partial_0 A_{12} - \partial_1 \partial_2 A_0 -C_{012} ) - (\hat{C}^{12}_0 + \partial_0 \hat{\Gamma}^{12}) \Gamma_{12}  \right. \\
    & \quad \quad  \left. - (\hat{C} + \partial_1 \partial_2 \hat{\Gamma}^{12}) \Gamma_0  - \partial_0 \hat{\Gamma}^{12} \, A_{12}  - \partial_1 \partial_2 \hat{\Gamma}^{12}\,  A_0 \right] \\
    &= \frac{iN}{2\pi}  \left[  \hat{\Gamma}^{12} C_{012}  - (\hat{C}^{12}_0 + \partial_0 \hat{\Gamma}^{12}) \Gamma_{12}   -  (\hat{C} + \partial_1 \partial_2 \hat{\Gamma}^{12}) \Gamma_0 \right] \,. \label{eanomaly}
\end{split}
\end{align}
If $\bm{C} = (C_{012})$ and $ \bm{\Gamma} = (\Gamma_0, \Gamma_{12})$ are absent or $\bm{\hat{C}} =(\hat{C}^{12}_0, \hat{C})$ and $\bm{\hat{\Gamma}} = (\hat{\Gamma}^{12})$ are absent, the partition function is invariant. So we can gauge the one side of the symmetry solely, but we cannot gauge both of the symmetries simultaneously. That is a mixed 't Hooft anomaly for the $\Z_N \times \Z_N$ subsystem symmetries \cite{Burnell:2021reh}.

\subsection{Exotic SSPT Phase for \texorpdfstring{$\Z_N \times \Z_N$}{ZN ZN} in 3+1 Dimensions} 
\label{subsection:exotic_sspt_phase}

We saw the exotic $BF$ theory in 2+1 dimensions has the mixed 't Hooft anomaly. This anomaly can be canceled by a classical field theory in one dimension higher, which is the continuum description of what is called an SSPT phase. In this section, we consider the SSPT phase for $\Z_N \times \Z_N$ in 3+1 dimensions. We first take Euclidean spacetime to be a (3+1)-dimensional torus $\T^{3+1}$ of lengths $l^0$, $l^1$, $l^2$ and $l^3$, and the coordinate $(x^0, x^1, x^2 ,x^3)$ on it.
The foliation structure \cite{Slagle:2018swq,Slagle:2020ugk,Hsin:2021mjn} of the SSPT phase is flat $x^1$ and $x^2$ foliations, so it is a fractonic system with two simultaneous foliations.\footnote{The foliation structure is similar to the lattice. For example, $x^1$ foliation on a three dimensional space is a decomposition into an infinite number of planes orthogonal to the $x^1$ direction, so the space has lattice-like structure in the $x^1$ direction.} The SSPT phase has the 90 degree discrete rotational symmetry $\Z_4$ for $(x^1,x^2)$ and the continuous rotational symmetry $SO(2)$ for $(x^0,x^3)$ as the spacetime rotational symmetry.

This theory has a background tensor two-form gauge field $\bm{C} = (C_{012}, C_{312},C_{03})$ and a background tensor one-form gauge field $\hat{C} = (\hat{C}^{12}_0, \hat{C}, \hat{C}^{12}_3)$, which are representations of $\Z_4 \times SO(2)$. $C_{012}$, $C_{312}$, $\hat{C}^{12}_0$ and $\hat{C}^{12}_3$ are in the representation $\bm{1}_2$ of $\Z_4$, and $\hat{C}$ is in the representation $\bm{1}_0$ of $\Z_4$. $(C_{012},C_{312})$ and $(\hat{C}^{12}_0,\hat{C}^{12}_3)$ are vectors of $SO(2)$, and $C_{03}$ is an anti-symmetric tensor of $SO(2)$. The background gauge transformations of $\bm{C} = (C_{012}, C_{312},C_{03})$ are
\begin{align}
    C_{012} &\sim C_{012} + \partial_0 \Gamma_{12} -\partial_1\partial_2 \Gamma_0 \,, \\
    C_{312} &\sim C_{312} + \partial_3 \Gamma_{12} -\partial_1\partial_2 \Gamma_3 \,, \\
    C_{03} &\sim C_{03} + \partial_0 \Gamma_3 -\partial_3 \Gamma_0 \,,
\end{align}
where $\bm{\Gamma} = (\Gamma_0, \Gamma_{12}, \Gamma_3)$ is a background tensor one-form gauge parameter.
The background gauge transformations of $\bm{\hat{C}} = (\hat{C}^{12}_0, \hat{C}, \hat{C}^{12}_3)$ are
\begin{align}
    \hat{C}^{12}_0 &\sim \hat{C}^{12}_0 + \partial_0  \hat{\Gamma}^{12} \,, \\
    \hat{C} &\sim \hat{C} + \partial_1 \partial_2  \hat{\Gamma}^{12} \,, \\
    \hat{C}^{12}_3 &\sim \hat{C}^{12}_3 + \partial_3  \hat{\Gamma}^{12} \,,
\end{align}
where $\bm{\hat{\Gamma}} = (\hat{\Gamma}^{12})$ is a background tensor zero-form gauge parameter.
To restrict these fields to $\Z_N$, we introduce a dynamical tensor one-form gauge field $\bm{\beta} = (\beta_0, \beta_{12}, \beta_3)$ and a dynamical tensor zero-form gauge field $\bm{\hat{\beta}} = (\hat{\beta}^{12})$. Their dynamical gauge transformations are
\begin{align}
    \beta_0 &\sim \beta_0 + \partial_0 s \,, \\
    \beta_{12} &\sim \beta_{12} + \partial_1 \partial_2 s \,, \\
    \beta_3 &\sim \beta_3 + \partial_3 s \,, \\
    \hat{\beta}^{12} &\sim \hat{\beta}^{12} + 2\pi \hat{s}^1 - 2\pi \hat{s}^2 \,, 
\end{align}
where $s$ is a gauge parameter in $\bm{1}_0$, and $\hat{s}^k$ is an $x^k$-dependent integer-valued gauge parameter. On the other hand, $\bm{\beta}$ and $\bm{\hat\beta}$ also have the background gauge transformations:
\begin{align}
    \beta_0 &\sim \beta_0 - \Gamma_0 \,, \\
    \beta_{12} &\sim \beta_{12} - \Gamma_{12} \,, \\
    \beta_3 &\sim \beta_3 - \Gamma_3 \,, \\
    \hat{\beta}^{12} &\sim \hat{\beta}^{12} - \hat{\Gamma}^{12} \,.
\end{align}

The SSPT phase is described by the Lagrangian
\begin{align}
\begin{split}
    &\L_{\text{SSPT,e}}\left[ \bm{C}, \bm{\hat{C}} \right] \\
    & \quad =  \frac{iN}{2\pi} \hat{\beta}^{12} \left( \partial_0 C_{312} - \partial_3 C_{012} + \partial_1 \partial_2 C_{03} \right) \\
    & \quad \quad +  \frac{iN}{2\pi} \left[ \beta_0 \left( \partial_3 \hat{C} - \partial_1 \partial_2 \hat{C}^{12}_3  \right) + \beta_{12} \left( \partial_3 \hat{C}^{12}_0 - \partial_0 \hat{C}^{12}_3  \right)  - \beta_3 \left( \partial_0 \hat{C} - \partial_1 \partial_2 \hat{C}^{12}_0  \right) \right] \\
    &\quad \quad  + \frac{iN}{2\pi}  \left( \hat{C}^{12}_3 C_{012}  - \hat{C}^{12}_0 C_{312}  + \hat{C} C_{03}   \right) \,. \label{sspte}
\end{split}
\end{align}
If the theory is on spacetime without a boundary, it is gauge invariant.\footnote{The gauge-invariant derivatives of $\bm{C}$ and $\bm{\hat{C}}$ are the components of $d_\text{e} \bm{C}$ and $d_\text{e} \bm{\hat{C}}$ in Section \ref{subsection:exotic_tensor_gauge_fields_z4so2_3d}.} However, if spacetime has a boundary, the partition function of the 3+1d SSPT phase is not invariant. From the anomaly inflow mechanism \cite{Callan:1984sa}, this variation is expected to be canceled by the anomaly of the 2+1d exotic $BF$ theory on the boundary \eqref{eanomaly}. To see this, we put the SSPT phase on the spacetime $\T^{2+1} \times \R_{x^3 \geq 0}$ with the boundary $x^3 = 0$. From the gauge invariance, the boundary conditions of the dynamical gauge parameters are
\begin{align}
    s |_{x^3 = 0} &= 0 \,, \\
    \hat{s}^k |_{x^3 = 0} &= 0  \,.
\end{align}
On the boundary, we put the 2+1d exotic $BF$ theory coupled to the background gauge fields \eqref{ebfc}, and the background gauge fields in the 3+1d SSPT phase are related to those in the 2+1d exotic $BF$ theory as
\begin{align}
    C_{\text{SSPT},012}\, |_{x^3 = 0} &= C_{BF,012} \,, \\
    \hat{C}^{12}_{\text{SSPT},0}\, |_{x^3 = 0} &= \hat{C}^{12}_{BF,0} \,, \\ 
    \hat{C}_{\text{SSPT}}\, |_{x^3 = 0} &= \hat{C}_{BF} \,.
\end{align}
Note that while the background gauge fields in the 3+1d SSPT phase are restricted to $\Z_N$ tensor gauge fields by the dynamical fields $\bm{\beta}$ and $\bm{\hat\beta}$, those in the 2+1d exotic $BF$ theory are restricted by the dynamical fields $\bm{\chi}$ and $\bm{\hat{\chi}}$.
Then, under the background gauge transformations, the Lagrangian transforms as
\begin{align}
\begin{split}
    &\delta_\text{g} \L_{\text{SSPT,e}}\left[ \bm{C}, \bm{\hat{C}} \right] \\
    &=  - \frac{iN}{2\pi} \hat{\Gamma}^{12} \left( \partial_0 C_{312} - \partial_3 C_{012} + \partial_1 \partial_2 C_{03} \right) \\
    & \quad +  \frac{iN}{2\pi} \left[ -\Gamma_0 \left( \partial_3 \hat{C} - \partial_1 \partial_2 \hat{C}^{12}_3  \right) - \Gamma_{12} \left( \partial_3 \hat{C}^{12}_0 - \partial_0 \hat{C}^{12}_3  \right)  + \Gamma_3 \left( \partial_0 \hat{C} - \partial_1 \partial_2 \hat{C}^{12}_0  \right) \right] \\
    &\quad  + \frac{iN}{2\pi}  \left[ \hat{C}^{12}_3 \left( \partial_0 \Gamma_{12} - \partial_1 \partial_2 \Gamma_0 \right)  - \hat{C}^{12}_0 \left( \partial_3 \Gamma_{12} -  \partial_1 \partial_2 \Gamma_3 \right)  + \hat{C} \left( \partial_0 \Gamma_3 - \partial_3 \Gamma_0 \right)  \right. \\
    &\quad + \partial_3 \hat{\Gamma}^{12}  \left(C_{012} + \partial_0 \Gamma_{12} - \partial_1 \partial_2 \Gamma_0 \right)  - \partial_0 \hat{\Gamma}^{12}  \left(C_{312} + \partial_3 \Gamma_{12} - \partial_1 \partial_2 \Gamma_3 \right)  \\
    &\quad \left. + \partial_1 \partial_2 \hat{\Gamma}^{12} \left(C_{03} + \partial_0 \Gamma_3 - \partial_3 \Gamma_0 \right)  \right]  \\
    &= \frac{iN}{2\pi} \partial_3 \left[ \hat{\Gamma}^{12} C_{012}  -( \hat{C}^{12}_0 + \partial_0 \hat{\Gamma}^{12} ) \Gamma_{12}  - ( \hat{C} + \partial_1 \partial_2 \hat{\Gamma}^{12} ) \Gamma_0   \right] \,,
\end{split}
\end{align}
where we have dropped total derivative terms with respect to $x^0$, $x^1$ and $x^2$. Thus on the boundary $x^3 = 0$, the term 
\begin{align}
    \delta_\text{g} S_{\text{SSPT,e}} \left[ \bm{C},\bm{\hat{C}} \right] = \int_{x^3 =0} d^3x \frac{iN}{2\pi}  \left[ -\hat{\Gamma}^{12} C_{012} + ( \hat{C}^{12}_0 + \partial_0 \hat{\Gamma}^{12} ) \Gamma_{12} + ( \hat{C} + \partial_1 \partial_2 \hat{\Gamma}^{12} ) \Gamma_0    \right] \label{eboundarysspt}
\end{align}
arises. This boundary term matches the 't Hooft anomaly of the 2+1d exotic $BF$ theory \eqref{eanomaly} on the boundary $x^3=0$. Therefore we can see that the 't Hooft anomaly of the 2+1d exotic $BF$ theory on the boundary is canceled by the gauge-variation of the 3+1d SSPT phase on the bulk.\footnote{The anomaly \eqref{eanomaly} and the boundary term \eqref{eboundarysspt} have the opposite sign, so we have to consider the product of the partition function of  the exotic $BF$ theory $Z_{BF,\text{e}} [ \bm{C},\bm{\hat{C}} ] =\int d\bm{\hat{\phi}} d\bm{A} d\chi d\hat{h}^{12} \,e^{-S_{BF,\text{e}}[ \bm{C},\bm{\hat{C}} ]}$ and the partition function of the exotic SSPT phase $Z_{\text{SSPT,e}}[ \bm{C},\bm{\hat{C}} ] =\int d\bm{\beta} d\bm{\hat{\beta}} \,e^{-S_{\text{SSPT,e}}[\bm{C},\bm{\hat{C}}]}$.}

\section{Anomaly in the 2+1d Foliated \texorpdfstring{$BF$}{BF} Theory}
\label{section:anomaly_foliated_bf}

In this section, we first review the foliated $BF$ theory in 2+1 dimensions \cite{Ohmori:2022rzz} and then consider its mixed 't Hooft anomaly. The mixed 't Hooft anomaly is considered to be the same one as the exotic $BF$ theory in 2+1 dimensions in Section \ref{section:2+1d_bfanomaly} from the foliated-exotic duality. To see this anomaly, we have to couple the subsystem symmetries of the foliated $BF$ theory to background gauge fields, but in the foliated form, its construction is non-trivial. In Section \ref{subsection:foliated_bf_field_correspondence}, we will establish field correspondences between background tensor gauge fields and background foliated gauge fields. Then, in Section \ref{subsection:coupling_background_foliated_gauge_fields}, we will construct the foliated $BF$ Lagrangian coupled to the foliated background gauge fields. Along the way, we will discuss a new type of symmetry-like operator. Next, in Section \ref{subsection:foliated_sspt}, we will construct the foliated SSPT phase in 3+1 dimensions from the exotic SSPT phase in Section \ref{subsection:exotic_sspt_phase} using the field correspondences. The foliated SSPT phase is the foliated form of the exotic SSPT phase, so we will have established the foliated-exotic duality of the 3+1d SSPT phase.

\subsection{Foliated \texorpdfstring{$BF$}{BF} Theory and Field Correspondences}
\label{subsection:foliated_bf_field_correspondence}

Again we take Euclidean spacetime to be a (2+1)-dimensional torus $\T^{2+1}$ of lengths $l^0$, $l^1$ and $l^2$, and the coordinate $(x^0, x^1, x^2)$ on it. We consider a $BF$ theory on the two-dimensional spatial manifold that is regarded as a stack of an infinite number of one-dimensional spatial submanifolds. Such a structure of a decomposition of a manifold is a codimension-one foliation and QFT on such a manifold is called foliated QFT \cite{Slagle:2018swq,Slagle:2020ugk,Hsin:2021mjn}. A codimension-one foliation is characterized by a one-form foliation field $e$, which is orthogonal to the leaves. We consider two simultaneous flat foliations $e^k = dx^k \, (k =1,2)$, where the index $k$ is the direction of the foliations. More explanations can be found in Section \ref{section:foliated_qft_foliated_gauge_fields}.

The foliated $BF$ theory in 2+1 dimensions contains two types of foliated gauge fields, which are regarded as gauge fields on the leaves, and bulk gauge fields mediating the foliated gauge fields on each leaf \cite{Slagle:2018swq,Slagle:2020ugk,Hsin:2021mjn,Ohmori:2022rzz}. The foliated gauge fields are $U(1)$ type-$A$ foliated (1+1)-form gauge field $\bm{A} = (A^k \wedge e^k, a)$ where $A^k \wedge e^k$ is a foliated (1+1)-form gauge field on the leaves of the $x^k$ foliation and $a$ is a one-form gauge field, and $U(1)$ type-$B$ foliated zero-form gauge field $\bm{\hat{\Phi}} = (\hat{\Phi}^k, \hat\Phi)$ where $\hat{\Phi}^k$ is a foliated zero-form gauge field on the leaves of the $x^k$ foliation and $\hat\Phi$ is a bulk one-form gauge field.\footnote{Here we omit the auxiliary field of the type-$B$ foliated gauge field.} The gauge transformations of the foliated gauge fields are
\begin{align}
    A^k \wedge e^k &\sim A^k \wedge e^k +d \lambda^k \wedge e^k  \,, \label{fgauge1} \\
    a &\sim a + d\lambda - \sum^{2}_{k=1} \lambda^k e^k\,, \label{fgauge3} \\
    \hat{\Phi}^k &\sim \hat{\Phi}^k +2\pi \hat{W}^k + \hat{\xi} \,, \label{fgauge2} \\
    \hat\Phi &\sim \hat\Phi + d\hat{\xi}\,, \label{fgauge4}
\end{align}
where $\bm{\lambda} = (\lambda^k e^k, \lambda)$ is a type-$A$ foliated (0+1)-form gauge parameter, $\hat{W}^k$ is an $x^k$-dependent integer-valued gauge parameter, and $\xi$ is a zero-form gauge parameter. 
The gauge parameter $\bm{\lambda}$ has its own gauge transformations $\lambda^k e^k \sim \lambda^k e^k + 2\pi d M^k$ and $\lambda \sim \lambda + 2\pi M^1 + 2\pi M^2$, where $M^k$ is an $x^k$-dependent integer-valued gauge parameter. 
Note that the constant modes of $M^k$ make $\lambda$ a $U(1)$-valued parameter. These foliated gauge fields and bulk gauge fields can have particular types of singularities and discontinuities. (See \cite{Ohmori:2022rzz} for more details.)

The foliated $BF$ Lagrangian\footnote{We use $\L$ and $L$ as letters for Lagrangians. $\L$ is often referred to as Lagrangian density and $L$ is $\L \, d^Dx$ in $D$-dimensional spacetime, so the action is written as $S = \int \L \, d^Dx = \int L$.} is
\begin{align}
\begin{split}
    L_{\text{f}} &=  \frac{iN}{2\pi} \bm{\hat{\Phi}} \wedge_\text{f} d_\text{f} \bm{A} \\
    &=  \frac{iN}{2\pi} \sum^{2}_{k=1} \hat{\Phi}^k\, d A^k \wedge e^k + \frac{iN}{2\pi} \hat{\Phi} \wedge \left( da + \sum^{2}_{k=1} A^k \wedge e^k \right) \,, \label{21folilag}
\end{split} 
\end{align}
where $N$ is an integer and $d_\text{f} \bm{A} = (dA^k \wedge e^k, da + \sum^2_{k=1}A^k \wedge e^k)$. Due to the $BF$ couplings, these $U(1)$ gauge fields are Higgsed down to $\Z_N$.

The equations of motion are
\begin{align}
     \frac{iN}{2\pi}(d\hat{\Phi}^k - \hat\Phi) \wedge e^k = 0\,, \label{feom1}\\
     \frac{iN}{2\pi}d\hat\Phi = 0\,, \label{feom2}\\
     \frac{iN}{2\pi}dA^k \wedge e^k = 0\,, \label{feom3}\\
     \frac{iN}{2\pi} \left( da + \sum^{2}_{k=1}  A^k \wedge e^k  \right)  =0\,. \label{feom4}
\end{align}

This foliated $BF$ theory is equivalent to the exotic $BF$ theory under field correspondences, which is called the foliated-exotic duality \cite{Ohmori:2022rzz,Spieler:2023wkz}. 
To derive the foliated $BF$ theory from the exotic $BF$ theory \eqref{elagrangian}, we assume the correspondences between the exotic tensor gauge fields and the foliated gauge fields as\footnote{The symbol $\simeq$ means that the correspondence between the gauge fields or parameters in exotic and foliated theory.} 
\begin{align}
    A_0 &\simeq a_0 \,,  \label{acorr_n}\\
    A_{12}  &\simeq A^1_2 + \partial_2 a_1 \,, \label{a12corr_n} \\
    \hat{\phi}^{12} &\simeq \hat{\Phi}^1 - \hat{\Phi}^2  \,, \label{bcorr_n}
\end{align}
and add to the exotic Lagrangian the term
\begin{align}
    \frac{iN}{2\pi} \hat{\Phi}' \wedge \left( da + \sum_{k=1}^2 A^k \wedge e^k  \right)  \,, \label{consttermbf}
\end{align}
where $\hat{\Phi}'$ is a Lagrange multiplier one-form field. This term imposes the constraint \eqref{feom4}, or in components,
\begin{align}
    \frac{iN}{2\pi} (\partial_1 a_2 - \partial_2 a_1 + A^2_1 - A^1_2 ) &= 0 \,, \label{aeomcomp1} \\
    \frac{iN}{2\pi} (\partial_2 a_0 - \partial_0 a_2 - A^2_0) &= 0 \,, \label{aeomcomp2} \\
    -\frac{iN}{2\pi} (\partial_0 a_1 - \partial_1 a_0 + A^1_0) &= 0 \,. \label{aeomcomp3}
\end{align} 
Note that the gauge parameters $\lambda^k e^k$ and $\xi$ on the right-hand sides of \eqref{acorr_n}--\eqref{bcorr_n} cancel out, so the degrees of freedom of the fields are consistent. We also have the correspondences between the gauge parameters
\begin{align}
    \Lambda &\simeq \lambda \ \,, \\
    \hat{w}^k  &\simeq \hat{W}^k \,, \ (k =1,2) \,.
\end{align}
Using these correspondences, we have the Lagrangian
\begin{align}
\begin{split}
    L_\text{e} &\simeq -\frac{iN}{2\pi} (\hat{\Phi}^1 - \hat{\Phi}^2) \left[ \partial_0 (A^1_2 + \partial_2 a_1 ) - \partial_1 \partial_2 a_0 \right] d^3x \\
    & \quad + \frac{iN}{2\pi} \hat{\Phi}' \wedge \left( da + \sum_{k=1}^2 A^k \wedge e^k  \right) \,.
\end{split}
\end{align}
Then, we consider a change of variables for the dynamical field from $\hat{\Phi}'$ to $\hat\Phi$:\footnote{Since we choose the correspondence \eqref{a12corr_n}, we consider $\hat{\Phi}_0' = \hat{\Phi}_0 - \partial_0 \hat\Phi^2$ to replace $A^1_2 + \partial_2 a_1$ with $A^2_1 + \partial_1 a_2$. Instead of that, we can adopt the correspondence $A_{12} \simeq A^2_1 + \partial_1 a_2$ and then we have to use the change of variable $\hat{\Phi}_0' = \hat{\Phi}_0 - \partial_0 \hat\Phi^1$.}
\begin{align}
    \hat{\Phi}_0' &= \hat{\Phi}_0 - \partial_0 \hat\Phi^2 \,, \\
    \hat{\Phi}_1' &= \hat{\Phi}_1 - \partial_1 \hat\Phi^2 \,, \\
    \hat{\Phi}_2' &= \hat{\Phi}_2 - \partial_2 \hat\Phi^1 \,.  
\end{align}
From the gauge transformation \eqref{fgauge2}, we have the gauge transformation of $\hat\Phi$ as \eqref{fgauge4}. After this change of variables, the Lagrangian becomes
\begin{align}
\begin{split}
    L_\text{e} &\simeq \frac{iN}{2\pi}  \left[ \hat\Phi^1 (\partial_2 A^1_0 - \partial_0 A^1_2) + \hat\Phi^2 ( \partial_0 A^2_1 - \partial_1 A^2_0 )   \right]d^3x \\
    & \quad + \frac{iN}{2\pi} \hat{\Phi} \wedge \left( da + \sum_{k=1}^2 A^k \wedge e^k  \right) \\
    &= L_\text{f} \,,
\end{split}
\end{align}
where we have dropped total derivative terms. Thus we have derived the foliated $BF$ Lagrangian from the exotic $BF$ Lagrangian using the field correspondences.

Conversely, we can also derive the exotic $BF$ theory from the foliated $BF$ theory. To show that, we must integrate out the time component of the fields $\hat\Phi_0$, $A^1_0$ and $A^2_0$ in the foliated $BF$ theory, then solve the equations of motion for $\hat\Phi_1$ and $\hat\Phi_2$, and plug them into the Lagrangian. However, from the form of the couplings, this manipulation leads to the same result as integrating out $\hat\Phi_1$ and $\hat\Phi_2$ instead of $A^1_0$ and $A^2_0$, then solving the equations for $A^1_0$ and $A^2_0$, and plugging them in. Here we integrate out $\hat\Phi$ for simplicity, and then we can use the equations of motion \eqref{aeomcomp1}--\eqref{aeomcomp3}. Then, the field correspondeces become
\begin{align}
    A_0 &\simeq a_0 \,,  \label{acorr}\\
    \partial_k A_0 &\simeq A^k_0 + \partial_0 a_k \,,\quad (k=1,2) \,, \label{ak0corr} \\
    A_{12}  &\simeq A^1_2 + \partial_2 a_1 = A^2_1 + \partial_1 a_2 \,, \label{a12corr} \\
    \hat{\phi}^{12} &\simeq \hat{\Phi}^1 - \hat{\Phi}^2  \,. \label{bcorr}
\end{align}
Using these correspondences, we can derive the exotic $BF$ Lagrangian from the foliated $BF$ Lagrangian.

Using the field correspondences, we can also obtain the symmetry operators and defects in the foliated $BF$ theory. 
The $\Z_N$ electric global symmetry is generated by the symmetry operator
\begin{align}
 V_m[x] = \exp \left[ i m (\hat{\Phi}^1 - \hat{\Phi}^2 ) \right] \,. \label{felesym}
\end{align}
The strip operators associated with the $\Z_N$ dipole global symmetries are
\begin{align}
    W_m^k \left[[x^k_1,x^k_2] \times \mathcal{C}^j \right] = \exp \left[ i m \int^{x^k_2}_{x^k_1} dx^k \oint_{\mathcal{C}^j} dx^j \left( A^k  \wedge e^k + d(a_k dx^k) \right) \right] \,, \quad (k=1,2) \,. \label{modstrip}
\end{align}
The quadrupole operator associated with the $\Z_N$ tensor time-like symmetry is
\begin{align}
    T_m \left[\mathcal{C}^{12,\text{rect}}(x^1_1,x^1_2,x^2_1,x^2_2)\right] = \exp\left[i m \Delta_{12} ( \hat{\Phi}^1 - \hat{\Phi}^2) (x^1_1,x^1_2,x^2_1,x^2_2) \right] \,. 
\end{align}
By using the equations of motion \eqref{feom1} and \eqref{feom2}, we can write this operator as
\begin{align}
    T_m \left[\mathcal{C}\right] = \exp\left[i m \oint_{\mathcal{C}} \hat{\Phi}\right] \,, \label{ftimesym}
\end{align}
where $\mathcal{C}$ is a one-dimensional loop.
The fracton defect is
\begin{align}
F_n[\mathcal{C}^0] = \exp \left[ i \oint_{\mathcal{C}^0} a \right]\,. \label{ffracton}
\end{align}

\subsection{Coupling to the Background Foliated Gauge Fields}
\label{subsection:coupling_background_foliated_gauge_fields}

In Section \ref{subsection:couple_background_gauge_fields}, we coupled the symmetries in the exotic $BF$ theory to the background tensor gauge fields. Since the foliated $BF$ theory is equivalent to the exotic $BF$ theory, it should be possible to couple the foliated $BF$ theory to the same set of the background gauge fields. However, in the case of the foliated theory, the structure of symmetry operators seems to be more complicated and it is non-trivial to find appropriate couplings. Here we construct the foliated $BF$ Lagrangian including background foliated and bulk gauge fields using the field correspondences in the foliated-exotic duality.

The exotic $BF$ Lagrangian coupled to background tensor gauge fields is
\begin{align}
    \L_{\text{e}} \left[ \bm{C}, \bm{\hat{C}} \right] &=  \L_{\text{e},BF}\left[ \bm{C}, \bm{\hat{C}} \right]  + \L_{\text{e}, \hat{h}}\left[ \bm{C}\right] + \L_{\text{e},\chi}\left[ \bm{\hat{C}} \right] \,,\\
    \L_{\text{e},BF} \left[ \bm{C}, \bm{\hat{C}} \right] &= \frac{iN}{2\pi} \left[ -\hat{\phi}^{12} (\partial_0 A_{12} - \partial_1 \partial_2 A_0 -C_{012} ) -  \hat{C}^{12}_0 A_{12} - \hat{C} A_0  \right] \,, \\ 
    \L_{\text{e},\hat{h}} \left[ \bm{C}\right] &= -\frac{iN}{2\pi}(2\pi \hat{h}^{12}) C_{012}\,, \\
    \L_{\text{e},\chi} \left[ \bm{\hat{C}} \right] &= -\frac{iN}{2\pi} \chi (\partial_0 \hat{C} - \partial_1 \partial_2 \hat{C}^{12}_0)  \,.
\end{align}
Firstly, we consider the $BF$ part $\L_{\text{e},BF}[ \bm{C}, \bm{\hat{C}} ]$. 
When coupled to background gauge fields, we assume that the background gauge transformations of the foliated gauge fields are
\begin{align}
    A^k\wedge e^k &\sim A^k \wedge e^k + \gamma^k \wedge e^k  \,,  \label{akbk} \\
    a &\sim a + \gamma \,, \label{abk} \\
    \hat{\Phi}^k &\sim \hat{\Phi}^k + \hat{\gamma}^k  \,,\label{bkbk} 
\end{align}
where $\gamma^k \wedge e^k$ is a background type-$A$ foliated (1+1)-form gauge parameter, $\gamma$ is a background type-$A$ bulk one-form gauge parameter and $\hat{\gamma}^k$ is a background type-$B$ foliated zero-form gauge parameter. From the background gauge invariance, we add a constraint term
\begin{align}
    \frac{iN}{2\pi} \hat{\Phi}' \wedge \left( da + \sum_{k=1}^2  A^k \wedge e^k - c \right) \,, 
\end{align}
instead of \eqref{consttermbf}, where $c$ is a two-form gauge field that has a background gauge transformation
\begin{align}
    c \sim c + d\gamma + \sum_{k=1}^2 \gamma^k \wedge e^k \,. \label{cbk}
\end{align}
Then, the field correspondences are the same as \eqref{acorr_n}--\eqref{bcorr_n}:
\begin{align}
    A_0 &\simeq a_0 \,, \\
    A_{12}  &\simeq A^1_2 + \partial_2 a_1  \,, \\
    \hat{\phi}^{12} &\simeq \hat{\Phi}^1 - \hat{\Phi}^2  \,.
\end{align}
The correspondences between the background gauge parameters are
\begin{align}
    \Gamma_0 &\simeq \gamma_0 \,, \\
    \Gamma_{12} &\simeq \gamma^1_2 + \partial_2 \gamma_1 \,, \\
    \hat{\Gamma}^{12} &\simeq \hat{\gamma}^1 - \hat{\gamma}^2 \,.
\end{align}
Note that $\gamma^2_1$ and $\gamma_2$ do not appear.\footnote{In more detail, the background gauge parameters also have their own gauge transformation. For example, $\hat{\Gamma}^{12}$ transforms as $\hat{\Gamma}^{12} \sim \hat{\Gamma}^{12} + 2\pi \hat{P}^1 - 2\pi \hat{P}^2$,  $\hat{\gamma}$ transforms as $\hat{\gamma} \sim \hat{\gamma} + d \hat{p}$, and $\hat{\gamma}^k$ transforms as $\hat{\gamma}^k \sim \hat{\gamma}^k + 2\pi \hat{p}^k  + \hat{p}$, where $\hat{p}$ is a type-$B$ bulk zero-form gauge parameter and $\hat{P}^k$ and $\hat{p}^k$ are $x^k$-dependent integer-valued gauge parameters. Then, as for the form $\hat{\gamma}^1 - \hat{\gamma}^2$, the parameter $\hat{p}$ cancels out and the gauge transformation is $2\pi \hat{p}^1 - 2\pi \hat{p}^2$, which is consistent with $\hat{\Gamma}^{12} \simeq \hat{\gamma}^1 - \hat{\gamma}^2$ under the correspondence $\hat{P}^k \simeq \hat{p}^k$.}

Using the field correspondences, $\L_{\text{e},BF}[\bm{C}, \bm{\hat{C}}]$ can be written as
\begin{align}
\begin{split}
    \L_{\text{e},BF} \left[ \bm{C}, \bm{\hat{C}} \right] &\simeq  \frac{iN}{2\pi} \left\{ -(\hat{\Phi}^1 - \hat{\Phi}^2) \left[ \partial_0 ( A^1_2 + \partial_2 a_1 ) -  \partial_1 \partial_2 a_0  - C_{012} \right] \right. \\
    & \left. \quad - \hat{C}^{12}_0 (A^1_2 + \partial_2 a_1)  -  \hat{C}a_0 \right\}   + \frac{iN}{2\pi} \hat{\Phi}' \wedge \left( da + \sum_{k=1}^2  A^k \wedge e^k - c \right) \,.
\end{split}
\end{align}
Since $\bm{A} = (A^k \wedge e^k,a)$ and $\hat{\Phi}^k$ are the foliated gauge fields in the foliated theory, we want to substitute some background foliated gauge fields for the background exotic tensor gauge fields $\bm{C} = (C_{012})$ and $\bm{\hat{C}} = (\hat{C}^{12}_0, \hat{C})$. Therefore we introduce a $U(1)$ background type-$A$ foliated (2+1)-form gauge field $C^k \wedge e^k\, (k=1,2)$, a $U(1)$ background type-$A$ bulk two-form gauge field $c$, a $U(1)$ background type-$B$ foliated one-form gauge field $\hat{C}^k \, (k=1,2)$ that obeys $\hat{C}^k_k = 0$, and the $x^1 x^2$-component of a $U(1)$ background type-$B$ bulk two-form gauge field $\hat{c}_{12}$. Then, we assume the correspondences between the background gauge fields as\footnote{Under the $90$ degree rotation $x^1 \rightarrow x^2$, $x^2 \rightarrow -x^1$, the $\bm{C}$ fields transform as $C_{012} \rightarrow -C_{012}$ and $C^1_{02} \leftrightarrow - C^2_{01}$, and the $\bm{\hat{C}}$ fields transform as $\hat{C}^{12}_0 \rightarrow -\hat{C}^{12}_0$, $\hat{C} \rightarrow \hat{C}$, $\hat{C}^1_0 \leftrightarrow \hat{C}^2_0$, $\hat{C}^1_2 \rightarrow -\hat{C}^2_1$ and $\hat{C}^2_1 \rightarrow \hat{C}^1_2$. Thus the rotational transformations on the both sides are compatible.}
\begin{align}
C_{012} &\simeq C^1_{02} + \partial_2 c_{01}  \,, \label{ccorr1_n}  \\
\hat{C}_0^{12} &\simeq \hat{C}^1_0 - \hat{C}^2_0 \,,  \label{chatcorr1} \\
\hat{C} &\simeq \partial_1 \hat{C}^1_2 - \partial_2 \hat{C}^2_1 + \hat{c}_{12} \,. \label{chatcorr2}
\end{align}
Note that the field correspondences of $\bm{\hat{C}}$ are the same form as type-$B$ foliated field in the foliated $BF$ theory with two foliations in 3+1 dimensions\cite{Spieler:2023wkz}. To impose a constraint on $C^k \wedge e^k$ and $c$, we must add to the Lagrangian the term 
\begin{align}
    -\frac{iN}{2\pi} \hat{\chi}' \left(  dc - \sum_{k=1}^2 C^k \wedge e^k \right) \,,
\end{align}
where $\hat{\chi}'$ is a zero-form dynamical field. The background gauge transformations of $C^k \wedge e^k$, $\hat{C}^k$ and $\hat{c}_{12}$ are
\begin{align}
    C^k \wedge e^k &\sim C^k \wedge e^k + d\gamma^k \wedge e^k    \,, \\
    \hat{C}^k &\sim \hat{C}^k + d\hat{\gamma}^k - \hat{\gamma}   \,, \\
    \hat{c}_{12} &\sim \hat{c}_{12} + (d\hat{\gamma})_{12} \,, 
\end{align}
where $\hat{\gamma}$ is a background type-$B$ bulk one-form gauge parameter.
Under the assumption, $\L_{\text{e},BF}[\bm{C},\bm{\hat{C}}]$ can be written as
\begin{align}
\begin{split}
    \L_{\text{e},BF}\left[\bm{C},\bm{\hat{C}}\right] &\simeq  \frac{iN}{2\pi} \left\{ -(\hat{\Phi}^1 - \hat{\Phi}^2) \left[ \partial_0 ( A^1_2 + \partial_2 a_1 ) -  \partial_1 \partial_2 a_0  - ( C^1_{02} + \partial_2 c_{01} )  \right] \right. \\
    & \left. \quad - (\hat{C}^1_0 - \hat{C}^2_0) (A^1_2 + \partial_2 a_1)  -  (\partial_1 \hat{C}^1_2 - \partial_2 \hat{C}^2_1 + \hat{c}_{12})a_0 \right\}  \\
    & \quad +\frac{iN}{2\pi} \hat{\Phi}' \wedge \left( da + \sum_{k=1}^2  A^k \wedge e^k - c \right) -\frac{iN}{2\pi} \hat{\chi}' \left(  dc - \sum_{k=1}^2 C^k \wedge e^k \right) \,.
\end{split}
\end{align}
Then, we consider changes of variables from $\hat{\Phi}'$ to $\hat\Phi$ and from $\hat{h}'$ to $\hat{h}''$:
\begin{align}
    \hat{\Phi}_0' &= \hat{\Phi}_0 - \partial_0 \hat\Phi^2 + \hat{C}^2_0 \,, \\
    \hat{\Phi}_1' &= \hat{\Phi}_1 - \partial_1 \hat\Phi^2 + \hat{C}^2_1 \,, \\
    \hat{\Phi}_2' &= \hat{\Phi}_2 - \partial_2 \hat\Phi^1 + \hat{C}^1_2 \,, \\
    \hat{\chi}' &= \hat{\chi}'' - \hat{\Phi}^2  \,.
\end{align}
We again have the dynamical gauge transformation of $\hat\Phi$ as \eqref{fgauge4}. The dynamical gauge transformation of $\hat{\chi}''$ is
\begin{align}
    \hat{\chi}'' \sim \hat{\chi}'' + 2\pi \hat{W}^2 + \hat\xi \,.
\end{align}
Also, we have the background gauge transformations of $\hat\Phi$ and $\hat{h}''$ as
\begin{align}
    \hat\Phi &\sim \hat\Phi + \hat{\gamma} \,, \label{bpass} \\
    \hat{\chi}'' &\sim \hat{\chi}'' + \hat{\gamma}^2 \,.
\end{align}    
After this change of variables, the Lagrangian becomes
\begin{align}
\begin{split}
    &L_{\text{e},BF}\left[\bm{C},\bm{\hat{C}}\right] \\
    &\simeq  \frac{iN}{2\pi} \left[  \hat{\Phi}^1 (\partial_2 A^1_0 -  \partial_0 A^1_2 + C^1_{02}) + \hat{\Phi}^2 (\partial_0 A^2_1 -  \partial_1 A^2_0 - C^2_{01})   \right. \\
    & \left. \quad -  \hat{C}^1_0 (A^1_2 + \partial_2 a_1) + \hat{C}^2_0 (A^2_1 + \partial_1 a_2)  - \hat{C}^2_1 (A^2_0 + \partial_0 a_2)  + \hat{C}^1_2 (A^1_0 + \partial_0 a_1)   -   \hat{c}_{12} a_0 \right] d^3x  \\
    & \quad - \frac{iN}{2\pi} \left(  \hat{C}^2_0 c_{12} - \hat{C}^2_1 c_{02}  + \hat{C}^1_2 c_{01}  \right) \\
    & \quad +\frac{iN}{2\pi} \hat{\Phi} \wedge \left( da + \sum_{k=1}^2  A^k \wedge e^k - c \right) - \frac{iN}{2\pi} \hat{\chi}'' \left(  dc - \sum_{k=1}^2 C^k \wedge e^k \right) \,,
\end{split}
\end{align}
where we have dropped total derivative terms. We will see later that the term 
\begin{align}
    -\frac{iN}{2\pi} \left( \hat{C}^2_0 c_{12}  -  \hat{C}^2_1 c_{02} +  \hat{C}^1_2 c_{01}  \right) d^3x \label{dropterm}
\end{align}
can be dropped by combining the SSPT phase as a counterterm, so we drop it here. Then, we obtain the foliated Lagrangian
\begin{align}
\begin{split}
    &L_{\text{f},BF}\left[\bm{C},\bm{\hat{C}}\right] \\
    &= \frac{iN}{2\pi} \sum^2_{k=1} \left[  \hat{\Phi}^k (d A^k - C^k) \wedge e^k + \hat{C}^k \wedge \left( A^k \wedge e^k + d(a_k dx^k )  \right) \right] - \frac{iN}{2\pi}  \hat{c}_{12} a_0 d^3x \\
    & \quad +\frac{iN}{2\pi} \hat{\Phi} \wedge \left( da + \sum_{k=1}^2  A^k \wedge e^k - c \right) - \frac{iN}{2\pi} \hat{\chi}'' \left(  dc - \sum_{k=1}^2 C^k \wedge e^k \right)  \,.
\end{split}
\end{align}

Next, we consider the remaining terms $\L_{\text{e},\hat{h}}[\bm{C}]$ and $\L_{\text{e},\chi}[\bm{\hat{C}}]$. As for $\L_{\text{e},\hat{h}}[\bm{C}]$, we assume the correspondence 
\begin{align}
    \hat{h}^{12} \simeq \hat{h}^1 - \hat{h}^2 \,,
\end{align}
where $\hat{h}^k\, (k=1,2)$ are $x^k$-dependent integer-valued dynamical fields. Their dynamical gauge transformations are
\begin{align}
    \hat{h}^k &\sim \hat{h}^k + \hat{W}^k + \hat{\tau} \,,
\end{align}
where $\hat{\tau}$ is an integer-valued constant gauge parameter. The gauge parameter $\hat{W}^k$ is canceled by the gauge transformations of $\hat{\Phi}^k$. 
Then, we have
\begin{align}
\begin{split}
    \L_{\text{e},\hat{h}}\left[\bm{C} \right] &= -\frac{iN}{2\pi} (2\pi \hat{h}^{12}) C_{012} \\ 
    &\simeq -\frac{iN}{2\pi} \left[ 2\pi (\hat{h}^1 - \hat{h}^2)\right] ( C^1_{02} + \partial_2 c_{01} ) \,.
\end{split}
\end{align}
We again consider a change of variables from $\hat{\chi}''$ to $\hat{\chi}$:
\begin{align}
    \hat{\chi}'' = \hat{\chi} + 2\pi \hat{h}^2 \,.
\end{align}
The dynamical gauge transformation of $\hat{\chi}$ is
\begin{align}
    \hat{\chi} \sim \hat{\chi} + \hat{\xi} - 2\pi \hat{\tau} \,, 
\end{align}
and the background gauge transformation of $\hat{\chi}$ is
\begin{align}
    \hat{\chi} \sim \hat{\chi} + \hat{\gamma}^2 \,. \label{chihatbkg_drop}
\end{align}
Adding the additional terms, $\L_{\text{e},\hat{h}}[\bm{C}]$ becomes
\begin{align}
\begin{split}
    L_{\text{f},\hat{h}}\left[C^k \wedge e^k, c \right]  
    &\simeq -\frac{iN}{2\pi}\left[ 2\pi \hat{h}^1 ( C^1_{02} + \partial_2 c_{01} )  - 2\pi \hat{h}^2 ( C^2_{01} + \partial_1 c_{02} - \partial_0 c_{12} ) \right] d^3x \\
    &= \frac{iN}{2\pi} \sum^2_{k=1} (2\pi \hat{h}^k) C^k \wedge e^k \,.
\end{split}
\end{align}
Note that we have dropped the total derivative terms and use $\partial_2 \hat{h}^1 = \partial_1 \hat{h}^2 = \partial_0 \hat{h}^2 = 0$.

As for $\L_{\text{e},\chi}[\bm{\hat{C}}]$, we use $\chi$ and introduce a dynamical type-$A$ foliated (0+1)-form field $\chi'^k e^k\, (k=1,2)$, and add a constraint term
\begin{align}
    \frac{iN}{2\pi}\left[ \hat{c}'_{01} ( \partial_2 \chi - \chi'^2) - \hat{c}'_{02} (\partial_1 \chi - \chi'^1) \right]  \,,
\end{align}
where $\hat{c}'_{01}$ and $\hat{c}'_{02}$ are dynamical fields.
Their dynamical gauge transformations are
\begin{align}
    \chi &\sim \chi + \lambda \,, \\
    \chi'^k &\sim \chi'^k + \partial_k \lambda \,.
\end{align}
Then, we can rewrite $\L_{\text{e},\chi}[\bm{\hat{C}}]$ as
\begin{align}
\begin{split}
    \L_{\text{e},\chi}\left[\bm{\hat{C}}\right] &= -\frac{iN}{2\pi} \chi (\partial_0 \hat{C} - \partial_1 \partial_2 \hat{C}^{12}_0) + \frac{iN}{2\pi}\left[ \hat{c}'_{01} ( \partial_2 \chi - \chi'^2) - \hat{c}'_{02} (\partial_1 \chi - \chi'^1) \right] \\
    &\simeq -\frac{iN}{2\pi} \chi \left[ \partial_0 (\partial_1 \hat{C}^1_2 - \partial_2 \hat{C}^2_1 + \hat{c}_{12})  - \partial_1 \partial_2 (\hat{C}^1_0 - \hat{C}^2_0) \right] \\
    & \quad  + \frac{iN}{2\pi}\left[ \hat{c}'_{01} ( \partial_2 \chi - \chi'^2) - \hat{c}'_{02} (\partial_1 \chi - \chi'^1) \right] \,.
\end{split}
\end{align}
Then, we consider changes of variables from $\hat{c}'_{01}$ and $\hat{c}'_{02}$ to $\hat{c}_{01}$ and $\hat{c}_{02}$:
\begin{align}
    \hat{c}'_{01} &= \hat{c}_{01} + \partial_0 \hat{C}^2_1 - \partial_1 \hat{C}^2_0 \,, \\
    \hat{c}'_{02} &= \hat{c}_{02} + \partial_0 \hat{C}^1_2 - \partial_2 \hat{C}^1_0 \,.
\end{align}
The background gauge transformations of $\hat{c}_{01}$ and $\hat{c}_{02}$ are
\begin{align}
    \hat{c}_{01} &\sim \hat{c}_{01} + (d\hat{\gamma})_{01} \,, \\
    \hat{c}_{02} &\sim \hat{c}_{02} + (d\hat{\gamma})_{02}  \,.
\end{align}
Performing this change of variables and dropping the total derivative terms, $\L_{\text{e},\chi}[\bm{\hat{C}}]$ becomes
\begin{align}
\begin{split}
    L_{\text{f},\chi}\left[\hat{C}^k, \hat{c}_{12}\right] 
    &= \frac{iN}{2\pi} \left[ \chi'^1 (\partial_0 \hat{C}^1_2 - \partial_2 \hat{C}^1_0 + \hat{c}_{02}) - \chi'^2 (\partial_0 \hat{C}^2_1 - \partial_1 \hat{C}^2_0 + \hat{c}_{01}) \right. \\
    &\left. \quad - \chi (\partial_0 \hat{c}_{12} - \partial_1 \hat{c}_{02} + \partial_2 \hat{c}_{01}) \right] d^3x  \\
    &= -\frac{iN}{2\pi} \left[ \sum_{k=1}^2  \chi'^k  e^k \wedge ( d\hat{C}^k + \hat{c})  + \chi d \hat{c}  \right]  \,.
\end{split}
\end{align}
Note that we combine the background gauge field $\hat{c}_{12}$ and the dynamical gauge fields $\hat{c}_{01}$ and $\hat{c}_{02}$ into $\hat{c}$.

Then, the full foliated $BF$ Lagrangian coupled to the background gauge fields is given by
\begin{align}
\begin{split}
    &L_{\text{f}} \left[ C^k \wedge e^k ,c, \hat{C}^k, \hat{c}_{12} \right] \\
    &= \frac{iN}{2\pi} \left\{ \sum^2_{k=1} \left[  \hat{\Phi}^k (d A^k - C^k) \wedge e^k + \hat{C}^k \wedge \left( A^k \wedge e^k + d(a_k dx^k )  \right) \right] - \hat{c}_{12} a_0 d^3x \right. \\
    & \qquad + \hat{\Phi} \wedge \left( da + \sum_{k=1}^2  A^k \wedge e^k + c \right) - \hat{\chi} \left(  dc - \sum_{k=1}^2 C^k \wedge e^k \right) \\  
    & \left. \qquad  + \sum^2_{k=1} (2\pi \hat{h}^k) C^k \wedge e^k - \sum_{k=1}^2  \chi'^k  e^k \wedge ( d\hat{C}^k + \hat{c})  - \chi d \hat{c} \right\} \,. 
\end{split}
\end{align}
The background foliated gauge field $C^k \wedge e^k$ is coupled to the $\Z_N$ electric symmetry \eqref{felesym}, the background bulk gauge field $c$ is coupled to the $\Z_N$ tensor time-like symmetry \eqref{ftimesym}, and the background foliated gauge field $\hat{C}^k$ is coupled to the $\Z_N$ dipole symmetries \eqref{modstrip}. From the term $\frac{iN}{2\pi} a_0 \hat{c}_{12} d^3 x$, we can say the background gauge fields $\hat{c}_{12}$ is coupled to the fracton defect $F_n[\mathcal{C}^0] = \exp \left[ in \oint_{\mathcal{C}^0} a \right]$. The symmetry action \eqref{tfctvw} can be written as
\begin{align}
    T_m \left[\mathcal{C}\right] \cdot F_n\left[\mathcal{C}^0\right] = e^{2\pi i m n /N} F_n\left[\mathcal{C}^0\right]
\end{align}
in the foliated form \cite{Slagle:2020ugk,Ohmori:2022rzz}. $T_m \left[\mathcal{C}\right]$ acts on the gauge field $\hat\Phi$, but we can formally interpret the background gauge transformation of $\hat\Phi$ \eqref{bpass} as being passively acted by the defect $F_n[\mathcal{C}^0]$. Even if $\hat{\gamma}$ were not a local transformation, this transformation would not be a symmetry transformation in the original 2+1d foliated $BF$ theory \eqref{21folilag}. Moreover, the fracton defect $F_n[\mathcal{C}^0]$ is not topological, so the defect is not a symmetry operator. However, this situation is similar to that of global symmetries, so we call the fracton defect a \textit{symmetry-like passive action operator}.

Moreover, we can rewrite the Lagrangian more simply. We perform a change of variables from $\chi'^k$ to $\chi^k$:
\begin{align}
    \chi'^k = \chi^k + a_k \,.
\end{align} 
The dynamical gauge transformation of $\chi^k e^k$ is
\begin{align}
    \chi^k e^k \sim \chi^k e^k + \lambda^k e^k \,,
\end{align}
and the background gauge transformation of
\begin{align}
    \chi^k e^k \sim \chi^k e^k -\gamma_k e^k \,, \label{chikbkg_drop}
\end{align}
The foliated Lagrangian becomes
\begin{align}
\begin{split}
    &L_{\text{f}} \left[ C^k \wedge e^k ,c, \hat{C}^k, \hat{c}_{12} \right] \\
    &= \frac{iN}{2\pi} \left\{ \sum^2_{k=1} \left[  \hat{\Phi}^k (d A^k - C^k) \wedge e^k + \hat{C}^k \wedge  A^k \wedge e^k  \right] - \hat{c} \wedge a \right. \\
    & \qquad + \hat{\Phi} \wedge \left( da + \sum_{k=1}^2  A^k \wedge e^k - c \right) - \hat{\chi} \left(  dc - \sum_{k=1}^2 C^k \wedge e^k \right) \\  
    & \left. \qquad  + \sum^2_{k=1} (2\pi \hat{h}^k)\, C^k \wedge e^k - \sum_{k=1}^2  \chi^k  e^k \wedge ( d\hat{C}^k + \hat{c})  - \chi d \hat{c}  \right\} \\
    &= \frac{iN}{2\pi} \left[ \bm{\hat{\Phi}} \wedge_\text{f} (d_\text{f} \bm{A} - \bm{C}) + \bm{\hat{C}} \wedge_\text{f} \bm{A} + d_\text{f}\bm{\hat{\chi}} \wedge_\text{f} \bm{C} - \bm{\chi} \wedge_\text{f} d_\text{f}\bm{\hat{C}}   \right]  \,, \label{fbfc}
\end{split}
\end{align}
where $\bm{\hat{\Phi}} = (\hat{\Phi}^k,\hat{\Phi})$ is a type-$B$ foliated zero-form gauge field, $\bm{A} = (A^k \wedge e^k,a)$ is a type-$A$ foliated (1+1)-form gauge field, $\bm{C} = (C^k \wedge e^k, c)$ is a type-$A$ foliated (2+1)-form gauge field, $\bm{\hat{C}} = (\hat{C}^k,\hat{c})$ is a type-$B$ foliated one-form field, $d_\text{f} \bm{\hat{\chi}} = (2\pi\hat{h}^k + \hat{\chi}, d\hat{\chi})$ is a type-$B$ foliated zero-form field\footnote{We can consider a field $\bm{\hat\chi} = (\hat{\chi}^k, \hat\chi)$ satisfying $d\hat{\chi}^k = 2\pi \hat{h}^k$, which is called a type-$B$ foliated $(-1)$-form gauge field} and $\bm{\chi} = (\chi^k e^k , \chi)$ is a type-$A$ foliated (0+1)-form field.\footnote{The fields $\bm{\chi}$ and $d_\text{f} \bm{\hat\chi}$ have different gauge transformations from the ordinary foliated gauge fields.} Note that $\hat{c}$ includes both the background gauge field $\hat{c}_{12}$ and the dynamical gauge fields $\hat{c}_{01}$ and $\hat{c}_{02}$. After all, we have constructed the 2+1d foliated $BF$ Lagrangian coupled to the background gauge fields.

To obtain the exotic theory from the foliated theory, we must integrate out $\hat{\Phi}$, $\hat{\chi}$, $\hat{c}_{01}$ and $\hat{c}_{02}$ to get the constraints
\begin{align}
    \frac{iN}{2\pi} \left( da + \sum_{k=1}^2  A^k \wedge e^k - c \right) &= 0 \,, \label{aconst_n} \\
    -\frac{iN}{2\pi} \left(  dc - \sum_{k=1}^2 C^k \wedge e^k \right) &= 0 \,, \label{cconst_n} 
\end{align}
or in components,
\begin{align}
    \frac{iN}{2\pi} (\partial_1 a_2 - \partial_2 a_1 + A^2_1 - A^1_2 - c_{12} ) &= 0 \,, \\
    \frac{iN}{2\pi} (\partial_2 a_0 - \partial_0 a_2 - A^2_0 + c_{02}) &= 0 \,, \\
    \frac{iN}{2\pi} (\partial_0 a_1 - \partial_1 a_0 + A^1_0 - c_{01}) &= 0 \,, \\
    -\frac{iN}{2\pi} ( \partial_0 c_{12} - \partial_1 c_{02} + \partial_2 c_{01} + C^1_{02} - C^2_{01} ) &= 0 \,,
\end{align}
and
\begin{align}
    \frac{iN}{2\pi}(\partial_k \chi - \chi^k - a_k) &= 0 \,, \, (k = 1,2) \label{chiconst_n} \,.
\end{align}
Then, we can use the dictionary
\begin{align}
    A_0 &\simeq a_0 \,, \\
    \partial_k A_0 &\simeq A^k_0 + \partial_0 a_k - c_{0k} \,, \quad (k=1,2)\,, \\
    A_{12}  &\simeq A^1_2 + \partial_2 a_1 = A^2_1 + \partial_1 a_2  - c_{12}  \,, \\
    \hat{\phi}^{12} &\simeq \hat{\Phi}^1 - \hat{\Phi}^2  \,, \\
    C_{012} &\simeq C^1_{02} + \partial_2 c_{01} = C^2_{01} + \partial_1 c_{02} - \partial_0 c_{12}  \,,   \\
    \hat{C}_0^{12} &\simeq \hat{C}^1_0 - \hat{C}^2_0 \,,  \\
    \hat{C} &\simeq \partial_1 \hat{C}^1_2 - \partial_2 \hat{C}^2_1 + \hat{c}_{12} \,, \\
    \hat{h}^{12} &\simeq \hat{h}^1 - \hat{h}^2 \,, \\
    \partial_k \chi &\simeq \chi^k + a_k \,, \quad (k=1,2) \,.
\end{align}
By applying these correspondences to the foliated Lagrangian \eqref{fbfc} and considering the dropped term \eqref{dropterm}, we can reproduce the exotic Lagrangian \eqref{ebfc}.

Finally, we consider the 't Hooft anomaly in the foliated theory. We will see later that the background gauge transformation of $\hat{\chi}$ and $\chi^k e^k$ given by \eqref{chihatbkg_drop} and \eqref{chikbkg_drop} can be dropped by combining the SSPT phase, so we ignore it here.
Then, under the background gauge transformations, the foliated $BF$ Lagrangian transforms as
\begin{align}
\begin{split}
    &\delta_\text{g} L_{\text{f}} \left[ C^k \wedge e^k ,c, \hat{C}^k, \hat{c}_{12} \right] \\
    &= \frac{iN}{2\pi}\left\{ \sum^2_{k=1} \left[ \hat{\gamma}^k (d A^k - C^k) \wedge e^k + (d\hat\gamma^k - \hat\gamma) \wedge A^k  \wedge e^k \right. \right.  \\
    & \left.\qquad  +  (\hat{C}^k + d\hat\gamma^k - \hat\gamma) \wedge \gamma^k \wedge e^k  \right] - \left[ d\hat\gamma \wedge a   + (\hat{c} + d\hat\gamma)\wedge \gamma \right]  \\
    & \left. \qquad + \hat{\gamma} \wedge \left( da + \sum_{k=1}^2  A^k \wedge e^k - c \right) \right\} \\ 
    &= -\frac{iN}{2\pi} \Bigg\{ \sum^2_{k=1} \left[  \hat{\gamma}^k\, C^k \wedge e^k - (\hat{C}^k + d\hat\gamma^k - \hat\gamma) \wedge \gamma^k \wedge e^k \right]   +(\hat{c} + d\hat\gamma ) \wedge \gamma + \hat{\gamma} \wedge  c \Bigg\} \,.  \label{fanomaly}
\end{split}
\end{align}
This variation is the mixed 't Hooft anomaly for the $\Z_N \times \Z_N$ subsystem symmetries in the foliated $BF$ theory.

\subsection{Foliated SSPT Phase for \texorpdfstring{$\Z_N \times \Z_N$}{ZN ZN} in 3+1 Dimensions}
\label{subsection:foliated_sspt}

In Section \ref{subsection:exotic_sspt_phase}, we saw that the mixed 't Hooft anomaly in the 2+1d exotic $BF$ theory are canceled by the 3+1d exotic SSPT phase. In this section, we construct a description of the foliated SSPT phase for $\Z_N \times \Z_N$ that is equivalent to the exotic one with two foliations by determining field correspondences. Again we take the coordinates $(x^0,x^1,x^2,x^3)$.

First, we introduce a $U(1)$ background type-$A$ foliated (2+1)-form gauge field $C^k \wedge e^k\, (k = 1,2)$, a $U(1)$ background type-$A$ bulk two-form gauge field $c$, a $U(1)$ background type-$B$ foliated one-form gauge field $\hat{C}^k\, (k = 1,2)$ that obeys $\hat{C}^k_k = 0$, and the $x^1x^2$-component of a $U(1)$ background type-$B$ bulk two-form gauge field $\hat{c}_{12}$. Their background gauge transformations are
\begin{align}
    C^k \wedge e^k &\sim C^k \wedge e^k + d\gamma^k \wedge e^k \,, \\
    c &\sim c + d\gamma + \sum_{k=1}^2 \gamma^k \wedge e^k \,, \\
    \hat{C}^k &\sim \hat{C}^k + d \hat{\gamma}^k - \hat{\gamma}  \,, \\
    \hat{c}_{12} &\sim \hat{c}_{12} + (d\hat{\gamma})_{12} \,,
\end{align}
where $\gamma^k \wedge e^k$ is a background type-$A$ foliated (1+1)-form gauge parameter, $\gamma$ is a background type-$A$ bulk one-form gauge parameter, $\hat{\gamma}^k$ is a background type-$B$ foliated zero-form gauge parameter, and $\hat{\gamma}$ is a background type-$B$ bulk one-form gauge parameter. To impose the constraints for $C^k \wedge e^k$ and $c$, we add to the Lagrangian the term
\begin{align}
    \frac{iN}{2\pi} \hat{\beta}' \wedge  \left( dc - \sum_{k=1}^2 C^k \wedge e^k \right)
\end{align}
where $\hat{\beta}'$ is a dynamical one-form field.

Next, we assume the correspondences between the background gauge fields $\bm{C}$ and $\bm{\hat{C}}$ in the exotic SSPT phase and the above background gauge fields:
\begin{align}
    C_{012} &\simeq C^1_{02} + \partial_2 c_{01} \,, \label{3cor1_n} \\
    C_{312} &\simeq C^1_{32} + \partial_2 c_{31}  \label{3cor2_n} \,, \\
    C_{03} &\simeq c_{03} \label{3cor3_n} \,, \\
    \hat{C}^{12}_0 &\simeq \hat{C}^1_0 - \hat{C}^2_0 \,,\label{3cor4_n} \\
    \hat{C} &\simeq \partial_1 \hat{C}^1_2 - \partial_2 \hat{C}^2_1 + \hat{c}_{12} \,, \label{3cor5_n}\\
    \hat{C}^{12}_3 &\simeq \hat{C}^1_3 - \hat{C}^2_3  \label{3cor6_n}\,.
\end{align}
These correspondences are consistent with the ones between the background gauge fields in the 2+1d exotic and foliated $BF$ theories. We also have the correspondences between background gauge parameters:
\begin{align}
    \Gamma_0 &\simeq \gamma_0 \,, \\
    \Gamma_3 &\simeq \gamma_3 \,, \\
    \Gamma_{12} &\simeq \gamma^1_2 + \partial_2 \gamma_1  \,, \\
    \hat{\Gamma}^{12} &\simeq \hat{\gamma}^1 - \hat{\gamma}^2 \,.
\end{align}

To restrict $\bm{C}$ and $\bm{\hat{C}}$ fields to $\Z_N$ in the exotic SSPT phase, we introduced the dynamical gauge fields $ \bm{\beta} = (\beta_0, \beta_{12}, \beta_3)$ and $\bm{\hat\beta} = (\hat{\beta}^{12})$. We also introduce dynamical gauge fields $\beta^k \wedge e^k\ (k=1,2)$, $\beta$ and $\hat{\beta}^k\ (k=1,2)$, and assume the correspondences 
\begin{align}
    \beta_{12} &\simeq \beta^1_2 + \partial_2 \beta_1 \,, \\
    \hat{\beta}^{12} &\simeq \hat{\beta}^1 - \hat{\beta}^2 \,, 
\end{align}
where $\beta^k\wedge e^k$ is a $A$-type foliated (1+1)-form gauge field and $\beta$ is a type-$A$ bulk one-form gauge field, and $\hat{\beta}^k$ is a type-$B$ foliated zero-form gauge field.\footnote{For simplicity, we use the same $\beta_0$ and $\beta_3$ in the exotic and foliated theiories.} They have dynamical gauge transformations:
\begin{align}
    \beta^k \wedge e^k &\sim \beta^k \wedge e^k + d u^k \wedge e^k   \,, \\
    \beta &\sim \beta + d u - \sum_{k=1}^2 u^k e^k \,, \\
    \hat{\beta}^k &\sim \hat{\beta}^k + 2\pi \hat{u}^k + \hat{u}   \,, 
\end{align}
where $u^k e^k$ is a type-$A$ foliated (0+1)-form gauge parameter, $u$ is a type-$A$ bulk zero-form gauge parameter, $\hat{u}^k$ is an $x^k$-dependent integer-valued gauge parameter, and $\hat{u}$ is a type-$B$ bulk zero-form gauge parameter. They correspond to the dynamical gauge parameters of $\bm{\hat{\beta}}$ and $\bm{\beta}$ as
\begin{align}
    s &\simeq u \,, \\
    \hat{s}^k &\simeq \hat{u}^k \,.
\end{align}
They also have background gauge transformations:
\begin{align}
    \beta^k \wedge e^k &\sim \beta^k \wedge e^k - \gamma^k \wedge e^k  \,, \\ 
    \beta &\sim \beta - \gamma  \,, \\
    \hat{\beta}^k &\sim \hat{\beta}^k - \hat{\gamma}^k  \,.
\end{align}
Moreover, to impose the constraint for $\beta^k\wedge e^k$, $\beta$ and $c$, we add the term
\begin{align}
    \frac{iN}{2\pi} (\hat{c}'\, |_{\hat{c}'_{12} = 0}) \wedge \left(d\beta + \sum_{k=1}^2 \beta^k \wedge e^k + c \right)  \,,
\end{align}
where $\hat{c}_{ij}\, ((i,j) = (0,1),(0,2),(0,3),(2,3),(3,1))$ is the $x^ix^j$-component of dynamical two-form field.

Let us construct the foliated Lagrangian describing the 3+1d SSPT phase. The exotic SSPT phase with two foliations is described by the Lagrangian \eqref{sspte}
\begin{align}
    \L_{\text{SSPT},\text{e}} \left[ \bm{C}, \bm{\hat{C}} \right] &= \L_{\text{SSPT},\text{e},\hat{\beta}}\left[ \bm{C}\right] + \L_{\text{SSPT},\text{e},\beta}\left[\bm{\hat{C}} \right] + \L_{\text{SSPT},\text{e}, C\hat{C}}\left[ \bm{C}, \bm{\hat{C}} \right] \,, \\
    \L_{\text{SSPT},\text{e}, \hat{\beta}}\left[ \bm{C} \right] &= \frac{iN}{2\pi} \hat{\beta}^{12} \left( \partial_0 C_{312} - \partial_3 C_{012} + \partial_1 \partial_2 C_{03} \right) \,, \\
    \L_{\text{SSPT},\text{e},\beta}\left[ \bm{\hat{C}} \right] &= \frac{iN}{2\pi} \left[ \beta_0 \left( \partial_3 \hat{C} - \partial_1 \partial_2 \hat{C}^{12}_3  \right) + \beta_{12} \left( \partial_3 \hat{C}^{12}_0 - \partial_0 \hat{C}^{12}_3  \right) \right.\notag \\
    & \qquad -\left. \beta_3 \left( \partial_0 \hat{C} - \partial_1 \partial_2 \hat{C}^{12}_0  \right) \right]  \,,  \\
    \L_{\text{SSPT},\text{e}, \hat{C}C}\left[ \bm{C}, \bm{\hat{C}} \right] &=   \frac{iN}{2\pi}  \left( \hat{C}^{12}_3 C_{012}  - \hat{C}^{12}_0 C_{312}  +  \hat{C} C_{03}   \right) \,.
\end{align}
Using the correspondences above, we can rewrite it in terms of the foliated fields:
\begin{align}
    \L_{\text{SSPT},\text{e}, \hat{\beta}} \left[ \bm{C} \right] \simeq  \frac{iN}{2\pi} (\hat{\beta}^1 - \hat{\beta}^2) \left[ \partial_0 ( C^1_{32} + \partial_2 c_{31} ) - \partial_3 ( C^1_{02} + \partial_2 c_{01} ) + \partial_1 \partial_2 c_{03} \right] \,,
\end{align}
and
\begin{align}
\begin{split}
    \L_{\text{SSPT},\text{e},\beta} \left[ \bm{\hat{C}} \right]
    & \simeq  \frac{iN}{2\pi} \left\{ \beta_0 \left[ \partial_3  \left( \partial_1 \hat{C}^1_2  - \partial_2 \hat{C}^2_1 + \hat{c}_{12} \right) - \partial_1 \partial_2 \left(\hat{C}^1_3 - \hat{C}^2_3  \right) \right] \right.   \\
    & \quad + (\beta^1_2 + \partial_2 \beta_1) \left[ \partial_3 \left( \hat{C}^1_0 - \hat{C}^2_0 \right) - \partial_0 \left( \hat{C}^1_3 - \hat{C}^2_3 \right) \right] \\
    & \quad \left. - \beta_3 \left[ \partial_0 \left( \partial_1 \hat{C}^1_2 - \partial_2 \hat{C}^2_1 + \hat{c}_{12} \right) - \partial_1 \partial_2 \left( \hat{C}^1_0 - \hat{C}^2_0 \right)  \right] \right\} \,,
\end{split}
\end{align}
and
\begin{align}
\begin{split}
    \L_{\text{SSPT},\text{e}, \hat{C}C} \left[ \bm{C}, \bm{\hat{C}} \right] & \simeq   \frac{iN}{2\pi}  \left[ ( \hat{C}^1_3 - \hat{C}^2_3 ) ( C^1_{02} + \partial_2 c_{01} )  -  ( \hat{C}^1_0 - \hat{C}^2_0 )( C^1_{32} + \partial_2 c_{31} )  \right. \\
    & \quad \left. +  ( \partial_1 \hat{C}^1_2  - \partial_2 \hat{C}^2_1 + \hat{c}_{12} ) c_{03}  \right] \,.
\end{split}
\end{align}
We consider changes of variables from $\hat{\beta}'$ to $\hat{\beta}$:
\begin{align}
    \hat{\beta}'_0 &= \hat{\beta}_0 - \partial_0 \hat{\beta}^2 - \hat{C}^2_0 \,, \\
    \hat{\beta}'_1 &= \hat{\beta}_1 - \partial_1 \hat{\beta}^2 - \hat{C}^2_1 \,, \\
    \hat{\beta}'_2 &= \hat{\beta}_2 - \partial_2 \hat{\beta}^1 - \hat{C}^1_2 \,, \\
    \hat{\beta}'_3 &= \hat{\beta}_3 - \partial_3 \hat{\beta}^2 - \hat{C}^2_3 \,,
\end{align}
and from $\hat{c}'_{ij}$ to $\hat{c}_{ij}$:
\begin{align}
    \hat{c}'_{01} &= \hat{c}_{01} + \partial_0 \hat{C}^2_1 - \partial_1 \hat{C}^2_0 \,, \\
    \hat{c}'_{02} &= \hat{c}_{02} + \partial_0 \hat{C}^1_2 - \partial_2 \hat{C}^1_0 \,, \\
    \hat{c}'_{03} &= \hat{c}_{03} + \partial_0 \hat{C}^2_3 - \partial_3 \hat{C}^2_0 \,, \\
    \hat{c}'_{23} &= \hat{c}_{23} + \partial_2 \hat{C}^1_3 - \partial_3 \hat{C}^1_2 \,, \\
    \hat{c}'_{31} &= \hat{c}_{31} + \partial_3 \hat{C}^2_1 - \partial_1 \hat{C}^2_3 \,.
\end{align}
The background gauge transformations of $\hat{\beta}$ and $\hat{c}$ are
\begin{align}
    \hat{\beta} &\sim \hat{\beta}  - \hat{\gamma}  \,, \\
    \hat{c}_{ij} &\sim \hat{c}_{ij} + (d\hat{\gamma})_{ij} \,,
\end{align}
and the dynamical gauge transformation of $\hat{\beta}$ are
\begin{align}
    \hat{\beta} &\sim \hat{\beta} + d \hat{u}   \,.
\end{align}
Then, adding the additional terms, we have
\begin{align}
\begin{split}
    \L_{\text{SSPT},\text{e}, \hat{\beta}} \left[ \bm{C} \right] &\simeq  \frac{iN}{2\pi} \left\{ \hat{\beta}^1(\partial_0 C^1_{32} - \partial_3 C^1_{02} + \partial_2 C^1_{03})- \hat{\beta}^2 (\partial_0 C^2_{31} - \partial_3 C^2_{01} + \partial_1 C^2_{03}) \right.\\
    &\quad - \partial_0 \left[ \hat{\beta}^2 (\partial_2 c_{31} + \partial_3 c_{12} + \partial_1 c_{23} + C^1_{32} - C^2_{31}) \right] \\
    &\quad - \partial_1 \left[ \hat{\beta}^2 ( -\partial_0 c_{23} + \partial_2 c_{03} - \partial_3 c_{02} - C^2_{03}) \right] \\
    &\quad - \partial_2 \left[ \hat{\beta}^1 ( -\partial_0 c_{31} - \partial_1 c_{03} + \partial_3 c_{01} + C^1_{03}) \right] \\
    &\left. \quad - \partial_3 \left[ \hat{\beta}^2 ( - \partial_2 c_{01} - \partial_0 c_{12} + \partial_1 c_{02} + C^2_{01} - C^1_{02}) \right] \right\} \,,
\end{split}
\end{align}
and
\begin{align}
\begin{split}
    \L_{\text{SSPT},\text{e},\beta} \left[ \bm{\hat{C}} \right]
    & \simeq  \frac{iN}{2\pi} \left[ (\beta^1_0 + \partial_0 \beta_1)(\partial_2 \hat{C}^1_3  - \partial_3 \hat{C}^1_2) + (\beta^2_0 + \partial_0 \beta_2)(\partial_3 \hat{C}^2_1  - \partial_1 \hat{C}^2_3)  \right.   \\
    & \quad + (\beta^1_2 + \partial_2 \beta_1) (\partial_3 \hat{C}^1_0  - \partial_0 \hat{C}^1_3 ) + (\beta^2_1 + \partial_1 \beta_2) (\partial_0 \hat{C}^2_3  - \partial_3 \hat{C}^2_0) \\
    & \quad + (\beta^1_3 + \partial_3 \beta_1)(\partial_0 \hat{C}^1_2  - \partial_2 \hat{C}^1_0) + (\beta^2_3 + \partial_3 \beta_2)(\partial_1 \hat{C}^2_0  - \partial_0 \hat{C}^2_1) \\
    & \left. \quad + \beta_0 \partial_3 \hat{c}_{12} - \beta_3 \partial_0 \hat{c}_{12}  \right] \\
    &= \frac{iN}{2\pi} \left\{ \beta^1_0 (\partial_2 \hat{C}^1_3  - \partial_3 \hat{C}^1_2) + \beta^2_0 (\partial_3 \hat{C}^2_1  - \partial_1 \hat{C}^2_3) + \beta^1_2  (\partial_3 \hat{C}^1_0  - \partial_0 \hat{C}^1_3 ) \right.   \\
    & \quad + \beta^2_1 (\partial_0 \hat{C}^2_3  - \partial_3 \hat{C}^2_0)  + \beta^1_3 (\partial_0 \hat{C}^1_2  - \partial_2 \hat{C}^1_0) + \beta^2_3 (\partial_1 \hat{C}^2_0  - \partial_0 \hat{C}^2_1) \\
    & \quad + \hat{c}_{12} (\partial_0 \beta_3 - \partial_3 \beta_0) \\
    &  \quad + \partial_0\left[ \beta_1 (\partial_2 \hat{C}^1_3  - \partial_3 \hat{C}^1_2) + \beta_2 (\partial_3 \hat{C}^2_1  - \partial_1 \hat{C}^2_3) -\beta_3 \hat{c}_{12} \right]\\
    &  \quad + \partial_1\left[ \beta_2 (\partial_0 \hat{C}^2_3  - \partial_3 \hat{C}^2_0)\right] + \partial_2 \left[\beta_1 (\partial_3 \hat{C}^1_0  - \partial_0 \hat{C}^1_3)  \right]\\
    & \left. \quad + \partial_3 \left[ \beta_1 (\partial_0 \hat{C}^1_2  - \partial_2 \hat{C}^1_0) + \beta_2 (\partial_1 \hat{C}^2_0  - \partial_0 \hat{C}^2_1) + \beta_0 \hat{c}_{12} \right]  \right\} \,,
\end{split}
\end{align}
and
\begin{align}
\begin{split}
    \L_{\text{SSPT},\text{e}, \hat{C}C} \left[ \bm{C}, \bm{\hat{C}} \right] & \simeq   \frac{iN}{2\pi}  \left[ \hat{C}^1_3 C^1_{02} - \hat{C}^1_0 C^1_{32} - \hat{C}^2_3 C^2_{01} + \hat{C}^2_0 C^2_{31}\right. \\
    & \quad - \hat{C}^1_2 C^1_{03} + \hat{C}^2_1 C^2_{03} + \hat{c}_{12} c_{03}    \\
    & \quad + \partial_0 (\hat{C}^2_1 c_{23} - \hat{C}^1_2 c_{13} + \hat{C}^2_3 c_{12}) \\
    & \quad + \partial_1 (\hat{C}^1_2 c_{03} - \hat{C}^2_0 c_{23} + \hat{C}^2_3 c_{02}) \\
    & \quad + \partial_2 (-\hat{C}^2_1 c_{03} - \hat{C}^1_0 c_{31} + \hat{C}^1_3 c_{01}) \\
    &  \left.\quad + \partial_3 (-\hat{C}^2_0 c_{12} + \hat{C}^2_1 c_{02} - \hat{C}^1_2 c_{01} ) \right] \,.
\end{split}
\end{align}
For later convenience, we have left the total derivative terms. We will put the SSPT on the region $x^3 \geq 0$ and the foliated $BF$ theory on the boundary $x^3 = 0$. Then, we can drop the total derivative terms except for the $x^3$-derivative term and the $x^3$-derivative term
\begin{align}
    \frac{iN}{2\pi} \left[ - \partial_3 \left(\hat{C}^2_0 c_{12} - \hat{C}^2_1 c_{02} S+ \hat{C}^1_2 c_{01} \right) \right] d^4 x  \label{bdcc}
\end{align}
cancels out the boundary term \eqref{dropterm}, so we drop this term.
As a result, we obtain the Lagrangian
\begin{align}
\begin{split}
    &L_{\text{SSPT},\text{f}} \left[ C^k \wedge e^k, c, \hat{C}^k, \hat{c}_{12} \right] \\
    &=  \frac{iN}{2\pi} \Bigg\{ -\sum_{k=1}^2 \hat{\beta}^k \, dC^k \wedge e^k + \sum_{k=1}^2 \beta^k \wedge e^k \wedge d\hat{C}^k + \hat{c}_{12}(d\beta)_{03}d^4x \\
    & \qquad + \sum_{k=1}^2 \hat{C}^k \wedge C^k \wedge e^k  + \hat{c}_{12} c_{03}d^4x \\
    &\qquad +  \hat{\beta} \wedge  \left( dc - \sum_{k=1}^2 C^k \wedge e^k \right) + (\hat{c}\, |_{\hat{c}_{12} = 0}) \wedge \left(d\beta + \sum_{k=1}^2 \beta^k \wedge e^k + c \right) \\
    &\qquad + \partial_3 \Big[ -\hat{\beta}^2 ( - \partial_2 c_{01} - \partial_0 c_{12} + \partial_1 c_{02} + C^2_{01} - C^1_{02}) \\
    &  \qquad + \beta_1 (\partial_0 \hat{C}^1_2  - \partial_2 \hat{C}^1_0) + \beta_2 (\partial_1 \hat{C}^2_0  - \partial_0 \hat{C}^2_1) + \beta_0 \hat{c}_{12} \Big] d^4x \Bigg\} \,.
\end{split}
\end{align}
We combine the background gauge field $\hat{c}_{12}$ and the dynamical gauge fields $\hat{c}_{ij}\, ((i,j) = (0,1),(0,2),(0,3),(2,3),(3,1))$ into $\hat{c}$. We can write
\begin{align}
\begin{split}
    &\left[ \hat{c}_{12} ( c_{03} + (d\beta)_{03} ) \right] d^4x + (\hat{c}\, |_{\hat{c}_{12} = 0}) \wedge \left(d\beta + \sum_{k=1}^2 \beta^k \wedge e^k + c \right) \\
    & = \hat{c} \wedge \left(d\beta + \sum_{k=1}^2 \beta^k \wedge e^k + c \right) \\
    & = \beta \wedge  d \hat{c}   + \sum_{k=1}^2  \beta^k \wedge e^k \wedge \hat{c} + \hat{c} \wedge c \\
    & \qquad + \partial_3 (- \hat{c}_{01} \beta_2 + \hat{c}_{02} \beta_1 - \hat{c}_{12} \beta_0 ) d^4x \,.
\end{split}
\end{align}
If we put this theory on the spacetime $\T^{2+1} \times \R_{x^3 \geq 0}$, we have the term
\begin{align}
    -\frac{iN}{2\pi} \Big[ \hat{\beta}^2 \left( dc - \sum_{k=1}^2 C^k \wedge e^k \right)_{012} + \beta_1 \left(d\hat{C}^k + \hat{c} \right)_{02} - \beta_2 \left(d\hat{C}^k + \hat{c} \right)_{01}  \Big] d^3x \label{droppedtermbeta}
\end{align}
on the boundary $x^3 =0$. In the foliated $BF$ theory on the boundary, we consider changes of variables from $\hat{\chi}$ to $\hat{\chi}'''$ and from $\chi^k$ to $\chi''^k$ as
\begin{align}
    \hat{\chi} &= \hat{\chi}''' + \hat{\beta}^2 \,, \\
    \chi^1 &= \chi''^1 + \beta_1 \,, \\
    \chi^2 &= \chi''^2 + \beta_2 \,.
\end{align}
The background gauge transformations of $\hat{\chi}$ and $\chi^k e^k$ given by \eqref{chihatbkg_drop} and \eqref{chikbkg_drop} vanish and the boundary term \eqref{droppedtermbeta} is canceled out. Thus, we omit the primes on $\hat{\chi}'''$ and $\chi''^k e^k$. Then, the foliated Lagrangian of the SSPT phase for $\Z_N \times \Z_N$ is
\begin{align}
\begin{split}
    &L_{\text{SSPT},\text{f}} \left[ C^k \wedge e^k, c, \hat{C}^k, \hat{c}_{12} \right] \\
    &=  \frac{iN}{2\pi} \Bigg[ -\sum_{k=1}^2 \hat{\beta}^k \, dC^k \wedge e^k +  \hat{\beta} \wedge  \left( dc - \sum_{k=1}^2 C^k \wedge e^k \right)    \\
    & \qquad + \sum_{k=1}^2 \beta^k \wedge e^k \wedge (d\hat{C}^k + \hat{c}) + \beta \wedge d\hat{c}  + \sum_{k=1}^2 \hat{C}^k \wedge C^k \wedge e^k + \hat{c} \wedge c \Bigg] \\
    &=  \frac{iN}{2\pi} \left[ -\bm{\hat\beta} \wedge_\text{f} d_\text{f} \bm{C} +  \bm{\beta} \wedge_\text{f} d_\text{f} \bm{\hat{C}} + \bm{\hat{C}} \wedge_\text{f} \bm{C}   \right] \,, \label{foliatedssptlag_n}
\end{split}
\end{align}
where $\bm{C} = (C^k \wedge e^k, c)$ is a type-$A$ foliated (2+1)-form gauge field, $\bm{\hat{C}} = (\hat{C}^k, \hat{c})$ is a type-$B$ foliated one-form gauge field, $\bm{\beta} = (\beta^k \wedge e^k, \beta)$ is a type-$A$ foliated (1+1)-form gauge field, and $\bm{\hat\beta} = (\hat{\beta}^k, \hat{\beta})$ is a type-$B$ foliated zero-form gauge field.

As in the case of the exotic SSPT phase, if the theory is on spacetime without a boundary, it is gauge invariant. However, if spacetime has a boundary, the partition function of the 3+1d SSPT phase is not invariant. From the anomaly inflow mechanism \cite{Callan:1984sa}, this variation is expected to be canceled by the anomaly of the 2+1d foliated $BF$ theory on the boundary \eqref{fanomaly}. To see this, we put the foliated SSPT phase on the spacetime $\T^{2+1} \times \R_{x^3 \geq 0}$ with the boundary $x^3 = 0$. From the gauge invariance, the boundary conditions of the dynamical gauge parameters are
\begin{align}
    u^k |_{x^3 = 0} &= 0 \,, \\ 
    u\, |_{x^3 = 0} &= 0 \,, \\
    \hat{u}^k |_{x^3 = 0}  &= 0 \,, \\
    \hat{u}\, |_{x^3 = 0} &= 0 \,,
\end{align}
which are consistent with the boundary conditions in the exotic SSPT phase. On the boundary, we put the 2+1d foliated $BF$ theory coupled to the background gauge fields \eqref{fbfc}, and the background gauge fields in the 3+1d SSPT phase are related to those in the 2+1d foliated $BF$ theory as\footnote{On the boundary, $\hat{c}_{\text{SSPT},01}$ and $\hat{c}_{\text{SSPT},02}$ do not appear. Then, these have no relation to $\hat{c}_{BF,01}$ and $\hat{c}_{BF,02}$.}
\begin{align}
    C^k_{\text{SSPT}} \wedge e^k \, |_{x^3 = 0} &= C^k_{BF} \wedge e^k \,, \\
    \hat{C}^k_{\text{SSPT}}\, |_{x^3 = 0} &= \hat{C}^k_{BF} \,, \\ 
    c_{\text{SSPT}}\, |_{x^3 = 0} &= c_{BF} \,, \\
    \hat{c}_{\text{SSPT},12}\, |_{x^3 = 0} &= \hat{c}_{BF,12} \,.
\end{align}
Note that while the background gauge fields in the 3+1d SSPT phase are restricted to $\Z_N$ tensor gauge fields by the dynamical fields $\hat{\beta}^k$, $\hat{\beta}$ $\beta^k \wedge e^k$ and $\beta$, those in the 2+1d foliated $BF$ theory are restricted by the dynamical fields $\hat{h}^k$, $\hat{\chi}$ $\chi^k e^k$ and $\chi$.
Then, under the background gauge transformations, the Lagrangian transforms as
\begin{align}
\begin{split}
    &\delta_\text{g} L_{\text{SSPT,f}} \left[ C^k \wedge e^k ,c, \hat{C}^k, \hat{c}_{12} \right] \\
    &=  \frac{iN}{2\pi}  \Bigg[ \sum_{k=1}^2 \hat{\gamma}^k d C^k  \wedge e^k - \hat{\gamma} \wedge \left( dc - \sum_{k=1}^2 C^k \wedge e^k \right)    \\
    & \qquad - \sum_{k=1}^2 \gamma^k \wedge e^k \wedge (d\hat{C}^k + \hat{c}) - \gamma \wedge d\hat{c}   + \sum_{k=1}^2 (d\hat{\gamma}^k - \hat{\gamma}) \wedge C^k \wedge e^k + d\hat{\gamma} \wedge c  \\
    & \qquad + \sum_{k=1}^2 (\hat{C}^k + d\hat{\gamma}^k - \hat{\gamma}) \wedge d\gamma^k \wedge e^k + (\hat{c} + d\hat{\gamma} ) \wedge \left(d\gamma + \sum_{k=1}^2 \gamma^k \wedge e^k \right) \Bigg] \\
    & =  \frac{iN}{2\pi} d \Bigg[ \sum_{k=1}^2 \left\{ \hat{\gamma}^k \,C^k \wedge e^k - (\hat{C} + d{\gamma}^k - \hat{\gamma}) \wedge \gamma^k \wedge e^k \right\} + (\hat{c} + d\hat{\gamma})\wedge \gamma + \hat{\gamma} \wedge c \Bigg] \,.
\end{split}
\end{align}
Thus on the boundary, the term 
\begin{align}
\begin{split}
    &\delta_\text{g} S_{\text{SSPT,f}}\left[ C^k \wedge e^k ,c, \hat{C}^k, \hat{c}_{12} \right]\\
     &= \int_{x^3=0} \frac{iN}{2\pi} \Bigg[ \sum_{k=1}^2 \left\{ \hat{\gamma}^k \,C^k \wedge e^k - (\hat{C} + d{\gamma}^k - \hat{\gamma}) \wedge \gamma^k \wedge e^k \right\} + (\hat{c} + d\hat{\gamma})\wedge \gamma + \hat{\gamma} \wedge c \Bigg]  \label{anomcan2}
\end{split}
\end{align}
arises. This boundary term of the 3+1d foliated SSPT phase matches the 't Hooft anomaly of the 2+1d foliated $BF$ theory \eqref{fanomaly} on the boundary $x^3=0$. Therefore we can see that the 't Hooft anomaly of the 2+1d foliated $BF$ theory on the boundary is canceled by the gauge-variation of the 3+1d foliated SSPT phase on the bulk.

To obtain the exotic SSPT phase from the foliated SSPT phase, we must integrate out $\hat{\beta}$ and $\hat{c}_{ij}\, ((i,j) = (0,1),(0,2),(0,3),(2,3),(3,1))$ to get the constraints
\begin{align}
    \frac{iN}{2\pi} \left( dc - \sum_{k=1}^2  C^k \wedge e^k \right) &= 0 \,, \label{cconst_n_sspt} \\
    \frac{iN}{2\pi} \left(  d\beta + \sum_{k=1}^2 \beta^k \wedge e^k + c \right)_{ij} &= 0 \,,\quad ((i,j) \neq (0,3)) \,, \label{betaconst_n_sspt} 
\end{align}
or in components,
\begin{align}
    \frac{iN}{2\pi} ( \partial_1 c_{23} + \partial_2 c_{31} + \partial_3 c_{12} + C^1_{32} - C^2_{31} ) &= 0 \,, \\
    \frac{iN}{2\pi} ( -\partial_0 c_{23} + \partial_2 c_{03} - \partial_3 c_{02} - C^2_{03} ) &= 0 \,, \\
    \frac{iN}{2\pi} ( -\partial_0 c_{31} - \partial_1 c_{03} + \partial_3 c_{01} + C^1_{03} ) &= 0 \,, \\
    \frac{iN}{2\pi} ( -\partial_0 c_{12} + \partial_1 c_{02} - \partial_2 c_{01} - C^1_{02} + C^2_{01} ) &= 0 \,, 
\end{align}
and
\begin{align}
    \frac{iN}{2\pi} (\partial_2 \beta_3 - \partial_3 \beta_2 - \beta^2_3 + c_{23}  ) &= 0 \,, \\
    \frac{iN}{2\pi} (-\partial_1 \beta_3 + \partial_3 \beta_1 + \beta^1_3 + c_{31}  ) &= 0 \,, \\ 
    \frac{iN}{2\pi} (\partial_1 \beta_2 - \partial_2 \beta_1  + \beta^2_1 - \beta^1_2 + c_{12}
     ) &= 0 \,, \\
    \frac{iN}{2\pi} (\partial_0 \beta_1 - \partial_1 \beta_0 + \beta^1_0 + c_{01}  ) &= 0 \,, \\
    \frac{iN}{2\pi} (\partial_0 \beta_2 - \partial_2 \beta_0 + \beta^2_0 + c_{02}  ) &= 0 \,. 
\end{align}
Then, we can use the dictionary
\begin{align}
    C_{012} &\simeq C^1_{02} + \partial_2 c_{01} = C^2_{01} + \partial_1 c_{02} - \partial_0 c_{12} \,, \label{3cor1} \\
    C_{312} &\simeq C^1_{32} + \partial_2 c_{31} = C^2_{31} + \partial_1 c_{32} - \partial_3 c_{12} \label{3cor2} \,, \\
    C_{03} &\simeq c_{03} \label{3cor3} \,, \\
    \partial_1 C_{03} &\simeq C^1_{03} + \partial_3 c_{01}  -  \partial_0 c_{31} \,, \label{3cor4}\\
    \partial_2 C_{03} &\simeq C^2_{03} + \partial_3 c_{02} - \partial_0 c_{32} \label{3cor5}   \,, \\
    \hat{C}^{12}_0 &\simeq \hat{C}^1_0 - \hat{C}^2_0 \,,\label{3cor6} \\
    \hat{C} &\simeq \partial_1 \hat{C}^1_2 - \partial_2 \hat{C}^2_1 + \hat{c}_{12} \,, \label{3cor7}\\
    \hat{C}^{12}_3 &\simeq \hat{C}^1_3 - \hat{C}^2_3  \label{3cor8}\,, \\
    \partial_k \beta_0 &\simeq \beta^k_0 + \partial_0 \beta_k + c_{0k} \,, \quad (k=1,2) \,, \\
    \partial_k \beta_3 &\simeq \beta^k_3 + \partial_3 \beta_k + c_{3k} \,, \quad (k=1,2) \,, \\
    \beta_{12} &\simeq \beta^1_2 + \partial_2 \beta_1 = \beta^2_1 + \partial_1 \beta_2 + c_{12} \,, \\
    \hat{\beta}^{12} &\simeq \hat{\beta}^1 - \hat{\beta}^2 \,. 
\end{align}
By applying these correspondences to the foliated SSPT Lagrangian \eqref{foliatedssptlag_n} and dropping the total derivative terms, we can reproduce the exotic SSPT Lagrangian \eqref{sspte}.

\section{Change of Foliation Structure}
\label{section:change_foliation}

In Section \ref{subsection:foliated_sspt}, we obtained the 3+1d foliated SSPT phase for $\Z_N \times \Z_N$ with two foliations:
\begin{align}
\begin{split}
    &L^2_{\text{SSPT},\text{f}} \left[ C^k \wedge e^k, c, \hat{C}^k, \hat{c}_{12} \right] \\
    &=  \frac{iN}{2\pi} \Bigg[ -\sum_{k=1}^2 \hat{\beta}^k \, dC^k \wedge e^k +  \hat{\beta} \wedge  \left( dc - \sum_{k=1}^2 C^k \wedge e^k \right)    \\
    & \qquad + \sum_{k=1}^2 \beta^k \wedge e^k \wedge (d\hat{C}^k + \hat{c}) + \beta \wedge d\hat{c}  + \sum_{k=1}^2 \hat{C}^k \wedge C^k \wedge e^k + \hat{c} \wedge c \Bigg] \,. \label{2foliated_sspt_cf}
\end{split}
\end{align}

Here we change the foliation structure from two foliations $e^k = dx^k\, (k=1,2)$ to three foliations $e^k = dx^k\, (k=1,2,3)$. By adding gauge fields with $k=3$ and modifying the gauge transformations, we can easily construct the 3+1d foliated SSPT phase with three foliations. The SSPT phase with a boundary also cancels the mixed 't Hooft anomaly of the 2+1d foliated $BF$ theory on the boundary. The situation where other (exotic) SSPT phases with different foliations cancel the same anomaly appears in \cite{Burnell:2021reh,Luo:2022mrj}.

Moreover, we will assume correspondences between the exotic and foliated fields in the SSPT phases with three foliations, and convert from the foliated SSPT phase with three foliations to the exotic SSPT phase with $S_4$ rotational invariance. While it is non-trivial to construct the exotic form of the 3+1d SSPT phase with $S_4$ rotational invariance from the one with $\Z_4$ rotational invariance, we can construct it via the foliated theory. It is a systematic way to construct the exotic theory using the foliated-exotic duality.

\subsection{Foliated SSPT Phase with Three Foliations}
\label{subsection:foliated_sspt_3foliations}

Firstly, we construct the 3+1d foliated SSPT phase with three foliations. We introduce a $U(1)$ background type-$A$ foliated (2+1)-form gauge field $C^k \wedge e^k\ (k = 1,2,3)$, a $U(1)$ background type-$A$ bulk two-form gauge field $c$, a $U(1)$ background type-$B$ foliated one-form gauge field $\hat{C}^k\ (k = 1,2,3)$ that obeys $\hat{C}^k_k = 0$, and the $x^ix^j$-component of a $U(1)$ background type-$B$ bulk two-form gauge field $\hat{c}_{ij}\, ((i,j) = (1,2), (2,3), (3,1))$. Their background gauge transformations are
\begin{align}
    C^k \wedge e^k &\sim C^k \wedge e^k + d\gamma^k \wedge e^k \,, \\
    c &\sim c + d\gamma + \sum_{k=1}^3 \gamma^k \wedge e^k \,, \\
    \hat{C}^k &\sim \hat{C}^k + d \hat{\gamma}^k - \hat{\gamma}  \,, \\
    \hat{c}_{ij} &\sim \hat{c}_{ij} + (d\hat{\gamma})_{ij} \,,
\end{align}
where $\gamma^k \wedge e^k$ is a background type-$A$ foliated (1+1)-form gauge parameter, $\gamma$ is a background type-$A$ bulk one-form gauge parameter, $\hat{\gamma}^k$ is a background type-$B$ foliated zero-form gauge parameter, and $\hat{\gamma}$ is a background type-$B$ bulk one-form gauge parameter.

To restrict $C^k \wedge e^k$, $c$, $\hat{C}^k$ and $\hat{c}_{ij}\, ((i,j) = (1,2), (2,3), (3,1))$ to $\Z_N$, we introduce a $U(1)$ dynamical type-$A$ foliated (1+1)-form gauge field $\beta^k \wedge e^k \ (k=1,2,3)$, a $U(1)$ dynamical type-$A$ bulk  one-form gauge field $\beta$, a $U(1)$ dynamical foliated type-$B$ zero-form gauge field $\hat{\beta}^k \ (k =1,2,3)$ and a $U(1)$ dynamical bulk type-$B$ one-form gauge field. We also introduce the $x^0x^i$-component of a $U(1)$ background type-$B$ bulk two-form gauge field $\hat{c}_{0i}\, (i = 1,2,3)$ for a constraint on $\beta^k \wedge e^k$ and $\beta$. They have background gauge transformations:
\begin{align}
    \beta^k \wedge e^k &\sim \beta^k \wedge e^k - \gamma^k \wedge e^k  \,, \\ 
    \beta &\sim \beta - \gamma  \,, \\
    \hat{\beta}^k &\sim \hat{\beta}^k - \hat{\gamma}^k  \,, \\
    \hat{\beta} &\sim \hat{\beta} - \hat{\gamma} \\
    \hat{c}_{0i} &\sim \hat{c}_{0i} + (d\hat{\gamma})_{0i} \,, \quad  (i = 1, 2, 3)  \,.
\end{align}
They also have dynamical gauge transformations:
\begin{align}
    \beta^k \wedge e^k &\sim \beta^k \wedge e^k + d u^k \wedge e^k   \,, \\
    \beta &\sim \beta + d u - \sum_{k=1}^3 u^k e^k \,, \\
    \hat{\beta}^k &\sim \hat{\beta}^k + 2\pi \hat{u}^k + \hat{u}  \,, \\
    \hat{\beta} &\sim \hat{\beta} + d \hat{u} \,, 
\end{align}
where $u^k e^k$ is a background type-$A$ foliated (0+1)-form gauge parameter, $u$ is a background type-$A$ bulk zero-form gauge parameter, $\hat{u}^k$ is an $x^k$-dependent integer-valued gauge parameter, $\hat{u}$ is a background type-$A$ bulk zero-form gauge parameter.
Then, the 3+1d foliated SSPT phase with three foliations is written as
\begin{align}
\begin{split}
    &L^3_{\text{SSPT},\text{f}} \left[ C^k \wedge e^k, c,  \hat{C}^k,  \hat{c}_{12}, \hat{c}_{23}, \hat{c}_{31} \right]  \\
    &=  \frac{iN}{2\pi} \Bigg[ -\sum_{k=1}^3 \hat{\beta}^k \, dC^k \wedge e^k +  \hat{\beta} \wedge  \left( dc - \sum_{k=1}^3 C^k \wedge e^k \right)    \\
    & \qquad + \sum_{k=1}^3 \beta^k \wedge e^k \wedge (d\hat{C}^k + \hat{c}) + \beta \wedge d\hat{c}  + \sum_{k=1}^3 \hat{C}^k \wedge C^k \wedge e^k + \hat{c} \wedge c \Bigg]  \,. \label{3folisspt4}
\end{split}
\end{align}
If the theory is on spacetime without a boundary, it is gauge invariant. As in the case of the SSPT phase with two foliations, if spacetime has a boundary, the partition function of the 3+1d SSPT phase with three foliations is not gauge invariant and the variation is canceled by the anomaly of the 2+1d foliated $BF$ theory on the boundary \eqref{fanomaly}. To see this, we put the foliated SSPT phase on the spacetime $\T^{2+1} \times \R_{x^3 \geq 0}$ with the boundary $x^3 = 0$. From the gauge invariance, the boundary conditions of the dynamical gauge parameters are
\begin{align}
    u^k |_{x^3 = 0} &= 0 \,, \\ 
    u\, |_{x^3 = 0} &= 0 \,, \\
    \hat{u}^k |_{x^3 = 0}  &= 0 \,, \\
    \hat{u}\, |_{x^3 = 0} &= 0 \,,
\end{align}
which are consistent with the boundary conditions in the exotic SSPT phase.
On the boundary, we put the 2+1d foliated $BF$ theory coupled to the background gauge fields \eqref{fbfc}, and the background gauge fields in the 3+1d SSPT phase are related to those in the 2+1d foliated $BF$ theory as
\begin{align}
    C^k_{\text{SSPT}} \wedge e^k \, |_{x^3 = 0} &= C^k_{BF} \wedge e^k \,, \quad (k=1,2) \,, \\
    \hat{C}^k_{\text{SSPT}}\, |_{x^3 = 0} &= \hat{C}^k_{BF} \,, \quad (k=1,2) \,, \\
    c_{\text{SSPT}} \, |_{x^3 = 0} &= c_{BF} \,, \\
    \hat{c}_{\text{SSPT},12}\, |_{x^3 = 0} &= \hat{c}_{BF,12} \,.
\end{align}
Then, under the background gauge transformations, the Lagrangian transforms as
\begin{align}
\begin{split}
    &\delta_\text{g} L^3_{\text{SSPT,f}} \left[ C^k \wedge e^k ,c, \hat{C}^k, \hat{c}_{12} \right]\\
    &=  \frac{iN}{2\pi}  \Bigg[ \sum_{k=1}^3 \hat{\gamma}^k d C^k  \wedge e^k - \hat{\gamma} \wedge \left( dc - \sum_{k=1}^3 C^k \wedge e^k \right)    \\
    & \qquad - \sum_{k=1}^3 \gamma^k \wedge e^k \wedge (d\hat{C}^k + \hat{c}) - \gamma \wedge d\hat{c}   + \sum_{k=1}^3 (d\hat{\gamma}^k - \hat{\gamma}) \wedge C^k \wedge e^k + d\hat{\gamma} \wedge c  \\
    & \qquad + \sum_{k=1}^3 (\hat{C}^k + d\hat{\gamma}^k - \hat{\gamma}) \wedge d\gamma^k \wedge e^k + (\hat{c} + d\hat{\gamma} ) \wedge \left(d\gamma + \sum_{k=1}^3 \gamma^k \wedge e^k \right) \Bigg] \\
    & =  \frac{iN}{2\pi} d \Bigg[ \sum_{k=1}^2 \left\{ \hat{\gamma}^k \,C^k \wedge e^k - (\hat{C} + d{\gamma}^k - \hat{\gamma}) \wedge \gamma^k \wedge e^k \right\} + (\hat{c} + d\hat{\gamma})\wedge \gamma + \hat{\gamma} \wedge c \Bigg] \,.
\end{split}
\end{align}
Note that on the boundary $x^3 = 0$, the terms containing $e^3 = dx^3$, such as the foliated fields with $k = 3$, do not appear. Then, we have 
\begin{align}
\begin{split}
    &\delta_\text{g} S^3_{\text{SSPT,f}} \left[ C^k \wedge e^k ,c, \hat{C}^k, \hat{c}_{12} \right] \\
    &= \int_{x^3=0} \frac{iN}{2\pi} \Bigg[ \sum_{k=1}^2 \left\{ \hat{\gamma}^k \,C^k \wedge e^k - (\hat{C} + d{\gamma}^k - \hat{\gamma}) \wedge \gamma^k \wedge e^k \right\} + (\hat{c} + d\hat{\gamma})\wedge \gamma + \hat{\gamma} \wedge c \Bigg] \,.
\end{split}
\end{align}
This expression is the same as the boundary term of the 3+1d foliated SSPT with two foliations \eqref{anomcan2}, and matches the 't Hooft anomaly of the 2+1d foliated $BF$ theory \eqref{fanomaly} on the boundary $x^3=0$. Therefore we can see that the 't Hooft anomaly of the 2+1d foliated $BF$ theory on the boundary is also canceled by the gauge-variation of the 3+1d foliated SSPT phase with three foliations on the bulk.

An 't Hooft anomaly of an ordinary global symmetry in a relativistic QFT corresponds to a classical field theory in one dimension higher from the anomaly inflow \cite{Callan:1984sa}, and such field theories are called invertible field theories \cite{Freed:2016rqq}. Invertible field theories are the low-energy effective field theories of the symmetry-protected topological (SPT) phases \cite{Gu:2009dr,Chen:2010gda}, which cannot be smoothly deformed into trivially gapped systems while preserving the symmetry.

In the case of the subsystem global symmetries, the 3+1d SSPT phases with two foliations and three foliations correspond to the same 't Hooft anomaly on the boundary. The SSPT phase with two foliations has the spatial rotational symmetry $\Z_4$ with respect to $(x^1,x^2)$, while the SSPT phase with three foliations has the spatial rotational symmetry $S_4$ with respect to $(x^1,x^2,x^3)$. However, on the boundary $x^3 =0$, both SSPT phases reproduce the 90 degree rotational symmetry $\Z_4$ in the foliated $BF$ theory in 2+1 dimensions. Furthermore, the SSPT phase with three foliations \eqref{3folisspt4} is smoothly deformed into the one with two foliations \eqref{foliatedssptlag_n} while preserving the rotational symmetry $\Z_4$ under the deformation $e^3 = dx^3 \rightarrow 0$, that is $\hat{\beta}^3$, $\beta^3 \wedge e^3$, $C^3 \wedge e^3$ and $\hat{C}^3$ go to $0$.\footnote{Precisely, we must consider the version where the bulk gauge fields $\hat{c}$ are not dynamical. This fact implies that we cannot apply the same discussion to the exotic SSPT phases.} Therefore, an 't Hooft anomaly of a subsystem symmetry in a fracton QFT is considered to correspond to a certain deformation class of the SSPT phases and foliation structures. This gives an implication for the characterization of 't Hooft anomalies of subsystem symmetry.

\subsection{Exotic SSPT Phase with Three Foliations}
\label{subsection:exotic_sspt_3foliations}

In Section \ref{subsection:foliated_sspt_3foliations}, we derived the foliated SSPT phase with three foliations simply by adding the $k = 3$ foliation terms. Here assuming field correspondences, we determine exotic tensor gauge fields in the SSPT phase with $S_4$ rotational symmetry and construct the exotic Lagrangian describing the SSPT phase with three foliations.

To derive the exotic form, we must integrate out $\hat{\beta}$, $\beta_0^k \, (k =1,2,3)$ and $\hat{c}_{0i} \, (i = 1,2,3)$ to get the constraints
\begin{align}
    \frac{iN}{2\pi} \left( dc - \sum_{k=1}^3  C^k \wedge e^k \right) &= 0 \,, \label{3cconst_n_sspt} \\
    \frac{iN}{2\pi} \left( d\hat{C}^k + \hat{c} \right)_{ij} &= 0 \,,\quad ((i,j,k) = (1,2,3), (2,3,1), (3,1,2)) \,, \label{3chatconst_n_sspt} \\
    \frac{iN}{2\pi} \left(  d\beta + \sum_{k=1}^3 \beta^k \wedge e^k + c \right)_{ij} &= 0 \,,\quad ((i,j) = (2,3),(3,1),(1,2)) \,, \label{3betaconst_n_sspt}
\end{align}
or in components,
\begin{align}
    \frac{iN}{2\pi} ( \partial_1 c_{23} + \partial_2 c_{31} + \partial_3 c_{12} - C^1_{23} - C^2_{31} - C^3_{12} ) &= 0 \,, \label{3cconst_1} \\
    \frac{iN}{2\pi} ( -\partial_0 c_{23} + \partial_2 c_{03} - \partial_3 c_{02} - C^2_{03} + C^3_{02} ) &= 0 \,, \\
    \frac{iN}{2\pi} ( -\partial_0 c_{31} - \partial_1 c_{03} + \partial_3 c_{01} + C^1_{03} - C^3_{01} ) &= 0 \,, \\
    \frac{iN}{2\pi} ( -\partial_0 c_{12} + \partial_1 c_{02} - \partial_2 c_{01} - C^1_{02} + C^2_{01} ) &= 0 \,,
\end{align}
and
\begin{align}
    \frac{iN}{2\pi} ( \partial_2 \hat{C}^1_3 - \partial_3 \hat{C}^1_2 + \hat{c}_{23}  ) &= 0 \,, \\
    \frac{iN}{2\pi} ( \partial_3 \hat{C}^2_1 - \partial_1 \hat{C}^2_3 + \hat{c}_{31}  ) &= 0 \,, \\
    \frac{iN}{2\pi} ( \partial_1 \hat{C}^3_2 - \partial_2 \hat{C}^3_1 + \hat{c}_{12}  ) &= 0 \,,
\end{align}
and
\begin{align}
    \frac{iN}{2\pi} (\partial_2 \beta_3 - \partial_3 \beta_2 + \beta^3_2 - \beta^2_3 + c_{23}  ) &= 0 \,, \\
    \frac{iN}{2\pi} (-\partial_1 \beta_3 + \partial_3 \beta_1 - \beta^3_1 + \beta^1_3  + c_{31}  ) &= 0 \,, \\
    \frac{iN}{2\pi} (\partial_1 \beta_2 - \partial_2 \beta_1 + \beta^2_1 - \beta^1_2 + c_{12}  ) &= 0 \,.
\end{align}

The fracton theory with three simultaneous foliations $e^k = dx^k\, (k=1,2,3)$ has the 90 degree spatial rotational symmetry $S_4$ in 3+1 dimensions. Then, exotic tensor gauge fields of such a theory are in representations of $S_4$. Irreducible representations of $S_4$ are $\bm{1}$, $\bm{1}'$, $\bm{2}$,  $\bm{3}$ and $\bm{3}'$, and we use the notation in \cite{Seiberg:2020wsg}. Firstly, we introduce $U(1)$ background exotic tensor two-form gauge field $\bm{C} = (C_{0ij},C_{[ij]k})$ in $(\bm{3}',\bm{2})$ and assume the correspondences
\begin{align}
    C_{012} &\simeq C^1_{02} + \partial_2 c_{01} = C^2_{01} + \partial_1 c_{02} - \partial_0 c_{12} \,, \label{33ccor1} \\
    C_{023} &\simeq C^2_{03} + \partial_3 c_{02} = C^3_{02} + \partial_2 c_{03} - \partial_0 c_{23} \,, \label{33ccor2} \\
    C_{031} &\simeq C^3_{01} + \partial_1 c_{03} = C^1_{03} + \partial_3 c_{01} - \partial_0 c_{31} \,, \label{33ccor3}
\end{align}
and
\begin{align}
    C_{[12]3} &\simeq C^3_{12} - \partial_1 c_{23}  \,, \label{33ccor4} \\
    C_{[23]1} &\simeq C^1_{23} - \partial_2 c_{31}  \,, \label{33ccor5} \\
    C_{[31]2} &\simeq C^2_{31} - \partial_3 c_{12}  \,. \label{33ccor6}
\end{align}
Note that from \eqref{3cconst_1}, $C_{[ij]k}$ satisfy the relation $C_{[12]3} + C_{[23]1} + C_{[31]2} = 0$ of the representation $\bm{2}$. Instead of this, we can use another basis $C_{i(jk)}$ in $\bm{2}$ with the relations\footnote{They satisfy the formula:
$\sum_{(i,j,k)} C_{i(jk)} \hat{C}^{i(jk)} = 3\sum_{(i,j,k)} C_{[ij]k} \hat{C}^{[ij]k}$.}
\begin{align}
    C_{i(jk)} &= C_{[ij]k} - C_{[ki]j} \,, \\
    C_{[ij]k} &= \frac{1}{3} \left( C_{i(jk)} - C_{j(ki)} \right) \,,
\end{align}
and then we have
\begin{align}
    C_{1(23)} &\simeq C^3_{12} - C^2_{31}  - \partial_1 c_{23} + \partial_3 c_{12} \,, \label{33ccor7} \\
    C_{2(31)} &\simeq C^1_{23} - C^3_{12}  - \partial_2 c_{31} + \partial_1 c_{23} \,, \label{33ccor8} \\
    C_{3(12)} &\simeq C^2_{31} - C^1_{23}  - \partial_3 c_{12} + \partial_2 c_{31} \,. \label{33ccor9}
\end{align}
We also assume the correspondences between the background gauge parameters:
\begin{align}
    \Gamma_0 &\simeq \gamma_0 \,, \\
    \Gamma_{12} &\simeq \gamma^1_2 + \partial_2 \gamma_1 \,, \\
    \Gamma_{23} &\simeq \gamma^2_3 + \partial_3 \gamma_2 \,, \\
    \Gamma_{31} &\simeq \gamma^3_1 + \partial_1 \gamma_3 \,,
\end{align}
where $\bm{\Gamma} = (\Gamma_0, \Gamma_{ij})$ is a background tensor one-form gauge parameter in $(\bm{1},\bm{3}')$. Then, we have background gauge transformations of $\bm{C}$ as
\begin{align}
    C_{0ij} &\sim C_{0ij} + \partial_0 \Gamma_{ij} - \partial_i \partial_j \Gamma_0 \,, \quad ((i,j) = (1,2),(2,3),(3,1)) \,, \\
    C_{[ij]k} &\sim C_{[ij]k} + \partial_i \Gamma_{jk} - \partial_j \Gamma_{ki} \,, \quad ((i,j,k) = (1,2,3),(2,3,1),(3,1,2)) \,, \\
    C_{k(ij)} &\sim C_{k(ij)} + 2 \partial_k \Gamma_{ij} - \partial_i \Gamma_{jk} - \partial_j \Gamma_{ki} \,, \quad ((i,j,k) = (1,2,3),(2,3,1),(3,1,2)) \,.
\end{align}

Next, we introduce a $U(1)$ background exotic tensor one-form gauge field \par
\noindent $\bm{\hat{C}} = (\hat{C}^{k(ij)}_0, \hat{C}^{ij})$ in the representation $(\bm{2},\bm{3'})$, and we assume the correspondences
\begin{align}
    \hat{C}_0^{3(12)} &\simeq \hat{C}_0^1 - \hat{C}_0^2 \,, \label{33chatcor1} \\
    \hat{C}_0^{1(23)} &\simeq \hat{C}_0^2 - \hat{C}_0^3 \,, \label{33chatcor2} \\
    \hat{C}_0^{2(31)} &\simeq \hat{C}_0^3 - \hat{C}_0^1 \,, \label{33chatcor3}
\end{align}
and
\begin{align}
    \hat{C}^{12} &\simeq \hat{C}_3^1 - \hat{C}_3^2 \,, \label{33chatcor4} \\
    \hat{C}^{23} &\simeq \hat{C}_1^2 - \hat{C}_1^3 \,, \label{33chatcor5} \\
    \hat{C}^{31} &\simeq \hat{C}_2^3 - \hat{C}_2^1 \,. \label{33chatcor6} 
\end{align}
Note that the gauge transformations $\hat{\gamma}$ on the right-hand sides cancel out, so the degrees of freedom of the fields are consistent. We can also use another basis $\hat{C}^{[ij]k}_0$ in $\bm{2}$ instead of $\hat{C}^{k(ij)}_0$, and then we have
\begin{align}
    \hat{C}^{[12]3}_0 &\simeq \frac{1}{3} \left( \hat{C}^1_0 + \hat{C}^2_0 +  \hat{C}^3_0 \right)  - \hat{C}^3_0 \,, \label{33chatcor7} \\
    \hat{C}^{[23]1}_0 &\simeq \frac{1}{3} \left( \hat{C}^2_0 + \hat{C}^3_0 +  \hat{C}^1_0 \right) - \hat{C}^1_0 \,, \label{33chatcor8} \\
    \hat{C}^{[31]2}_0 &\simeq \frac{1}{3} \left( \hat{C}^3_0 + \hat{C}^1_0 +  \hat{C}^2_0 \right) - \hat{C}^2_0 \,. \label{33chatcor9}
\end{align}
We also assume cornerspondences between the background gauge parameters:
\begin{align}
    \hat{\Gamma}^{3(12)} &\simeq \hat{\gamma}^1 - \hat{\gamma}^2 \,, \\
    \hat{\Gamma}^{1(23)} &\simeq \hat{\gamma}^2 - \hat{\gamma}^3 \,, \\
    \hat{\Gamma}^{2(31)} &\simeq \hat{\gamma}^3 - \hat{\gamma}^1 \,, 
\end{align}
and
\begin{align}
    \hat{\Gamma}^{[12]3} &\simeq \frac{1}{3} \left( \hat{\gamma}^1 + \hat{\gamma}^2 + \hat{\gamma}^3 \right) - \hat{\gamma}^3 \,, \\
    \hat{\Gamma}^{[23]1} &\simeq \frac{1}{3} \left( \hat{\gamma}^2 + \hat{\gamma}^3 + \hat{\gamma}^1 \right) - \hat{\gamma}^1 \,, \\
    \hat{\Gamma}^{[31]2} &\simeq \frac{1}{3} \left( \hat{\gamma}^3 + \hat{\gamma}^1 + \hat{\gamma}^2 \right) - \hat{\gamma}^2 \,,
\end{align}
where $\bm{\hat{\Gamma}} = (\hat{\Gamma}^{k(ij)}) = (\hat{\Gamma}^{[ij]k})$ is a background tensor zero-form gauge parameter in $\bm{2}$. Then, we have background gauge transformations of $\bm{\hat{C}}$ as
\begin{align}
    \hat{C}_0^{k(ij)} &\sim \hat{C}_0^{k(ij)} + \partial_0 \hat{\Gamma}^{k(ij)}  \,, \quad ((i,j,k) = (1,2,3),(2,3,1),(3,1,2)) \,, \\
    \hat{C}^{ij} &\sim \hat{C}^{ij} + \partial_k \hat{\Gamma}^{k(ij)} \,, \quad ((i,j,k) = (1,2,3),(2,3,1),(3,1,2)) \,, 
\end{align}
or
\begin{align}
    \hat{C}_0^{[ij]k} &\sim \hat{C}_0^{[ij]k} + \partial_0 \hat{\Gamma}^{[ij]k}  \,, \quad ((i,j,k) = (1,2,3),(2,3,1),(3,1,2)) \,, \\
    \hat{C}^{ij} &\sim \hat{C}^{ij} + \partial_k \hat{\Gamma}^{[ki]j} - \partial_k \hat{\Gamma}^{[jk]i} \,, \quad ((i,j,k) = (1,2,3),(2,3,1),(3,1,2)) \,. 
\end{align}
In addition, since we have the background gauge fields $\hat{c}_{12}$, $\hat{c}_{23}$ and $\hat{c}_{31}$, we introduce an auxiliary background gauge field $\hat{C}^i$ in the representation $\bm{3}$ satisfying the correspondences
\begin{align}
    \hat{C}^1 &\simeq \partial_2 \hat{C}^2_3 - \partial_3 \hat{C}^3_2 + \hat{c}_{23} \,, \label{33chatcor10} \\
    \hat{C}^2 &\simeq \partial_3 \hat{C}^3_1 - \partial_1 \hat{C}^1_3 + \hat{c}_{31} \,, \label{33chatcor11} \\
    \hat{C}^3 &\simeq \partial_1 \hat{C}^1_2 - \partial_2 \hat{C}^2_1 + \hat{c}_{12} \,. \label{33chatcor12}
\end{align}
Their background gauge transformations are
\begin{align}
    \hat{C}^k &\sim \hat{C}^k + \partial_i \partial_j \hat{\Gamma}^{k(ij)} \,, \quad ((i,j,k) = (1,2,3),(2,3,1),(3,1,2)) \,,
\end{align}
or
\begin{align}
    \hat{C}^k &\sim \hat{C}^k + \partial_i \partial_j \hat{\Gamma}^{[ki]j} - \partial_i \partial_j \hat{\Gamma}^{[jk]i}  \,, \quad ((i,j,k) = (1,2,3),(2,3,1),(3,1,2)) \,.
\end{align}
After integrating out $\beta^k_0$, these fields satisfy
\begin{align}
    \partial_2 \hat{C}^{12} + \partial_3 \hat{C}^{31} + \hat{C}^1 &= 0 \,, \\
    \partial_3 \hat{C}^{23} + \partial_1 \hat{C}^{12} + \hat{C}^2 &= 0 \,, \\
    \partial_1 \hat{C}^{31} + \partial_2 \hat{C}^{23} + \hat{C}^3 &= 0 \,, \label{chat3_const3}
\end{align}
from the constraint \eqref{3chatconst_n_sspt}.

We also introduce dynamical gauge fields $\bm{\beta}$ and $\bm{\hat{\beta}}$ to restrict $\bm{C}$ and $\bm{\hat{C}}$ fields to $\Z_N$. We use $\beta_0$ and assume the correspondences 
\begin{align}
    \beta_{12} &\simeq \beta^1_2 + \partial_2 \beta_1 = \beta^2_1 + \partial_1 \beta_2 + c_{12}  \,, \\
    \beta_{23} &\simeq \beta^2_3 + \partial_3 \beta_2 = \beta^3_2 + \partial_2 \beta_3 + c_{23}  \,, \\
    \beta_{31} &\simeq \beta^3_1 + \partial_1 \beta_3 = \beta^1_3 + \partial_3 \beta_1 + c_{31}  \,, 
\end{align}
and
\begin{align}
    \hat{\beta}^{3(12)} &\simeq \hat{\beta}^1 - \hat{\beta}^2 \,, \\
    \hat{\beta}^{1(23)} &\simeq \hat{\beta}^2 - \hat{\beta}^3 \,, \\
    \hat{\beta}^{2(31)} &\simeq \hat{\beta}^3 - \hat{\beta}^1 \,,
\end{align}
or
\begin{align}
    \hat{\beta}^{[12]3} &\simeq \frac{1}{3} \left( \hat{\beta}^1 + \hat{\beta}^2 + \hat{\beta}^3 \right) - \hat{\beta}^3 \,, \\
    \hat{\beta}^{[23]1} &\simeq \frac{1}{3} \left( \hat{\beta}^2 + \hat{\beta}^3 + \hat{\beta}^1 \right) - \hat{\beta}^1 \,, \\
    \hat{\beta}^{[31]2} &\simeq \frac{1}{3} \left( \hat{\beta}^3 + \hat{\beta}^1 + \hat{\beta}^2 \right) - \hat{\beta}^2 \,,
\end{align}
where $\bm{\beta} = (\beta_0, \beta_{ij})$ in $(\bm{1},\bm{3'})$ is a dynamical exotic tensor one-form gauge field and $\bm{\hat{\beta}} = (\hat{\beta}^{k(ij)}) = (\hat{\beta}^{[ij]k})$ is a dynamical exotic tensor zero-form gauge field.
Their background gauge transformations are
\begin{align}
    \beta_0 &\sim \beta_0 - \Gamma_0 \,, \\
    \beta_{ij} &\sim \beta_{ij} - \Gamma_{ij} \,, \quad ((i,j) = (1,2),(2,3),(3,1)) \,, \\
    \hat{\beta}^{k(ij)} &\sim \hat{\beta}^{k(ij)} - \hat{\Gamma}^{k(ij)} \,, \quad ((i,j,k) = (1,2,3),(2,3,1),(3,1,2)) \,, \\
    \hat{\beta}^{[ij]k} &\sim \hat{\beta}^{[ij]k} - \hat{\Gamma}^{[ij]k} \,, \quad ((i,j,k) = (1,2,3),(2,3,1),(3,1,2)) \,.
\end{align}
They also have dynamical gauge transformations
\begin{align}
    \beta_0 &\sim \beta_0 + \partial_0 s \,, \\
    \beta_{ij} &\sim \beta_{ij} + \partial_i \partial_j s \,, \quad ((i,j) = (1,2),(2,3),(3,1)) \,, \\
    \hat{\beta}^{k(ij)} &\sim \hat{\beta}^{k(ij)} + 2\pi \hat{s}^i - 2\pi \hat{s}^j \,, \quad ((i,j,k) = (1,2,3),(2,3,1),(3,1,2)) \,, \\
    \hat{\beta}^{[ij]k} &\sim \hat{\beta}^{[ij]k} + \frac{2\pi}{3} ( \hat{s}^i + \hat{s}^j - 2 \hat{s}^k ) \,, \quad ((i,j,k) = (1,2,3),(2,3,1),(3,1,2)) \,,
\end{align}
where $\hat{s}^k$ is a $x^k$-dependent integer-valued gauge parameter, and $\bm{s} =(s)$ is a dynamical tensor zero-form gauge parameter in $\bm{1}$, and we have
\begin{align}
    s &\simeq u \,, \\
    \hat{s}^k &\simeq \hat{u}^k 
     \,.
\end{align}

Using these field correspondences, we construct the exotic Lagrangian describing the 3+1d SSPT phase with three foliations. As for $\L^3_{\text{SSPT,f},\hat{\beta}}[\bm{C}]$, we have
\begin{align}
\begin{split}
   L^3_{\text{SSPT,f},\hat{\beta}}\left[\bm{C}\right] &= -\frac{iN}{2\pi}   \sum_{k=1}^3  \hat{\beta}^k\, d C^k \wedge e^k \\
   &\simeq -\frac{iN}{2\pi} \hat{\beta}^1 \left( \partial_0 C_{[23]1} - \partial_2 C_{031} + \partial_3 C_{012}   \right) d^4x \\
   & \quad -\frac{iN}{2\pi} \hat{\beta}^2 \left( \partial_0 C_{[31]2} - \partial_3 C_{012} + \partial_1 C_{023}   \right)d^4x \\
   & \quad -\frac{iN}{2\pi} \hat{\beta}^3 \left( \partial_0 C_{[12]3} - \partial_1 C_{023} + \partial_2 C_{031}   \right)d^4x \\
   &\simeq \frac{iN}{2\pi} \sum_{(i,j,k)} \hat{\beta}^{[ij]k} \left( \partial_0 C_{[ij]k} - \partial_i C_{0jk} + \partial_j C_{0ki} \right)d^4x \\
   &= \frac{iN}{2\pi} \sum_{(i,j,k)}\frac{1}{3} \hat{\beta}^{k(ij)}\left(\partial_0 C_{k(ij)} - 2\partial_k C_{0ij} + \partial_i C_{0jk} +  \partial_j C_{0ki} \right)d^4x \,,
\end{split}
\end{align}
where the sum for $(i,j,k)$ is over $(i,j,k) = (1,2,3),(2,3,1),(3,1,2)$. As for $\L^3_{\text{SSPT,f},\beta}[\bm{\hat{C}}]$, we have
\begin{align}
\begin{split}
    & L^3_{\text{SSPT,f},\beta} \left[ \bm{\hat{C}} \right] \\
    &= \frac{iN}{2\pi}  \left[ \sum_{k=1}^3  \beta^k \wedge  e^k \wedge  ( d \hat{C}^k + \hat{c} ) +  \beta \wedge d\hat{c} \right] \\
    &= \frac{iN}{2\pi}  \left[ \beta^1_2 ( \partial_3 \hat{C}^1_0 - \partial_0 \hat{C}^1_3 )  + \beta^2_1 ( - \partial_3 \hat{C}^2_0 + \partial_0 \hat{C}^2_3 ) + \beta^2_3 ( - \partial_0 \hat{C}^2_1 + \partial_1 \hat{C}^2_0 ) \right. \\
    &\quad + \beta^3_2 (  \partial_0 \hat{C}^3_1 - \partial_1 \hat{C}^3_0 ) + \beta^3_1 ( - \partial_0 \hat{C}^3_2 + \partial_2 \hat{C}^3_0 ) + \beta^1_3 (  \partial_0 \hat{C}^1_2 - \partial_2 \hat{C}^1_0 )  \\
    &\quad \left.  + \beta_0 ( \partial_1 \hat{c}_{23} + \partial_2 \hat{c}_{31} + \partial_3 \hat{c}_{12} ) \right]d^4x \\
    &\simeq -\frac{iN}{2\pi} \left[ \sum_{(i,j,k)} \beta_{ij} \left( \partial_0 \hat{C}^{ij} - \partial_k \hat{C}^{k(ij)} \right) +  \beta_0 \sum_{(i,j)}  \partial_i \partial_j \hat{C}^{ij}  \right] d^4x \\
    &\quad + \frac{iN}{2\pi} \left[ - c_{12} ( -\partial_3 \hat{C}^2_0 + \partial_0 \hat{C}^2_3 ) - c_{23} ( \partial_0 \hat{C}^3_1 - \partial_1 \hat{C}^3_0 ) - c_{31}( \partial_0 \hat{C}^1_2 - \partial_2 \hat{C}^1_0 ) \right]d^4x \,, \label{3ssptbeta}
\end{split}
\end{align}
where we have dropped the total derivative terms. Also, we used the constraint \eqref{3chatconst_n_sspt} to get
\begin{align}
\begin{split}
    \partial_1 \hat{c}_{23} + \partial_2 \hat{c}_{31} + \partial_3 \hat{c}_{12} &= -\partial_1 \left( \partial_2 \hat{C}^1_3  - \partial_3 \hat{C}^1_2 \right) -\partial_2 \left( \partial_3 \hat{C}^2_1 - \partial_1 \hat{C}^2_3 \right) \\
    & \quad -\partial_3 \left( \partial_1 \hat{C}^3_2 - \partial_2 \hat{C}^3_1   \right) \\
    &\simeq -\sum_{(i,j)}  \partial_i \partial_j \hat{C}^{ij}  \,.
\end{split}
\end{align} 
Regarding $\L^3_{\text{SSPT},\text{f}, \hat{C}C} [ \bm{C}, \bm{\hat{C}}]$, we can write as
\begin{align}
\begin{split}
    &L^3_{\text{SSPT},\text{f}, \hat{C}C} \left[ \bm{C}, \bm{\hat{C}}\right]  \\
    &= \frac{iN}{2\pi}  \sum_{k=1}^3 \hat{C}^k \wedge  C^k \wedge e^k + \frac{iN}{2\pi} \hat{c} \wedge c  \\
    &= \frac{iN}{2\pi} \left(  \hat{C}^1_3 C^1_{02} -  \hat{C}^1_2 C^1_{03} + \hat{C}^1_0 C^1_{23}  - \hat{C}^2_3 C^2_{01}  + \hat{C}^2_1  C^2_{03}   \right. \\
    & \quad \left.  +  \hat{C}^2_0 C^2_{31} -  \hat{C}^3_1 C^3_{02} +  \hat{C}^3_2 C^3_{01} +  \hat{C}^3_0 C^3_{12} + \hat{c}_{23} c_{01}  + \hat{c}_{31} c_{02}  + \hat{c}_{12} c_{03} \right)d^4x \\
    &\simeq \frac{iN}{2\pi} \sum_{(i,j,k)} \left[  \hat{C}^{ij} C_{0ij} - \hat{C}^{[ij]k}_0 C_{[ij]k}  \right] d^4x \\
    & \quad +  \frac{iN}{2\pi} \big[ -\hat{C}^1_3  \partial_2 c_{01}  + \hat{C}^1_2 ( \partial_3 c_{01} - \partial_0 c_{31} )  + \hat{C}^1_0 \partial_2 c_{31}  + \hat{C}^2_3  ( \partial_1 c_{02} - \partial_0 c_{12} )  \\ 
    & \quad  - \hat{C}^2_1  \partial_3 c_{02}  + \hat{C}^2_0 \partial_3 c_{12}  + \hat{C}^3_1 ( \partial_2 c_{03} - \partial_0 c_{23} )  - \hat{C}^3_2 \partial_1 c_{03} + \hat{C}^3_0 \partial_1 c_{23}   \\ 
    & \quad  -   (\partial_2 \hat{C}^1_3 - \partial_3 \hat{C}^1_2 )c_{01} -  (\partial_3 \hat{C}^2_1 - \partial_1 \hat{C}^2_3 )c_{02} - (\partial_1 \hat{C}^3_2 - \partial_2 \hat{C}^3_1 )c_{03} \big]d^4x \,. \label{3ssptcc}
\end{split}
\end{align}
Combining the last terms of \eqref{3ssptbeta} and \eqref{3ssptcc} and dropping the total derivative terms with respect to $x^0$, $x^1$, and $x^2$, we have
\begin{align}
    \frac{iN}{2\pi} \partial_3 \left(\hat{C}^2_0 c_{12}  - \hat{C}^2_1 c_{02}  +  \hat{C}^1_2 c_{01}   \right) \,.
\end{align}
As in the case of the SSPT phase with two foliations, we have to add the term \eqref{bdcc} when converting the SSPT phase from the foliated form to the exotic form, and this term cancels out. Then, we can write
\begin{align}
\begin{split}
    \L^3_{\text{SSPT},\text{e}, \hat{C}C} \left[ \bm{C}, \bm{\hat{C}}\right]  
    &= \frac{iN}{2\pi} \sum_{(i,j,k)} \left[  \hat{C}^{ij} C_{0ij} - \hat{C}^{[ij]k}_0 C_{[ij]k}  \right]  \\
    &= \frac{iN}{2\pi} \sum_{(i,j,k)} \left[  \hat{C}^{ij} C_{0ij} - \frac{1}{3}\hat{C}^{k(ij)}_0 C_{k(ij)} \right] \,.
\end{split}
\end{align}

After all, we have constructed the exotic SSPT phase with three foliations in 3+1 dimensions:
\begin{align}
\begin{split}
    &\L^3_{\text{SSPT},\text{e}} \left[ \bm{C}, \bm{\hat{C}} \right]  \\
    & = \frac{iN}{2\pi} \sum_{(i,j,k)} \hat{\beta}^{[ij]k} \left( \partial_0 C_{[ij]k} - \partial_i C_{0jk} + \partial_j C_{0ki} \right) \\ 
    & \quad -\frac{iN}{2\pi} \left[ \sum_{(i,j,k)} \beta_{ij} \left( \partial_0 \hat{C}^{ij} - \partial_k \hat{C}^{[ki]j} + \partial_k \hat{C}^{[jk]i} \right) +  \beta_0 \sum_{(i,j)}  \partial_i \partial_j \hat{C}^{ij}  \right] \\
    &\quad  + \frac{iN}{2\pi} \sum_{(i,j,k)} \left[  \hat{C}^{ij} C_{0ij} - \hat{C}^{[ij]k}_0 C_{[ij]k}  \right] \,, \label{3foliesspt1}
\end{split}
\end{align}
or
\begin{align}
\begin{split}
    &\L^3_{\text{SSPT},\text{e}} \left[ \bm{C}, \bm{\hat{C}} \right]  \\
    &= \frac{iN}{2\pi} \sum_{(i,j,k)}\frac{1}{3} \hat{\beta}^{k(ij)}\left(\partial_0 C_{k(ij)} - 2\partial_k C_{0ij} + \partial_i C_{0jk} +  \partial_j C_{0ki} \right) \\ 
    & \quad -\frac{iN}{2\pi} \left[ \sum_{(i,j,k)} \beta_{ij} \left( \partial_0 \hat{C}^{ij} - \partial_k \hat{C}^{k(ij)} \right) +  \beta_0 \sum_{(i,j)}  \partial_i \partial_j \hat{C}^{ij}  \right] \\
    &\quad  + \frac{iN}{2\pi} \sum_{(i,j,k)} \left[  \hat{C}^{ij} C_{0ij} - \frac{1}{3}\hat{C}^{k(ij)}_0 C_{k(ij)}  \right] \,. \label{3foliesspt2}
\end{split}
\end{align}

As in the other cases, if the theory is on spacetime without a boundary, it is gauge invariant.\footnote{The gauge-invariant derivatives of $\bm{C}$ and $\bm{\hat{C}}$ are the components of $d_\text{e} \bm{C}$ and $d_\text{e} \bm{\hat{C}}$ in Section \ref{subsection:exotic_tensor_gauge_fields_s4_3d}.} If spacetime has a boundary, the partition function is not gauge invariant and the variation is canceled by the anomaly of the 2+1d exotic $BF$ theory on the boundary \eqref{eanomaly}. To see this, we put the exotic SSPT phase with three foliations on the spacetime $\T^{2+1} \times \R_{x^3 \geq 0}$ with the boundary $x^3 = 0$. From the gauge invariance, the boundary conditions of the dynamical gauge parameters are
\begin{align}
    s |_{x^3 = 0} &= 0 \,, \\
    \hat{s}^k |_{x^3 = 0} &= 0  \,.
\end{align}
On the boundary, we put the 2+1d exotic $BF$ theory coupled to the background gauge fields \eqref{ebfc}, and the background gauge fields in the 3+1d SSPT phase with three foliations are related to those in the 2+1d exotic $BF$ theory as
\begin{align}
    C_{\text{SSPT},012}\, |_{x^3 = 0} &= C_{BF,012} \,, \label{3ecbou1} \\
    \hat{C}^{3(12)}_{\text{SSPT},0}\, |_{x^3 = 0} &= \hat{C}^{12}_{BF,0} \,, \label{3ecbou2} \\ 
    \hat{C}^3_{\text{SSPT}}\, |_{x^3 = 0} &= \hat{C}_{BF} \,. \label{3ecbou3}
\end{align}
Then, under the background gauge transformations, the Lagrangian transforms as
\begin{align}
\begin{split}
    &\delta_{\text{g}} \L^3_{\text{SSPT},\text{e}} \left[ \bm{C}, \bm{\hat{C}} \right]  \\
    &= -\frac{iN}{2\pi} \sum_{(i,j,k)}\frac{1}{3} \hat{\Gamma}^{k(ij)}\left(\partial_0 C_{k(ij)} - 2\partial_k C_{0ij} + \partial_i C_{0jk} +  \partial_j C_{0ki} \right) \\ 
    & \quad +\frac{iN}{2\pi} \left[ \sum_{(i,j,k)} \Gamma_{ij} \left( \partial_0 \hat{C}^{ij} - \partial_k \hat{C}^{k(ij)} \right) +  \Gamma_0 \sum_{(i,j)}  \partial_i \partial_j \hat{C}^{ij}  \right] \\
    &\quad  + \frac{iN}{2\pi} \sum_{(i,j,k)} \left[  \partial_k \hat{\Gamma}^{k(ij)} C_{0ij} - \frac{1}{3} \partial_0 \hat{\Gamma}^{k(ij)} C_{k(ij)}  \right] \\
    &\quad  + \frac{iN}{2\pi} \sum_{(i,j,k)} \left[  \left(\hat{C}^{ij} + \partial_k \hat{\Gamma}^{k(ij)} \right) \left(\partial_0 \Gamma_{ij} - \partial_i \partial_j \Gamma_0  \right) \right. \\
    & \left. \qquad - \frac{1}{3}\left(\hat{C}^{k(ij)}_0 + \partial_0 \hat{\Gamma}^{k(ij)} \right) \left(2\partial_k \Gamma_{ij} - \partial_i  \Gamma_{jk} - \partial_j \Gamma_{ki}  \right) \right] \\
    &= \frac{iN}{2\pi} \sum_{(i,j,k)} \left\{ \partial_k \left[ \hat{\Gamma}^{k(ij)} C_{0ij} - \Gamma_{ij} \hat{C}^{k(ij)}_0 + \hat{\Gamma}^{k(ij)} (\partial_0 \Gamma_{ij} - \partial_i \partial_j \Gamma_0) \right] \right. \\
    & \left. \qquad + \Gamma_0 \partial_i \partial_j \hat{C}^{ij} - \hat{C}^{ij}\partial_i \partial_j \Gamma_0 \right\} \,,
\end{split}
\end{align}
where we used formulae such as
\begin{align}
\begin{split}
    \sum_{(i,j,k)} \hat{C}_0^{k(ij)} ( \partial_k \Gamma_{ij} + \partial_i \Gamma_{jk} + \partial_j \Gamma_{ki} )  &= \sum_{(i,j,k)}  (\hat{C}_0^{k(ij)} + \hat{C}_0^{j(ki)} + \hat{C}_0^{i(jk)} ) \partial_k \Gamma_{ij}  \\
    &= 0 \,.
\end{split}
\end{align}
Dropping the total derivative terms with respect to $x^0$, $x^1$, and $x^2$, we have
\begin{align}
\begin{split}
    &\delta_{\text{g}} \L^3_{\text{SSPT},\text{e}} \left[ \bm{C}, \bm{\hat{C}} \right]  \\
    &=  \frac{iN}{2\pi} \partial_3 \left[ \hat{\Gamma}^{3(12)} C_{012} - (\hat{C}^{3(12)}_0 + \partial_0  \hat{\Gamma}^{3(12)}) \Gamma_{12}   - (-\partial_2 \hat{C}^{23} - \partial_1 \hat{C}^{31} +\partial_1 \partial_2  \hat{\Gamma}^{3(12)} ) \Gamma_0   \right] \,.
\end{split}
\end{align}
Thus, using the equation \eqref{chat3_const3}, the term
\begin{align}
\begin{split}
    &\delta_\text{g} S^3_{\text{SSPT,e}} \left[ \bm{C},\bm{\hat{C}} \right] \\
    & \quad = \int_{x^3 =0} d^3x \frac{iN}{2\pi}  \left[ -\hat{\Gamma}^{3(12)} C_{012} + (\hat{C}^{3(12)}_0 + \partial_0  \hat{\Gamma}^{3(12)}) \Gamma_{12}   + (\hat{C}^3 +\partial_1 \partial_2  \hat{\Gamma}^{3(12)} ) \Gamma_0   \right] 
\end{split}
\end{align}
arises on the boundary $x^3 = 0$. From the boundary conditions \eqref{3ecbou1}--\eqref{3ecbou3}, the background gauge parameters also satisfy
\begin{align}
    \Gamma_{\text{SSPT},0}\, |_{x^3 = 0} &= \Gamma_{BF,0} \,, \\
    \Gamma_{\text{SSPT},12}\, |_{x^3 = 0} &= \Gamma_{BF,12} \,, \\
    \hat{\Gamma}^{3(12)}_{\text{SSPT},0}\, |_{x^3 = 0} &= \hat{\Gamma}^{12}_{BF,0}  
\end{align}
on the boundary, and then it matches the 't Hooft anomaly of the 2+1d exotic $BF$ theory \eqref{eanomaly}. Therefore we can see that the 't Hooft anomaly of the 2+1d exotic $BF$ theory on the boundary is also canceled by the gauge-variation of the
3+1d exotic SSPT phase with three foliations on the bulk.

In the foliated form, the SSPT phase with two foliations \eqref{2foliated_sspt_cf} are related to the one with three foliations \eqref{3folisspt4} in a rather simple way. On the other hand, in the exotic form, relation between the SSPT phase with two foliations \eqref{sspte} and three foliations \eqref{3foliesspt1} is non-trivial. To convert the SSPT phase from the foliated form to the exotic form, we must consider the bulk gauge fields $\hat{c}_{01}$, $\hat{c}_{02}$, $\hat{c}_{03}$, $\hat{c}_{23}$ and $\hat{c}_{31}$ as dynamical fields in the case with two foliations, while we must consider the bulk gauge fields $\hat{c}_{23}$ and $\hat{c}_{31}$ as background fields in the case with three foliations. Thus we cannot naively deform the exotic SSPT phase with two foliations into the one with three foliations, which is an obscure point of the deformation.

\section{Summary of this Chapter}

In this chapter, we have discussed the mixed 't Hooft anomaly of subsystem global symmetry in the exotic and foliated $BF$ theories in 2+1 dimensions and constructed the SSPT phases in 3+1 dimensions that cancel it via the anomaly inflow. Here is the summary of this chapter.

\subsection*{The Fractonic $BF$ theory in 2+1 dimensions}

\subsubsection*{The Exotic $BF$ theory in 2+1 dimensions}

The exotic Lagrangian is
\begin{align}
    \L_{\text{e}} = -\frac{i N}{2\pi}\hat{\phi}^{12}(\partial_0 A_{12} - \partial_{1}\partial_{2} A_0) \,. 
\end{align}
where $\bm{\hat\phi} = (\hat{\phi}^{12})$ is a dynamical tensor zero-form gauge field and $\bm{A} = (A_0,A_{12})$ is a dynamical tensor one-form gauge field of $\Z_4$.
Their gauge transformations are
\begin{align}
    \hat{\phi}^{12} &\sim \hat{\phi}^{12} + 2\pi \hat{w}^1 - 2\pi \hat{w}^2 \,, \\
    A_{0} &\sim A_{0} + \partial_0 \Lambda \,, \\
    A_{12} &\sim A_{12} + \partial_1\partial_2 \Lambda \,, 
\end{align}
where $\hat{w}^k$ is an $x^k$-dependent integer-valued gauge parameter, and $\bm{\Lambda} = (\Lambda)$ is a tensor zero-form gauge parameter.

The exotic Lagrangian coupled to the background gauge fields is
\begin{align}
\begin{split}
    \L_{\text{e}} \left[ \bm{C}, \bm{\hat{C}} \right] = &\frac{iN}{2\pi} \left[ -\hat{\phi}^{12} (\partial_0 A_{12} - \partial_1 \partial_2 A_0 -C_{012} ) - \hat{C}^{12}_0 A_{12}  - \hat{C} A_0  \right] \\ 
    &- \frac{iN}{2\pi} \chi (\partial_0 \hat{C} - \partial_1 \partial_2 \hat{C}^{12}_0) - \frac{iN}{2\pi}( 2\pi\hat{h}^{12} ) C_{012} \,,
\end{split}
\end{align}
where $\bm{C} = (C_{012})$ is a background tensor two-form gauge field and $\bm{\hat{C}} = (\hat{C}^{12}_0, \hat{C})$ is a background tensor one-form gauge field, $\chi$ and $\hat{h}^{12}$ are dynamical gauge fields.
The dynamical gauge transformations of $\chi$ and $\hat{h}^{12}$ are
\begin{align}
    \chi &\sim \chi + \Lambda \,, \\
    \hat{h}^{12} &\sim \hat{h}^{12} + \hat{w}^1 - \hat{w}^2 \,.
\end{align}
The background gauge transformations are
\begin{align}
    \hat{\phi}^{12} &\sim \hat{\phi}^{12} +  \hat{\Gamma}^{12} \,, \\
    A_0 &\sim A_0 + \Gamma_0 \,, \\
    A_{12} &\sim A_{12} + \Gamma_{12} \,,
\end{align}
and
\begin{align}
    C_{012} &\sim C_{012} +  \partial_0 \Gamma_{12} - \partial_1 \partial_2 \Gamma_0 \,, \\
    \hat{C}^{12}_0 &\sim \hat{C}^{12}_0 + \partial_0 \hat{\Gamma}^{12} \,, \\
    \hat{C} &\sim \hat{C} + \partial_1 \partial_2 \hat{\Gamma}^{12} \,,
\end{align}
where $\bm{\Gamma} = (\Gamma_0, \Gamma_{12})$ is a background tensor one-form gauge parameter, and $\bm{\hat{\Gamma}} = (\hat{\Gamma}^{12})$ is a background tensor zero-form gauge parameter.

\subsubsection*{The Foliated $BF$ theory in 2+1 dimensions}

The foliated Lagrangian is
\begin{align}
\begin{split}
    L_{\text{f}} &=  \frac{iN}{2\pi} \bm{\hat{\Phi}} \wedge_\text{f} d_\text{f} \bm{A} \\
    &=  \frac{iN}{2\pi} \sum^{2}_{k=1} \hat{\Phi}^k\, d A^k \wedge e^k + \frac{iN}{2\pi} \hat{\Phi} \wedge \left( da + \sum^{2}_{k=1} A^k \wedge e^k \right) \,,
\end{split} 
\end{align}
where $\bm{A} = (A^k \wedge e^k, a)$ is a dynamical type-$A$ foliated (1+1)-form gauge field and $\bm{\hat\Phi} = (\hat{\Phi}^k, \hat{\Phi})$ is a dynamical type-$B$ foliated zero-form gauge field.
The dynamical gauge transformations of $\bm{\hat\Phi} = (\hat{\Phi}^k, \hat{\Phi})$, $\bm{A} = (A^k \wedge e^k, a)$ are
\begin{align}
    A^k \wedge e^k &\sim A^k \wedge e^k +d \lambda^k \wedge e^k  \,,  \\
    a &\sim a + d\lambda - \sum^{2}_{k=1} \lambda^k e^k\,,  \\
    \hat{\Phi}^k &\sim \hat{\Phi}^k +2\pi \hat{W}^k + \hat{\xi} \,,  \\
    \hat\Phi &\sim \hat\Phi + d\hat{\xi}\,,
\end{align}
where $\bm{\lambda} = (\lambda^k e^k, \lambda)$ is a type-$A$ foliated (0+1)-form gauge parameter, $\hat{W}^k$ is an $x^k$-dependent integer-valued gauge parameter, and $\xi$ is a zero-form gauge parameter.

The foliated Lagrangian coupled to the background gauge fields is
\begin{align}
\begin{split}
    &L_{\text{f}} \left[ C^k \wedge e^k ,c, \hat{C}^k, \hat{c}_{12} \right] \\
    &= \frac{iN}{2\pi} \left[ \bm{\hat{\Phi}} \wedge_\text{f} (d_\text{f} \bm{A} - \bm{C}) + \bm{\hat{C}} \wedge_\text{f} \bm{A} + d_\text{f}\bm{\hat{\chi}} \wedge_\text{f} \bm{C} - \bm{\chi} \wedge_\text{f} d_\text{f}\bm{\hat{C}}   \right] \\
    &= \frac{iN}{2\pi} \left\{ \sum^2_{k=1} \left[  \hat{\Phi}^k (d A^k - C^k) \wedge e^k + \hat{C}^k \wedge  A^k \wedge e^k  \right] - \hat{c} \wedge a \right. \\
    & \qquad + \hat{\Phi} \wedge \left( da + \sum_{k=1}^2  A^k \wedge e^k - c \right) - \hat{\chi} \left(  dc - \sum_{k=1}^2 C^k \wedge e^k \right) \\  
    & \left. \qquad  + \sum^2_{k=1} (2\pi \hat{h}^k)\, C^k \wedge e^k - \sum_{k=1}^2  \chi^k  e^k \wedge ( d\hat{C}^k + \hat{c})  - \chi d \hat{c}  \right\}   \,,
\end{split}
\end{align}
where $\bm{C} = (C^k \wedge e^k, c)$ is a type-$A$ foliated (2+1)-form gauge field, $\bm{\hat{C}} = (\hat{C}^k,\hat{c})$ is a type-$B$ foliated one-form field, $d_\text{f} \bm{\hat{\chi}} = (2\pi\hat{h}^k + \hat{\chi}, d\hat{\chi})$ is a dynamical type-$B$ foliated zero-form field, and $\bm{\chi} = (\chi^k e^k , \chi)$ is a dynamical type-$A$ foliated (0+1)-form field. The gauge fields $C^k \wedge e^k$, $c$, $\hat{C}^k$ and $\hat{c}_{12}$ are background gauge fields, and the gauge fields $\hat{c}_{01}$ and $\hat{c}_{02}$ are dynamical gauge fields.
The dynamical gauge transformations of $\bm{\chi} = (\chi^k e^k, \chi)$, $\hat{h}^k$ and $\hat\chi$ are
\begin{align}
    \chi &\sim \chi + \lambda \,, \\
    \chi^k e^k &\sim \chi^k e^k + \lambda^k e^k \,, \\
    \hat{h}^k &\sim \hat{h}^k + \hat{W}^k + \hat{\tau} \,, \\
    \hat{\chi} &\sim \hat{\chi} + \hat{\xi} - 2\pi \hat{\tau} \,.
\end{align}
The background gauge transformations of the fields are
\begin{align}
    A^k\wedge e^k &\sim A^k \wedge e^k + \gamma^k \wedge e^k  \,,  \\
    a &\sim a + \gamma \,, \\
    \hat{\Phi}^k &\sim \hat{\Phi}^k + \hat{\gamma}^k  \,, \\
    \hat\Phi &\sim \hat\Phi + \hat{\gamma} \,,
\end{align}
and
\begin{align}
    C^k \wedge e^k &\sim C^k \wedge e^k + d\gamma^k \wedge e^k    \,, \\
    c &\sim c + d\gamma + \sum_{k=1}^2 \gamma^k \wedge e^k \,, \\
    \hat{C}^k &\sim \hat{C}^k + d\hat{\gamma}^k - \hat{\gamma}   \,, \\
    \hat{c} &\sim \hat{c} + d\hat{\gamma} \,. 
\end{align}

\subsubsection*{Correspondences in the fractonic $BF$ theory in 2+1 dimensions}

The field correspondences in the fractonic $BF$ theories are
\begin{align}
    A_0 &\simeq a_0 \,, \label{summary:cor1} \\
    A_{12}  &\simeq A^1_2 + \partial_2 a_1 \,,  \\
    \hat{\phi}^{12} &\simeq \hat{\Phi}^1 - \hat{\Phi}^2  \,. \label{summary:cor3}
\end{align}
The correspondences between the dynamical gauge parameters are
\begin{align}
    \Lambda &\simeq \lambda \ \,, \\
    \hat{w}^k  &\simeq \hat{W}^k \,.
\end{align}

The field correspondences in the fractonic $BF$ theory coupled to the background gauge fields are \eqref{summary:cor1}--\eqref{summary:cor3} and
\begin{align}
C_{012} &\simeq C^1_{02} + \partial_2 c_{01}  \,,   \\
\hat{C}_0^{12} &\simeq \hat{C}^1_0 - \hat{C}^2_0 \,,  \\
\hat{C} &\simeq \partial_1 \hat{C}^1_2 - \partial_2 \hat{C}^2_1 + \hat{c}_{12} \,,
\end{align}
and
\begin{align}
    \hat{h}^{12} \simeq \hat{h}^1 - \hat{h}^2 \,.
\end{align}
The correspondences between the background gauge parameters are
\begin{align}
    \Gamma_0 &\simeq \gamma_0 \,, \\
    \Gamma_{12} &\simeq \gamma^1_2 + \partial_2 \gamma_1 \,, \\
    \hat{\Gamma}^{12} &\simeq \hat{\gamma}^1 - \hat{\gamma}^2 \,.
\end{align}

\subsection*{The SSPT phases for $\Z_N \times \Z_N$ with two foliations in 3+1 dimensions}

\subsubsection*{The Exotic SSPT phase for $\Z_N \times \Z_N$ with two foliations in 3+1 dimensions}

The exotic Lagrangian is
\begin{align}
\begin{split}
    &\L_{\text{SSPT,e}}\left[ \bm{C}, \bm{\hat{C}} \right] \\
    & \quad =  \frac{iN}{2\pi} \hat{\beta}^{12} \left( \partial_0 C_{312} - \partial_3 C_{012} + \partial_1 \partial_2 C_{03} \right) \\
    & \quad \quad +  \frac{iN}{2\pi} \left[ \beta_0 \left( \partial_3 \hat{C} - \partial_1 \partial_2 \hat{C}^{12}_3  \right) + \beta_{12} \left( \partial_3 \hat{C}^{12}_0 - \partial_0 \hat{C}^{12}_3  \right)  - \beta_3 \left( \partial_0 \hat{C} - \partial_1 \partial_2 \hat{C}^{12}_0  \right) \right] \\
    &\quad \quad  + \frac{iN}{2\pi}  \left( \hat{C}^{12}_3 C_{012}  - \hat{C}^{12}_0 C_{312}  + \hat{C} C_{03}   \right) \,,
\end{split}
\end{align}
where $\bm{C} = (C_{012}, C_{312}, C_{03})$ is a background tensor two-form gauge field and $\bm{\hat{C}} = (\hat{C}^{12}_0, \hat{C}, \hat{C}^{12}_3)$ is a background tensor one-form gauge field, $\bm{\beta} = (\beta_0, \beta_{12},\beta_3)$ is a dynamical tensor one-form gauge field and $\bm{\hat{\beta}} = (\hat{\beta}^{12})$ is a dynamical tensor zero-form gauge field of $\Z_4 \times SO(2)$.
The dynamical gauge transformations of $\bm{\beta}$ and $\bm{\hat{\beta}}$ are
\begin{align}
    \beta_0 &\sim \beta_0 + \partial_0 s \,, \\
    \beta_{12} &\sim \beta_{12} + \partial_1 \partial_2 s \,, \\
    \beta_3 &\sim \beta_3 + \partial_3 s \,, \\
    \hat{\beta}^{12} &\sim \hat{\beta}^{12} + 2\pi \hat{s}^1 - 2\pi \hat{s}^2 \,,
\end{align}
where $s$ is a dynamical tensor zero-form gauge parameter, and $\hat{s}^k$ is an $x^k$-dependent integer-valued gauge parameter.
The background gauge transformations of the fields are
\begin{align}
    C_{012} &\sim C_{012} + \partial_0 \Gamma_{12} -\partial_1\partial_2 \Gamma_0 \,, \\
    C_{312} &\sim C_{312} + \partial_3 \Gamma_{12} -\partial_1\partial_2 \Gamma_3 \,, \\
    C_{03} &\sim C_{03} + \partial_0 \Gamma_3 -\partial_3 \Gamma_0 \,,
\end{align}
and
\begin{align}
    \hat{C}^{12}_0 &\sim \hat{C}^{12}_0 + \partial_0  \hat{\Gamma}^{12} \,, \\
    \hat{C} &\sim \hat{C} + \partial_1 \partial_2  \hat{\Gamma}^{12} \,, \\
    \hat{C}^{12}_3 &\sim \hat{C}^{12}_3 + \partial_3  \hat{\Gamma}^{12} \,,
\end{align}
and
\begin{align}
    \beta_0 &\sim \beta_0 - \Gamma_0 \,, \\
    \beta_{12} &\sim \beta_{12} - \Gamma_{12} \,, \\
    \beta_3 &\sim \beta_3 - \Gamma_3 \,, \\
    \hat{\beta}^{12} &\sim \hat{\beta}^{12} - \hat{\Gamma}^{12} \,,
\end{align}
where $\bm{\Gamma} = (\Gamma_0, \Gamma_{12}, \Gamma_3)$ is a background tensor one-form gauge parameter, and $\bm{\hat{\Gamma}} = (\hat{\Gamma}^{12})$ is a background tensor zero-form gauge parameter.

\subsubsection*{The Foliated SSPT phase for $\Z_N \times \Z_N$ with two foliations in 3+1 dimensions}

The foliated Lagrangian is
\begin{align}
\begin{split}
    &L_{\text{SSPT},\text{f}} \left[ C^k \wedge e^k, c, \hat{C}^k, \hat{c}_{12} \right] \\
    &=  \frac{iN}{2\pi} \left[ -\bm{\hat\beta} \wedge_\text{f} d_\text{f} \bm{C} +  \bm{\beta} \wedge_\text{f} d_\text{f} \bm{\hat{C}} + \bm{\hat{C}} \wedge_\text{f} \bm{C}   \right]  \\
    &=  \frac{iN}{2\pi} \Bigg[ -\sum_{k=1}^2 \hat{\beta}^k \, dC^k \wedge e^k +  \hat{\beta} \wedge  \left( dc - \sum_{k=1}^2 C^k \wedge e^k \right)    \\
    & \qquad + \sum_{k=1}^2 \beta^k \wedge e^k \wedge (d\hat{C}^k + \hat{c}) + \beta \wedge d\hat{c}  + \sum_{k=1}^2 \hat{C}^k \wedge C^k \wedge e^k + \hat{c} \wedge c \Bigg] \,,
\end{split}
\end{align}
where $\bm{C} = (C^k \wedge e^k, c)$ is a type-$A$ foliated (2+1)-form gauge field, $\bm{\hat{C}} = (\hat{C}^k, \hat{c})$ is a type-$B$ foliated one-form gauge field, $\bm{\beta} = (\beta^k \wedge e^k, \beta)$ is a dynamical type-$A$ foliated (1+1)-form gauge field and $\bm{\hat\beta} = (\hat{\beta}^k, \hat{\beta})$ is a dynamical type-$B$ foliated zero-form gauge field. The gauge fields $C^k \wedge e^k$, $c$, $\hat{C}^k$ and $\hat{c}_{12}$ are background gauge fields, and the gauge fields $\hat{c}_{ij} \ ((i,j) = (0,1), (0,2), (0,3), (2,3), (3,1))$ are dynamical gauge fields.
The dynamical gauge transformations of $\bm{\beta}$ and $\bm{\hat\beta}$ are
\begin{align}
    \beta^k \wedge e^k &\sim \beta^k \wedge e^k + d u^k \wedge e^k   \,, \\
    \beta &\sim \beta + d u - \sum_{k=1}^2 u^k e^k \,, \\
    \hat{\beta}^k &\sim \hat{\beta}^k + 2\pi \hat{u}^k + \hat{u}   \,, \\
    \hat{\beta} &\sim \hat{\beta} + d \hat{u}   \,,
\end{align}
where  $\bm{u}=(u^k e^k,u)$ is a dynamical type-$A$ foliated (0+1)-form gauge parameter, $\hat{u}^k$ is an $x^k$-dependent integer-valued gauge parameter, and $\hat{u}$ is a dynamical type-$B$ bulk zero-form gauge parameter.
The background gauge transformations of the fields are
\begin{align}
    C^k \wedge e^k &\sim C^k \wedge e^k + d\gamma^k \wedge e^k \,, \\
    c &\sim c + d\gamma + \sum_{k=1}^2 \gamma^k \wedge e^k \,, \\
    \hat{C}^k &\sim \hat{C}^k + d \hat{\gamma}^k - \hat{\gamma}  \,, \\
    \hat{c} &\sim \hat{c} + d\hat{\gamma} \,,
\end{align}
and
\begin{align}
    \beta^k \wedge e^k &\sim \beta^k \wedge e^k - \gamma^k \wedge e^k  \,, \\ 
    \beta &\sim \beta - \gamma  \,, \\
    \hat{\beta}^k &\sim \hat{\beta}^k - \hat{\gamma}^k  \,, \\
    \hat{\beta} &\sim \hat{\beta}  - \hat{\gamma} \,,
\end{align}
where $\bm{\gamma} = (\gamma^k \wedge e^k, \gamma)$ is a background type-$A$ foliated (0+1)-form gauge parameter, and $\bm{\hat{\gamma}} = (\hat{\gamma}^k, \hat{\gamma})$ is a background type-$B$ foliated zero-form gauge parameter.

\subsubsection*{Correspondences in the SSPT phases for $\Z_N \times \Z_N$ with two foliations in 3+1 dimensions}

The field correspondences in the SSPT phase are
\begin{align}
    C_{012} &\simeq C^1_{02} + \partial_2 c_{01} \,,  \\
    C_{312} &\simeq C^1_{32} + \partial_2 c_{31}   \,, \\
    C_{03} &\simeq c_{03}  \,, \\
    \hat{C}^{12}_0 &\simeq \hat{C}^1_0 - \hat{C}^2_0 \,, \\
    \hat{C} &\simeq \partial_1 \hat{C}^1_2 - \partial_2 \hat{C}^2_1 + \hat{c}_{12} \,, \\
    \hat{C}^{12}_3 &\simeq \hat{C}^1_3 - \hat{C}^2_3  \,,
\end{align}
and
\begin{align}
    \beta_{12} &\simeq \beta^1_2 + \partial_2 \beta_1 \,, \\
    \hat{\beta}^{12} &\simeq \hat{\beta}^1 - \hat{\beta}^2 \,.
\end{align}
The correspondences between the dynamical gauge parameters are
\begin{align}
    s &\simeq u \,, \\
    \hat{s}^k &\simeq \hat{u}^k \,.
\end{align}
The correspondences between the background gauge parameters are
\begin{align}
    \Gamma_0 &\simeq \gamma_0 \,, \\
    \Gamma_3 &\simeq \gamma_3 \,, \\
    \Gamma_{12} &\simeq \gamma^1_2 + \partial_2 \gamma_1  \,, \\
    \hat{\Gamma}^{12} &\simeq \hat{\gamma}^1 - \hat{\gamma}^2 \,.
\end{align}

\subsection*{The SSPT phases for $\Z_N \times \Z_N$ with three foliations in 3+1 dimensions}

\subsubsection*{The foliated SSPT phase for $\Z_N \times \Z_N$ with three foliations in 3+1 dimensions}

The foliated Lagrangian is
\begin{align}
\begin{split}
    &L^3_{\text{SSPT},\text{f}} \left[ C^k \wedge e^k, c, \hat{C}^k, \hat{c}_{12},\hat{c}_{23},\hat{c}_{31} \right] \\
    &=  \frac{iN}{2\pi} \left[ -\bm{\hat\beta} \wedge_\text{f} d_\text{f} \bm{C} +  \bm{\beta} \wedge_\text{f} d_\text{f} \bm{\hat{C}} + \bm{\hat{C}} \wedge_\text{f} \bm{C}   \right]  \\
    &=  \frac{iN}{2\pi} \Bigg[ -\sum_{k=1}^3 \hat{\beta}^k \, dC^k \wedge e^k +  \hat{\beta} \wedge  \left( dc - \sum_{k=1}^3 C^k \wedge e^k \right)    \\
    & \qquad + \sum_{k=1}^3 \beta^k \wedge e^k \wedge (d\hat{C}^k + \hat{c}) + \beta \wedge d\hat{c}  + \sum_{k=1}^3 \hat{C}^k \wedge C^k \wedge e^k + \hat{c} \wedge c \Bigg] \,,
\end{split}
\end{align}
where $\bm{C} = (C^k \wedge e^k, c)$ is a type-$A$ foliated (2+1)-form gauge field, $\bm{\hat{C}} = (\hat{C}^k, \hat{c})$ is a type-$B$ foliated one-form gauge field, $\bm{\beta} = (\beta^k \wedge e^k, \beta)$ is a type-$A$ foliated (0+1)-form gauge field, and $\bm{\hat\beta} = (\hat{\beta}^k, \hat{\beta})$ is a type-$B$ foliated zero-form gauge field. The gauge fields $C^k \wedge e^k$, $c$, $\hat{C}^k$, $\hat{c}_{12}$, $\hat{c}_{23}$ and $\hat{c}_{31}$ are background gauge fields, and the gauge fields $\hat{c}_{01}$, $\hat{c}_{02}$ and $\hat{c}_{03}$ are dynamical gauge fields.
The dynamical gauge transformations of $\bm{\beta}$ and $\bm{\hat\beta}$ are
\begin{align}
    \beta^k \wedge e^k &\sim \beta^k \wedge e^k + d u^k \wedge e^k   \,, \\
    \beta &\sim \beta + d u - \sum_{k=1}^3 u^k e^k \,, \\
    \hat{\beta}^k &\sim \hat{\beta}^k + 2\pi \hat{u}^k + \hat{u}   \,, \\
    \hat{\beta} &\sim \hat{\beta} + d \hat{u}   \,,
\end{align}
where $u^k e^k$ is a type-$A$ foliated (0+1)-form gauge parameter, $u$ is a type-$A$ bulk zero-form gauge parameter, $\hat{u}^k$ is an $x^k$-dependent integer-valued gauge parameter, and $\hat{u}$ is a type-$B$ bulk zero-form gauge parameter.
The background gauge transformations of the fields are
\begin{align}
    C^k \wedge e^k &\sim C^k \wedge e^k + d\gamma^k \wedge e^k \,, \\
    c &\sim c + d\gamma + \sum_{k=1}^3 \gamma^k \wedge e^k \,, \\
    \hat{C}^k &\sim \hat{C}^k + d \hat{\gamma}^k - \hat{\gamma}  \,, \\
    \hat{c} &\sim \hat{c} + d\hat{\gamma} \,,
\end{align}
and
\begin{align}
    \beta^k \wedge e^k &\sim \beta^k \wedge e^k - \gamma^k \wedge e^k  \,, \\ 
    \beta &\sim \beta - \gamma  \,, \\
    \hat{\beta}^k &\sim \hat{\beta}^k - \hat{\gamma}^k  \,, \\
    \hat{\beta} &\sim \hat{\beta}  - \hat{\gamma} \,,
\end{align}
where $\bm{\gamma} = (\gamma^k \wedge e^k, \gamma)$ is a background type-$A$ foliated (0+1)-form gauge parameter, and $\bm{\hat{\gamma}} = (\hat{\gamma}^k, \hat{\gamma})$ is a background type-$B$ foliated zero-form gauge parameter.

\subsubsection*{The exotic SSPT phase for $\Z_N \times \Z_N$ with three foliations in 3+1 dimensions}

The exotic Lagrangian is
\begin{align}
\begin{split}
    &\L^3_{\text{SSPT},\text{e}} \left[ \bm{C}, \bm{\hat{C}} \right]  \\
    & = \frac{iN}{2\pi} \sum_{(i,j,k)} \hat{\beta}^{[ij]k} \left( \partial_0 C_{[ij]k} - \partial_i C_{0jk} + \partial_j C_{0ki} \right) \\ 
    & \quad -\frac{iN}{2\pi} \left[ \sum_{(i,j,k)} \beta_{ij} \left( \partial_0 \hat{C}^{ij} - \partial_k \hat{C}^{[ki]j} + \partial_k \hat{C}^{[jk]i} \right) +  \beta_0 \sum_{(i,j)}  \partial_i \partial_j \hat{C}^{ij}  \right] \\
    &\quad  + \frac{iN}{2\pi} \sum_{(i,j,k)} \left[  \hat{C}^{ij} C_{0ij} - \hat{C}^{[ij]k}_0 C_{[ij]k}  \right] \,, 
\end{split}
\end{align}
or
\begin{align}
\begin{split}
    &\L^3_{\text{SSPT},\text{e}} \left[ \bm{C}, \bm{\hat{C}} \right]  \\
    &= \frac{iN}{2\pi} \sum_{(i,j,k)}\frac{1}{3} \hat{\beta}^{k(ij)}\left(\partial_0 C_{k(ij)} - 2\partial_k C_{0ij} + \partial_i C_{0jk} +  \partial_j C_{0ki} \right) \\ 
    & \quad -\frac{iN}{2\pi} \left[ \sum_{(i,j,k)} \beta_{ij} \left( \partial_0 \hat{C}^{ij} - \partial_k \hat{C}^{k(ij)} \right) +  \beta_0 \sum_{(i,j)}  \partial_i \partial_j \hat{C}^{ij}  \right] \\
    &\quad  + \frac{iN}{2\pi} \sum_{(i,j,k)} \left[  \hat{C}^{ij} C_{0ij} - \frac{1}{3}\hat{C}^{k(ij)}_0 C_{k(ij)}  \right] \,,
\end{split}
\end{align}
where $\bm{C} = (C_{0ij}, C_{[ij]k}) = (C_{0ij}, C_{k(ij)})$ is a background tensor two-form gauge field, $\bm{\hat{C}} = (\hat{C}_0^{k(ij)}, \hat{C}^{ij}) = (\hat{C}_0^{[ij]k},\hat{C}^{ij})$ is a background tensor one-form gauge field, $\bm{\beta} = (\beta_0, \beta_{ij})$ is a dynamical tensor one-form gauge field, $\bm{\hat{\beta}} =(\hat{\beta}^{k(ij)}) = (\hat{\beta}^{[ij]k})$ is a dynamical tensor zero-form gauge field of $S_4$, and $(i,j,k) = (1,2,3), (2,3,1), (3,1,2)$.
The dynamical gauge transformations of $\bm{\beta} = (\beta_0, \beta_{ij})$ and $\bm{\hat\beta} =(\hat{\beta}^{k(ij)}) = (\hat{\beta}^{[ij]k})$ are
\begin{align}
    \beta_0 &\sim \beta_0 + \partial_0 s \,, \\
    \beta_{ij} &\sim \beta_{ij} + \partial_i \partial_j s \,, \\
    \hat{\beta}^{k(ij)} &\sim \hat{\beta}^{k(ij)} + 2\pi \hat{s}^i - 2\pi \hat{s}^j \,, \\
    \hat{\beta}^{[ij]k} &\sim \hat{\beta}^{[ij]k} + \frac{2\pi}{3} ( \hat{s}^i + \hat{s}^j - 2 \hat{s}^k )  \,,
\end{align}
where $s$ is a dynamical tensor zero-form gauge parameter and $\hat{s}^k$ is a $x^k$-dependent integer-valued gauge parameter.
The background gauge transformations of the fields $\bm{C} = (C_{0ij}, C_{[ij]k}) = (C_{0ij}, C_{k(ij)})$ are
\begin{align}
    C_{0ij} &\sim C_{0ij} + \partial_0 \Gamma_{ij} - \partial_i \partial_j \Gamma_0 \,,  \\
    C_{[ij]k} &\sim C_{[ij]k} + \partial_i \Gamma_{jk} - \partial_j \Gamma_{ki} \,, \\
    C_{k(ij)} &\sim C_{k(ij)} + 2 \partial_k \Gamma_{ij} - \partial_i \Gamma_{jk} - \partial_j \Gamma_{ki} \,,
\end{align}
where $\bm{\Gamma} = (\Gamma_0, \Gamma_{ij})$ is a background tensor one-form gauge parameter, and the background gauge transformations of the fields $\bm{\hat{C}} = (\hat{C}_0^{k(ij)}, \hat{C}^{ij}) = (\hat{C}_0^{[ij]k},\hat{C}^{ij})$ are
\begin{align}
    \hat{C}_0^{k(ij)} &\sim \hat{C}_0^{k(ij)} + \partial_0 \hat{\Gamma}^{k(ij)}   \,, \\
    \hat{C}^{ij} &\sim \hat{C}^{ij} + \partial_k \hat{\Gamma}^{k(ij)} \,, 
\end{align}
or equivalently,
\begin{align}
    \hat{C}_0^{[ij]k} &\sim \hat{C}_0^{[ij]k} + \partial_0 \hat{\Gamma}^{[ij]k}   \,, \\
    \hat{C}^{ij} &\sim \hat{C}^{ij} + \partial_k \hat{\Gamma}^{[ki]j} - \partial_k \hat{\Gamma}^{[jk]i} \,,
\end{align}
where $\bm{\hat{\Gamma}} = (\hat{\Gamma}^{k(ij)}) = (\hat{\Gamma}^{[ij]k})$ is a background tensor zero-form gauge parameter.
The background gauge transformations of the fields $\bm{\beta} = (\beta_0, \beta_{ij})$ and $\bm{\hat\beta} =(\hat{\beta}^{k(ij)}) = (\hat{\beta}^{[ij]k})$ are
\begin{align}
    \beta_0 &\sim \beta_0 - \Gamma_0 \,, \\
    \beta_{ij} &\sim \beta_{ij} - \Gamma_{ij} \,, \\
    \hat{\beta}^{k(ij)} &\sim \hat{\beta}^{k(ij)} - \hat{\Gamma}^{k(ij)} \,,  \\
    \hat{\beta}^{[ij]k} &\sim \hat{\beta}^{[ij]k} - \hat{\Gamma}^{[ij]k} \,.
\end{align}

\subsubsection*{Correspondences in the SSPT phases for $\Z_N \times \Z_N$ with three foliations in 3+1 dimensions}

The correspondences in $\bm{C}$ in the SSPT phase are
\begin{align}
    C_{012} &\simeq C^1_{02} + \partial_2 c_{01} \,,\\
    C_{023} &\simeq C^2_{03} + \partial_3 c_{02}  \,,  \\
    C_{031} &\simeq C^3_{01} + \partial_1 c_{03} \,,
\end{align}
and
\begin{align}
    C_{[12]3} &\simeq C^3_{12} - \partial_1 c_{23}  \,,  \\
    C_{[23]1} &\simeq C^1_{23} - \partial_2 c_{31}  \,,  \\
    C_{[31]2} &\simeq C^2_{31} - \partial_3 c_{12}  \,,
\end{align}
or quivalently,
\begin{align}
    C_{1(23)} &\simeq C^3_{12} - C^2_{31}  - \partial_1 c_{23} + \partial_3 c_{12} \,,  \\
    C_{2(31)} &\simeq C^1_{23} - C^3_{12}  - \partial_2 c_{31} + \partial_1 c_{23} \,,  \\
    C_{3(12)} &\simeq C^2_{31} - C^1_{23}  - \partial_3 c_{12} + \partial_2 c_{31} \,.
\end{align}
The correspondences in $\bm{\hat{C}}$ are
\begin{align}
    \hat{C}_0^{3(12)} &\simeq \hat{C}_0^1 - \hat{C}_0^2 \,,  \\
    \hat{C}_0^{1(23)} &\simeq \hat{C}_0^2 - \hat{C}_0^3 \,,  \\
    \hat{C}_0^{2(31)} &\simeq \hat{C}_0^3 - \hat{C}_0^1 \,, 
\end{align}
or equivalently,
\begin{align}
    \hat{C}^{[12]3}_0 &\simeq \frac{1}{3} \left( \hat{C}^1_0 + \hat{C}^2_0 +  \hat{C}^3_0 \right)  - \hat{C}^3_0 \,,  \\
    \hat{C}^{[23]1}_0 &\simeq \frac{1}{3} \left( \hat{C}^2_0 + \hat{C}^3_0 +  \hat{C}^1_0 \right) - \hat{C}^1_0 \,, \\
    \hat{C}^{[31]2}_0 &\simeq \frac{1}{3} \left( \hat{C}^3_0 + \hat{C}^1_0 +  \hat{C}^2_0 \right) - \hat{C}^2_0 \,,
\end{align}
and
\begin{align}
    \hat{C}^{12} &\simeq \hat{C}_3^1 - \hat{C}_3^2 \,, \\
    \hat{C}^{23} &\simeq \hat{C}_1^2 - \hat{C}_1^3 \,, \\
    \hat{C}^{31} &\simeq \hat{C}_2^3 - \hat{C}_2^1 \,. 
\end{align}
The correspondences in $\bm{\beta}$ and $\bm{\hat{\beta}}$ are
\begin{align}
    \beta_{12} &\simeq \beta^1_2 + \partial_2 \beta_1 \,, \\
    \beta_{23} &\simeq \beta^2_3 + \partial_3 \beta_2\,, \\
    \beta_{31} &\simeq \beta^3_1 + \partial_1 \beta_3 \,, 
\end{align}
and
\begin{align}
    \hat{\beta}^{3(12)} &\simeq \hat{\beta}^1 - \hat{\beta}^2 \,, \\
    \hat{\beta}^{1(23)} &\simeq \hat{\beta}^2 - \hat{\beta}^3 \,, \\
    \hat{\beta}^{2(31)} &\simeq \hat{\beta}^3 - \hat{\beta}^1 \,,
\end{align}
or equivalently,
\begin{align}
    \hat{\beta}^{[12]3} &\simeq \frac{1}{3} \left( \hat{\beta}^1 + \hat{\beta}^2 + \hat{\beta}^3 \right) - \hat{\beta}^3 \,, \\
    \hat{\beta}^{[23]1} &\simeq \frac{1}{3} \left( \hat{\beta}^2 + \hat{\beta}^3 + \hat{\beta}^1 \right) - \hat{\beta}^1 \,, \\
    \hat{\beta}^{[31]2} &\simeq \frac{1}{3} \left( \hat{\beta}^3 + \hat{\beta}^1 + \hat{\beta}^2 \right) - \hat{\beta}^2 \,,
\end{align}
The correspondences in the dynamical gauge parameters are
\begin{align}
    s &\simeq u \,, \\
    \hat{s}^k &\simeq \hat{u}^k 
     \,.
\end{align}
The correspondences in the background gauge parameters are
\begin{align}
    \Gamma_0 &\simeq \gamma_0 \,, \\
    \Gamma_{12} &\simeq \gamma^1_2 + \partial_2 \gamma_1 \,, \\
    \Gamma_{23} &\simeq \gamma^2_3 + \partial_3 \gamma_2 \,, \\
    \Gamma_{31} &\simeq \gamma^3_1 + \partial_1 \gamma_3 \,,
\end{align}
and
\begin{align}
    \hat{\Gamma}^{3(12)} &\simeq \hat{\gamma}^1 - \hat{\gamma}^2 \,, \\
    \hat{\Gamma}^{1(23)} &\simeq \hat{\gamma}^2 - \hat{\gamma}^3 \,, \\
    \hat{\Gamma}^{2(31)} &\simeq \hat{\gamma}^3 - \hat{\gamma}^1 \,, 
\end{align}
or equivalently,
\begin{align}
    \hat{\Gamma}^{[12]3} &\simeq \frac{1}{3} \left( \hat{\gamma}^1 + \hat{\gamma}^2 + \hat{\gamma}^3 \right) - \hat{\gamma}^3 \,, \\
    \hat{\Gamma}^{[23]1} &\simeq \frac{1}{3} \left( \hat{\gamma}^2 + \hat{\gamma}^3 + \hat{\gamma}^1 \right) - \hat{\gamma}^1 \,, \\
    \hat{\Gamma}^{[31]2} &\simeq \frac{1}{3} \left( \hat{\gamma}^3 + \hat{\gamma}^1 + \hat{\gamma}^2 \right) - \hat{\gamma}^2 \,.
\end{align}

\chapter{Fractonic \texorpdfstring{$\phi$}{phi}-Theory in 2+1 Dimensions and Their Anomalies}
\label{chapter_phianomaly}

This chapter is based on the work in \cite{Ohmori:2025fuy}.

Building on Chapter \ref{chapter:2+1d_bfanomaly_and_3+1d_sspt} where we constructed a foliated SSPT phase from an anomalous boundary theory via the foliated-exotic duality, we now reverse the approach. 
In this chapter, we discuss a method to derive a boundary theory from an SSPT phase in one dimension higher, leading to a construction of a foliated $\phi$-theory. The foliated $\phi$-theory can be shown to be equivalent to the exotic $\phi$-theory \cite{Seiberg:2020bhn}, a gapless scalar field theory with $U(1) \times U(1)$ anomalous subsystem symmetry \cite{Gorantla:2021bda}, by tuning parameters and applying correspondences between the fields.
The resulting foliated theory contains a 1+1d compact scalar field $\Phi^k$ on each layer of the foliation in the $x^k$ direction $(k=1,2)$, coupled to a 2+1d compact scalar field $\Phi$.
Finally, we consider the $T$-duality-type duality between the exotic $\phi$-theory and the exotic $\hat\phi$-theory \cite{Seiberg:2020bhn,Spieler:2024fby}. We also construct a foliated $\hat\phi$-theory from the foliated SSPT phase. The foliated $\hat\phi$-theory contains 1+1d compact scalar fields $\hat\Phi^k$ on each layer of the foliation in the $x^k$ direction $(k=1,2)$, coupled to a 2+1d one-form gauge field $\hat\Phi$.

\begin{figure}[H]
   \begin{center}
    \includegraphics[width=1\hsize]{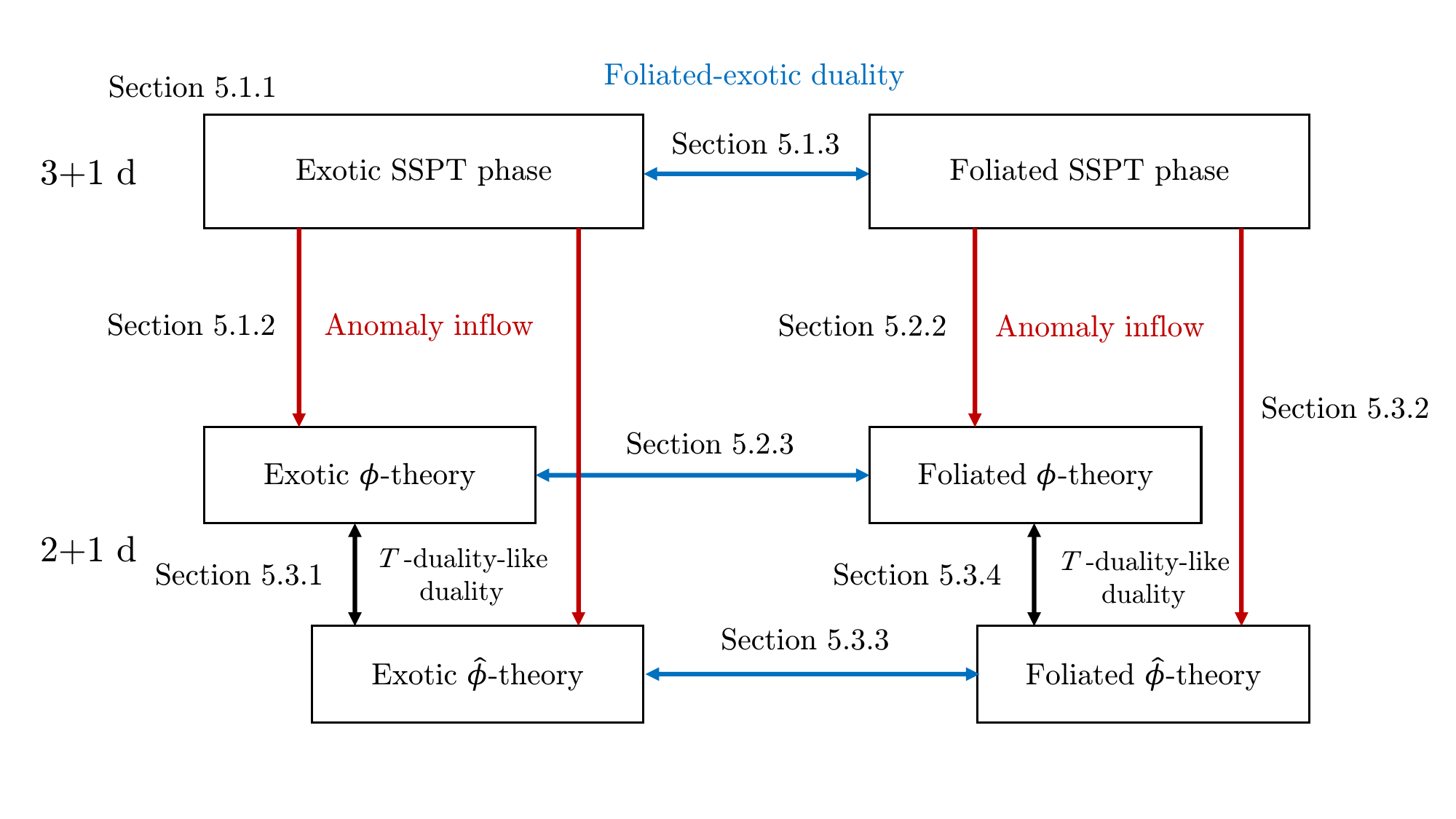} 
    \end{center}
    \vspace{-1cm}
      \caption{The structure of the anomaly inflow and dualities.}
      \label{fig_gapless}
\end{figure}

The organization of this chapter is as follows and is also summarized in Figure \ref{fig_gapless}.

In Section \ref{subsection:u1u1_exotic_sspt}, we will review the exotic description of the SSPT phase for $U(1) \times U(1)$ subsystem symmetry in 3+1 dimensions \cite{Burnell:2021reh}.
This exotic SSPT phase cancels the 't Hooft anomaly of $U(1) \times U(1)$ subsystem symmetry in the exotic $\phi$-theory in 2+1 dimensions \cite{Seiberg:2020bhn}, which we will review in Section \ref{subsection:exotic_phi_theory}.
In Section \ref{subsection:field_correspondences_and_foliated_sspt}, we establish correspondences between exotic tensor gauge fields and foliated gauge fields, and then construct the foliated description of the SSPT phase, which is equivalent to the exotic SSPT phase for $U(1) \times U(1)$ subsystem symmetry.
This foliated-exotic duality is almost identical to the duality in the fractonic $BF$ theory with two flat foliations \cite{Spieler:2023wkz}.

Section \ref{subsection:construction_of_boundary_theory} describes a way to construct an anomalous theory on a boundary from an SPT phase in one dimension higher.
As a relativistic example, we will construct a compact scalar field theory in 1+1 dimensions from the SPT phase for $U(1) \times U(1)$ global symmetry in 2+1 dimensions.
In Section \ref{subsection:construction_of_foliated_phi_theory}, we will construct a foliated $\phi$-theory from the foliated SSPT phase.
In Section \ref{subsection:foliated_exotic_duality_phi}, we discuss the foliated-exotic duality between the exotic $\phi$-theory and the foliated $\phi$-theory.

In Section \ref{section:duality_foliated_phihat}, we investigate the relation between the foliated-exotic duality and the $T$-duality-like duality between the $\phi$-theory and the $\hat\phi$-theory.
In Section \ref{subsection:exotic_phihat_theory}, we will review the (self-)duality between the exotic $\phi$-theory and the exotic $\hat\phi$-theory \cite{Seiberg:2020bhn,Spieler:2024fby}.
In Section \ref{subsection:construction_of_foliated_phihat_theory}, we will construct a foliated $\hat\phi$-theory from the foliated SSPT phase in the same way as in Section \ref{subsection:construction_of_foliated_phi_theory}.
Section \ref{subsection:foliated_exotic_duality_phihat} discusses the foliated-exotic duality between the exotic $\hat\phi$-theory and the foliated $\hat\phi$-theory.
 In Section \ref{subsection:foliated_duality_phi}, we will describe the duality between the foliated $\phi$-theory and the foliated $\hat\phi$-theory, which is parallel to the discussion in Section \ref{subsection:exotic_phihat_theory}.

\section{Exotic and Foliated SSPT Phase for \texorpdfstring{$U(1) \times U(1)$}{U(1) × U(1)} Subsystem Symmetry}
\label{section:exotic_and_foliated_sspt_phase}

In this section, we first review the exotic description of the SSPT phase for $U(1) \times U(1)$ subsystem symmetry in 3+1 dimensions \cite{Burnell:2021reh} and an anomalous $\phi$-theory on (2+1)-dimensional boundary \cite{Seiberg:2020bhn}. This SSPT phase describes an 't Hooft anomaly of a $U(1) \times U(1)$ subsystem symmetry in the exotic $\phi$-theory via the anomaly inflow mechanism \cite{Gorantla:2021bda}. Next, we will consider the field correspondences between exotic tensor gauge fields and foliated gauge fields, and construct the foliated description of the SSPT phase.

\subsection{Exotic SSPT Phase for \texorpdfstring{$U(1) \times U(1)$}{U(1) × U(1)} in 3+1 Dimensions}
\label{subsection:u1u1_exotic_sspt}

We review the exotic SSPT phase for $U(1) \times U(1)$ subsystem symmetry in 3+1 dimensions \cite{Burnell:2021reh}. We take Euclidean spacetime to be a (3+1)-dimensional torus $\T^{3+1}$ of lengths $l^0$, $l^1$, $l^2$ and $l^3$, and the coordinate $(x^0, x^1, x^2 ,x^3)$ on it. First, we consider the exotic SSPT phase on the torus, and then we will restrict the space to the region $x^3 \geq 0$ and take the boundary $x^3 = 0$. The SSPT phase has the 90-degree discrete rotational symmetry $\Z_4$ for $(x^1,x^2)$ and the continuous rotational symmetry $SO(2)$ for $(x^0,x^3)$. Related to that, this theory has $U(1)$ exotic tensor gauge fields in representations of $\Z_4 \times SO(2)$.
This theory has a background tensor one-form gauge fields $\bm{C}=(C_0, C_{12}, C_3)$ and $\bm{\hat{C}}=(\hat{C}^{12}_{0}, \hat{C}, \hat{C}^{12}_{3})$.\footnote{We denote the tensor gauge fields by $\bm{C}$ and $\bm{\hat{C}}$, but they are different fields from the ones in the previous chapter.} For $\Z_4$ on subspace $(x^1,x^2)$, the tensor gauge fields $C_{12}$, $\hat{C}^{12}_{0}$ and $\hat{C}^{12}_{3}$ are in the representation $\bm{1}_2$, and $C_0$, $C_3$ and $\hat{C}$ are in the representation $\bm{1}_0$. For $SO(2)$ on subspace $(x^0,x^3)$, the fields $(C_0, C_3)$ and $(\hat{C}_0^{12}, \hat{C}_3^{12})$ are in the vector representation, and $C_{12}$ and $\hat{C}$ are in the trivial representation.  
The background gauge transformations of $\bm{C}$ and $\bm{\hat{C}}$ are 
\begin{align}
    C_0 &\sim C_0 + \partial_0 \Gamma \,, \label{bkgugec0} \\
    C_{12} &\sim C_{12} + \partial_1\partial_2 \Gamma \,, \label{bkgugec12} \\ 
    C_3 &\sim C_3 + \partial_3 \Gamma  \,, \label{bkgugec3}
\end{align}
and
\begin{align}
    \hat{C}^{12}_0 &\sim \hat{C}^{12}_0 + \partial_0  \hat{\Gamma}^{12} \,, \label{bkgugechat120}\\
    \hat{C} &\sim \hat{C} + \partial_1 \partial_2  \hat{\Gamma}^{12} \,, \label{bkgugechat}\\
    \hat{C}^{12}_3 &\sim \hat{C}^{12}_3 + \partial_3  \hat{\Gamma}^{12} \,, \label{bkgugechat123}
\end{align}
where $\bm{\Gamma} =(\Gamma)$ and $\bm{\hat{\Gamma}} = (\hat{\Gamma}^{12})$ are tensor zero-form gauge parameters in $\bm{1}_0$ and $\bm{1}_2$, respectively.

The 3+1d SSPT phase for $U(1) \times U(1)$ subsystem symmetry is described by the exotic Lagrangian
\begin{align}
\begin{split}
    &\L_{\text{SSPT,e}}\left[ \bm{C},\bm{\hat{C}} \right] \\ 
     & \quad = \frac{i}{2\pi} \left[ \hat{C} ( \partial_0 C_3 - \partial_3 C_0 ) - \hat{C}_0^{12} ( \partial_3 C_{12} - \partial_1 \partial_2 C_3 )  + \hat{C}_3^{12} ( \partial_0 C_{12} - \partial_1 \partial_2 C_0 )  \right] \,. \label{exoticsspt}
\end{split}
\end{align}
The Lagrangian is gauge invariant on spacetime without a boundary. However, if spacetime has a boundary, a gauge variation arises on the boundary, which will match a mixed 't Hooft anomaly of $U(1) \times U(1)$ subsystem symmetry. Then, we consider the spacetime $\T^{2+1} \times \R_{x^3 \geq 0}$ with the boundary $x^3 = 0$. The gauge variation of the Lagrangian is
\begin{align}
\begin{split}
    &\delta_\text{g} \L_{\text{SSPT,e}} \left[ \bm{C},\bm{\hat{C}} \right] \\ 
      &= \frac{i}{2\pi} \left[ \partial_1 \partial_2 \hat{\Gamma}^{12} ( \partial_0 C_3 - \partial_3 C_0 ) - \partial_0 \hat{\Gamma}^{12} ( \partial_3 C_{12} - \partial_1 \partial_2 C_3 ) + \partial_3 \hat{\Gamma}^{12} ( \partial_0 C_{12} - \partial_1 \partial_2 C_0 )  \right] \\
     & = \partial_3 \left[ \frac{i}{2\pi} \hat{\Gamma}^{12}  ( \partial_0 C_{12} - \partial_1 \partial_2 C_0 ) \right] \,,
\end{split}
\end{align}
where we have dropped the total derivative terms with respect to $x^0$, $x^1$ and $x^2$, so the variation of the action is
\begin{align}
    \delta_\text{g}  S_{\text{SSPT,e}} \left[ \bm{C},\bm{\hat{C}} \right] = \int_{x^3 = 0} d^3x \left[ -\frac{i}{2\pi} \hat{\Gamma}^{12}  ( \partial_0 C_{12} - \partial_1 \partial_2 C_0 ) 
    \right] \,. \label{variexoticsspt}
\end{align}

\subsection{Exotic \texorpdfstring{$\phi$}{φ}-Theory in 2+1 Dimensions}
\label{subsection:exotic_phi_theory}

Next, we review the exotic $\phi$-theory in 2+1 dimensions \cite{Seiberg:2020bhn}, which is the continuum QFT of the XY-plaquette model \cite{Paramekanti:2002iup}. We take Euclidean spacetime to be a (2+1)-dimensional torus $\T^{2+1}$ of lengths $l^0$, $l^1$ and $l^2$, and the coordinate $(x^0, x^1, x^2)$ on it. This theory has two types of $U(1)$ subsystem symmetries: the momentum dipole global symmetry and the winding dipole global symmetry. These symmetries have a mixed 't Hooft anomaly, so they cannot be gauged simultaneously \cite{Gorantla:2021bda}. The anomaly is described by the SSPT phase in section \ref{subsection:u1u1_exotic_sspt}, so the $\phi$-theory is considered as a boundary theory of it \cite{Burnell:2021reh}.

The Lagrangian describing the exotic $\phi$-theory is
\begin{align}
    \L_{\phi,\text{e}} = \frac{\mu_0}{2} (\partial_0 \phi )^2 + \frac{1}{2 \mu_{12}} (\partial_1 \partial_2 \phi )^2 \,,  \label{exoticphi}
\end{align}
where $\mu_0$ and $\mu_{12}$ are parameters with mass dimension one. The field $\phi$ is a dynamical compact scalar field with the periodicity $\phi \sim \phi + 2\pi$, and we have a gauge transformation
\begin{align}
    \phi \sim \phi + 2\pi w^1 + 2\pi w^2 \,.
\end{align}
$w^k$ is an $x^k$-dependent integer-valued gauge parameter. it can have step function discontinuities in the $x^k$ direction.
Then, $\phi$ can also have the discontinuities and these configurations have contributions to the theory due to the term of the second-order derivative.
This theory has the discrete spatial rotational symmetry $\Z_4$, and the field $\phi$ is a $U(1)$ exotic tensor zero-form gauge field $\bm{\phi} = (\phi)$ described in Section \ref{subsection:exotic_tensor_gauge_fields_z4_2d}.

The equation of motion is
\begin{align}
    - \mu_0 \partial_0^2 \phi + \frac{1}{\mu_{12}} \partial^2_1 \partial_2^2 \phi = 0 \,. \label{phieom}
\end{align}

Let us discuss subsystem symmetry. The currents of the momentum dipole symmetry are
\begin{align}
    J_0 &= i \mu_0 \partial_0 \phi \,,  \label{momsym1} \\
    J_{12} &=  \frac{i}{\mu_{12}} \partial_1 \partial_2 \phi \,, \label{momsym2}
\end{align}
and the conservation law is
\begin{align}
    \partial_0 J_0 - \partial_1 \partial_2 J_{12} = 0 \,,
\end{align}
from the equation of motion \eqref{phieom}. The conserved charges are
\begin{align}
    Q^1(x^1) = \oint dx^2 J_0 \,, \\
    Q^2(x^2) = \oint dx^1 J_0 \,. 
\end{align}
By integrating them over the fixed width and exponentiating them, we have the symmetry operators
\begin{align}
    U_{\alpha}^1 \left[ [x_1^1, x_2^1] \times \mathcal{C}^2  \right] = \exp \left[ i \alpha \int_{x_1^1}^{x_2^1} dx^1 \oint_{\mathcal{C}^2} dx^2  \mu_0 \partial_0 \phi  \right] \,, \\
    U_{\alpha}^2 \left[ [x_1^2, x_2^2] \times \mathcal{C}^1 \right] = \exp \left[ i \alpha  \int_{x_1^2}^{x_2^2} dx^2 \oint_{\mathcal{C}^1} dx^1  \mu_0 \partial_0 \phi  \right] \,,
\end{align}
where $\alpha$ is $2\pi$-periodic: $e^{i\alpha} \in U(1)$, and $\mathcal{C}^k$ is a one-dimensional loop along the $x^k$ direction. The momentum dipole symmetry acts on the charged operator
\begin{align}
    V_n [x] = e^{in\phi}  \,,
\end{align}
where $n$ is an integer, as
\begin{align}
    U_{\alpha}^1 \left[ [x_1^1, x_2^1] \times \mathcal{C}^2  \right] \, V_n [x] \, U_{\alpha}^1 \left[ [x_1^1, x_2^1] \times \mathcal{C}^2  \right]^{-1} = e^{in\alpha} \,V_n [x] \,, \quad x_1^1 < x^1 < x_2^1 \,, \\
    U_{\alpha}^2 \left[ [x_1^2, x_2^2] \times \mathcal{C}^1 \right] \, V_n [x] \, U_{\alpha}^2 \left[ [x_1^2, x_2^2] \times \mathcal{C}^1 \right]^{-1} = e^{in\alpha} \,V_n [x] \,, \quad x_1^2 < x^2 < x_2^2 \,.
\end{align}
For the field action, we have
\begin{align}
    \phi &\rightarrow \phi + \Gamma^1(x^1) + \Gamma^2(x^2) \,,
\end{align}
where $\Gamma^k(x^k)$ is a $2\pi$-periodic scalar that can have step function discontinuities in the $x^k$ direction.

On the other hand, the currents of the winding dipole symmetry are
\begin{align}
    \hat{J}^{12}_0 &= \frac{1}{2 \pi} \partial_1 \partial_2 \phi \,, \label{winsym1} \\
    \hat{J} &= \frac{1}{2 \pi} \partial_0 \phi \,,  \label{winsym2}
\end{align}
and the conservation law is
\begin{align}
    \partial_0 \hat{J}^{12}_0 - \partial_1 \partial_2 \hat{J} = 0 \,.
\end{align}
The conserved charges are
\begin{align}
    \hat{Q}^1(x^1) = \oint dx^2 \hat{J}^{12}_0 \,, \\
    \hat{Q}^2(x^2) = \oint dx^1 \hat{J}^{12}_0 \,. 
\end{align}
Then, we have the symmetry operators
\begin{align}
    \hat{U}_{\hat{\alpha}}^1 \left[ [x_1^1, x_2^1] \times \mathcal{C}^2 \right] = \exp \left[ i \hat{\alpha}  \int_{x_1^1}^{x_2^1} dx^1 \oint_{\mathcal{C}^2} dx^2  \frac{1}{2 \pi} \partial_1 \partial_2 \phi  \right] \,, \\
    \hat{U}_{\hat{\alpha}}^2 \left[  [x_1^2, x_2^2] \times \mathcal{C}^1 \right] = \exp \left[ i \hat{\alpha}  \int_{x_1^2}^{x_2^2} dx^2 \oint_{\mathcal{C}^1} dx^1   \frac{1}{2 \pi} \partial_1 \partial_2 \phi  \right] \,,
\end{align}
where $\hat{\alpha}$ is $2\pi$-periodic: $e^{i\hat{\alpha}} \in U(1)$. The winding dipole symmetry acts on a charged operator, but it is written by the dual field $\hat{\phi}^{12}$ not $\phi$. We will consider the duality in Section \ref{section:duality_foliated_phihat}.

Let us see that the $U(1) \times U(1)$ subsystem symmetry has a mixed 't Hooft anomaly. To see this, we couple the subsystem symmetries to background gauge fields. Then, the symmetry action on the field becomes the local transformation:
\begin{align}
    \phi &\sim \phi + \Gamma \,.
\end{align}
where $\Gamma$ is a $2\pi$-periodic local parameter and it is absorbed into the gauge transformations of the background gauge fields.

The momentum dipole symmetry is coupled to the background tensor one-form gauge field $\bm{C} = (C_0 ,C_{12})$ in the representation $(\bm{1}_0, \bm{1}_2)$. Their background gauge transformations are
\begin{align}
    C_0 &\sim C_0 + \partial_0 \Gamma \,,  \\
    C_{12} &\sim C_{12} + \partial_1\partial_2 \Gamma \,. 
\end{align}
Then, $\bm{\Gamma} = (\Gamma)$ is a tensor zero-form gauge parameter in $\bm{1}_0$.
The winding dipole symmetry is coupled to the background tensor one-form gauge field $\bm{\hat{C}} =  (\hat{C}_0^{12} ,\hat{C})$ in the representation $(\bm{1}_2, \bm{1}_0)$. Their background gauge transformations are
\begin{align}
    \hat{C}^{12}_0 &\sim \hat{C}^{12}_0 + \partial_0  \hat{\Gamma}^{12} \,,  \\
    \hat{C} &\sim \hat{C} + \partial_1 \partial_2  \hat{\Gamma}^{12} \,,
\end{align}
where $\bm{\hat{\Gamma}} = (\hat{\Gamma}^{12})$ is a tensor zero-form gauge parameter in $\bm{1}_2$.

Then, the Lagrangian of the $\phi$-theory including the background exotic tensor gauge fields is
\begin{align}
\begin{split}
    \L_{\phi,\text{e}}\left[ \bm{C}, \bm{\hat{C}} \right] &= \frac{\mu_0}{2} (\partial_0 \phi - C_0 )^2 + \frac{1}{2 \mu_{12}} (\partial_1 \partial_2 \phi - C_{12})^2 \\
    & \quad + \frac{i}{2\pi} \hat{C}^{12}_0 ( \partial_1 \partial_2 \phi - C_{12}) + \frac{i}{2\pi} \hat{C} ( \partial_0 \phi - C_0 ) \,.  \label{exoticphic}
\end{split}
\end{align}
Under the background gauge transformation, it varies as
\begin{align}
\begin{split}
    \delta_\text{g} \L_{\phi,\text{e}}\left[ \bm{C}, \bm{\hat{C}} \right] &=  \frac{i}{2\pi} \left[ \partial_0 \hat{\Gamma}^{12} ( \partial_1 \partial_2 \phi - C_{12} ) + \partial_1 \partial_2 \hat{\Gamma}^{12} ( \partial_0 \phi - C_0 )   \right] \\
    &= \frac{i}{2\pi} \hat{\Gamma}^{12} ( \partial_0 C_{12} - \partial_1 \partial_2 C_0 ) \,, \label{variexoticphi}
\end{split}
\end{align}
where we have dropped the total derivative terms.
Therefore, the partition function is not invariant and we cannot gauge the $U(1) \times U(1)$ subsystem symmetry, indicating a mixed 't Hooft anomaly. This anomaly is canceled by combining the SSPT phase \eqref{exoticsspt} in one dimension higher. We put the $\phi$-theory on the boundary $x^3 = 0$, and set the boundary conditions
\begin{align}
    C_{\text{SSPT},0} |_{x^3 = 0} &= C_{\phi,0} \,, \\
    C_{\text{SSPT},12} |_{x^3 = 0} &= C_{\phi,12} \,, \\
    \hat{C}^{12}_{\text{SSPT},0} |_{x^3 = 0} &= \hat{C}^{12}_{\phi,0} \,, \\
    \hat{C}_{\text{SSPT}} |_{x^3 = 0} &= \hat{C}_{\phi} \,,
\end{align}
so the variation of the SSPT phase \eqref{variexoticsspt} cancels the anomaly \eqref{variexoticphi}.

\subsection{Field Correspondences and Foliated SSPT Phase for \texorpdfstring{$U(1) \times U(1)$}{U(1) × U(1)} in 3+1 Dimensions}
\label{subsection:field_correspondences_and_foliated_sspt}

In Section \ref{subsection:u1u1_exotic_sspt}, we reviewed the exotic description of the SSPT phase with the rotational symmetry $\Z_4 \times SO(2)$ in 3+1 dimensions. In this section, we use the foliated-exotic duality between the exotic and foliated SSPT phase, and construct the foliated description of the SSPT phase for $U(1) \times U(1)$ subsystem symmetry with two flat foliations $e^k = dx^k\, (k =1,2)$. In this duality, the field correspondences are almost the same as those for the fractonic $BF$ theory with two flat foliations in 3+1 dimensions \cite{Spieler:2023wkz}. As for the foliated gauge fields, further details are explained in Section \ref{section:foliated_qft_foliated_gauge_fields}.

The foliated SSPT phase has a $U(1)$ background type-$A$ foliated (1+1)-form gauge field $C^k \wedge e^k\, (k = 1,2)$, a $U(1)$ background type-$A$ bulk one-form gauge field $c$, a $U(1)$ background type-$B$ foliated one-form gauge field $\hat{C}^k\, (k = 1,2)$ that obeys $\hat{C}^k_k = 0$, and the $x^1x^2$-component of a $U(1)$ background type-$B$ bulk two-form gauge field $\hat{c}_{12}$.
We assume the correspondences between the background gauge fields as
\begin{align}
    C_0 &\simeq c_0 \label{corr0-0} \,, \\
    C_{12} &\simeq C^1_2 + \partial_2 c_1 \,,  \label{corr12-1} \\
    C_3 &\simeq c_3 \label{corr3-3} \,,
\end{align}
and
\begin{align}
    \hat{C}_0^{12} &\simeq \hat{C}_0^1 - \hat{C}_0^2 \,, \label{ssptcorr8} \\
    \hat{C} &\simeq \partial_1 \hat{C}_2^1 - \partial_2 \hat{C}_1^2 + \hat{c}_{12}  \,, \\
    \hat{C}_3^{12} &\simeq \hat{C}_3^1 - \hat{C}_3^2 \,. \label{ssptcorr10}
\end{align}
In addition, to impose constraints to the fields $C^k \wedge e^k$ and $c$, we add to the Lagrangian the term
\begin{align}
    \frac{i}{2\pi} \hat{c}' \wedge \left. \left(dc + \sum^{2}_{k=1} C^k \wedge e^k  \right)\right|_{\hat{c}'_{12} = 0} \,, \label{constterm}
\end{align}
or in components,
\begin{align}
\begin{split}
    &\frac{i}{2\pi} \left[ \hat{c}'_{01} (\partial_2 c_3 - \partial_3 c_2 - C^2_3) + \hat{c}'_{02} (\partial_3 c_1 - \partial_1 c_3  + C^1_3)  \right. \\
    & \left.  \quad + \hat{c}'_{03} (\partial_1 c_2 - \partial_2 c_1 + C^2_1 - C^1_2)  + \hat{c}'_{23} (\partial_0 c_1 - \partial_1 c_0 + C^1_0) + \hat{c}'_{31} (\partial_0 c_2 - \partial_2 c_0 + C^2_0) \right] \,,
\end{split}
\end{align}
where $\hat{c}'_{ij} \, ((i,j) = (0,1), (0,2), (0,3), (2,3), (3,1))$ are the $x^ix^j$-component of a $U(1)$ dynamical type-$B$ bulk two-form gauge field.

The background gauge transformations of the foliated gauge fields are
\begin{align}
    C^k \wedge e^k &\sim C^k \wedge e^k +d \gamma^k \wedge e^k  \, , \\
    c &\sim c + d\gamma - \sum^{2}_{k=1} \gamma^k e^k\,,  \\
    \hat{C}^k &\sim \hat{C}^k +d \hat{\gamma}^k - \hat{\gamma} \,, \\
    \hat{c}_{12} &\sim \hat{c}_{12} + (d \hat{\gamma})_{12} \,, 
\end{align}
where $\gamma^k e^k$ is a background type-$A$ foliated (0+1)-form gauge parameter, $\gamma$ is a background type-$A$ bulk zero-form gauge parameter, $\hat{\gamma}^k$ is a background type-$B$ foliated zero-form gauge parameter, and $\hat{\gamma}$ is a background type-$B$ bulk one-form gauge parameter. From the background gauge transformations of the tensor gauge field \eqref{bkgugec0}--\eqref{bkgugec3} and \eqref{bkgugechat120}--\eqref{bkgugechat123}, we have the correspondences between the background gauge parameters
\begin{align}
    \Gamma &\simeq \gamma \,, \label{ssptcorr11}\\
    \hat{\Gamma}^{12} &\simeq \hat{\gamma}^1 - \hat{\gamma}^2 \,. \label{ssptcorr12}
\end{align}
Note that the gauge parameters $\gamma^k e^k$ and $\hat{\gamma}$ do not appear on the exotic side, and the degrees of freedom of the fields are the same on both sides due to their own gauge transformations.

Using the correspondences \eqref{corr0-0}--\eqref{corr3-3} and \eqref{ssptcorr8}--\eqref{ssptcorr10}, we can construct the foliated description of the SSPT phase for $U(1) \times U(1)$ subsystem symmetry. The Lagrangian can be written as
\begin{align}
\begin{split}
    &\L_{\text{SSPT,e}} \left[ \bm{C}, \bm{\hat{C}} \right] \\
     &\simeq  \frac{i}{2\pi} \left[ (\partial_1 \hat{C}^1_2 - \partial_2 \hat{C}^2_1 + \hat{c}_{12}) ( \partial_0 c_3 - \partial_3 c_0 ) - ( \hat{C}^1_0 - \hat{C}^2_0 ) \left\{ \partial_3 ( C^1_2 + \partial_2 c_1 ) - \partial_1 \partial_2 c_3   \right\} \right. \\
    & \left. \quad  + ( \hat{C}^1_3 - \hat{C}^2_3 ) \left\{ \partial_0  ( C^1_2 + \partial_2 c_1 )   - \partial_1 \partial_2 c_0 \right\}  \right] \,. \label{ssptetof1}
\end{split}
\end{align}
Integrating by parts, dropping the $x^0$-, $x^1$- and $x^2$-derivative terms, and then adding the term for constraints \eqref{constterm}, we have
\begin{align}
\begin{split}
    &\L_{\text{SSPT,f}} \left[ C^k \wedge e^k, c, \hat{C}^k, \hat{c}_{12} \right] \\
    &= \frac{i}{2\pi} \left[ - \hat{C}^1_2  \partial_1 ( \partial_0 c_3 - \partial_3 c_0 ) + \hat{C}^2_1  \partial_2 ( \partial_0 c_3 - \partial_3 c_0 ) - ( \hat{C}^1_0 - \hat{C}^2_0 ) \left\{ \partial_3 ( C^1_2 + \partial_2 c_1 ) - \partial_1 \partial_2 c_3   \right\} \right. \\
    & \left. \quad \quad + ( \hat{C}^1_3 - \hat{C}^2_3 ) \left\{ \partial_0  ( C^1_2 + \partial_2 c_1 )   - \partial_1 \partial_2 c_0 \right\} + \hat{c}_{12} ( \partial_0 c_3 - \partial_3 c_0 )  \right] \\
    & \quad + \frac{i}{2\pi} \left[ \hat{c}'_{01} (\partial_2 c_3 - \partial_3 c_2 - C^2_3) + \hat{c}'_{02} (\partial_3 c_1 - \partial_1 c_3  + C^1_3)  \right. \\
    & \left. \quad \quad + \hat{c}'_{03} (\partial_1 c_2 - \partial_2 c_1 + C^2_1 - C^1_2)  + \hat{c}'_{23} (\partial_0 c_1 - \partial_1 c_0 + C^1_0) + \hat{c}'_{31} (\partial_0 c_2 - \partial_2 c_0 + C^2_0) \right]\,. \label{ssptlag2}
\end{split}
\end{align}
Next, let us consider changes of variables for the dynamical fields from $\hat{c}'_{ij}$ to $\hat{c}_{ij}$:\footnote{Since we choose the correspondence \eqref{corr12-1}, we consider $\hat{c}_{03} \rightarrow \hat{c}_{03} + \partial_0 \hat{C}^2_3 - \partial_3 \hat{C}^2_0$ to replace $C^1_2 + \partial_2 c_1$ with $C^2_1 + \partial_1 c_2$. Instead of that, we can adopt the correspondence $C_{12} \simeq C^2_1 + \partial_1 c_2$ and then we have to use the change of variable $\hat{c}_{03} \rightarrow \hat{c}_{03} + \partial_0 \hat{C}^1_3 - \partial_3 \hat{C}^1_0$.}
\begin{align}
    \hat{c}'_{01} &= \hat{c}_{01} + \partial_0 \hat{C}^2_1 - \partial_1 \hat{C}^2_0 \,, \\
    \hat{c}'_{02} &= \hat{c}_{02} + \partial_0 \hat{C}^1_2 - \partial_2 \hat{C}^1_0 \,, \\
    \hat{c}'_{03} &= \hat{c}_{03} + \partial_0 \hat{C}^2_3 - \partial_3 \hat{C}^2_0 \,, \\
    \hat{c}'_{23} &= \hat{c}_{23} + \partial_2 \hat{C}^1_3 - \partial_3 \hat{C}^1_2 \,, \\
    \hat{c}'_{31} &= \hat{c}_{31} + \partial_3 \hat{C}^2_1 - \partial_1 \hat{C}^2_3 \,.
\end{align}
The term $\partial_i \hat{C}^k_j - \partial_j \hat{C}^k_i$ has the background gauge transformation 
\begin{align}
    \partial_i \hat{C}^k_j - \partial_j \hat{C}^k_i \sim \partial_i \hat{C}^k_j - \partial_j \hat{C}^k_i - \partial_i \hat{\gamma}_j + \partial_j \hat{\gamma}_i \,,
\end{align}
so the dynamical fields $\hat{c}_{ij}\, ((i,j) = (0,1), (0,2), (0,3), (2,3), (3,1) )$ also have the background gauge transformations
\begin{align}
    \hat{c}_{ij} &\sim \hat{c}_{ij} + (d \hat{\gamma} )_{ij}   \,. \label{dychatbkg}
\end{align}
Under the changes of variables we have additional terms
\begin{align}
\begin{split}
    &\frac{i}{2\pi} \Big[ - \hat{C}^1_2 \left\{ -\partial_1 ( \partial_0 c_3 - \partial_3 c_0 ) + ( \partial_0 C^1_3 - \partial_3 C^1_0 )  \right\} + \hat{C}^2_1 \left\{ -\partial_2 ( \partial_0 c_3 - \partial_3 c_0 ) + ( \partial_0 C^2_3 - \partial_3 C^2_0 )  \right\} 
     \\
    & \quad  - \hat{C}^1_0 \left\{ \partial_1 \partial_2 c_3 - \partial_2 ( C^1_3 + \partial_3 c_1 ) \right\} + \hat{C}^2_0 \left\{ -\partial_3 (C^1_2 + \partial_2 c_1 ) + \partial_1 \partial_2 c_3 + ( \partial_3 C^2_1 - \partial_1 C^2_3 ) \right\} \\
    & \quad + \hat{C}^1_3 \left\{ \partial_1 \partial_2 c_0 - \partial_2 ( C^1_0 + \partial_0 c_1 ) \right\} - \hat{C}^2_3 \left\{ -\partial_0 (C^1_2 + \partial_2 c_1 ) + \partial_1 \partial_2 c_0  +  (\partial_0 C^2_1 - \partial_1  C^2_0 )  \right\}  \\
    & \quad + \partial_3 \left\{ -\hat{C}^2_0 ( \partial_1 c_2 - \partial_2 c_1 + C^2_1 - C^1_2) - \hat{C}^1_2 (\partial_0 c_1 - \partial_1 c_0 + C^1_0) + \hat{C}^2_1 (  \partial_0 c_2 - \partial_2 c_0 + C^2_0) \right\} \Big] \,, \label{ssptaddterm}
\end{split}
\end{align}
where we have dropped the total derivative terms with respect to $x^0$, $x^1$ and $x^2$. The total derivative term with respect to $x^3$ has an effect when the spacetime has a boundary $x^3 = 0$ and will be considered in the boundary theory, so we drop it here. Adding \eqref{ssptaddterm} to \eqref{ssptlag2}, we can derive the foliated Lagrangian of the SSPT phase
\begin{align}
\begin{split}
    &L_{\text{SSPT,f}} \left[ C^k \wedge e^k, c, \hat{C}^k, \hat{c}_{12} \right] \\
    & = \frac{i}{2\pi} \left[  \sum_{k = 1}^2 \hat{C}^k \wedge d C^k \wedge e^k + \hat{c} \wedge \left( dc + \sum_{k=1}^2 C^k \wedge e^k \right) \right] \\
    &= \frac{i}{2\pi} \bm{\hat{C}} \wedge_\text{f} d_\text{f} \bm{C}  \,, \label{folissptlag}
\end{split}
\end{align}
where $\bm{\hat{C}} = (\hat{C}^k, \hat{c})$ is a type-$B$ foliated one-form gauge field, $\bm{C} = (C^k \wedge e^k , c)$ is a type-$A$ foliated (1+1)-form gauge field, and $d_\text{f}\bm{C} =(d C^k \wedge e^k , dc + \sum^{2}_{k=1} C^k \wedge e^k )$ is the exterior derivative of $\bm{C}$.
Note that we combine the background gauge field $\hat{c}^{12}$ and the dynamical fields $\hat{c}_{ij} \ ((i,j) = (0,1), (0,2), (0,3), (2,3), (3,1))$ into $\hat{c}$.

Conversely, to obtain the exotic theory from the foliated theory, we must integrate out the dynamical fields $\hat{c}_{ij} \ ((i,j) = (0,1), (0,2), (0,3), (2,3), (3,1))$ as Lagrange multipliers. Then, we have a constraint
\begin{align}
    \left( dc + \sum^{2}_{k=1} C^k \wedge e^k \right)_{ij} = 0 \,,\quad (i,j) = (2,3), (3,1), (1,2), (0,1), (0,2)  \,, 
\end{align}
or in components,
\begin{align}
    \partial_2 c_3 - \partial_3 c_2 -C^2_3 &= 0 \,, \label{constraint1} \\
    \partial_3 c_1 - \partial_1 c_3 + C^1_3  &= 0 \,, \label{constraint2} \\
    \partial_1 c_2 - \partial_2 c_1 + C^2_1 - C^1_2  &= 0  \,, \\
    \partial_0 c_1 - \partial_1 c_0 + C^1_0  &= 0 \,, \\
    \partial_0 c_2 - \partial_2 c_0 + C^2_0 &= 0 \,. \label{constraint5}
\end{align}
Combining these constraints with the correspondences \eqref{corr0-0}--\eqref{corr3-3} and \eqref{ssptcorr8}--\eqref{ssptcorr10}, we can derive the dictionary 
\begin{align}
    C_0 &\simeq c_0 \,, \label{ssptdic1} \\
    C_{12} &\simeq C^1_2 + \partial_2 c_1 = C^2_1 + \partial_1 c_2  \,,  \\
    C_3 &\simeq c_3 \,, \\
    \partial_1 C_0 &\simeq C^1_0 + \partial_0 c_1 \,, \\
    \partial_2 C_0 &\simeq C^2_0 + \partial_0 c_2 \,, \\
    \partial_1 C_3 &\simeq C^1_3 + \partial_3 c_1 \,, \\
    \partial_2 C_3 &\simeq C^2_3 + \partial_3 c_2 \,, \label{ssptdic7} \\
    \hat{C}_0^{12} &\simeq \hat{C}_0^1 - \hat{C}_0^2 \,,  \\
    \hat{C} &\simeq \partial_1 \hat{C}_2^1 - \partial_2 \hat{C}_1^2 + \hat{c}_{12}  \,, \\
    \hat{C}_3^{12} &\simeq \hat{C}_3^1 - \hat{C}_3^2 \,. \label{ssptdic10}
\end{align}
Using these correspondences, integrating by parts and dropping the total derivative terms with respect to $x^0$, $x^1$ and $x^2$, we can reproduce the exotic Lagrangian \eqref{exoticsspt}.

As in the case of the exotic SSPT phase, we put the foliated SSPT phase on the spacetime $\T^{2+1} \times \R_{x^3 \geq 0}$ with the boundary $x^3 = 0$. Then, under the background gauge transformations, the gauge variation is
\begin{align}
\begin{split}
    &\delta_\text{g} L_{\text{SSPT,f}}\left[ C^k \wedge e^k, c, \hat{C}^k, \hat{c}_{12} \right] \\
    &=  \frac{i}{2\pi} \left[  \sum^2_{k=1} ( d\hat{\gamma}^k - \hat{\gamma} ) \wedge d C^k \wedge e^k + d\hat{\gamma} \wedge \left( dc + \sum^2_{k=1}   C^k \wedge e^k    \right) \right] \\
    &= \frac{i}{2\pi}  d \left[ \sum^2_{k=1} \hat{\gamma}^k \ d C^k \wedge e^k + \hat{\gamma} \wedge \left( dc +  \sum^2_{k=1} C^k \wedge e^k  \right) \right] \,.
\end{split}
\end{align}
The variation of the action is
\begin{align}
\begin{split}
    &\delta_\text{g} S_{\text{SSPT,f}} \left[ C^k \wedge e^k, c, \hat{C}^k, \hat{c}_{12} \right] \\
    &= \int_{x^3 = 0} \frac{i}{2\pi} \left[  \sum^2_{k=1} \hat{\gamma}^k \ d C^k \wedge e^k + \hat{\gamma} \wedge \left( dc + \sum^2_{k=1} C^k \wedge e^k  \right) \right] \,. \label{varifolisspt}
\end{split}
\end{align}

\section{Boundary Theory for Foliated SSPT Phase}
\label{section:boundary_theory_for_foliated_sspt}

In the previous section, we reviewed the anomaly inflow mechanism between the exotic SSPT phase and the exotic $\phi$-theory, and constructed the foliated SSPT phase for $U(1) \times U(1)$ subsystem symmetry using the foliated-exotic duality. In this section, we construct a foliated $\phi$-theory that has an 't Hooft anomaly of the $U(1) \times U(1)$ subsystem symmetry from the foliated SSPT phase \eqref{folissptlag}. This foliated $\phi$-theory is considered as a boundary theory of the foliated SSPT phase. Furthermore, by tuning the parameters in the foliated Lagrangian, we will derive the foliated $\phi$-theory that is equivalent to the exotic $\phi$-theory \eqref{exoticphi}, and construct the foliated-exotic duality in the gapless theory.

\subsection{Construction of Boundary Theory}
\label{subsection:construction_of_boundary_theory}

First, we show a way to construct a boundary theory from an SPT phase in a relativistic case. We consider the SPT phase for a relativistic $U(1) \times U(1)$ symmetry in 2+1 dimensions, and construct the compact scalar field theory in 1+1 dimensions, which is a boundary theory of it.

As in Section \ref{section:exotic_and_foliated_sspt_phase}, we take Euclid spacetime to be a (2+1)-dimensional torus $\T^{2+1}$. The SPT phase for a $U(1) \times U(1)$ global symmetry is described by the Lagrangian
\begin{align}
    L_{\text{SPT}}\left[ C, \hat{C} \right] = \frac{i}{2\pi} \hat{C} \wedge d C \,, \label{spt1}
\end{align}
or
\begin{align}
    L'_{\text{SPT}} \left[ C, \hat{C} \right] = \frac{i}{2\pi} d \hat{C} \wedge  C \,, \label{spt2}
\end{align}
where $C$ and $\hat{C}$ are $U(1)$ background one-form gauge fields. These two Lagrangians differ only by a total derivative term, so they are the same on the spacetime without a boundary. The background gauge transformations of $C$ and $\hat{C}$ are
\begin{align}
    C &\sim C + d\gamma \,, \label{sptgauge1} \\
    \hat{C} &\sim \hat{C} + d\hat{\gamma} \,, \label{sptgauge2}
\end{align}
where $\gamma$ and $\hat{\gamma}$ are background zero-form gauge parameters. Under these gauge transformations, the Lagrangian is invariant if the spacetime does not have a boundary. However, when the spacetime has a boundary, we have a variation on the boundary. Let us consider the SPT phase \eqref{spt1} in the spacetime $\T^{1+1} \times \R_{x^2 \geq 0}$ with the boundary $x^2 = 0$. Then, under the gauge transformations, we have
\begin{align}
\begin{split}
    \delta_\text{g} L_{\text{SPT}} \left[ C, \hat{C} \right] &= \frac{i}{2\pi} d \hat{\gamma} \wedge d C \\
    &= d \left[ \frac{i}{2\pi}  \hat{\gamma} \ d C \right] \,,
\end{split}
\end{align}
and the variation of action is
\begin{align}
    \delta_\text{g} S_{\text{SPT}}\left[ C, \hat{C} \right] = \int_{x^2 = 0} \left[ -\frac{i}{2\pi} \hat{\gamma} d C \right] \,. \label{varispt}
\end{align}
From the anomaly inflow mechanism, this variation should match an 't Hooft anomaly of some boundary theory. In this situation, we have one way to construct a boundary theory.

First, we choose a background gauge field $C$ and introduce a one-form lower dynamical gauge field that cancels the background gauge transformation \eqref{sptgauge1}. Since the gauge field $C$ is one-form, we introduce a zero-form gauge field $\phi$, whose dynamical gauge transformation is
\begin{align}
    \phi \sim \phi + 2\pi w \,,
\end{align}
where $w$ is a locally constant function valued in an integer. This gauge equivalence makes $\phi$ a compact scalar: $\phi \sim \phi + 2\pi$.\footnote{The compact scalar field $\phi$ in this section is different from the compact scalar field $\phi$ in the exotic $\phi$-theory in Section \ref{subsection:exotic_phi_theory}.} We set the background gauge transformation of $\phi$ as
\begin{align}
    \phi \sim \phi + \gamma \,,
\end{align}
and then, the term $C - d\phi$ becomes invariant under the background gauge transformations. Next, we consider the Lagrangian \eqref{spt2} on $\T^{1+1} \times \R_{x^2 \geq 0}$ in which $\hat{C}$ has an exterior derivative. Then, $L'_{\text{SPT}}[C,\hat{C}]$ is not gauge invariant due to \eqref{sptgauge1}, so we substitute $C - d\phi$ for $C$ in the $L'_{\text{SPT}}[C,\hat{C}]$.\footnote{We can choose the gauge field $\hat{C}$ and introduce a compact scalar $\hat{\phi}$. Then, we can derive the dual theory of $\hat{\phi}$. We will see it in Section \ref{section:duality_foliated_phihat}.} Then, we have the gauge-invariant Lagrangian
\begin{align}
\begin{split}
    L''_{\text{SPT}}\left[C,\hat{C}\right] &= \frac{i}{2\pi} d \hat{C} \wedge ( C  - d\phi ) \\
    &= d \left[ \frac{i}{2\pi} \hat{C} \wedge ( C  - d\phi ) \right] + \frac{i}{2\pi} \hat{C} \wedge d C  \,. \label{spt''}
\end{split}
\end{align}
The boundary term in the action is
\begin{align}
    \int_{x^2 = 0} \left[ -\frac{i}{2\pi} \hat{C} \wedge ( C  - d\phi ) \right] \,,
\end{align}
on the boundary $x^2 = 0$, and the other term is the SPT phase \eqref{spt1}.\footnote{Instead, we can use $L''_{\text{SPT}}[C, \hat{C}] = d \left[ - \frac{i}{2\pi} \hat{C} \wedge  d\phi \right] + \frac{i}{2\pi} d \hat{C} \wedge C $. Then, the boundary term is $\int_{x^3 = 0} \frac{i}{2\pi} \hat{C} \wedge  d\phi$ and the Lagrangian of the SPT phase is \eqref{spt2}.} Finally, we add a gauge-invariant quadratic term to the boundary, and the result is
\begin{align}
\begin{split}
    L_{\phi, \text{rel}} \left[ C, \hat{C} \right] &= \frac{R^2}{2} ( d\phi - C ) \wedge \ast ( d\phi - C ) + \frac{i}{2\pi} \hat{C} \wedge ( d\phi - C )  \\
    &= \frac{R^2}{2} \left[ (\partial_0 \phi - C_0 ) ^2 + (\partial_1 \phi - C_1 )^2  \right]d^2x \\
    &\quad  + \frac{i}{2\pi} \left[ \hat{C}_0 ( \partial_1 \phi - C_1 ) - \hat{C}_1 ( \partial_0 \phi - C_0 )  \right] d^2x \,, \label{phicc1}
\end{split}
\end{align}
where $R$ is a parameter with mass dimension zero. Since the Lagrangian \eqref{spt''} is gauge invariant, the gauge variation of the action $S_{\phi, \text{rel}}[C,\hat{C}] = \int L_{\phi, \text{rel}}[C,\hat{C}]$ matches the variation of the SPT phase \eqref{varispt}. Therefore, this scalar field theory is an anomalous theory coupled to background gauge fields, whose 't Hooft anomaly is described by the SPT phase \eqref{spt1}.

We can identify the currents of the global symmetries. Coupling between background gauge fields and currents of global symmetry is in the form $i C \wedge \ast J$. To derive the conservation law, we just have to consider the gauge transformation of $C$ only, and set $d \gamma \wedge \ast J$ to be zero. Since $\gamma$ is arbitrary, we have the conservation law $d \ast J = 0 $.\footnote{From the Noether theorem, we can derive the conservation law by considering a local transformation of $\phi$ in the theory without the background gauge field $C$, which is equivalent to considering only the gauge transformation of $C$ in the theory coupled to $C$.} The coupling in \eqref{phicc1} is\footnote{In $D$-dimensional Euclidean spacetime, we have $\ast\ast A = (-1)^{n(D-n)}A$ for an $n$-form $A$. Thus, in 1+1 dimensions, we have $\ast\ast d\phi = -d\phi$.}
\begin{align}
    i C \wedge \ast \left( i R^2 d\phi \right) + i \hat{C} \wedge \ast \left( - \frac{1}{2\pi} \ast d\phi \right) \,.
\end{align}
Thus, we have the currents
\begin{align}
    J &= i R^2 d\phi \,, \\
    \hat{J} &= - \frac{1}{2\pi} \ast d\phi \,,
\end{align}
and the conservation laws
\begin{align}
    d \ast J &= 0 \,, \\
    d \ast \hat{J} &= 0 \,.
\end{align}
The global symmetry with $J$ is called the $U(1)$ momentum global symmetry, and the global symmetry with $\hat{J}$ is called the $U(1)$ winding global symmetry. In 1+1 dimensions, both of the global symmetries are zero-form symmetry.

By setting the background gauge fields and parameters to zero, we can derive the compact scalar field theory
\begin{align}
\begin{split}
    L_{\phi, \text{rel}}  &= \frac{R^2}{2}  d\phi \wedge \ast  d\phi \\
    &= \frac{R^2}{2} \left[ (\partial_0 \phi) ^2 + (\partial_1 \phi )^2  \right] d^2x \,,
\end{split}
\end{align}

Applying this method to the exotic SSPT phase \eqref{exoticsspt}, we can also derive the boundary exotic $\phi$-theory \eqref{exoticphic} and \eqref{exoticphi}. In this situation, the exotic $\phi$-theory does not have a continuous rotational symmetry, so we can introduce different parameters $\mu_0$ and $\mu_{12}$ as the coefficients of the kinetic terms.

\subsection{Construction of Foliated \texorpdfstring{$\phi$}{φ}-Theory}
\label{subsection:construction_of_foliated_phi_theory}

Let us construct a foliated $\phi$-theory in 2+1 dimensions by using the method in Section \ref{subsection:construction_of_boundary_theory}. We take the foliated SSPT phase \eqref{folissptlag}, and choose the background gauge field $\bm{C} = (C^k \wedge e^k,c)$. Then, we have to consider the Lagrangian integrated by parts
\begin{align}
\begin{split}
    &L'_{\text{SSPT,f}} \left[ C^k \wedge e^k, c, \hat{C}^k, \hat{c}_{12} \right] \\
    & \quad = \frac{i}{2\pi} \left[  \sum_{k = 1}^2 ( d \hat{C}^k + \hat{c} ) \wedge C^k \wedge e^k   - d\hat{c} \wedge c  \right]  \,,
\end{split}
\end{align}
on the spacetime $\T^{2+1} \times \R_{x^3 \geq 0}$. Next, we introduce a $U(1)$ type-$A$ foliated (0+1)-form gauge field $\Phi^k e^k\ (k = 1,2)$ and a $U(1)$ type-$A$ bulk zero-form gauge field $\Phi$, whose dynamical gauge transformations are
\begin{align}
    \Phi^k  e^k &\sim \Phi^k e^k + 2\pi d W^k \,, \label{dyngaugephik} \\
    \Phi &\sim \Phi + 2\pi W^1 + 2\pi W^2 \,,
\end{align}
where $W^k$ is an $x^k$-dependent integer-valued gauge parameter and it can have step function discontinuities in the $x^k$ direction. Since the background gauge transformations of the foliated gauge fields are
\begin{align}
    C^k \wedge e^k &\sim C^k \wedge e^k +d \gamma^k \wedge e^k  \, , \label{bkggaugetr1} \\
    c &\sim c + d\gamma - \sum^{2}_{k=1} \gamma^k e^k\,, \label{bkggaugetr3} \\
    \hat{C}^k &\sim \hat{C}^k +d \hat{\gamma}^k - \hat{\gamma} \,, \label{bkggaugetr2}\\
    \hat{c} &\sim \hat{c} + d \hat{\gamma} \,, \label{bkggaugetr4} 
\end{align}
$d\hat{C}^k + \hat{c}$ and $d\hat{c}$ are gauge invariant, but the parts of $C^k \wedge e^k$ and $c$ are not on the spacetime with boundary. As in the case of the relativistic theory in Section \ref{subsection:construction_of_boundary_theory}, we set the background gauge transformations of $\Phi^k e^k$ and $\Phi$ as
\begin{align}
    \Phi^k e^k &\sim \Phi^k e^k  + \gamma^k e^k  \,, \label{bkggaugephik} \\
    \Phi &\sim \Phi + \gamma \,, \label{bkggaugephi}
\end{align}
and substitute the gauge-invariant combinations $ (C^k - d\Phi^k ) \wedge e^k $ and $c - d\Phi + \sum_{k = 1}^2 \Phi^k e^k$ for $C^k \wedge e^k$ and $c$ in $L'_{\text{SSPT,f}}[ C^k \wedge e^k, c, \hat{C}^k, \hat{c}_{12} ]$, respectively.\footnote{We omit the auxiliary fields of the type-$B$ foliated gauge field $\bm{\hat{C}}=(\hat{C}^k,\hat{c}) $ and the gauge parameter of it explained in Section \ref{subsection:foliated_gauge_fields}.} Then, we have
\begin{align}
\begin{split}
    &L''_{\text{SSPT,f}} \left[ C^k \wedge e^k, c, \hat{C}^k, \hat{c}_{12} \right] \\
    &  = \frac{i}{2\pi} \left[  \sum_{k = 1}^2 ( d \hat{C}^k + \hat{c} ) \wedge ( C^k - d\Phi^k ) \wedge e^k - d\hat{c} \wedge \left( c - d\Phi + \sum_{k = 1}^2 \Phi^k e^k \right) \right]   \\
    &  = \frac{i}{2\pi} \left[ \sum_{k = 1}^2  \hat{C}^k \wedge d C^k \wedge e^k + \hat{c} \wedge \left( \sum_{k = 1}^2 C^k \wedge e^k + dc \right) \right] \\
    &\quad +  d \left[ \frac{i}{2\pi}\sum_{k = 1}^2  \hat{C}^k \wedge ( C^k - d\Phi^k ) \wedge e^k - \frac{i}{2\pi} \hat{c} \wedge \left( c - d\Phi +  \sum_{k = 1}^2 \Phi^k e^k  \right)  \right] \,.
\end{split}
\end{align}
The boundary term in the action is
\begin{align}
\begin{split}
    &\int_{x^3 = 0} \frac{i}{2\pi} \left[ - \sum_{k = 1}^2  \hat{C}^k \wedge ( d\Phi^k - C^k ) \wedge e^k + \hat{c} \wedge \left( d\Phi - \sum_{k = 1}^2 \Phi^k e^k - c \right) \right] \\
    & = \int_{x^3 = 0} \frac{i}{2\pi} \left[ \hat{C}^1_0 ( 
 \partial_2 \Phi^1 - C^1_2 ) - \hat{C}^1_2 ( 
 \partial_0 \Phi^1 - C^1_0 ) - \hat{C}^2_0 ( 
 \partial_1 \Phi^2 - C^2_1 ) + \hat{C}^2_1 ( 
 \partial_0 \Phi^2 - C^2_0 )  \right. \\
 & \qquad \qquad \left.  + \hat{c}_{12} ( \partial_0 \Phi - c_0 )  - \hat{c}_{02} ( \partial_1 \Phi - \Phi^1 - c_1  ) + \hat{c}_{01} ( \partial_2 \Phi - \Phi^2 - c_2  )  \right] d^3 x  \,.
\end{split}
\end{align}
Note that the gauge fields $\hat{c}_{01}$ and $\hat{c}_{02}$ on the boundary are dynamical, so integrating out these fields, we have the constraints
\begin{align}
    (  \partial_2 c_3 - \partial_3 c_2 - C^2_3 ) +  ( \partial_2 \Phi - \Phi^2 - c_2  ) \delta(x^3) &= 0 \,, \\
    (  \partial_3 c_1 - \partial_1 c_3 + C^1_3 ) - ( \partial_1 \Phi - \Phi^1 - c_1  ) \delta(x^3) &= 0 \,.
\end{align}
Then, we have the constraints \eqref{constraint1} and \eqref{constraint2} in the (3+1)-dimensional spacetime, and we have the constraints
\begin{align}
    \partial_2 \Phi - \Phi^2 - c_2  &= 0 \label{constraintphi2} \,, \\
    \partial_1 \Phi - \Phi^1 - c_1  &= 0 \,, \label{constraintphi1} 
\end{align}
on the boundary.\footnote{The first terms $C^1_3 + \partial_3 c_1 - \partial_1 c_3$ and $- C^2_3 + \partial_2 c_3 - \partial_3 c_2$ do not have a contribution proportional to $\delta(x^3)$ because the theory does not have a foliation in the $x^3$ direction.} In addition, since the background gauge fields are restricted to the (2+1)-dimensional boundary from the (3+1)-dimensional spacetime, we have constraints on the background gauge fields $C^k \wedge e^k$ and $c$ as
\begin{align}
    \partial_1 c_2 - \partial_2 c_1 + C^2_1 - C^1_2 &= 0 \label{constc12} \,, \\
    \partial_0 c_1 - \partial_1 c_0 + C^1_0 &= 0 \,, \label{constc01} \\
    \partial_0 c_2 - \partial_2 c_0 + C^2_0  &= 0 \,. \label{constc02}
\end{align}
Therefore, we add the term
\begin{align}
\begin{split}
    & \frac{i}{2\pi} \hat{\chi} \wedge \left(  dc + \sum^{2}_{k=1} C^k \wedge e^k \right) \\
    &= \frac{i}{2\pi} \left[ \hat{\chi}_0 ( \partial_1 c_2 - \partial_2 c_1 + C^2_1 - C^1_2 ) \right. \\
    & \quad \left. - \hat{\chi}_1 ( \partial_0 c_2 - \partial_2 c_0 +  C^2_0) + \hat{\chi}_2 (\partial_0 c_1 - \partial_1 c_0 +  C^1_0 ) \right] d^3 x \,,
\end{split}
\end{align}
on the boundary, where $\hat{\chi}$ is a dynamical one-form field.
Finally, we add background gauge-invariant and $\Z_4$ rotational invariant quadratic terms,\footnote{Quadratic terms $ ( \partial_1 \Phi - \Phi^1 - c_1 )^2$ and $ ( \partial_2 \Phi - \Phi^2 - c_2 )^2$ are absorbed to changes of the variables $\hat{c}_{01}$ and $\hat{c}_{02}$ on the boundary.} and we have
\begin{align}
\begin{split}
    &\L_{\phi,\text{f}} \left[ C^k \wedge e^k, c, \hat{C}^k, \hat{c}_{12} \right] \\
    & \quad = \frac{\mu_0}{2} ( \partial_0 \Phi - c_0 )^2 + \frac{1}{4\mu_{12}} ( \partial_2 \Phi^1 - C^1_2 )^2 + \frac{1}{4\mu_{12}} ( \partial_1 \Phi^2 - C^2_1 )^2 \\
    & \qquad + \frac{1}{2\mu_{012}} ( \partial_0 \Phi^1 - C^1_0 )^2  + \frac{1}{2\mu_{012}} ( \partial_0 \Phi^2 - C^2_0 )^2 \\
    & \qquad + \frac{i}{2\pi} \left[ \hat{C}^1_0 ( 
    \partial_2 \Phi^1 - C^1_2 ) - \hat{C}^1_2 ( 
    \partial_0 \Phi^1 - C^1_0 ) - \hat{C}^2_0 ( 
    \partial_1 \Phi^2 - C^2_1 )  \right. \\
    & \qquad \quad \left. + \hat{C}^2_1 ( 
    \partial_0 \Phi^2 - C^2_0 ) + \hat{c}_{12} ( \partial_0 \Phi - c_0 ) \right] \\
    & \qquad + \frac{i}{2\pi} \left[ - \hat{c}_{02} ( \partial_1 \Phi - \Phi^1 - c_1  ) + \hat{c}_{01} ( \partial_2 \Phi - \Phi^2 - c_2  ) \right] \\
    & \qquad + \frac{i}{2\pi} \left[ \hat{\chi}_0 ( C^2_1 - C^1_2 + \partial_1 c_2 - \partial_2 c_1 ) \right. \\
    & \qquad \quad \left. - \hat{\chi}_1 ( C^2_0 + \partial_0 c_2 - \partial_2 c_0) + \hat{\chi}_2 ( C^1_0 + \partial_0 c_1 - \partial_1 c_0 ) \right] \,, \label{genefoliatedphi}
\end{split}
\end{align}
where $\mu_0$, $\mu_{12}$ and $\mu_{012}$ are parameters with mass dimension one. We can check the $\Z_4$ rotational invariance of the Lagrangian using the transformation rules in Section \ref{appendix23}.\footnote{We can change the term $\frac{1}{4\mu_{12}} ( \partial_2 \Phi^1 - C^1_2 )^2 + \frac{1}{4\mu_{12}} ( \partial_1 \Phi^2 - C^2_1 )^2$ to $\frac{1}{2\mu^1_2} ( \partial_2 \Phi^1 - C^1_2 )^2 + \frac{1}{2\mu^2_1} ( \partial_1 \Phi^2 - C^2_1 )^2$, where $\mu^1_2$ and $\mu^2_1$ are parameters satisfying $1/\mu_{12} = 1/\mu^1_2 + 1/\mu^2_1$. Using the equations of motion \eqref{constraintphi2}, \eqref{constraintphi1} and \eqref{constc12}, this term becomes $\Z_4$ rotational invariant and the parameters $\mu^1_2$ and $\mu^2_1$ appear only in the form $1/\mu_{12} = 1/\mu^1_2 + 1/\mu^2_1$.} We will consider global symmetry in Section \ref{subsection:foliated_exotic_duality_phi}. 

We can also write the foliated Lagrangian in differential form by tuning the metric:
\begin{align}
\begin{split}
    &L_{\phi,\text{f}} \left[ C^k \wedge e^k, c, \hat{C}^k, \hat{c}_{12}\right] \\
    & = \frac{1}{2} \left( d \Phi - \sum_{k = 1}^2 \Phi^k e^k - c \right) \wedge \ast \left( d \Phi - \sum_{k = 1}^2 \Phi^k e^k - c \right) \\
    & \quad + \frac{1}{2} \sum_{k=1}^2  \left( d \Phi^k \wedge e^k - C^k \wedge e^k  \right) \wedge \ast \left( d \Phi^k \wedge e^k - C^k \wedge e^k  \right)  \\
    & \quad + \frac{i}{2\pi} \left[ - \sum_{k = 1}^2  \hat{C}^k \wedge ( d\Phi^k - C^k ) \wedge e^k + \hat{c} \wedge \left( d\Phi - \sum_{k = 1}^2 \Phi^k e^k - c \right) \right]  \\
    & \quad + \frac{i}{2\pi} \hat{\chi} \wedge \left( dc + \sum^{2}_{k=1} C^k \wedge e^k  \right) \\
    &= \frac{1}{2} \left( d_\text{f} \bm{\Phi} - \bm{C} \right)^2 - \frac{i}{2\pi} \bm{\hat{C}} \wedge_\text{f} \left(d_\text{f} \bm{\Phi} - \bm{C} \right) + \frac{i}{2\pi} \bm{\hat{\chi}} \wedge_\text{f} d_\text{f} \bm{C}  \,, \label{foliphideff}
\end{split}
\end{align}
where $\bm{\Phi} = (\Phi^k e^k, \Phi)$ is a type-$A$ foliated (0+1)-form gauge field, $\bm{C} = (C^k \wedge e^k, c)$ is a type-$A$ foliated (1+1)-form gauge field, $\bm{\hat{C}} = (\hat{C}^k, \hat{c})$ is a type-$B$ foliated one-form gauge field and $\bm{\hat{\chi}} = (0,\hat{\chi})$ is a type-$B$ foliated zero-form field.

\subsection{Foliated-Exotic Duality in \texorpdfstring{$\phi$}{φ}-Theory}
\label{subsection:foliated_exotic_duality_phi}

In this section, we compare the exotic $\phi$-theory \eqref{exoticphic} with the foliated $\phi$-theory \eqref{genefoliatedphi} in 2+1 dimensions. We establish the field correspondences in the $\phi$-theory, and by tuning the parameters, we will see that the foliated $\phi$-theory can be equivalent to the exotic $\phi$-theory, which is a gapless example of the foliated-exotic duality.

In Section \ref{subsection:exotic_phi_theory}, we reviewed the exotic $\phi$-theory coupled with the background tensor gauge fields
\begin{align}
\begin{split}
    \L_{\phi,\text{e}}\left[ \bm{C}, \bm{\hat{C}} \right] &= \frac{\mu_0}{2} (\partial_0 \phi - C_0 )^2 + \frac{1}{2 \mu_{12}} (\partial_1 \partial_2 \phi - C_{12})^2 \\
    & \quad + \frac{i}{2\pi} \hat{C}^{12}_0 ( \partial_1 \partial_2 \phi - C_{12}) + \frac{i}{2\pi} \hat{C} ( \partial_0 \phi - C_0 ) \,.  \label{exoticphic2}
\end{split}
\end{align}
First, we assume the correspondences between the scalar field $\phi$ in the exotic $\phi$-theory and the type-$A$ bulk zero-form gauge field $\Phi$ in a foliated $\phi$-theory as
\begin{align}
    \phi \simeq \Phi \,. \label{corrphi}
\end{align}
Since the dynamical gauge transformations of $\phi$ and $\Phi$ are
\begin{align}
    \phi &\sim \phi + 2 \pi w^1 + 2 \pi w^2 \,, \\
    \Phi &\sim \Phi + 2 \pi W^1 + 2 \pi W^2 \,,
\end{align}
the correspondences between the dynamical gauge parameters are
\begin{align}
    w^k \simeq W^k \,, \quad (k = 1,2) \,. 
\end{align}
In addition, to impose relations to the fields $\Phi$ and 
type-$A$ foliated (0+1)-form gauge fields $\Phi^k  e^k \ ( k = 1,2 )$, we add to the Lagrangian the term
\begin{align}
    \frac{i}{2\pi} \left[ -\hat{h}'_{02} ( \partial_1 \Phi - \Phi^1 - c_1 )  + \hat{h}'_{01} ( \partial_2 \Phi - \Phi^2 - c_2 ) \right] \,,
\end{align}
where $\hat{h}'_{01}$ and $\hat{h}'_{02}$ are dynamical fields. The dynamical gauge transformation of $\Phi^k e^k$ is \eqref{dyngaugephik}, and the background gauge transformations of $\Phi^k e^k$ and $\Phi$ are \eqref{bkggaugephik} and \eqref{bkggaugephi}.

For the background gauge fields, the background gauge fields in a foliated theory are $U(1)$ type-$A$ foliated (1+1)-form gauge fields $C^k \wedge e^k \ (k = 1,2)$, a $U(1)$ type-$A$ bulk one-form gauge field $c$, $U(1)$ type-$B$ foliated one-form gauge fields $\hat{C}^k \ (k = 1,2)$ and the $x^1x^2$-component of a $U(1)$ type-$B$ bulk two-form gauge field $\hat{c}_{12}$. Then, we assume correspondences
\begin{align}
    C_0 &\simeq c_0 \label{bcorr0-0} \,, \\
    C_{12} &\simeq C^1_2 + \partial_2 c_1 \,,  \label{bcorr12-1} 
\end{align}
and
\begin{align}
    \hat{C}_0^{12} &\simeq \hat{C}_0^1 - \hat{C}_0^2 \,, \label{bcorr012} \\
    \hat{C} &\simeq \partial_1 \hat{C}_2^1 - \partial_2 \hat{C}_1^2 + \hat{c}_{12}  \,. \label{bcorr1221}
\end{align}
Their background gauge transformations are \eqref{bkggaugetr1}--\eqref{bkggaugetr4}, and then the correpondences between the background gauge parameters are
\begin{align}
    \Gamma &\simeq \gamma \,, \\
    \hat{\Gamma}^{12} &\simeq \hat{\gamma}^1 - \hat{\gamma}^2 \,.
\end{align}
As in the previous section, we add to the Lagrangian the term
\begin{align}
\begin{split}
    & \frac{i}{2\pi} \hat{\chi}' \wedge \left( dc + \sum^{2}_{k=1} C^k \wedge e^k \right) \\
    &= \frac{i}{2\pi} \left[ \hat{\chi}'_0 ( \partial_1 c_2 - \partial_2 c_1 +C^2_1 - C^1_2 ) \right. \\
    & \quad \left. - \hat{\chi}'_1 ( \partial_0 c_2 - \partial_2 c_0 +  C^2_0) + \hat{\chi}'_2 (\partial_0 c_1 - \partial_1 c_0 +  C^1_0) \right] d^3 x \,.
\end{split}
\end{align}

Using these correspondences, the Lagrangian of the $\phi$-theory can be written as
\begin{align}
\begin{split}
    &\L_{\phi,\text{e}} \left[ \bm{C}, \bm{\hat{C}} \right] \\
    &\simeq \frac{\mu_0}{2} (\partial_0 \Phi - c_0 )^2 + \frac{1}{2 \mu_{12}} (\partial_1 \partial_2 \Phi - C^1_2 - \partial_2 c_1 )^2 \\
    & \quad + \frac{i}{2\pi} ( \hat{C}^{1}_0 - \hat{C}^2_0 ) ( \partial_1 \partial_2 \Phi - C^1_2 - \partial_2 c_1 ) \\
    & \quad - \frac{i}{2\pi} \hat{C}_2^1 ( \partial_0 \partial_1 \Phi - \partial_1 c_0 ) + \frac{i}{2\pi} \hat{C}_1^2 ( \partial_0 \partial_2 \Phi - \partial_2 c_0 )  + \frac{i}{2\pi} \hat{c}_{12} ( \partial_0 \Phi - c_0 ) \\
    & \quad + \frac{i}{2\pi} \left[ -\hat{h}'_{02} ( \partial_1 \Phi - \Phi^1 - c_1 )  + \hat{h}'_{01} ( \partial_2 \Phi - \Phi^2 - c_2 ) \right] \\
    & \quad + \frac{i}{2\pi} \left[ \hat{\chi}'_0 ( C^2_1 - C^1_2 + \partial_1 c_2 - \partial_2 c_1 ) \right. \\
    & \qquad \left. - \hat{\chi}'_1 ( C^2_0 + \partial_0 c_2 - \partial_2 c_0) + \hat{\chi}'_2 ( C^1_0 + \partial_0 c_1 - \partial_1 c_0 ) \right] \,,
\end{split}
\end{align}
where we have integrated by parts and dropped the total derivative terms.  

Next, we consider changes of variables of the dynamical fields from $\hat{h}'_{01}$, $\hat{h}'_{02}$ and $\hat{\chi}'$ to $\hat{h}_{01}$, $\hat{h}_{02}$ and $\hat{\chi}$ as
\begin{align}
    \hat{h}'_{02} &= \hat{h}_{02} + \frac{i\pi}{2\mu_{12}}  \partial_2 \left\{ \partial_2 ( \partial_1 \Phi + \Phi^1 + c_1 ) - 2 ( C^1_2 +  \partial_2 c_1 )   \right\} \\
    \hat{h}'_{01} &= \hat{h}_{01} - \frac{i\pi}{2\mu_{12}}  \partial_1 \left\{ \partial_1 ( \partial_2 \Phi + \Phi^2 + c_2 ) - 2 ( C^1_2 +  \partial_2 c_1 )   \right\} \\
    \hat{\chi}'_0 &= \hat{\chi}_0 - \frac{i\pi}{2\mu_{12}} \left\{ C^2_1 + \partial_1 c_2 + C^1_2 + \partial_2 c_1 - 2 \partial_1 ( \Phi^2 + c_2 ) \right\} \,, \\
    \hat{\chi}'_1 &= \hat{\chi}_1 \,, \\
    \hat{\chi}'_2 &= \hat{\chi}_2 \,.
\end{align}
The additional terms are background gauge invariant, so the background gauge transformations of the dynamical fields are not added. Then, we can see 
\begin{align}
\begin{split}
    &\frac{1}{2 \mu_{12}} (\partial_1 \partial_2 \Phi - C^1_2 - \partial_2 c_1 )^2 \\
    & \quad \rightarrow \frac{1}{4 \mu_{12}} ( \partial_2 \Phi^1 - C^1_2 )^2 + \frac{1}{4 \mu_{12}} ( \partial_1 \Phi^2 - C^2_1 )^2 \,.
\end{split}
\end{align}
Furthermore, we consider changes of variables from $\hat{h}_{01}$ and $\hat{h}_{02}$ to $\hat{c}_{01}$ and $\hat{c}_{02}$ as
\begin{align}
    \hat{h}_{02} &= \hat{c}_{02} + \partial_0 \hat{C}^1_2 - \partial_2 \hat{C}^1_0  \,, \\
    \hat{h}_{01} &= \hat{c}_{01} + \partial_0 \hat{C}^2_1 - \partial_1 \hat{C}^2_0  \,,
\end{align}
where $\hat{c}_{01}$ and $\hat{c}_{02}$ get the background gauge transformations
\begin{align}
    \hat{c}_{01} &\sim \hat{c}_{01} + (d \hat{\gamma})_{01}  \,, \\ 
    \hat{c}_{02} &\sim \hat{c}_{02} + (d \hat{\gamma})_{02} \,.
\end{align}
In addition, we add the boundary term of \eqref{ssptaddterm}:
\begin{align}
     \frac{i}{2\pi} \left[ \hat{C}^2_0 ( \partial_1 c_2 - \partial_2 c_1 + C^2_1 - C^1_2) + \hat{C}^1_2 (\partial_0 c_1 - \partial_1 c_0 +  C^1_0 ) - \hat{C}^2_1 ( \partial_0 c_2 - \partial_2 c_0 + C^2_0  )   \right] \,.
\end{align}
Then, we can see
\begin{align}
\begin{split}
    & \frac{i}{2\pi} ( \hat{C}^{1}_0 - \hat{C}^2_0 ) ( \partial_1 \partial_2 \Phi - C^1_2 - \partial_2 c_1 ) \\
    & \quad - \frac{i}{2\pi} \hat{C}_2^1 ( \partial_0 \partial_1 \Phi - \partial_1 c_0 ) + \frac{i}{2\pi} \hat{C}_1^2 ( \partial_0 \partial_2 \Phi - \partial_2 c_0 ) \\
    & \rightarrow  \frac{i}{2\pi} \left[ \hat{C}^{1}_0 ( \partial_2  \Phi^1 - C^1_2 ) - \hat{C}^{2}_0 ( \partial_1  \Phi^2 - C^2_1 ) \right. \\
    & \qquad\quad \left. -  \hat{C}_2^1 ( \partial_0  \Phi^1 - C^1_0 ) + \hat{C}_1^2 ( \partial_0 \Phi^2 - C^2_0 ) \right] \,.
\end{split}
\end{align}
After all, we derive the foliated $\phi$-theory that is equivalent to the exotic $\phi$-theory \eqref{exoticphic2}:
\begin{align}
\begin{split}
    &\L_{\phi,\text{e} \rightarrow \text{f}} \left[ C^k \wedge e^k, c, \hat{C}^k, \hat{c}_{12} \right] \\
    &= \frac{\mu_0}{2} ( \partial_0 \Phi - c_0 )^2 + \frac{1}{4\mu_{12}} ( \partial_2 \Phi^1 - C^1_2 )^2 + \frac{1}{4\mu_{12}} ( \partial_1 \Phi^2 - C^2_1 )^2 \\
    & \quad + \frac{i}{2\pi} \left[ \hat{C}^1_0 ( 
    \partial_2 \Phi^1 - C^1_2 ) - \hat{C}^1_2 ( 
    \partial_0 \Phi^1 - C^1_0 ) - \hat{C}^2_0 ( 
    \partial_1 \Phi^2 - C^2_1 )  \right. \\
    & \qquad  \left. + \hat{C}^2_1 ( 
    \partial_0 \Phi^2 - C^2_0 ) + \hat{c}_{12} ( \partial_0 \Phi - c_0 ) \right] \\
    & \qquad + \frac{i}{2\pi} \left[ -\hat{c}_{02} ( \partial_1 \Phi - \Phi^1 - c_1 )  + \hat{c}_{01} ( \partial_2 \Phi - \Phi^2 - c_2 ) \right] \\
    & \qquad + \frac{i}{2\pi} \left[ \hat{\chi}_0 ( C^2_1 - C^1_2 + \partial_1 c_2 - \partial_2 c_1 ) \right. \\
    & \qquad \quad \left. - \hat{\chi}_1 ( C^2_0 + \partial_0 c_2 - \partial_2 c_0) + \hat{\chi}_2 ( C^1_0 + \partial_0 c_1 - \partial_1 c_0 ) \right] \,. \label{etofoliatedphi}
\end{split}
\end{align}
If we set the background gauge fields to zero, we can obtain the foliated $\phi$-theory without the background gauge fields\footnote{When we set the background gauge fields to zero, we also set the background gauge parameters to zero, and then $\hat{c}_{01}$ and $\hat{c}_{02}$ become $\hat{h}_{01}$ and $\hat{h}_{02}$.}
\begin{align}
\begin{split}
    &\L_{\phi,\text{e} \rightarrow \text{f}}  = \frac{\mu_0}{2} ( \partial_0 \Phi )^2 + \frac{1}{4\mu_{12}} ( \partial_2 \Phi^1 )^2 + \frac{1}{4\mu_{12}} ( \partial_1 \Phi^2 )^2 \\
    & \qquad + \frac{i}{2\pi} \left[ -\hat{h}_{02} ( \partial_1 \Phi - \Phi^1 )  + \hat{h}_{01} ( \partial_2 \Phi - \Phi^2 ) \right]  \,. \label{etofoliatedphiczero}
\end{split}
\end{align}

Comparing the foliated Lagrangian \eqref{etofoliatedphi} with \eqref{genefoliatedphi}, the difference is the terms
\begin{align}
    \frac{1}{2\mu_{012}} ( \partial_0 \Phi^1 - C^1_0 )^2  + \frac{1}{2\mu_{012}} ( \partial_0 \Phi^2 - C^2_0 )^2 \,,
\end{align}
so we have only to take the limit $\mu_{012} \rightarrow \infty$ to get \eqref{etofoliatedphi} from \eqref{genefoliatedphi}. In the exotic description, these terms correspond to
\begin{align}
    \frac{1}{2\mu_{012}} \left\{ \partial_1 ( \partial_0 \phi - C_0 ) \right\}^2  + \frac{1}{2\mu_{012}} \left\{ \partial_2 ( \partial_0 \phi - C_0 ) \right\}^2 \,.
\end{align}
Note that, due to the UV/IR mixing, these terms have contribution to the energy of the same order as the leading term \cite{Seiberg:2020bhn}.

Conversely, to obtain the exotic $\phi$-theory \eqref{exoticphic2} from the foliated $\phi$-theory \eqref{etofoliatedphi}, we integrate out the dynamical fields $\hat{c}_{01}$, $\hat{c}_{02}$ and $\hat{\chi}_i \ (i = 0,1,2)$ as Lagrange multipliers. Then, we have the constraints \eqref{constraintphi1}, \eqref{constraintphi2} and \eqref{constc12}--\eqref{constc02}, and we can use the dictionary
\begin{align}
    \phi &\simeq \Phi \,,  \label{phidic1} \\
    \partial_1 \phi &\simeq \Phi^1 + c_1 \,, \\
    \partial_2 \phi &\simeq \Phi^2 + c_2 \,, \\
    C_0 &\simeq c_0 \,, \\
    C_{12} &\simeq C^1_2 + \partial_2 c_1 =  C^2_1 + \partial_1 c_2 \,, \\
    \partial_1 C_0 &\simeq C^1_0 + \partial_0 c_1 \,, \\
    \partial_2 C_0 &\simeq C^2_0 + \partial_0 c_2 \,, \\ 
    \hat{C}_0^{12} &\simeq \hat{C}_0^1 - \hat{C}_0^2 \,,  \\
    \hat{C} &\simeq \partial_1 \hat{C}_2^1 - \partial_2 \hat{C}_1^2 + \hat{c}_{12}  \,. \label{phidic9}
\end{align}
By applying these correspondences, we can recover the exotic $\phi$-theory \eqref{exoticphic2}.

Let us consider subsystem symmetry. From the background gauge transformations of $C^k \wedge e^k$ and $c$, which are \eqref{bkggaugetr1} and \eqref{bkggaugetr3}, in the Lagrangian with the background gauge fields \eqref{etofoliatedphi}, we have
\begin{gather}
    \partial_0 \left( i \mu_0 \partial_0 \Phi \right) + \partial_1 \left( \frac{1}{2\pi} \hat{h}_{02} \right) - \partial_2 \left( \frac{1}{2\pi} \hat{h}_{01} \right) = 0 \,, \\
    \partial_2 \left( \frac{i}{2\mu_{12}} \partial_2 \Phi^1 \right) + \frac{1}{2\pi} \hat{h}_{02} = 0 \,, \\
    \partial_1 \left( \frac{i}{2\mu_{12}} \partial_1 \Phi^2 \right) - \frac{1}{2\pi} \hat{h}_{01} = 0 \,,
\end{gather}
where $\hat{\chi}$ is canceled out and the background gauge parameters are turned off. These equations can also be obtained as the equations of motion of $\Phi$, $\Phi^1$ and $\Phi^2$ in the theory where the background gauge fields are zero \eqref{etofoliatedphiczero}.
Combining these equations, we derive a conservation law
\begin{align}
    \partial_0 J_0 - \partial_1 \partial_2 J_{12}  = 0 \,,
\end{align}
where the currents are\footnote{In the theory \eqref{genefoliatedphi}, the $x^0$ component of the current $J_0$ is modified to $J_0 = i \mu_0 \partial_0 \Phi -  \frac{i}{\mu_{012}} \partial_0 \partial_1 \Phi^1 - \frac{i}{\mu_{012}} \partial_0 \partial_2 \Phi^2$.}
\begin{align}
    J_0 &= i \mu_0 \partial_0 \Phi \,, \\
    J_{12} &= \frac{i}{2\mu_{12}} ( \partial_2 \Phi^1 + \partial_1 \Phi^2 ) \,.
\end{align}
Using the constraints derived by integrating out the dynamical fields $\hat{c}_{01}$ and $\hat{c}_{02}$
\begin{align}
    \partial_1 \Phi - \Phi^1 &= 0 \,, \label{constphinoc1} \\
    \partial_2 \Phi - \Phi^2 &= 0 \,, \label{constphinoc2}
\end{align}
the $x^1 x^2$-component of the current $J_{12}$ becomes
\begin{align}
    J_{12} = \frac{i}{\mu_{12}} \partial_1 \partial_2 \Phi \,.
\end{align}
These currents are equivalent to the currents \eqref{momsym1} and \eqref{momsym2} under the correspondence \eqref{corrphi}, and thus generate the momentum dipole symmetry. We also have conservation laws related to winding symmetry. From the background gauge transformations of $\hat{C}^k$ and $\hat{c}$, which are \eqref{bkggaugetr2} and \eqref{bkggaugetr4}, we have relations
\begin{gather}
    \partial_0 \left( \frac{1}{2\pi} \partial_2 \Phi^1  \right) - \partial_2 \left( \frac{1}{2\pi} \partial_0 \Phi^1  \right) = 0 \,, \\
    \partial_0 \left( -\frac{1}{2\pi} \partial_1 \Phi^2  \right) + \partial_1 \left( \frac{1}{2\pi} \partial_0 \Phi^2  \right) = 0 \,, \\
    \frac{1}{2\pi} \partial_2 \Phi^1 - \frac{1}{2\pi}\partial_1 \Phi^2 + \partial_2 \left\{ \frac{1}{2\pi} ( \partial_1 \Phi - \Phi^1 ) \right\} - \partial_1 \left\{ \frac{1}{2\pi} ( \partial_2 \Phi - \Phi^2 ) \right\} = 0 \,, \\
    \frac{1}{2\pi} \partial_0 \Phi^2 - \partial_2 \left( \frac{1}{2\pi} \partial_0 \Phi \right) + \partial_0 \left\{ \frac{1}{2\pi} ( \partial_2 \Phi - \Phi^2 ) \right\} = 0 \,, \\
    -\frac{1}{2\pi} \partial_0 \Phi^1 + \partial_1 \left( \frac{1}{2\pi} \partial_0 \Phi \right) - \partial_0 \left\{ \frac{1}{2\pi} ( \partial_1 \Phi - \Phi^1 ) \right\} = 0 \,,
\end{gather}
which are locally trivial. Under the constraints \eqref{constphinoc1} and \eqref{constphinoc2}, these relations become a conservation law
\begin{align}
    \partial_0 \hat{J}^{12}_0 - \partial_1 \partial_2 \hat{J} = 0 \,, 
\end{align}
where the currents are
\begin{align}
    \hat{J}^{12}_0 &= \frac{1}{2\pi} \partial_1 \partial_2 \Phi \,, \\
    \hat{J} &=  \frac{1}{2\pi} \partial_0 \Phi \,.
\end{align}
These currents are equivalent to the currents \eqref{winsym1} and \eqref{winsym2} under the correspondence \eqref{corrphi}, and thus generate the winding dipole symmetry.

\section{Duality and Foliated \texorpdfstring{$\hat{\phi}$}{φhat}-Theory}
\label{section:duality_foliated_phihat}

The exotic $\phi$-theory in 2+1 dimensions has the momentum dipole symmetry and the winding dipole symmetry. While the momentum dipole symmetry acts on the scalar field $\phi$, the winding dipole symmetry acts on the dual field $\hat{\phi}^{12}$ in the $\hat\phi$-theory \cite{Seiberg:2020bhn,Spieler:2024fby}. In this section, we consider the $T$-duality-like duality between the $\phi$-theory and the $\hat{\phi}$-theory, which is a self-duality in the $\phi$-theory, and construct the foliated description of the $\hat{\phi}$-theory.

\subsection{\texorpdfstring{$T$}{T}-Duality-like Duality and Exotic \texorpdfstring{$\hat{\phi}$}{φhat}-Theory}
\label{subsection:exotic_phihat_theory}

In this section, we review the duality between the exotic $\phi$-theory and the exotic $\hat{\phi}$-theory\cite{Seiberg:2020bhn}, which is similar to the $T$-duality in the compact scalar field theory.

We consider the exotic Lagrangian including the background tensor gauge fields
\begin{align}
\begin{split}
    \L_{\phi,\text{e}}\left[ \bm{C}, \bm{\hat{C}} \right] &= \frac{\mu_0}{2} ( E_0 )^2 + \frac{1}{2 \mu_{12}} ( B_{12} )^2  \\
    & \quad +\frac{i}{2\pi} \hat{E}^{12}_0(\partial_1 \partial_2 \phi - C_{12} - B_{12}) + \frac{i}{2\pi} \hat{B} (\partial_0 \phi - C_0 - E_0 )  \\
    & \quad + \frac{i}{2\pi} \hat{C}^{12}_0 ( \partial_1 \partial_2 \phi - C_{12}) + \frac{i}{2\pi} \hat{C} ( \partial_0 \phi - C_0 ) \,, \label{exoticphic-eb}
\end{split}
\end{align}
where $E_0$, $B_{12}$, $\hat{E}^{12}_0$ and $\hat{B}$ are dynamical tensor fields in $\bm{1}_0$, $\bm{1}_2$, $\bm{1}_2$ and $\bm{1}_0$, respectively. Integrating out $\hat{E}^{12}_0$ and $\hat{B}$ leads to the equations
\begin{align}
    B_{12} &= \partial_1 \partial_2 \phi - C_{12} \,, \label{e12re} \\
    E_0 &= \partial_0 \phi - C_0 \,, \label{b0re}
\end{align}
and then, we recover the Lagrangian of the exotic $\phi$-theory \eqref{exoticphic}. On the other hand, integrating out $E_0$ and $B_{12}$ leads to the equations
\begin{align}
    E_0 &= \frac{i}{2\pi \mu_0} \hat{B} \,, \label{b0ehat} \\
    B_{12} &= \frac{i\mu_{12}}{2\pi} \hat{E}^{12}_0 \,, \label{e12bhat012}
\end{align}
so we can write the Lagrangian as
\begin{align}
\begin{split}
    \L_{\phi,\text{e}}\left[ \bm{C}, \bm{\hat{C}} \right] &= \frac{\mu_{12}}{8\pi^2} ( \hat{E}^{12}_0 )^2 + \frac{1}{8 \pi^2 \mu_0} ( \hat{B} )^2  \\
    & \quad +\frac{i}{2\pi} (\hat{E}^{12}_0 + \hat{C}^{12}_0 )(\partial_1 \partial_2 \phi - C_{12}) + \frac{i}{2\pi} (\hat{B} + \hat{C} ) (\partial_0 \phi - C_0)  \,. \label{exoticphic-eb2}
\end{split}
\end{align}
Next, we integrate out $\phi$ to have the equation
\begin{align}
    \partial_1 \partial_2 (\hat{E}^{12}_0 + \hat{C}^{12}_0) = \partial_0 (\hat{B} + \hat{C}) \,,
\end{align}
which we can solve locally in terms of a tensor gauge field $\hat{\phi}^{12}$ in $\bm{1}_2$:
\begin{align}
    \hat{E}^{12}_0 &= \partial_0  \hat{\phi}^{12} - \hat{C}^{12}_0 \,, \label{bhat120re} \\
    \hat{B} &= \partial_1 \partial_2 \hat{\phi}^{12} - \hat{C} \,. \label{ehatre}
\end{align}
The dual field $\hat{\phi}^{12}$ is a dynamical compact scalar field with the periodicity $\hat{\phi}^{12} \sim \hat{\phi}^{12} + 2\pi$, and we have dynamical gauge transformation of $\hat{\phi}^{12}$ as
\begin{align}
    \hat{\phi}^{12} \sim \hat{\phi}^{12}  + 2\pi \hat{w}^1 - 2\pi \hat{w}^2 \,.
\end{align}
$\hat{w}^k$ is an $x^k$-dependent integer-valued gauge parameter.
We also have background gauge transformation as
\begin{align}
    \hat{\phi}^{12} &\sim \hat{\phi}^{12} + \hat{\Gamma}^{12} \,.
\end{align}
Then, we derive the Lagrangian of the exotic $\hat{\phi}$-theory including the background tensor gauge fields:
\begin{align}
\begin{split}
    \L_{\hat{\phi},\text{e}}\left[ \bm{C}, \bm{\hat{C}} \right] &= \frac{\hat{\mu}_0}{2} ( \partial_0  \hat{\phi}^{12} - \hat{C}^{12}_0 )^2 + \frac{1}{2 \hat{\mu}_{12}} ( \partial_1 \partial_2 \hat{\phi}^{12} - \hat{C} )^2  \\
    & \quad - \frac{i}{2\pi} C_{0} \partial_1 \partial_2 \hat{\phi}^{12} - \frac{i}{2\pi} C_{12} \partial_0  \hat{\phi}^{12}   \,, \label{exoticphihatc1}
\end{split}
\end{align}
where $\hat{\mu}_0 = \mu_{12}/(4\pi^2)$ and $\hat{\mu}_{12} = 4\pi^2 \mu_0$. We can obtain relations between $\phi$ and $\hat{\phi}^{12}$ from \eqref{e12re}, \eqref{b0re}, \eqref{b0ehat}, \eqref{e12bhat012}, \eqref{bhat120re} and \eqref{ehatre}:
\begin{align}
    \partial_0 \phi - C_0 &=  \frac{i}{2\pi \mu_0} (\partial_1 \partial_2 \hat{\phi}^{12} - \hat{C} ) \,, \label{phiphihatcre1} \\
    \partial_1 \partial_2 \phi - C_{12} &= \frac{i\mu_{12}}{2\pi} ( \partial_0 \hat{\phi}^{12} - \hat{C}_0^{12} ) \,. \label{phiphihatcre2}
\end{align}

We can consider the duality between the exotic $\phi$-theory and the exotic $\hat{\phi}$-theory as a self-duality. The exotic $\hat{\phi}$-theory is also a boundary theory of the exotic SSPT phase \eqref{exoticsspt} in the spacetime $\T^{2+1} \times \R_{x^3 \geq 0}$. Then, we add the terms
\begin{align}
    \frac{i}{2\pi} ( C_0 \hat{C} + C_{12} \hat{C}^{12}_0 ) \,, 
\end{align}
on the boundary and the terms
\begin{align}
    \partial_3 \left[ \frac{i}{2\pi} ( C_0 \hat{C} + C_{12} \hat{C}^{12}_0 )  \right] \,,
\end{align}
in the SSPT phase in 3+1 dimensions, the sum of which is zero. The boundary $\hat{\phi}$-theory becomes
\begin{align}
\begin{split}
    \L'_{\hat{\phi},\text{e}}\left[ \bm{C}, \bm{\hat{C}} \right] &= \frac{\hat{\mu}_0}{2} ( \partial_0  \hat{\phi}^{12} - \hat{C}^{12}_0 )^2 + \frac{1}{2 \hat{\mu}_{12}} ( \partial_1 \partial_2 \hat{\phi}^{12} - \hat{C} )^2  \\
    & \quad - \frac{i}{2\pi} C_{0} ( \partial_1 \partial_2 \hat{\phi}^{12} - \hat{C} ) - \frac{i}{2\pi} C_{12} ( \partial_0  \hat{\phi}^{12} - \hat{C}^{12}_0 )   \,,
\end{split}
\end{align}
and the SSPT phase \eqref{exoticsspt} becomes
\begin{align}
\begin{split}
    &\L'_{\text{SSPT,e}}\left[ \bm{C}, \bm{\hat{C}} \right] \\
    &\quad = \frac{i}{2\pi} \left[ - C_{12} ( \partial_0 \hat{C}^{12}_3 - \partial_3 \hat{C}^{12}_0 ) + C_0 (  \partial_3 \hat{C} - \partial_1 \partial_2 \hat{C}^{12}_3 ) - C_3 ( \partial_0 \hat{C} - \partial_1 \partial_2 \hat{C}^{12}_0 )  \right] \,. \label{exoticsspt2}
\end{split}
\end{align}
Therefore, considering transformations
\begin{align}
    \hat{\phi}^{12} &\rightarrow \phi \,, \\
    \hat{C}^{12}_0 &\rightarrow C_0 \,, \\
    \hat{C} &\rightarrow C_{12} \,, \\
    \hat{C}^{12}_3 &\rightarrow C_3 \,, \\
    C_0 &\rightarrow -\hat{C}^{12}_0 \,, \\
    C_{12} &\rightarrow -\hat{C} \,, \\
    C_3 &\rightarrow -\hat{C}^{12}_3 \,, 
\end{align}
and replacing the $\Z_4$ rotations with the $\Z_4$ rotations multiplied by the charge conjugation, under which all the fields are multiplied by $-1$, we can derive the original theory \eqref{exoticphic} and \eqref{exoticsspt}, when the parameters satisfy $\mu_0 = \hat{\mu_0}$ and $\mu_{12} = \hat{\mu}_{12}$, i.e., $4\pi^2 \mu_0 = \mu_{12}$ \cite{Seiberg:2020bhn,Spieler:2024fby}.

Let us consider the subsystem symmetries. We consider the exotic $\hat{\phi}$-theory without the background tensor gauge fields
\begin{align}
    \L_{\hat{\phi},\text{e}} &= \frac{\hat{\mu}_0}{2} ( \partial_0  \hat{\phi}^{12}  )^2 + \frac{1}{2 \hat{\mu}_{12}} ( \partial_1 \partial_2 \hat{\phi}^{12} )^2  \,.
\end{align}
Then, the relations \eqref{phiphihatcre1} and \eqref{phiphihatcre2} become
\begin{align}
    \partial_0 \phi  &=  \frac{i}{2\pi \mu_0} \partial_1 \partial_2 \hat{\phi}^{12}  \,, \label{phiphihatre1} \\
    \partial_1 \partial_2 \phi  &= \frac{i\mu_{12}}{2\pi}  \partial_0 \hat{\phi}^{12}  \,. \label{phiphihatre2}
\end{align}
Thus, the currents of the momentum dipole symmetry \eqref{momsym1} and \eqref{momsym2} in the exotic $\phi$-theory are
\begin{align}
    J_0 &= i\mu_0 \partial_0 \phi = -\frac{1}{2\pi} \partial_1 \partial_2 \hat{\phi}^{12} \,, \label{hatwinsym1} \\
    J_{12} &=  \frac{i}{\mu_{12}} \partial_1 \partial_2 \phi =  -\frac{1}{2\pi} \partial_0 \hat{\phi}^{12} \,, \label{hatwinsym2}
\end{align}
which are the currents of the winding dipole symmetry in the exotic $\hat{\phi}$-theory.\footnote{The minus signs are compatible with the coupling terms with $C_0$ and $C_{12}$ in \eqref{exoticphihatc1}.} On the other hand, the currents of the winding dipole symmetry \eqref{winsym1} and \eqref{winsym2} in the exotic $\phi$-theory are
\begin{align}
    \hat{J}^{12}_0 &= \frac{1}{2 \pi} \partial_1 \partial_2 \phi = i \hat{\mu}_0 \partial_0 \hat{\phi}^{12} \,, \label{hatmomsym1} \\
    \hat{J} &= \frac{1}{2 \pi} \partial_0 \phi = \frac{i}{\hat{\mu}_{12}} \partial_1 \partial_2 \hat{\phi}^{12} \,, \label{hatmomsym2}
\end{align}
which are the currents of the momentum dipole symmetry in the exotic $\hat{\phi}$-theory. Then, the charged operator is
\begin{align}
    \hat{V}_n [x] = e^{in\hat{\phi}^{12}} \,, 
\end{align}
For the field action, the momentum dipole symmetry acts on $\hat{\phi}^{12}$ as
\begin{align}
    \hat{\phi}^{12} &\rightarrow \hat{\phi}^{12} + \hat{\Gamma}^1(x^1) - \hat{\Gamma}^2(x^2) \,.
\end{align}
where $\Gamma^k(x^k)$ is a $2\pi$-periodic scalar that can have step function discontinuities in the $x^k$ direction.

\subsection{Construction of Foliated \texorpdfstring{$\hat{\phi}$}{φhat}-Theory}
\label{subsection:construction_of_foliated_phihat_theory}

As in the case of Section \ref{subsection:construction_of_foliated_phi_theory}, we can construct a foliated $\hat{\phi}$-theory in 2+1 dimensions by using the method in Section \ref{subsection:construction_of_boundary_theory}. We take the foliated SSPT phase \eqref{folissptlag}, and choose the gauge fields $\hat{C}^k$ and $\hat{c}$. Then, we have to consider the Lagrangian
\begin{align}
\begin{split}
    &L_{\text{SSPT,f}} \left[ C^k \wedge e^k, c, \hat{C}^k, \hat{c}_{12} \right] \\
    & \quad = \frac{i}{2\pi} \left[  \sum_{k = 1}^2 \hat{C}^k \wedge d C^k \wedge e^k + \hat{c} \wedge \left( dc + \sum_{k=1}^2 C^k \wedge e^k \right) \right]  \,, 
\end{split}
\end{align}
on the spacetime $\T^{2+1} \times \R_{x^3 \geq 0}$. We introduce a $U(1)$ type-$B$ foliated zero-form gauge field $\hat{\Phi}^k\ (k = 1,2)$ and a $U(1)$ type-$B$ bulk one-form gauge field $\hat{\Phi}$, whose dynamical gauge transformations are
\begin{align}
    \hat{\Phi}^k  &\sim \hat{\Phi}^k + 2\pi \hat{W}^k + \hat{\xi} \,, \\
    \hat{\Phi} &\sim \hat{\Phi} + d \hat{\xi} \,,
\end{align}
where $\hat{W}^k$ is an $x^k$-dependent integer-valued gauge parameter, and $\hat{\xi}$ is a type-$B$ bulk zero-form gauge parameter. We set the background gauge transformations of $\hat{\Phi}^k$ and $\hat{\Phi}$ as
\begin{align}
    \hat{\Phi}^k  &\sim \hat{\Phi}^k + \hat{\gamma}^k \,, \label{bkggaugephihatk} \\
    \hat{\Phi} &\sim \hat{\Phi} + \hat{\gamma} \,,
\end{align}
and substitute the form $(\hat{C}^k - d\hat{\Phi}^k + \hat{\Phi})\wedge e^k$ and $\hat{c} - d\hat{\Phi}$, which are the gauge-invariant combinations from \eqref{bkggaugetr1}--\eqref{bkggaugetr4}, for $\hat{C}^k \wedge e^k$ and $\hat{c}$ in $L_{\text{SSPT,f}}$, respectively. Then, we have
\begin{align}
\begin{split}
    &L'''_{\text{SSPT,f}} \left[ C^k \wedge e^k, c, \hat{C}^k, \hat{c}_{12} \right] \\
    & = \frac{i}{2\pi} \left[  \sum_{k = 1}^2 (\hat{C}^k - d\hat{\Phi}^k + \hat{\Phi}) \wedge d C^k \wedge e^k + (\hat{c} - d\hat{\Phi}) \wedge \left( dc + \sum_{k=1}^2 C^k \wedge e^k \right) \right] \\
    &  = \frac{i}{2\pi} \left[  \sum_{k = 1}^2  \hat{C}^k \wedge d C^k \wedge e^k + \hat{c} \wedge \left( dc + \sum_{k = 1}^2 C^k \wedge e^k  \right)  \right] \\
    & \quad + d \left[  \frac{i}{2\pi} \sum_{k = 1}^2 ( d\hat{\Phi}^k - \hat{\Phi}) \wedge  C^k \wedge e^k - \frac{i}{2\pi} d \hat{\Phi} \wedge c  \right]\,, 
\end{split}
\end{align}
The boundary term in the action is
\begin{align}
\begin{split}
    &\int_{x^3 = 0} \frac{i}{2\pi} \left[  \sum_{k = 1}^2 (d\hat{\Phi}^k - \hat{\Phi}) \wedge  C^k \wedge e^k - d \hat{\Phi} \wedge c  \right] \\
    & = \int_{x^3 = 0} \frac{i}{2\pi} \left[ C^1_0 ( \partial_2 \hat{\Phi}^1 - \hat{\Phi}_2 ) - C^2_0 ( \partial_1 \hat{\Phi}^2 - \hat{\Phi}_1 ) - C^1_2 ( \partial_0 \hat{\Phi}^1 - \hat{\Phi}_0 ) + C^2_1 ( \partial_0 \hat{\Phi}^2 - \hat{\Phi}_0 )       \right. \\
    & \qquad \qquad \left. - c_0 ( \partial_1 \hat{\Phi}_2 - \partial_2 \hat{\Phi}_1 )  + c_1 ( \partial_0 \hat{\Phi}_2 - \partial_2 \hat{\Phi}_0 ) - c_2 ( \partial_0 \hat{\Phi}_1 - \partial_1 \hat{\Phi}_0 )    \right] d^3 x 
\end{split}
\end{align}
As in the case of Section \ref{subsection:construction_of_foliated_phi_theory}, we add the term 
\begin{align}
\begin{split}
    & \frac{i}{2\pi} \hat{\chi} \wedge \left( dc + \sum^{2}_{k=1} C^k \wedge e^k  \right) \\
    &= \frac{i}{2\pi} \left[ \hat{\chi}_0 (\partial_1 c_2 - \partial_2 c_1 + C^2_1 - C^1_2  ) \right. \\
    & \quad \left. - \hat{\chi}_1 ( \partial_0 c_2 - \partial_2 c_0 +  C^2_0) + \hat{\chi}_2 (\partial_0 c_1 - \partial_1 c_0 +  C^1_0 ) \right] d^3 x \,,
\end{split}
\end{align}
on the boundary, where $\hat{\chi}$ is a dynamical one-form field. Finally, we add background gauge-invariant and $\Z_4$ rotational invariant quadratic terms, and we have
\begin{align}
\begin{split}
    &\L_{\hat{\phi},\text{f}} \left[ C^k \wedge e^k, c, \hat{C}^k, \hat{c}_{12} \right] \\
    & = \hat{\mu}_0 ( \partial_0 \hat{\Phi}^1 - \hat{\Phi}_0 -\hat{C}^1_0 )^2 + \hat{\mu}_0 ( \partial_0 \hat{\Phi}^2 - \hat{\Phi}_0 -\hat{C}^2_0 )^2 \\ 
    & \quad +  \frac{\hat{\mu}^{12}}{2}( \partial_2 \hat{\Phi}^1 - \hat{\Phi}_2 - \hat{C}^1_2 )^2 +   \frac{\hat{\mu}^{12}}{2}( \partial_1 \hat{\Phi}^2 - \hat{\Phi}_1 -\hat{C}^2_1 )^2 \\
    & \quad + \frac{1}{2\hat{\mu}_{12}}( \partial_1 \hat{\Phi}_2 - \partial_2 \hat{\Phi}_1 - \hat{c}_{12} )^2 \\
    & \quad + \frac{1}{2\hat{\mu}_{012}}( \partial_0 \hat{\Phi}_1 - \partial_1 \hat{\Phi}_0 - \hat{c}_{01} )^2 + \frac{1}{2\hat{\mu}_{012}}( \partial_0 \hat{\Phi}_2 - \partial_2 \hat{\Phi}_0 - \hat{c}_{02} )^2 \\
    & \quad + \frac{i}{2\pi} \left[ C^1_0 ( \partial_2 \hat{\Phi}^1 - \hat{\Phi}_2 ) - C^2_0 ( \partial_1 \hat{\Phi}^2 - \hat{\Phi}_1 ) \right. \\
    &  \qquad - C^1_2 ( \partial_0 \hat{\Phi}^1 - \hat{\Phi}_0 )  + C^2_1 ( \partial_0 \hat{\Phi}^2 - \hat{\Phi}_0 )  \\
    &  \qquad \left. - c_0 ( \partial_1 \hat{\Phi}_2 - \partial_2 \hat{\Phi}_1 )  + c_1 ( \partial_0 \hat{\Phi}_2 - \partial_2 \hat{\Phi}_0 ) - c_2 ( \partial_0 \hat{\Phi}_1 - \partial_1 \hat{\Phi}_0 )    \right] \\
    & \quad + \frac{i}{2\pi} \left[ \hat{\chi}_0 ( C^2_1 - C^1_2 + \partial_1 c_2 - \partial_2 c_1 ) \right. \\
    & \qquad \left. - \hat{\chi}_1 ( C^2_0 + \partial_0 c_2 - \partial_2 c_0) + \hat{\chi}_2 ( C^1_0 + \partial_0 c_1 - \partial_1 c_0 ) \right] \,, \label{genefoliatedphihat1}
\end{split}
\end{align}
where $\hat{\mu}_0$, $\hat{\mu}^{12}$ and $\hat{\mu}_{012}$  are parameters with mass dimension one.

We can also write the foliated Lagrangian in differential form by tuning the metric:
\begin{align}
\begin{split}
    &L_{\hat{\phi},\text{f}} \left[ C^k \wedge e^k, c, \hat{C}^k, \hat{c}_{12}\right] \\
    &  = \frac{1}{2} \sum_{k=1}^2 \left\{ \left(  d \hat{\Phi}^k - \hat{\Phi} - \hat{C}^k \right) \wedge e^k \right\} \wedge \ast \left\{ \left( d \hat{\Phi}^k - \hat{\Phi} - \hat{C}^k \right) \wedge e^k \right\} \\
    & \quad + \frac{1}{2} \left( d \hat{\Phi} -\hat{c}  \right) \wedge \ast \left( d \hat{\Phi} -\hat{c}  \right)  \\
    & \quad + \frac{i}{2\pi} \left[  \sum_{k = 1}^2  \left( d\hat{\Phi}^k - \hat{\Phi} \right)  \wedge C^k \wedge e^k -  d\hat{\Phi} \wedge c   \right]  \\
    & \quad + \frac{i}{2\pi} \hat{\chi} \wedge \left( dc + \sum^{2}_{k=1} C^k \wedge e^k  \right) \\
    &= \frac{1}{2} (d_\text{f} \bm{\hat{\Phi}} - \bm{\hat{C}} )^2 + \frac{i}{2\pi} d_\text{f} \bm{\hat{\Phi}} \wedge_\text{f} \bm{C} + \hat{\bm{\chi}} \wedge_\text{f} d_\text{f} \bm{C}   \,, \label{foliatedphihatdeff}
\end{split}
\end{align}
where $\bm{\hat{\Phi}} = (\hat{\Phi}^k, \hat{\Phi})$ is a type-$B$ foliated zero-form gauge field, $\bm{\hat{C}} = (\hat{C}^k, \hat{c})$ is a type-$B$ foliated one-form gauge field, $\bm{C} = (C^k \wedge e^k, c)$ is a type-$A$ foliated (1+1)-form gauge field and $\bm{\hat{\chi}} = (0,\hat{\chi})$ is a type-$B$ foliated zero-form field.

We can integrate out the dynamical gauge fields $\hat{c}_{01}$ and $\hat{c}_{02}$ on the boundary, and then we have
\begin{align}
    \hat{c}_{01} &=  \partial_0 \hat{\Phi}_1 - \partial_1 \hat{\Phi}_0 \,, \\
    \hat{c}_{02} &= \partial_0 \hat{\Phi}_2 - \partial_2 \hat{\Phi}_0 \,.
\end{align}
Furthermore, we consider a change of variable from $\hat{\chi}$ to $\hat{\kappa}$ as
\begin{align}
    \hat{\chi} = \hat{\kappa} + \hat{\Phi} \,.
\end{align}
Then, the field $\hat{\kappa}$ has a dynamical gauge transformation
\begin{align}
    \hat{\kappa} \sim \hat{\kappa} - d\hat{\xi} \,,
\end{align}
and a background gauge transformation
\begin{align}
    \hat{\kappa} \sim \hat{\kappa} - \hat{\gamma} \,.
\end{align}
Finally, we have the foliated Lagrangian
\begin{align}
    &\L_{\hat{\phi},\text{f}} \left[ C^k \wedge dx^k, c, \hat{C}^k, \hat{c}_{12}\right] \\
    & = \hat{\mu}_0 ( \partial_0 \hat{\Phi}^1 - \hat{\Phi}_0 -\hat{C}^1_0 )^2 + \hat{\mu}_0 ( \partial_0 \hat{\Phi}^2 - \hat{\Phi}_0 -\hat{C}^2_0 )^2 \\ 
    & \quad +  \frac{\hat{\mu}^{12}}{2}( \partial_2 \hat{\Phi}^1 - \hat{\Phi}_2 -\hat{C}^1_2 )^2 +   \frac{\hat{\mu}^{12}}{2}( \partial_1 \hat{\Phi}^2 - \hat{\Phi}_1 -\hat{C}^2_1 )^2  \\
    & \quad + \frac{1}{2\hat{\mu}_{12}}( \partial_1 \hat{\Phi}_2 - \partial_2 \hat{\Phi}_1 - \hat{c}_{12} )^2 \\
    & \quad + \frac{i}{2\pi} \left( C^1_0  \partial_2 \hat{\Phi}^1  - C^2_0  \partial_1 \hat{\Phi}^2 - C^1_2 \partial_0 \hat{\Phi}^1  + C^2_1  \partial_0 \hat{\Phi}^2 \right)  \\
    & \quad + \frac{i}{2\pi} \left[ \hat{\kappa}_0 ( C^2_1 - C^1_2 + \partial_1 c_2 - \partial_2 c_1 ) \right. \\
    & \qquad \left. - \hat{\kappa}_1 ( C^2_0 + \partial_0 c_2 - \partial_2 c_0) + \hat{\kappa}_2 ( C^1_0 + \partial_0 c_1 - \partial_1 c_0 ) \right] \,. \label{genefoliatedphihat2}
\end{align}

\subsection{Foliated-Exotic Duality in \texorpdfstring{$\hat{\phi}$}{φhat}-Theory}
\label{subsection:foliated_exotic_duality_phihat}

In this section, we establish the field correspondences in the $\hat{\phi}$-theory and obtain the foliated $\hat{\phi}$-theory, which is equivalent to the exotic $\hat{\phi}$-theory \eqref{exoticphihatc1}. We can also derive the foliated $\hat{\phi}$-theory by tuning the parameters in the general foliated $\hat{\phi}$-theory \eqref{genefoliatedphihat2}.

In section \ref{subsection:exotic_phihat_theory}, we have derived the exotic $\hat{\phi}$-theory coupled with the background tensor gauge fields
\begin{align}
\begin{split}
    \L_{\hat{\phi},\text{e}}\left[ \bm{C}, \bm{\hat{C}}  \right] &= \frac{\hat{\mu}_0}{2} ( \partial_0  \hat{\phi}^{12} - \hat{C}^{12}_0 )^2 + \frac{1}{2 \hat{\mu}_{12}} ( \partial_1 \partial_2 \hat{\phi}^{12} - \hat{C} )^2  \\
    & \quad - \frac{i}{2\pi} C_{0} \partial_1 \partial_2 \hat{\phi}^{12} - \frac{i}{2\pi} C_{12} \partial_0  \hat{\phi}^{12}   \,. \label{exoticphihatc2} 
\end{split}
\end{align}

First, we assume the correspondences between the tensor gauge field $\hat{\phi}^{12}$ in the exotic $\hat{\phi}$-theory and the type-$B$ foliated zero-form gauge fields $\hat{\Phi}^k\ (k=1,2)$ in a foliated $\hat{\phi}$-theory as
\begin{align}
    \hat{\phi}^{12} \simeq \hat{\Phi}^1 - \hat{\Phi}^2 \,. \label{corrphihat}
\end{align}
Since the dynamical gauge transformations of $\hat{\phi}^{12}$ and $\hat{\Phi}^k$ are
\begin{align}
    \hat{\phi}^{12} &\sim \hat{\phi}^{12} + 2 \pi \hat{w}^1 - 2 \pi \hat{w}^2 \,, \\
    \hat{\Phi}^k &\sim \hat{\Phi}^k + 2 \pi \hat{W}^k + \hat{\xi} \,,
\end{align}
the correspondences between the dynamical gauge parameters are
\begin{align}
    \hat{w}^k \simeq \hat{W}^k \,, \quad (k = 1,2) \,. 
\end{align}
The background gauge transformation of $\hat{\Phi}^k$ is \eqref{bkggaugephihatk}.

As in the case of Section \ref{subsection:foliated_exotic_duality_phi}, the background gauge fields in a foliated theory are $U(1)$ type-$A$ foliated (1+1)-form gauge fields $C^k \wedge e^k \ (k = 1,2)$, a $U(1)$ type-$A$ bulk one-form gauge field $c$, $U(1)$ type-$B$ foliated one-form gauge fields $\hat{C}^k \ (k = 1,2)$ and the $x^1x^2$-component of a $U(1)$ type-$B$ bulk two form gauge field $\hat{c}_{12}$. We assume correspondences
\begin{align}
    C_0 &\simeq c_0 \label{bcorr0-0,2} \,, \\
    C_{12} &\simeq C^1_2 + \partial_2 c_1 \,,  \label{bcorr12-1,2} 
\end{align}
and
\begin{align}
    \hat{C}_0^{12} &\simeq \hat{C}_0^1 - \hat{C}_0^2 \,, \label{bcorr012,2} \\
    \hat{C} &\simeq \partial_1 \hat{C}_2^1 - \partial_2 \hat{C}_1^2 + \hat{c}_{12}  \,. \label{bcorr1221,2}
\end{align}
Their background gauge transformations are \eqref{bkggaugetr1}--\eqref{bkggaugetr4}. We also add to the Lagrangian the term
\begin{align}
\begin{split}
    & \frac{i}{2\pi} \hat{\chi} \wedge \left( dc + \sum^{2}_{k=1} C^k \wedge e^k  \right) \\
    &= \frac{i}{2\pi} \left[ \hat{\chi}_0 ( \partial_1 c_2 - \partial_2 c_1 + C^2_1 - C^1_2 ) \right. \\
    & \quad \left. - \hat{\chi}_1 (\partial_0 c_2 - \partial_2 c_0 +  C^2_0) + \hat{\chi}_2 (  \partial_0 c_1 - \partial_1 c_0 + C^1_0 ) \right] d^3 x \,.
\end{split}
\end{align}

Using these correspondences, the Lagrangian of the $\hat{\phi}$-theory can be written as
\begin{align}
\begin{split}
    &\L_{\hat{\phi},\text{e}}\left[ \bm{C}, \bm{\hat{C}}  \right] \\
    &\simeq \frac{\hat{\mu}_0}{2} \left[ ( \partial_0 \hat{\Phi}^1 - \hat{C}^1_0 ) - ( \partial_0 \hat{\Phi}^2 - \hat{C}^2_0 ) \right]^2 \\
    & \quad + \frac{1}{2\hat{\mu}_{12}} \left[ \partial_1 ( \partial_2 \hat{\Phi}^1 - \hat{C}^1_2 ) - \partial_2 ( \partial_1 \hat{\Phi}^2 - \hat{C}^2_1 ) - \hat{c}_{12} \right]^2 \\
    & \quad + \frac{i}{2\pi} ( \partial_1 c_0 \partial_2 \hat{\Phi}^1 - \partial_2 c_0 \partial_1 \hat{\Phi}^2 )  - \frac{i}{2\pi} ( C^1_2 + \partial_2 c_1 ) \partial_0 ( \hat{\Phi}^1 - \hat{\Phi}^2 ) \\
    & \quad + \frac{i}{2\pi} \left[ \hat{\chi}_0 ( \partial_1 c_2 - \partial_2 c_1 + C^2_1 - C^1_2 ) \right. \\
    & \qquad \left. - \hat{\chi}_1 (\partial_0 c_2 - \partial_2 c_0 +  C^2_0) + \hat{\chi}_2 (  \partial_0 c_1 - \partial_1 c_0 + C^1_0 ) \right] \,,
\end{split}
\end{align}
where we have integrated by parts and dropped the total derivative terms. As in the case of Section \ref{subsection:foliated_exotic_duality_phi}, we have to add the boundary term of \eqref{ssptaddterm}:
\begin{align}
     \frac{i}{2\pi} \left[ \hat{C}^2_0 ( \partial_1 c_2 - \partial_2 c_1 +C^2_1 - C^1_2 ) - \hat{C}^2_1 ( \partial_0 c_2 - \partial_2 c_0 + C^2_0) + \hat{C}^1_2 ( \partial_0 c_1 - \partial_1 c_0 + C^1_0)   \right] \,.
\end{align}

Next, we consider changes of variables from $\hat{\chi}$ to $\hat{\kappa}$ as
\begin{align}
    \hat{\chi}_0 &= \hat{\kappa}_0 + \partial_0 \hat{\Phi}^2  - \hat{C}^2_0 \,, \\
    \hat{\chi}_1 &= \hat{\kappa}_1 + \partial_1 \hat{\Phi}^2  - \hat{C}^2_1 \,, \\
    \hat{\chi}_2 &= \hat{\kappa}_2 + \partial_2 \hat{\Phi}^1  - \hat{C}^1_2 \,, 
\end{align}
and then, the field $\hat{\kappa}$ has a dynamical gauge transformation
\begin{align}
    \hat{\kappa} \sim \hat{\kappa} - d\hat{\xi} \,,
\end{align}
and a background gauge transformation
\begin{align}
    \hat{\kappa} \sim \hat{\kappa} - \hat{\chi} \,,
\end{align}
as in the previous section. The result is
\begin{align}
\begin{split}
    & \frac{i}{2\pi} ( \partial_1 c_0 \partial_2 \hat{\Phi}^1 - \partial_2 c_0 \partial_1 \hat{\Phi}^2 )  - \frac{i}{2\pi} ( C^1_2 + \partial_2 c_1 ) \partial_0 ( \hat{\Phi}^1 - \hat{\Phi}^2 ) \\
    & \quad + \frac{i}{2\pi} \left[ \hat{C}^2_0 ( \partial_1 c_2 - \partial_2 c_1 +C^2_1 - C^1_2 ) - \hat{C}^2_1 ( \partial_0 c_2 - \partial_2 c_0 + C^2_0) + \hat{C}^1_2 ( \partial_0 c_1 - \partial_1 c_0 + C^1_0)   \right] \\
    & \rightarrow \frac{i}{2\pi} \left[  ( C^1_0 + \partial_0 c_1 ) \partial_2 \hat{\Phi}^1 - ( C^2_0 + \partial_0 c_2 ) \partial_1 \hat{\Phi}^2 - ( C^1_2 + \partial_2 c_1 ) \partial_0 \hat{\Phi}^1 + ( C^2_1 + \partial_1 c_2 ) \partial_0 \hat{\Phi}^2 \right] \\
    &\quad =  \frac{i}{2\pi} \left(  C^1_0  \partial_2 \hat{\Phi}^1 -  C^2_0  \partial_1 \hat{\Phi}^2 -  C^1_2  \partial_0 \hat{\Phi}^1 +  C^2_1  \partial_0 \hat{\Phi}^2 \right) \,,
\end{split}
\end{align}
where we have integrated by parts and dropped the total derivative terms. Finally, we obtain the foliated $\hat{\phi}$-theory, which is equivalent to the exotic $\hat{\phi}$-theory \eqref{exoticphihatc2}:
\begin{align}
\begin{split}
    &\L_{\hat{\phi},\text{e} \rightarrow \text{f}} \left[ C^k \wedge e^k, c, \hat{C}^k, \hat{c}_{12} \right] \\
    & = \frac{\hat{\mu}_0}{2} \left[ ( \partial_0 \hat{\Phi}^1 - \hat{C}^1_0 ) - ( \partial_0 \hat{\Phi}^2 - \hat{C}^2_0 ) \right]^2 \\
    & \quad + \frac{1}{2\hat{\mu}_{12}} \left[ \partial_1 ( \partial_2 \hat{\Phi}^1 - \hat{C}^1_2 ) - \partial_2 ( \partial_1 \hat{\Phi}^2 - \hat{C}^2_1 ) - \hat{c}_{12} \right]^2 \\
    & \quad + \frac{i}{2\pi} \left(  C^1_0  \partial_2 \hat{\Phi}^1 -  C^2_0  \partial_1 \hat{\Phi}^2 -  C^1_2  \partial_0 \hat{\Phi}^1 +  C^2_1  \partial_0 \hat{\Phi}^2 \right) \\
    & \quad + \frac{i}{2\pi} \left[ \hat{\kappa}_0 ( \partial_1 c_2 - \partial_2 c_1 + C^2_1 - C^1_2 ) \right. \\
    & \qquad \left. - \hat{\kappa}_1 (\partial_0 c_2 - \partial_2 c_0 +  C^2_0) + \hat{\kappa}_2 (  \partial_0 c_1 - \partial_1 c_0 + C^1_0 ) \right]  \,. \label{foliphihatc}
\end{split}
\end{align}

Conversely, to obtain the exotic $\hat{\phi}$-theory \eqref{exoticphihatc2} from the foliated $\phi$-theory \eqref{foliphihatc}, we integrate out the dynamical fields $\hat{\kappa}_i \ (i = 0,1,2)$ as Lagrange multipliers. Then, we have the constraints \eqref{constc12}--\eqref{constc02} and we can use the dictionary
\begin{align}
    \hat{\phi}^{12} &\simeq \hat{\Phi}^1 - \hat{\Phi}^2 \,, \label{phihatdic1} \\ 
    C_0 &\simeq c_0 \,, \\
    C_{12} &\simeq C^1_2 + \partial_2 c_1 =  C^2_1 + \partial_1 c_2 \,, \\
    \partial_1 C_0 &\simeq C^1_0 + \partial_0 c_1 \,, \\
    \partial_2 C_0 &\simeq C^2_0 + \partial_0 c_2 \,, \\
    \hat{C}_0^{12} &\simeq \hat{C}_0^1 - \hat{C}_0^2 \,,  \\
    \hat{C} &\simeq \partial_1 \hat{C}_2^1 - \partial_2 \hat{C}_1^2 + \hat{c}_{12} \,. \label{phihatdic7}
\end{align}
By applying them, we can recover the exotic $\hat{\phi}$-theory \eqref{exoticphihatc2}.

We can also obtain the foliated Lagrangian \eqref{foliphihatc} by tuning the parameters in the general foliated Lagrangian \eqref{genefoliatedphihat2}. First, the terms
\begin{align}
    \frac{\hat{\mu}^{12}}{2}( \partial_2 \hat{\Phi}^1 - \hat{\Phi}_2 -\hat{C}^1_2 )^2 +   \frac{\hat{\mu}^{12}}{2}( \partial_1 \hat{\Phi}^2 - \hat{\Phi}_1 -\hat{C}^2_1 )^2 \label{term12}
\end{align}
can be written as
\begin{align}
     \frac{1}{8\pi^2 \hat{\mu}^{12}} (h_{01})^2 + \frac{i}{2\pi} h_{01} ( \partial_2 \hat{\Phi}^1 - \hat{\Phi}_2 -\hat{C}^1_2 ) + \frac{1}{8\pi^2 \hat{\mu}^{12}} (h_{02})^2 - \frac{i}{2\pi} h_{02} ( \partial_1 \hat{\Phi}^2 - \hat{\Phi}_1 -\hat{C}^2_1 ) 
\end{align}
by using dynamical fields $h_{01}$ and $h_{02}$. Integrating out $h_{01}$ and $h_{02}$, we have the equations of motion
\begin{align}
     h_{01}  &=  2\pi i \hat{\mu}^{12} (\partial_2 \hat{\Phi}^1 - \hat{\Phi}_2 -\hat{C}^1_2 ) \,, \\
     h_{02}  &=  - 2\pi i \hat{\mu}^{12} (\partial_1 \hat{\Phi}^2 - \hat{\Phi}_1 -\hat{C}^2_1 ) \,.
\end{align}
Using them, we can recover the terms \eqref{term12}. Then, let us take the limit $\hat{\mu}^{12} \rightarrow \infty$, which leads to the terms
\begin{align}
    \frac{i}{2\pi} h_{01} ( \partial_2 \hat{\Phi}^1 - \hat{\Phi}_2 -\hat{C}^1_2 ) - \frac{i}{2\pi} h_{02} ( \partial_1 \hat{\Phi}^2 - \hat{\Phi}_1 -\hat{C}^2_1 )  \,.
\end{align}
Then, the foliated Lagrangian \eqref{genefoliatedphihat2} becomes
\begin{align}
\begin{split}
    &\L_{\hat{\phi},\text{f}} \left[ C^k \wedge e^k, c, \hat{C}^k, \hat{c}_{12} \right] \\
    &  = \hat{\mu}_0 ( \partial_0 \hat{\Phi}^1 - \hat{\Phi}_0 -\hat{C}^1_0 )^2 + \hat{\mu}_0 ( \partial_0 \hat{\Phi}^2 - \hat{\Phi}_0 -\hat{C}^2_0 )^2 \\ 
    & \quad + \frac{1}{2\hat{\mu}_{12}}( \partial_1 \hat{\Phi}_2 - \partial_2 \hat{\Phi}_1 - \hat{c}_{12} )^2 \\
    & \quad + \frac{i}{2\pi} \left( C^1_0  \partial_2 \hat{\Phi}^1  - C^2_0  \partial_1 \hat{\Phi}^2 - C^1_2 \partial_0 \hat{\Phi}^1  + C^2_1  \partial_0 \hat{\Phi}^2 \right)  \\
    & \quad + \frac{i}{2\pi} \left[ h_{01} ( \partial_2 \hat{\Phi}^1 - \hat{\Phi}_2 -\hat{C}^1_2 ) - h_{02} ( \partial_1 \hat{\Phi}^2 - \hat{\Phi}_1 -\hat{C}^2_1 ) \right] \\
    & \quad + \frac{i}{2\pi} \left[ \hat{\kappa}_0 ( C^2_1 - C^1_2 + \partial_1 c_2 - \partial_2 c_1 ) \right. \\
    & \qquad  \left. - \hat{\kappa}_1 ( C^2_0 + \partial_0 c_2 - \partial_2 c_0) + \hat{\kappa}_2 ( C^1_0 + \partial_0 c_1 - \partial_1 c_0 ) \right] \,. \label{genefoliatedphihat3}
\end{split}
\end{align}
In this case, integrating out $h_{01}$ and $h_{02}$, we have the constraints
\begin{align}
    \partial_2 \hat{\Phi}^1 - \hat{\Phi}_2 -\hat{C}^1_2 &= 0 \,, \label{hatconst1} \\
    \partial_1 \hat{\Phi}^2 - \hat{\Phi}_1 -\hat{C}^2_1 &= 0 \,. \label{hatconst2}
\end{align}
Using them, we have
\begin{align}
    \frac{1}{2\hat{\mu}_{12}}( \partial_1 \hat{\Phi}_2 - \partial_2 \hat{\Phi}_1 - \hat{c}_{12} )^2 = \frac{1}{2\hat{\mu}_{12}}\left[ \partial_1 ( \partial_2 \hat{\Phi}^1 -\hat{C}^1_2 )  - \partial_2 ( \partial_1 \hat{\Phi}^2 -\hat{C}^2_1 ) - \hat{c}_{12} \right]^2 \,.
\end{align}
Next, integrating out $\hat{\Phi}_0$, we have
\begin{align}
    \hat{\Phi}_0 =  \frac{1}{2} \left( \partial_0 \hat{\Phi}^1 - \hat{C}^1_0 + \partial_0 \hat{\Phi}^2 - \hat{C}^2_0 \right) \,. \label{phihat0formula}
\end{align}
Then we can write
\begin{align}
    \hat{\mu}_0 ( \partial_0 \hat{\Phi}^1 - \hat{\Phi}_0 -\hat{C}^1_0 )^2 + \hat{\mu}_0 ( \partial_0 \hat{\Phi}^2 - \hat{\Phi}_0 -\hat{C}^2_0 )^2 = \frac{\hat{\mu}_0}{2} \left[ ( \partial_0 \hat{\Phi}^1 -\hat{C}^1_0 ) - ( \partial_0 \hat{\Phi}^2 -\hat{C}^2_0  ) \right]^2 \,,
\end{align}
and we can obtain the foliated Lagrangian \eqref{foliphihatc}, which is equivalent to the exotic $\hat{\phi}$-theory \eqref{exoticphihatc2}.\footnote{We can change the term $\hat{\mu}_0 ( \partial_0 \hat{\Phi}^1 - \hat{\Phi}_0 -\hat{C}^1_0 )^2 + \hat{\mu}_0 ( \partial_0 \hat{\Phi}^2 - \hat{\Phi}_0 -\hat{C}^2_0 )^2$ to $\frac{\hat{\mu}^1_0}{2} ( \partial_0 \hat{\Phi}^1 - \hat{\Phi}_0 -\hat{C}^1_0 )^2 + \frac{\hat{\mu}^2_0}{2} ( \partial_0 \hat{\Phi}^2 - \hat{\Phi}_0 -\hat{C}^2_0 )^2$, where $\hat{\mu}^1_0$ and $\hat{\mu}^2_0$ are parameters satisfying $1/\hat{\mu}_{0} = 1/\hat{\mu}^1_0 + 1/\hat{\mu}^2_0$. In this case, we have
\begin{align}
\begin{split}
    \frac{\hat{\mu}^1_0}{2} ( \partial_0 \hat{\Phi}^1 - \hat{\Phi}_0 -\hat{C}^1_0 )^2 + \frac{\hat{\mu}^2_0}{2} ( \partial_0 \hat{\Phi}^2 - \hat{\Phi}_0 -\hat{C}^2_0 )^2 &= \frac{\hat{\mu}^1_0 \hat{\mu}^2_0}{2 ( \hat{\mu}^1_0 + \hat{\mu}^2_0) } \left[ ( \partial_0 \hat{\Phi}^1 -\hat{C}^1_0 ) - ( \partial_0 \hat{\Phi}^2 -\hat{C}^2_0  ) \right]^2 \\
    &= \frac{\hat{\mu}_0}{2} \left[ ( \partial_0 \hat{\Phi}^1 -\hat{C}^1_0 ) - ( \partial_0 \hat{\Phi}^2 -\hat{C}^2_0  ) \right]^2 \,.
\end{split}
\end{align}
This term becomes $\Z_4$ rotational invariant and the parameters $\hat{\mu}^1_0$ and $\hat{\mu}^2_0$ appear only in the form $1/\hat{\mu}_{0} = 1/\hat{\mu}^1_0 + 1/\hat{\mu}^2_0$.}

If we set the background gauge fields to zero, we can have the foliated $\hat{\phi}$-theory without the background gauge fields
\begin{align}
\begin{split}
    \L_{\hat{\phi},\text{e} \rightarrow \text{f}} &= \hat{\mu}_0 ( \partial_0 \hat{\Phi}^1 - \hat{\Phi}_0 )^2 + \hat{\mu}_0 ( \partial_0 \hat{\Phi}^2 - \hat{\Phi}_0 )^2  + \frac{1}{2\hat{\mu}_{12}}( \partial_1 \hat{\Phi}_2 - \partial_2 \hat{\Phi}_1 )^2 \\
    & + \frac{i}{2\pi} \left[ h_{01} ( \partial_2 \hat{\Phi}^1 - \hat{\Phi}_2 ) - h_{02} ( \partial_1 \hat{\Phi}^2 - \hat{\Phi}_1 ) \right] \,, \label{foliatedphihatczero}
\end{split}
\end{align}
or
\begin{align}
\begin{split}
    \L_{\hat{\phi},\text{e} \rightarrow \text{f}} &=
    \frac{\hat{\mu}_0}{2} \left( \partial_0 \hat{\Phi}^1 -  \partial_0 \hat{\Phi}^2  \right)^2 + \frac{1}{2\hat{\mu}_{12}} \left( \partial_1  \partial_2 \hat{\Phi}^1  - \partial_2  \partial_1 \hat{\Phi}^2 \right)^2 \,. 
\end{split}
\end{align}

As in the case of the foliated $\phi$-theory, let us consider subsystem symmetry. From the background gauge transformations of $\hat{C}^k$ and $\hat{c}$, which are \eqref{bkggaugetr2} and \eqref{bkggaugetr4}, in the Lagrangian with the background gauge fields \eqref{genefoliatedphihat3}, we have
\begin{gather}
    2 \partial_0 \left\{ i \hat{\mu}_0 ( \partial_0 \hat{\Phi}^1 - \hat{\Phi}_0 ) \right\} - \partial_2 \left( \frac{1}{2\pi} h_{01} \right) = 0 \,, \\
    2 \partial_0 \left\{ i \hat{\mu}_0 ( \partial_0 \hat{\Phi}^2 - \hat{\Phi}_0 ) \right\} + \partial_1 \left( \frac{1}{2\pi} h_{02} \right) = 0 \,, \\
    \hat{\mu}_0 ( \partial_0 \hat{\Phi}^1 - \hat{\Phi}_0  ) + \hat{\mu}_0 ( \partial_0 \hat{\Phi}^2 - \hat{\Phi}_0  ) = 0 \,, \\
    \frac{1}{2\pi} h_{02} - \partial_2 \left\{ \frac{i}{\hat{\mu}_{12}} ( \partial_1 \hat{\Phi}_2 - \partial_2 \hat{\Phi}_1 ) \right\} = 0 \,, \\
    -\frac{1}{2\pi} h_{01} + \partial_1 \left\{ \frac{i}{\hat{\mu}_{12}} ( \partial_1 \hat{\Phi}_2 - \partial_2 \hat{\Phi}_1 ) \right\} = 0 \,.
\end{gather}
These equations can also be obtained as the equations of motion of $\hat{\Phi}^1$, $\hat{\Phi}^2$, $\hat{\Phi}_0$ $\hat{\Phi}_1$ and $\hat{\Phi}_2$ in the theory where the background gauge fields are zero \eqref{foliatedphihatczero}. Combining these equations, we derive a conservation law
\begin{align}
    \partial_0 \hat{J}^{12}_0 - \partial_1 \partial_2 \hat{J}  = 0 \,,
\end{align}
where the currents are
\begin{align}
    \hat{J}^{12}_0 &= i \hat{\mu}_0 \partial_0 ( \hat{\Phi}^1 - \hat{\Phi}^2 ) \,, \\
    \hat{J} &= -\frac{i}{\hat{\mu}_{12}} ( \partial_1  \hat{\Phi}_2 - \partial_2 \hat{\Phi}_1 ) \,.
\end{align}
Using the constraints derived by integrating out the dynamical fields $h_{01}$ and $h_{02}$
\begin{align}
    \partial_2 \hat{\Phi}^1 - \hat{\Phi}_2 &= 0 \,, \label{donstphinoc1} \\
     \partial_1 \hat{\Phi}^2 - \hat{\Phi}_1 &= 0 \,, \label{donstphinoc2}
\end{align}
the $x^1 x^2$-component of the current $\hat{J}_{12}$ becomes
\begin{align}
    \hat{J} &= \frac{i}{\hat{\mu}_{12}} \partial_1 \partial_2 ( \hat{\Phi}^1 - \hat{\Phi}^2 ) \,.
\end{align}
These currents are equivalent to the currents \eqref{hatmomsym1} and \eqref{hatmomsym2} under the correspondence \eqref{corrphihat}, and thus generate the momentum dipole symmetry in the $\hat{\phi}$-theory. We also have conservation laws related to winding symmetry. From the background gauge transformations of $C^k \wedge e^k$ and $c$, which are \eqref{bkggaugetr1} and \eqref{bkggaugetr3}, we have relations
\begin{gather}
    \partial_0 \left(  \frac{1}{2\pi} \partial_2 \hat{\Phi}^1 \right)  - \partial_2 \left( \frac{1}{2\pi} \partial_0 \hat{\Phi}^1 \right) = 0 \,, \\
    \partial_0 \left(  - \frac{1}{2\pi} \partial_1 \hat{\Phi}^2 \right) + \partial_1 \left( \frac{1}{2\pi} \partial_0 \hat{\Phi}^2 \right) = 0 \,, 
\end{gather}
which are locally trivial. Combining these equations, we have a conservation law
\begin{align}
    \partial_0 J_0 - \partial_1 \partial_2 J_{12} = 0 \,, 
\end{align}
where the currents are
\begin{align}
    J_0 &= - \frac{1}{2\pi} \partial_1 \partial_2 ( \hat{\Phi}^1 - \hat{\Phi}^2 ) \,, \\
    J_{12} &=   \frac{1}{2\pi} \partial_0 (  \hat{\Phi}^1 - \hat{\Phi}^2 ) \,.
\end{align}
These currents are equivalent to the currents \eqref{hatwinsym1} and \eqref{hatwinsym2} under the correspondence \eqref{corrphihat}, and thus generate the winding dipole symmetry in the $\hat{\phi}$-theory.

\subsection{\texorpdfstring{$T$}{T}-Duality-like Duality in Foliated \texorpdfstring{$\phi$}{phi}-Theory}
\label{subsection:foliated_duality_phi}

In this section, we directly show the duality between the foliated $\phi$-theory \eqref{etofoliatedphi} and the foliated $\hat{\phi}$-theory \eqref{foliphihatc} in the same way as Section \ref{subsection:exotic_phihat_theory}. In the exotic description, the $\phi$-theory is dual to the $\hat{\phi}$-theory, which is considered as a self-duality, but in the foliated description, the self-duality is non-trivial.

We consider the foliated Lagrangian including the background foliated gauge fields
\begin{align}
\begin{split}
    &\L_{\phi,\text{f}} \left[ C^k \wedge e^k, c, \hat{C}^k, \hat{c}_{12} \right] \\
    & \quad = \frac{\mu_0}{2} ( E_0 )^2 + \frac{1}{4\mu_{12}} ( B^1_2 )^2 +  \frac{1}{4\mu_{12}} ( B^2_1 )^2 \\
    & \qquad + \frac{i}{2\pi} \hat{E}^1_0 ( \partial_2 \Phi^1 - C^1_2 - B^1_2 ) + \frac{i}{2\pi} \hat{E}^2_0 ( \partial_1 \Phi^2 - C^2_1 - B^2_1 ) \\
    & \qquad + \frac{i}{2\pi} ( \partial_1 \hat{B}^1_2 - \partial_2 \hat{B}^2_1 + \hat{B}_{12} ) (\partial_0 \Phi - c_0 - E_0 ) \\
    & \qquad + \frac{i}{2\pi} \left[ \hat{C}^1_0 ( 
    \partial_2 \Phi^1 - C^1_2 ) - \hat{C}^1_2 ( 
    \partial_0 \Phi^1 - C^1_0 ) - \hat{C}^2_0 ( 
    \partial_1 \Phi^2 - C^2_1 )  \right. \\
    & \qquad \quad \left. + \hat{C}^2_1 ( 
    \partial_0 \Phi^2 - C^2_0 ) + \hat{c}_{12} ( \partial_0 \Phi - c_0 ) \right] \\
    & \qquad + \frac{i}{2\pi} \left[ -\hat{c}_{02} ( \partial_1 \Phi - \Phi^1 - c_1 )  + \hat{c}_{01} ( \partial_2 \Phi - \Phi^2 - c_2 ) \right] \\
    & \qquad + \frac{i}{2\pi} \left[ \hat{\chi}_0 ( C^2_1 - C^1_2 + \partial_1 c_2 - \partial_2 c_1 ) \right. \\
    & \qquad \quad \left. - \hat{\chi}_1 ( C^2_0 + \partial_0 c_2 - \partial_2 c_0) + \hat{\chi}_2 ( C^1_0 + \partial_0 c_1 - \partial_1 c_0 ) \right] \,,
\end{split}
\end{align}
where the fields $E_0$, $B^1_2$, $B^2_1$, $\hat{E}^1_0$, $\hat{E}^2_0$, $\hat{B}^1_2$, $\hat{B}^2_1$ and $\hat{B}_{12}$ are gauge-invariant dynamical fields. Integrating out $\hat{E}^1_0$, $\hat{E}^2_0$ and $\hat{B}_{12}$, we have equations
\begin{align}
    B^1_2 &= \partial_2 \Phi^1 - C^1_2 \,, \label{dualeq1} \\
    B^2_1 &= \partial_1 \Phi^2 - C^2_1 \,, \label{dualeq2}  \\
    E_0 &= \partial_0 \Phi - c_0 \,,  \label{dualeq3}
\end{align}
and then we can obtain the Lagrangian of the foliated $\phi$-theory \eqref{etofoliatedphi}.

On the other hand, integrating out $E_0$, $B^1_2$ and $B^2_1$, we have equations
\begin{gather}
    \mu_0 E_0 = \frac{i}{2\pi}  ( \partial_1 \hat{B}^1_2 - \partial_2 \hat{B}^2_1 + \hat{B}_{12} ) \,, \label{dualeq4}\\
    \frac{1}{2\mu_{12}} B^1_2 = \frac{i}{2\pi} \hat{E}^1_0 \,, \label{dualeq5}\\
    \frac{1}{2\mu_{12}} B^2_1 = \frac{i}{2\pi} \hat{E}^2_0 \,. \label{dualeq6}
\end{gather}
Using them, we have
\begin{align}
\begin{split}
    &\L_{\phi,\text{f}} \left[ C^k \wedge e^k, c, \hat{C}^k, \hat{c}_{12} \right] \\
    & \quad = \frac{1}{8\pi^2 \mu_0} ( \partial_1 \hat{B}^1_2 - \partial_2 \hat{B}^2_1 + \hat{B}_{12} )^2 + \frac{\mu_{12}}{4\pi^2} ( \hat{E}^1_0 )^2 +  \frac{\mu_{12}}{4\pi^2} ( \hat{E}^2_0 )^2 \\
    & \qquad + \frac{i}{2\pi} ( \hat{E}^1_0 + \hat{C}^1_0 ) ( \partial_2 \Phi^1 - C^1_2 ) + \frac{i}{2\pi} ( \hat{E}^2_0 + \hat{C}^2_0 ) ( \partial_1 \Phi^2 - C^2_1  ) \\
    & \qquad + \frac{i}{2\pi} ( \partial_1 \hat{B}^1_2 - \partial_2 \hat{B}^2_1 + \hat{B}_{12} + \hat{c}_{12} ) (\partial_0 \Phi - c_0 ) \\
    & \qquad + \frac{i}{2\pi} \left[ - \hat{C}^1_2 ( 
    \partial_0 \Phi^1 - C^1_0 ) + \hat{C}^2_1 ( 
    \partial_0 \Phi^2 - C^2_0 )  \right] \\
    & \qquad + \frac{i}{2\pi} \left[ -\hat{c}_{02} ( \partial_1 \Phi - \Phi^1 - c_1 )  + \hat{c}_{01} ( \partial_2 \Phi - \Phi^2 - c_2 ) \right] \\
    & \qquad + \frac{i}{2\pi} \left[ \hat{\chi}_0 ( C^2_1 - C^1_2 + \partial_1 c_2 - \partial_2 c_1 ) \right. \\
    & \qquad \quad \left. - \hat{\chi}_1 ( C^2_0 + \partial_0 c_2 - \partial_2 c_0) + \hat{\chi}_2 ( C^1_0 + \partial_0 c_1 - \partial_1 c_0 ) \right] \,.
\end{split}
\end{align}
Next, integrating out $\Phi$, $\Phi^1$ and $\Phi^2$, we can have equations
\begin{align}
    \partial_1 \hat{c}_{02} - \partial_2 \hat{c}_{01} &= \partial_0 ( \partial_1 \hat{B}^1_2 - \partial_2 \hat{B}^2_1 + \hat{B}_{12} + \hat{c}_{12} ) \,, \\
    \hat{c}_{02} &= \partial_2 ( \hat{E}^1_0 + \hat{C}^1_0 ) - \partial_0 \hat{C}^1_2 \,, \\
    \hat{c}_{01} &= \partial_1 ( \hat{E}^2_0 + \hat{C}^2_0 ) - \partial_0 \hat{C}^2_1 \,.
\end{align}
Combining them, we can obtain
\begin{align}
    \partial_1 \partial_2 ( \hat{E}^1_0 + \hat{C}^1_0 ) - \partial_2 \partial_1 ( \hat{E}^2_0 + \hat{C}^2_0 ) = \partial_0 \partial_1 ( \hat{B}^1_2 + \hat{C}^1_2 ) - \partial_0 \partial_2 ( \hat{B}^2_1 + \hat{C}^2_1 ) + \partial_0 (\hat{B}_{12} + \hat{c}_{12}) \,.
\end{align}
We can locally solve this equation as
\begin{align}
    \hat{E}^1_0 &= \partial_0 \hat{\Phi}^1 - \hat{\Phi}_0 - \hat{C}^1_0 \,, \label{dualeq7} \\
    \hat{E}^2_0 &= \partial_0 \hat{\Phi}^2 - \hat{\Phi}_0 - \hat{C}^2_0 \,, \label{dualeq8} \\
    \hat{B}^1_2 &= \partial_2 \hat{\Phi}^1 - \hat{\Phi}_2 - \hat{C}^1_2 \,, \label{dualeq9} \\
    \hat{B}^2_1 &= \partial_1 \hat{\Phi}^2 - \hat{\Phi}_1 - \hat{C}^2_1 \,, \label{dualeq10}\\
    \hat{B}_{12} &=  \partial_1 \hat{\Phi}_2 - \partial_2 \hat{\Phi}_1  - \hat{c}_{12} \,,  \label{dualeq11}
\end{align}
where $\hat{\Phi}^k\ (k = 1,2)$ is a $U(1)$ type-$B$ foliated zero-form gauge field and $\hat{\Phi}$ is a $U(1)$ type-$B$ bulk one-form gauge field. Their dynamical gauge transformations are
\begin{align}
    \hat{\Phi}^k  &\sim \hat{\Phi}^k + 2\pi \hat{W}^k + \hat{\xi} \,, \\
    \hat{\Phi} &\sim \hat{\Phi} + d \hat{\xi} \,,
\end{align}
where $\hat{W}^k$ is an $x^k$-dependent integer-valued gauge parameter, and $\hat{\xi}$ is a type-$B$ bulk zero-form gauge parameter. Their background gauge transformations are also
\begin{align}
    \hat{\Phi}^k  &\sim \hat{\Phi}^k + \hat{\gamma}^k \,,  \\
    \hat{\Phi} &\sim \hat{\Phi} + \hat{\gamma} \,.
\end{align}
The dynamical gauge fields $\hat{c}_{01}$ and $\hat{c}_{02}$ are written as
\begin{align}
    \hat{c}_{01} &= \partial_1 ( \partial_0 \hat{\Phi}^2 - \hat{\Phi}_0 ) - \partial_0 \hat{C}^2_1 \,, \\
    \hat{c}_{02} &= \partial_2 ( \partial_0 \hat{\Phi}^1 - \hat{\Phi}_0 ) - \partial_0 \hat{C}^1_2 \,.
\end{align}

Then, we can derive
\begin{align}
\begin{split}
    &\L_{\phi,\text{f}} \left[ C^k \wedge e^k, c, \hat{C}^k, \hat{c}_{12} \right] \\
    & \quad = \hat{\mu}_{0} ( \partial_0 \hat{\Phi}^1 - \hat{\Phi}_0 - \hat{C}^1_0 )^2 +  \hat{\mu}_{0} ( \partial_0 \hat{\Phi}^2 - \hat{\Phi}_0 - \hat{C}^2_0 )^2 \\
    & \qquad + \frac{1}{2\hat{\mu}_{12}} \left[ \partial_1 ( \partial_2 \hat{\Phi}^1 - \hat{C}^1_2 ) - \partial_2 ( \partial_1 \hat{\Phi}^2 - \hat{C}^2_1)  - \hat{c}_{12} \right]^2 \\
    & \qquad - \frac{i}{2\pi} C^1_2 ( \partial_0 \hat{\Phi}^1 - \hat{\Phi}_0  ) + \frac{i}{2\pi}  C^2_1 ( \partial_0 \hat{\Phi}^2 - \hat{\Phi}_0 )  \\
    & \qquad - \frac{i}{2\pi} c_0 \left[ \partial_1 ( \partial_2 \hat{\Phi}^1 - \hat{C}^1_2 ) - \partial_2 ( \partial_1 \hat{\Phi}^2 - \hat{C}^2_1 ) \right] + \frac{i}{2\pi} \left[ C^1_0 \hat{C}^1_2  - C^2_0 \hat{C}^2_1  \right] \\
    & \qquad + \frac{i}{2\pi} \left\{ c_1 \left[ \partial_2 ( \partial_0 \hat{\Phi}^1 - \hat{\Phi}_0 ) - \partial_0 \hat{C}^1_2 \right]  - c_2 \left[ \partial_1 ( \partial_0 \hat{\Phi}^2 - \hat{\Phi}_0 ) - \partial_0 \hat{C}^2_1 \right] \right\} \\
    & \qquad + \frac{i}{2\pi} \left[ \hat{\chi}_0 ( C^2_1 - C^1_2 + \partial_1 c_2 - \partial_2 c_1 ) \right. \\
    & \qquad \quad \left. - \hat{\chi}_1 ( C^2_0 + \partial_0 c_2 - \partial_2 c_0) + \hat{\chi}_2 ( C^1_0 + \partial_0 c_1 - \partial_1 c_0 ) \right] \,,
\end{split}
\end{align}
where $\hat{\mu}_0 = \mu_{12}/(4\pi^2)$ and $\hat{\mu}_{12} = 4\pi^2 \mu_0$. We consider changes of variables from $\hat{\chi}$ to $\hat{\kappa}$ as
\begin{align}
    \hat{\chi}_0 &= \hat{\kappa}_0 + \hat{\Phi}_0 \,, \\
    \hat{\chi}_1 &= \hat{\kappa}_1 + \partial_1 \hat{\Phi}^2 - \hat{C}^2_1 \,, \\
    \hat{\chi}_2 &= \hat{\kappa}_2 + \partial_2 \hat{\Phi}^1 - \hat{C}^1_2 \,, 
\end{align}
and then, the field $\kappa$ has a dynamical gauge transformation
\begin{align}
    \hat{\kappa} \sim \hat{\kappa} - d\hat{\xi} \,,
\end{align}
and a background gauge transformation
\begin{align}
    \hat{\kappa} \sim \hat{\kappa} - \hat{\gamma} \,,
\end{align}
as in the previous section. The result is
\begin{align}
\begin{split}
    &- \frac{i}{2\pi} C^1_2 ( \partial_0 \hat{\Phi}^1 - \hat{\Phi}_0  ) + \frac{i}{2\pi}  C^2_1 ( \partial_0 \hat{\Phi}^2 - \hat{\Phi}_0 )  \\
    & \quad - \frac{i}{2\pi} c_0 \left[ \partial_1 ( \partial_2 \hat{\Phi}^1 - \hat{C}^1_2 ) - \partial_2 ( \partial_1 \hat{\Phi}^2 - \hat{C}^2_1 ) \right] + \frac{i}{2\pi} \left[ C^1_0 \hat{C}^1_2  - C^2_0 \hat{C}^2_1  \right] \\
    &  \quad + \frac{i}{2\pi} \left\{ c_1 \left[ \partial_2 ( \partial_0 \hat{\Phi}^1 - \hat{\Phi}_0 ) - \partial_0 \hat{C}^1_2 \right]  - c_2 \left[ \partial_1 ( \partial_0 \hat{\Phi}^2 - \hat{\Phi}_0 ) - \partial_0 \hat{C}^2_1 \right] \right\} \\
    & \rightarrow  \frac{i}{2\pi} \left(  C^1_0  \partial_2 \hat{\Phi}^1 -  C^2_0  \partial_1 \hat{\Phi}^2 -  C^1_2  \partial_0 \hat{\Phi}^1 +  C^2_1  \partial_0 \hat{\Phi}^2 \right) \,.
\end{split}
\end{align}
By integrating out $\hat{\Phi}_0$, we finally obtain the foliated $\hat{\phi}$-theory
\begin{align}
\begin{split}
    &\L_{\hat{\phi},\text{e} \rightarrow \text{f}} \left[ C^k \wedge e^k, c, \hat{C}^k, \hat{c}_{12} \right] \\
    & = \frac{\hat{\mu}_0}{2} \left[ ( \partial_0 \hat{\Phi}^1 - \hat{C}^1_0 ) - ( \partial_0 \hat{\Phi}^2 - \hat{C}^2_0 ) \right]^2 \\
    & \quad + \frac{1}{2\hat{\mu}_{12}} \left[ \partial_1 ( \partial_2 \hat{\Phi}^1 - \hat{C}^1_2 ) - \partial_2 ( \partial_1 \hat{\Phi}^2 - \hat{C}^2_1 ) - \hat{c}_{12} \right]^2 \\
    & \quad + \frac{i}{2\pi} \left(  C^1_0  \partial_2 \hat{\Phi}^1 -  C^2_0  \partial_1 \hat{\Phi}^2 -  C^1_2  \partial_0 \hat{\Phi}^1 +  C^2_1  \partial_0 \hat{\Phi}^2 \right) \\
    & \quad + \frac{i}{2\pi} \left[ \hat{\kappa}_0 ( C^2_1 - C^1_2 + \partial_1 c_2 - \partial_2 c_1 ) \right. \\
    & \qquad \left. - \hat{\kappa}_1 ( C^2_0 + \partial_0 c_2 - \partial_2 c_0) + \hat{\kappa}_2 ( C^1_0 + \partial_0 c_1 - \partial_1 c_0 ) \right] \,,
\end{split}
\end{align}
which is the same as \eqref{foliphihatc}. We can find relations among the fields using the equations \eqref{dualeq1}--\eqref{dualeq3},\eqref{dualeq4}--\eqref{dualeq6} and \eqref{dualeq7}--\eqref{dualeq11}. The relations are
\begin{gather}
    \mu_0 (\partial_0 \Phi - c_0 ) = \frac{i}{2\pi} \left\{ \partial_1 \partial_2 (\hat\Phi^1 - \hat\Phi^2 ) - (\partial_1 \hat{C}^1_2 - \partial_2 \hat{C}^2_1 + \hat{c}_{12}  )  \right\} \,, \\
    \frac{1}{2\mu_{12}} ( \partial_2 \Phi^1 - C^1_2 ) = \frac{i}{2\pi} ( \partial_0 \hat\Phi^1 - \hat\Phi_0 - \hat{C}^1_0 ) \,, \\
    \frac{1}{2\mu_{12}} ( \partial_1 \Phi^2 - C^2_1 ) = \frac{i}{2\pi} ( \partial_0 \hat\Phi^2 - \hat\Phi_0 - \hat{C}^2_0 ) \,.
\end{gather}
Turning off the background gauge fields and using the relations  \eqref{phihat0formula}, \eqref{donstphinoc1} and \eqref{donstphinoc2}, these relations become
\begin{gather}
    \mu_0 \partial_0 \Phi  = \frac{i}{2\pi} \left( \partial_1  \hat\Phi_2 -  \partial_2 \hat\Phi_1  \right) \,, \\
    \frac{1}{2\mu_{12}} \partial_2 \Phi^1  =  \frac{i}{2\pi} ( \partial_0 \hat\Phi^1 - \hat\Phi_0 ) \,,  \\
    \frac{1}{2\mu_{12}} \partial_1 \Phi^2  =-  \frac{i}{2\pi} ( \partial_0 \hat\Phi^2 - \hat\Phi_0 ) \,,
\end{gather}
which indicates that $\hat\Phi$ is the one-form $T$-dual field of the zero-form field $\Phi$ in 2+1 dimensions, and $\hat\Phi^k$ is the zero-form $T$-dual field of the zero-form field $\Phi^k$ on the (1+1)-dimensional layers.

\section{Summary of this Chapter}

In this chapter, we have discussed the SSPT phase for $U(1) \times U(1)$ subsystem symmetry in 3+1 dimensions in terms of the exotic and foliated descriptions. This exotic SSPT phase cancels the mixed 't Hooft anomaly of subsystem global symmetry in the exotic $\phi$-theory in 2+1 dimensions via the anomaly inflow. Then, we have constructed the boundary theory, which is a foliated $\phi$-theory, from the foliated SSPT phase, and we also have constructed the foliated-exotic duality between the exotic $\phi$-theory and the foliated $\phi$-theory in 2+1 dimensions. Furthermore, we have discussed the $T$-duality-like duality and the dual $\hat{\phi}$-theory, and we have constructed the foliated $\hat{\phi}$-theory, the foliated-exotic duality in the $\hat{\phi}$-theory, and the $T$-duality-like duality in the foliated $\phi$-theory. Here is the summary of this chapter.

\subsection*{The SSPT phases for $U(1) \times U(1)$ in 3+1 dimensions}

\subsubsection*{The Exotic SSPT phase for $U(1) \times U(1)$ in 3+1 dimensions}

The exotic Lagrangian is
\begin{align}
\begin{split}
    &\L_{\text{SSPT,e}}\left[ \bm{C},\bm{\hat{C}} \right] \\ 
     & \quad = \frac{i}{2\pi} \left[ \hat{C} ( \partial_0 C_3 - \partial_3 C_0 ) - \hat{C}_0^{12} ( \partial_3 C_{12} - \partial_1 \partial_2 C_3 )  + \hat{C}_3^{12} ( \partial_0 C_{12} - \partial_1 \partial_2 C_0 )  \right] \,,
\end{split}
\end{align}
where $\bm{C}=(C_0, C_{12}, C_3)$ and $\bm{\hat{C}}=(\hat{C}^{12}_{0}, \hat{C}, \hat{C}^{12}_{3})$. are background tensor one-form gauge fields of $\Z_4 \times SO(2)$. Their gauge transformations are
\begin{align}
    C_0 &\sim C_0 + \partial_0 \Gamma \,, \\
    C_{12} &\sim C_{12} + \partial_1\partial_2 \Gamma \,,  \\ 
    C_3 &\sim C_3 + \partial_3 \Gamma  \,, 
\end{align}
and
\begin{align}
    \hat{C}^{12}_0 &\sim \hat{C}^{12}_0 + \partial_0  \hat{\Gamma}^{12} \,,\\
    \hat{C} &\sim \hat{C} + \partial_1 \partial_2  \hat{\Gamma}^{12} \,, \\
    \hat{C}^{12}_3 &\sim \hat{C}^{12}_3 + \partial_3  \hat{\Gamma}^{12} \,, 
\end{align}
where $\bm{\Gamma} =(\Gamma)$ and $\bm{\hat{\Gamma}} = (\hat{\Gamma}^{12})$ are tensor zero-form gauge parameters.

\subsubsection*{The Foliated SSPT phase for $U(1) \times U(1)$ in 3+1 dimensions}

The foliated Lagrangian is
\begin{align}
\begin{split}
    &L_{\text{SSPT,f}} \left[ C^k \wedge e^k, c, \hat{C}^k, \hat{c}_{12} \right] \\
    &= \frac{i}{2\pi} \bm{\hat{C}} \wedge_\text{f} d_\text{f} \bm{C} \\
    & = \frac{i}{2\pi} \left[  \sum_{k = 1}^2 \hat{C}^k \wedge d C^k \wedge e^k + \hat{c} \wedge \left( dc + \sum_{k=1}^2 C^k \wedge e^k \right) \right]   \,, 
\end{split}
\end{align}
where $\bm{C} = (C^k \wedge e^k , c)$ is a type-$A$ foliated (1+1)-form gauge field and $\bm{\hat{C}} = (\hat{C}^k, \hat{c})$ is a type-$B$ foliated one-form gauge field. The gauge fields $C^k \wedge e^k$, $c$, $\hat{C}^k$ and $\hat{c}_{12}$ are background gauge fields, and the gauge fields $\hat{c}_{ij} \ ((i,j) = (0,1), (0,2), (0,3), (2,3), (3,1))$ are dynamical gauge fields. Their background gauge transformations are
\begin{align}
    C^k \wedge e^k &\sim C^k \wedge e^k +d \gamma^k \wedge e^k  \, , \\
    c &\sim c + d\gamma - \sum^{2}_{k=1} \gamma^k e^k\,,  \\
    \hat{C}^k &\sim \hat{C}^k +d \hat{\gamma}^k - \hat{\gamma} \,, \\
    \hat{c} &\sim \hat{c} + d \hat{\gamma}\,, 
\end{align}
where $\bm{\gamma} = (\gamma^k e^k, \gamma)$ is a background type-$A$ foliated (0+1)-form gauge parameter and $\hat{\bm{\gamma}} = (\hat{\gamma}^k, \hat{\gamma})$ is a background type-$B$ foliated zero-form gauge parameter.

\subsubsection*{Correspondences in the SSPT phases for $U(1) \times U(1)$ in 3+1 dimensions}

The field correspondences in the SSPT phases are
\begin{align}
    C_0 &\simeq c_0  \,, \\
    C_{12} &\simeq C^1_2 + \partial_2 c_1 \,,   \\
    C_3 &\simeq c_3  \,,
\end{align}
and
\begin{align}
    \hat{C}_0^{12} &\simeq \hat{C}_0^1 - \hat{C}_0^2 \,, \\
    \hat{C} &\simeq \partial_1 \hat{C}_2^1 - \partial_2 \hat{C}_1^2 + \hat{c}_{12}  \,, \\
    \hat{C}_3^{12} &\simeq \hat{C}_3^1 - \hat{C}_3^2 \,.
\end{align}
The correspondences between the background gauge parameters are
\begin{align}
    \Gamma &\simeq \gamma \,, \\
    \hat{\Gamma}^{12} &\simeq \hat{\gamma}^1 - \hat{\gamma}^2 \,.
\end{align}

\subsection*{The $\phi$-theory in 2+1 dimensions}

\subsubsection*{The Exotic $\phi$-theory in 2+1 dimensions}

The exotic Lagrangian is
\begin{align}
    \L_{\phi,\text{e}} = \frac{\mu_0}{2} (\partial_0 \phi )^2 + \frac{1}{2 \mu_{12}} (\partial_1 \partial_2 \phi )^2 \,,
\end{align}
where $\bm{\phi} = (\phi)$ is a dynamical tensor zero-form gauge field. Its dynamical gauge transformation is
\begin{align}
    \phi \sim \phi + 2\pi w^1 + 2\pi w^2 \,,
\end{align}
where $w^k$ is an $x^k$-dependent integer-valued gauge parameter.

The exotic Lagrangian coupled to the background gauge fields is
\begin{align}
\begin{split}
    \L_{\phi,\text{e}}\left[ \bm{C}, \bm{\hat{C}} \right] &= \frac{\mu_0}{2} (\partial_0 \phi - C_0 )^2 + \frac{1}{2 \mu_{12}} (\partial_1 \partial_2 \phi - C_{12})^2 \\
    & \quad + \frac{i}{2\pi} \hat{C}^{12}_0 ( \partial_1 \partial_2 \phi - C_{12}) + \frac{i}{2\pi} \hat{C} ( \partial_0 \phi - C_0 ) \,,  
\end{split}
\end{align}
where $\bm{C} = (C_0, C_{12})$ and $\bm{\hat{C}} = (\hat{C}^{12}_0, \hat{C})$ is background tensor one-form gauge fields of $\Z_4$. The background gauge transformations of the fields are
\begin{align}
    \phi &\sim \phi + \Gamma \,,
\end{align}
and
\begin{align}
    C_0 &\sim C_0 + \partial_0 \Gamma \,, \label{summary2:c0gauge}  \\
    C_{12} &\sim C_{12} + \partial_1\partial_2 \Gamma \,, \\
    \hat{C}^{12}_0 &\sim \hat{C}^{12}_0 + \partial_0  \hat{\Gamma}^{12} \,,  \\
    \hat{C} &\sim \hat{C} + \partial_1 \partial_2  \hat{\Gamma}^{12} \,, \label{summary2:chatgauge}
\end{align}
where $\bm{\Gamma} = (\Gamma)$ and $\bm{\hat{\Gamma}} = (\hat{\Gamma}^{12})$ are background tensor zero-form gauge parameters.

\subsubsection*{The Foliated $\phi$-theory in 2+1 dimensions}

The foliated $\phi$-theory equivalent to the exotic $\phi$-theory is
\begin{align}
\begin{split}
    &\L_{\phi,\text{e} \rightarrow \text{f}}  = \frac{\mu_0}{2} ( \partial_0 \Phi )^2 + \frac{1}{4\mu_{12}} ( \partial_2 \Phi^1 )^2 + \frac{1}{4\mu_{12}} ( \partial_1 \Phi^2 )^2 \\
    & \qquad + \frac{i}{2\pi} \left[ -\hat{h}_{02} ( \partial_1 \Phi - \Phi^1 )  + \hat{h}_{01} ( \partial_2 \Phi - \Phi^2 ) \right]  \,,
\end{split}
\end{align}
where $\bm{\Phi} = (\Phi^k e^k, \Phi)$ is a dynamical type-$A$ foliated (0+1)-form gauge field, and $\hat{h}_{01}$ and $\hat{h}_{02}$ are dynamical fields. Their dynamical gauge transformations are
\begin{align}
    \Phi^k  e^k &\sim \Phi^k e^k + 2\pi d W^k \,,  \\
    \Phi &\sim \Phi + 2\pi W^1 + 2\pi W^2 \,,
\end{align}
where $W^k$ is an $x^k$-dependent integer-valued gauge parameter.

The foliated $\phi$-theory that is equivalent to the exotic $\phi$-theory is
\begin{align}
\begin{split}
    &\L_{\phi,\text{e} \rightarrow \text{f}} \left[ C^k \wedge e^k, c, \hat{C}^k, \hat{c}_{12} \right] \\
    &= \frac{\mu_0}{2} ( \partial_0 \Phi - c_0 )^2 + \frac{1}{4\mu_{12}} ( \partial_2 \Phi^1 - C^1_2 )^2 + \frac{1}{4\mu_{12}} ( \partial_1 \Phi^2 - C^2_1 )^2 \\
    & \quad + \frac{i}{2\pi} \left[ \hat{C}^1_0 ( 
    \partial_2 \Phi^1 - C^1_2 ) - \hat{C}^1_2 ( 
    \partial_0 \Phi^1 - C^1_0 ) - \hat{C}^2_0 ( 
    \partial_1 \Phi^2 - C^2_1 )  \right. \\
    & \qquad  \left. + \hat{C}^2_1 ( 
    \partial_0 \Phi^2 - C^2_0 ) + \hat{c}_{12} ( \partial_0 \Phi - c_0 ) \right] \\
    & \qquad + \frac{i}{2\pi} \left[ -\hat{c}_{02} ( \partial_1 \Phi - \Phi^1 - c_1 )  + \hat{c}_{01} ( \partial_2 \Phi - \Phi^2 - c_2 ) \right] \\
    & \qquad + \frac{i}{2\pi} \left[ \hat{\chi}_0 ( C^2_1 - C^1_2 + \partial_1 c_2 - \partial_2 c_1 ) \right. \\
    & \qquad \quad \left. - \hat{\chi}_1 ( C^2_0 + \partial_0 c_2 - \partial_2 c_0) + \hat{\chi}_2 ( C^1_0 + \partial_0 c_1 - \partial_1 c_0 ) \right] \,.
\end{split}
\end{align}
where $\bm{C} = (C^k \wedge e^k, c)$ is a type-$A$ foliated (1+1)-form gauge field, $\bm{\hat{C}} = (\hat{C}^k, \hat{c})$ is a type-$B$ foliated one-form gauge field, and $\hat{\chi}$ is a dynamical one-form field. The gauge fields $C^k \wedge e^k$, $c$, $\hat{C}^k$ and $\hat{c}_{12}$ are background gauge fields, and the gauge fields $\hat{c}_{01}$ and $\hat{c}_{02}$ are dynamical gauge fields. The background gauge transformations of the fields are
\begin{align}
    \Phi^k e^k &\sim \Phi^k e^k  + \gamma^k e^k  \,, \\
    \Phi &\sim \Phi + \gamma \,,
\end{align}
and
\begin{align}
    C^k \wedge e^k &\sim C^k \wedge e^k +d \gamma^k \wedge e^k  \,, \label{summary2:ckcorr} \\
    c &\sim c + d\gamma - \sum^{2}_{k=1} \gamma^k e^k\,,\\
    \hat{C}^k &\sim \hat{C}^k +d \hat{\gamma}^k - \hat{\gamma} \,, \\
    \hat{c} &\sim \hat{c} + d \hat{\gamma} \,, \label{summary2:smallchatcorr}
\end{align}
where $\bm{\gamma} = (\gamma^k e^k, \gamma)$ is a background type-$A$ foliated (0+1)-form gauge parameter and $\hat{\bm{\gamma}} = (\hat{\gamma}^k, \hat{\gamma})$ is a background type-$B$ foliated zero-form gauge parameter.

The general foliated Lagrangian is
\begin{align}
\begin{split}
    &L_{\phi,\text{f}} \left[ C^k \wedge e^k, c, \hat{C}^k, \hat{c}_{12}\right] \\
    &= \frac{1}{2} \left( d_\text{f} \bm{\Phi} - \bm{C} \right)^2 - \frac{i}{2\pi} \bm{\hat{C}} \wedge_\text{f} \left(d_\text{f} \bm{\Phi} - \bm{C} \right) + \frac{i}{2\pi} \bm{\hat{\chi}} \wedge_\text{f} d_\text{f} \bm{C} \\
    & = \frac{1}{2} \left( d \Phi - \sum_{k = 1}^2 \Phi^k e^k - c \right) \wedge \ast \left( d \Phi - \sum_{k = 1}^2 \Phi^k e^k - c \right) \\
    & \quad + \frac{1}{2} \sum_{k=1}^2  \left( d \Phi^k \wedge e^k - C^k \wedge e^k  \right) \wedge \ast \left( d \Phi^k \wedge e^k - C^k \wedge e^k  \right)  \\
    & \quad + \frac{i}{2\pi} \left[ - \sum_{k = 1}^2  \hat{C}^k \wedge ( d\Phi^k - C^k ) \wedge e^k + \hat{c} \wedge \left( d\Phi - \sum_{k = 1}^2 \Phi^k e^k - c \right) \right]  \\
    & \quad + \frac{i}{2\pi} \hat{\chi} \wedge \left( dc + \sum^{2}_{k=1} C^k \wedge e^k  \right)
      \,, 
\end{split}
\end{align}
where $\bm{\hat{\chi}} = (0,\hat{\chi})$ is a type-$B$ foliated zero-form field.

\subsubsection*{Correspondences in the $\phi$-theory in 2+1 dimensions}

The field correspondence in the $\phi$-theory is
\begin{align}
    \phi \simeq \Phi \,. \label{summary2:phicorr}
\end{align}
The correspondences between the dynamical gauge parameter is
\begin{align}
    w^k \simeq W^k \,. 
\end{align}

The field correspondences in the $\phi$-theory coupled to the background gauge fields are \eqref{summary2:phicorr} and
\begin{align}
    C_0 &\simeq c_0  \,, \label{summary2:c0corr} \\
    C_{12} &\simeq C^1_2 + \partial_2 c_1 \,,  \\
    \hat{C}_0^{12} &\simeq \hat{C}_0^1 - \hat{C}_0^2 \,,  \\
    \hat{C} &\simeq \partial_1 \hat{C}_2^1 - \partial_2 \hat{C}_1^2 + \hat{c}_{12}  \,. \label{summary2:chatcorr}
\end{align}
The correspondences between the background gauge parameters are
\begin{align}
    \Gamma &\simeq \gamma \,, \\
    \hat{\Gamma}^{12} &\simeq \hat{\gamma}^1 - \hat{\gamma}^2 \,.
\end{align}

\subsection*{The $\hat{\phi}$-theory in 2+1 dimensions}

\subsubsection*{The Exotic $\hat{\phi}$-theory in 2+1 dimensions}

The exotic Lagrangian is
\begin{align}
    \L_{\hat{\phi},\text{e}} &= \frac{\hat{\mu}_0}{2} ( \partial_0  \hat{\phi}^{12}  )^2 + \frac{1}{2 \hat{\mu}_{12}} ( \partial_1 \partial_2 \hat{\phi}^{12} )^2  \,.
\end{align}
where $\bm{\hat{\phi}} = (\hat{\phi}^{12})$ is a dynamical tensor zero-form gauge field. Its dynamical gauge transformation is
\begin{align}
    \hat{\phi}^{12} \sim \hat{\phi}^{12} + 2\pi \hat{w}^1 - 2\pi \hat{w}^2 \,,
\end{align}
where $\hat{w}^k$ is an $x^k$-dependent integer-valued gauge parameter.

The exotic Lagrangian coupled to the background gauge fields is
\begin{align}
\begin{split}
    \L_{\hat{\phi},\text{e}}\left[ \bm{C}, \bm{\hat{C}} \right] &= \frac{\hat{\mu}_0}{2} ( \partial_0  \hat{\phi}^{12} - \hat{C}^{12}_0 )^2 + \frac{1}{2 \hat{\mu}_{12}} ( \partial_1 \partial_2 \hat{\phi}^{12} - \hat{C} )^2  \\
    & \quad - \frac{i}{2\pi} C_{0} \partial_1 \partial_2 \hat{\phi}^{12} - \frac{i}{2\pi} C_{12} \partial_0  \hat{\phi}^{12}   \,. 
\end{split}
\end{align}
The background gauge transformations of the fields are
\begin{align}
    \hat{\phi}^{12} &\sim \hat{\phi}^{12} + \hat{\Gamma}^{12} \,,
\end{align}
and \eqref{summary2:c0gauge}--\eqref{summary2:chatgauge}.

\subsubsection*{The Foliated $\hat{\phi}$-theory in 2+1 dimensions}

The foliated $\hat{\phi}$-theory equivalent to the exotic $\hat{\phi}$-theory is
\begin{align}
\begin{split}
    \L_{\hat{\phi},\text{e} \rightarrow \text{f}} &= \hat{\mu}_0 ( \partial_0 \hat{\Phi}^1 - \hat{\Phi}_0 )^2 + \hat{\mu}_0 ( \partial_0 \hat{\Phi}^2 - \hat{\Phi}_0 )^2  + \frac{1}{2\hat{\mu}_{12}}( \partial_1 \hat{\Phi}_2 - \partial_2 \hat{\Phi}_1 )^2 \\
    & + \frac{i}{2\pi} \left[ h_{01} ( \partial_2 \hat{\Phi}^1 - \hat{\Phi}_2 ) - h_{02} ( \partial_1 \hat{\Phi}^2 - \hat{\Phi}_1 ) \right] \,, 
\end{split}
\end{align}
where $\bm{\hat{\Phi}} = (\hat{\Phi}^k, \hat{\Phi})$ is a dynamical type-$B$ foliated zero-form gauge field, and $h_{01}$ and $h_{02}$ are dynamical fields. Their dynamical gauge transformations are
\begin{align}
    \hat{\Phi}^k  &\sim \hat{\Phi}^k + 2\pi \hat{W}^k + \hat{\xi} \,, \\
    \hat{\Phi} &\sim \hat{\Phi} + d \hat{\xi} \,,
\end{align}
where $\hat{W}^k$ is an $x^k$-dependent integer-valued gauge parameter, and $\hat{\xi}$ is a type-$B$ bulk zero-form gauge parameter.

The foliated $\hat{\phi}$-theory that is equivalent to the exotic $\hat{\phi}$-theory is
\begin{align}
\begin{split}
    &\L_{\hat{\phi},\text{f}} \left[ C^k \wedge e^k, c, \hat{C}^k, \hat{c}_{12} \right] \\
    &  = \hat{\mu}_0 ( \partial_0 \hat{\Phi}^1 - \hat{\Phi}_0 -\hat{C}^1_0 )^2 + \hat{\mu}_0 ( \partial_0 \hat{\Phi}^2 - \hat{\Phi}_0 -\hat{C}^2_0 )^2 \\ 
    & \quad + \frac{1}{2\hat{\mu}_{12}}( \partial_1 \hat{\Phi}_2 - \partial_2 \hat{\Phi}_1 - \hat{c}_{12} )^2 \\
    & \quad + \frac{i}{2\pi} \left( C^1_0  \partial_2 \hat{\Phi}^1  - C^2_0  \partial_1 \hat{\Phi}^2 - C^1_2 \partial_0 \hat{\Phi}^1  + C^2_1  \partial_0 \hat{\Phi}^2 \right)  \\
    & \quad + \frac{i}{2\pi} \left[ h_{01} ( \partial_2 \hat{\Phi}^1 - \hat{\Phi}_2 -\hat{C}^1_2 ) - h_{02} ( \partial_1 \hat{\Phi}^2 - \hat{\Phi}_1 -\hat{C}^2_1 ) \right] \\
    & \quad + \frac{i}{2\pi} \left[ \hat{\kappa}_0 ( C^2_1 - C^1_2 + \partial_1 c_2 - \partial_2 c_1 ) \right. \\
    & \qquad  \left. - \hat{\kappa}_1 ( C^2_0 + \partial_0 c_2 - \partial_2 c_0) + \hat{\kappa}_2 ( C^1_0 + \partial_0 c_1 - \partial_1 c_0 ) \right] \,, 
\end{split}
\end{align}
or
\begin{align}
\begin{split}
    &\L_{\hat{\phi},\text{e} \rightarrow \text{f}} \left[ C^k \wedge e^k, c, \hat{C}^k, \hat{c}_{12} \right] \\
    & = \frac{\hat{\mu}_0}{2} \left[ ( \partial_0 \hat{\Phi}^1 - \hat{C}^1_0 ) - ( \partial_0 \hat{\Phi}^2 - \hat{C}^2_0 ) \right]^2 \\
    & \quad + \frac{1}{2\hat{\mu}_{12}} \left[ \partial_1 ( \partial_2 \hat{\Phi}^1 - \hat{C}^1_2 ) - \partial_2 ( \partial_1 \hat{\Phi}^2 - \hat{C}^2_1 ) - \hat{c}_{12} \right]^2 \\
    & \quad + \frac{i}{2\pi} \left(  C^1_0  \partial_2 \hat{\Phi}^1 -  C^2_0  \partial_1 \hat{\Phi}^2 -  C^1_2  \partial_0 \hat{\Phi}^1 +  C^2_1  \partial_0 \hat{\Phi}^2 \right) \\
    & \quad + \frac{i}{2\pi} \left[ \hat{\kappa}_0 ( \partial_1 c_2 - \partial_2 c_1 + C^2_1 - C^1_2 ) \right. \\
    & \qquad \left. - \hat{\kappa}_1 (\partial_0 c_2 - \partial_2 c_0 +  C^2_0) + \hat{\kappa}_2 (  \partial_0 c_1 - \partial_1 c_0 + C^1_0 ) \right]  \,,
\end{split}
\end{align}
where $h_{01}$ and $h_{02}$ are dynamical fields, and $\hat{\kappa}$ is a dynamical one-form gauge field. The dynamical gauge transformation of $\hat{\kappa}$ is
\begin{align}
    \hat{\kappa} \sim \hat{\kappa} - d\hat{\xi} \,.
\end{align}
The background gauge transformations of the fields are
\begin{align}
    \hat{\Phi}^k  &\sim \hat{\Phi}^k + \hat{\gamma}^k \,,  \\
    \hat{\Phi} &\sim \hat{\Phi} + \hat{\gamma} \,,
\end{align}
and
\begin{align}
    \hat{\kappa} \sim \hat{\kappa} - \hat{\gamma} \,,
\end{align}
and \eqref{summary2:ckcorr}--\eqref{summary2:smallchatcorr}.

The general foliated Lagrangian is
\begin{align}
\begin{split}
    &L_{\hat{\phi},\text{f}} \left[ C^k \wedge e^k, c, \hat{C}^k, \hat{c}_{12}\right] \\
    &= \frac{1}{2} (d_\text{f} \bm{\hat{\Phi}} - \bm{\hat{C}} )^2 + \frac{i}{2\pi} d_\text{f} \bm{\hat{\Phi}} \wedge_\text{f} \bm{C} + \hat{\bm{\chi}} \wedge_\text{f} d_\text{f} \bm{C}  \\
    &  = \frac{1}{2} \sum_{k=1}^2 \left\{ \left(  d \hat{\Phi}^k - \hat{\Phi} - \hat{C}^k \right) \wedge e^k \right\} \wedge \ast \left\{ \left( d \hat{\Phi}^k - \hat{\Phi} - \hat{C}^k \right) \wedge e^k \right\} \\
    & \quad + \frac{1}{2} \left( d \hat{\Phi} -\hat{c}  \right) \wedge \ast \left( d \hat{\Phi} -\hat{c}  \right)  \\
    & \quad + \frac{i}{2\pi} \left[  \sum_{k = 1}^2  \left( d\hat{\Phi}^k - \hat{\Phi} \right)  \wedge C^k \wedge e^k -  d\hat{\Phi} \wedge c   \right]  \\
    & \quad + \frac{i}{2\pi} \hat{\chi} \wedge \left( dc + \sum^{2}_{k=1} C^k \wedge e^k  \right) \,,
\end{split}
\end{align}
where $\bm{\hat{\chi}} = (0,\hat{\chi})$ is a type-$B$ foliated zero-form field.

\subsubsection*{Correspondences in the $\hat{\phi}$-theory in 2+1 dimensions}

The field correspondence in the $\hat{\phi}$-theory is
\begin{align}
    \hat{\phi}^{12} \simeq \hat{\Phi}^1 - \hat{\Phi}^2 \,. \label{summary2:phihatcorr}
\end{align}
The correspondence between the dynamical gauge parameters is
\begin{align}
    \hat{w}^k \simeq \hat{W}^k  \,. 
\end{align}

The field correspondences in the $\hat{\phi}$-theory coupled to the background gauge fields are \eqref{summary2:phihatcorr} and \eqref{summary2:c0corr}--\eqref{summary2:chatcorr}.

\chapter{Conclusion and Discussion}
\label{chapter:conclusion}

In this dissertation, we have discussed the foliated-exotic duality and its application to 't Hooft anomalies of subsystem symmetries and SSPT phases. 
We have analyzed the mixed 't Hooft anomaly of $\Z_N \times \Z_N$ subsystem symmetry in the exotic and foliated $BF$ theories in 2+1 dimensions and the SSPT phases for $\Z_N \times \Z_N$ subsystem symmetry in 3+1 dimensions that cancel the anomaly via the anomaly inflow mechanism.
Then, we have constructed the exotic and foliated SSPT phases with two and three foliations, respectively, by using the foliated-exotic duality. 
We have also seen that both of the SSPT phase with two foliations and three foliations match the same 't Hooft anomaly of the exotic/foliated $BF$ theory, and have pointed out that this fact may be a clue for characterizing 't Hooft anomalies of subsystem symmetry.
We have also studied the SSPT phases for $U(1) \times U(1)$ subsystem symmetry in 3+1 dimensions and the boundary theory in 2+1 dimensions that has the 't Hooft anomaly of $U(1) \times U(1)$ subsystem symmetry.
We have constructed the boundary foliated $\phi$ theory in 2+1 dimensions and the foliated-exotic duality in the boundary $\phi$-theory, which is the first example of the foliated-exotic duality in gapless fractonic theories. 
We have also constructed the foliated $\hat{\phi}$-theory in 2+1 dimensions, which is dual to the foliated $\phi$-theory, and the foliated-exotic duality in the $\hat{\phi}$-theory.
Then, the field correspondences between the exotic tensor gauge fields and the foliated gauge fields in this dissertation can be applied to general fractonic QFTs with the same discrete spatial rotational symmetries and foliation structures.
Since exotic form and foliated form have different manifest structure, clarifying the correspondences will lead to a deeper understanding of fractonic QFTs and subsystem symmetries.

There are several future directions.
By using the foliated-exotic duality, we can convert exotic QFT to foliated QFT, which is constructed from the foliated fields on the lower dimensional leaves.
This decomposition is easier to handle and enables us to apply established tools from standard QFT to fractonic systems.
For example, we might be able to provide a framework for constructing fermionic fracton QFT \cite{Yamaguchi:2021qrx,Honda:2022shd,Katsura:2022xkg,Kawakami:2025vox} by establishing a boson-fermion duality in foliated QFT, which is considered easier than in the case of exotic QFT.\footnote{The boson-fermion duality in lattice models with subsystem symmetry is studied in \cite{Cao:2022lig}.}

For the other direction, we have not formulated the symmetric tensor form and the coupling product corresponding to the wedge product in differential form in Section \ref{section:exotic_qft_tensor_gauge_fields}.
We may consider the basis such as $dx^1 \odot dx^2$, which is symmetric tensor, and an exotic exterior derivative $d_\text{e}$ that acts on the exotic tensor gauge fields systematically.
Then, it is considered necessary to apply a projection to some representation of the spatial rotation group to obtain the exterior derivative of fields. 
The formulation would help us to construct exotic QFTs systematically and clarify the structure of the exotic tensor gauge fields.

\chapter*{Acknowledgements}
First and foremost, I would like to express my deepest gratitude to my primary advisor Kantaro Ohmori for his continuous support and guidance throughout the course of my graduate research.
Throughout my graduate studies, I had many discussions with him about theoretical physics, which greatly enriched my understanding and broadened my perspective; I count myself most fortunate to have studied under his supervision.

I am also deeply grateful to Prof. Yutaka Matsuo, who is my official supervisor at the University of Tokyo.
He provided a lot of valuable support during my graduate school life.
Had he not accepted me into his research group, I would not have been able to experience such a fulfilling five years of graduate school.

I would also like to thank Prof. Yuji Tachikawa, who is my co-advisor and collaborator, for his insightful comments and suggestions on my research.

I also thank financial support and valuable education from the World-leading INnovative Graduate Study Program for Frontiers of Mathematical Sciences and Physics (WINGS-FMSP), The University of Tokyo.
The program provided me with stimulating insights into both mathematical and social domains.

Next, I am grateful to the other members of the particle physics groups in Hongo, especially to  Hajime Fukuda, Kohki Kawabata, Go Noshita, Shinichiro Yahagi, Masahito Yamazaki and Masataka Watanabe. I had many interesting conversations with them, which helped me to broaden my perspective on various topics.

I would also like to thank Masaki Okada and Takashi Tsuda, my peers in the string theory community, for many invaluable and stimulating discussions and conversations.
I am truly grateful to have had the opportunity to spend time with them; their company sparked my curiosity across a wide range of topics and deepened my understanding.

Furthermore, I owe my deepest thanks to Ayaka Takamura. Thanks to her, my years in graduate school --- and my life more broadly --- were filled with dialogue and color; her presence became the driving force that sustained me each day.

Finally, I would like to express my deepest gratitude to my family for their unwavering support and constant encouragement. I could not have reached this point without them.

\bibliography{ref.bib}
\bibliographystyle{ytamsalpha}

\end{document}